\documentclass[preprint,10pt]{elsarticle}

\usepackage{geometry}
\newgeometry{ hmargin={25mm,25mm}}

\usepackage{algorithm}
\usepackage[noend]{algpseudocode}
\usepackage{amsmath}
\usepackage{amssymb}
\usepackage{bm}
\usepackage{booktabs}
\usepackage{caption,subcaption}

\usepackage{multicol}
\usepackage[numbers]{natbib}
\usepackage{graphicx}

\DeclareMathOperator*{\argmin}{argmin}

\DeclareMathOperator{\bp}{\bf p}

\DeclareMathOperator{\br}{\bf r}

\DeclareMathOperator{\bX}{\bf X}
\DeclareMathOperator{\bx}{\bf x}
\DeclareMathOperator{\bY}{\bf Y}
\DeclareMathOperator{\by}{\bf y}

\DeclareMathOperator{\bkappa}{\bm \kappa}

\DeclareMathOperator{\bomega}{\bm \omega}

\DeclareMathOperator{\bsigma}{\bm \sigma}

\DeclareMathOperator{\btheta}{\bm \theta}

\newcommand{\toi}[2][t]{%
  \mathop{
    \mathrm{#2}^{( #1 )}
  }
}

\journal{arXiv}
\begin{document}

\begin{frontmatter}

\title{Idiosyncratic Market Effects in Global Time Series Momentum}

\author{Andrew Butler, Maciej Zawadzki, Adam Butler}

\address{ReSolve Asset Management}

\begin{abstract}
Time series momentum is a pervasive explanatory factor in financial markets whereby a market's recent historical average return positively predicts
future returns. Recent applications of statistical learning augment the traditional linear model definition with a more descriptive global nonlinear factor model. 
We extend the work of \citet{MoskowitzNonlinear} and demonstrate that models fit at the individual market level exhibit response curves that are distinct from 
the global factor model. We highlight the heterogeneity of forward market returns as a function of time series momentum and provide a framework 
that enables each feature-market pair to express its unique prediction behaviour. Empirical results demonstrate that, despite substantially smaller sample sizes, 
fitting nonlinear models to individual markets is a robust and often overlooked modelling approach that can result in a marginal improvement in out-of-sample 
performance.
\end{abstract}


\end{frontmatter}


\section{Introduction} \label{sec:intro}

Forecasting global financial markets is a challenging problem in quantitative finance. The efficient market hypothesis posits that the sum total 
of investor decisions leads to a rational and complete price discovery. Under this hypothesis, all currently available information is immediately 
reflected in the asset price \citep{Fama1970}. That is to say, asset prices follow a martingale, whose future variations are unpredictable given 
current information. In contrast, many academics and practitioners reject the hypothesis that financial agents act in a perfectly rational manner
 with no friction. Instead, factors such as investor biases \citep{Hong1999}, social influence \citep{Kirman1991} and limits to arbitrage
  \citep{Shleifer1997} can result in pricing inefficiencies which can persist for a meaningful period of time.

The use of market `factors' to forecast and explain asset returns has a long and established history in the financial literature. Notably, the 
capital asset pricing model (CAPM), first introduced by \citet{Treynor1962, Sharpe1963} and \citet{Lintner1965}, models stock returns as a linear 
function of the stock's sensitivity to the corresponding equity market.   The seminal work of Fama and French \citep{Fama1988, Fama1992, Fama1993, Fama2015}
expands upon CAPM by introducing other explanatory factors such as market capitalization, value and profitability, in order to forecast stock returns.

The observation that asset returns generally exhibit non-zero autocorrelation has also been documented in the financial literature and suggests that 
asset prices do not follow a perfect martingale process.  For example, the generalized autoregressive conditional heteroskedasticity (GARCH) model 
proposed by \citet{Engle1982} and \citet{Bollerslev1986} demonstrates that the absolute value of returns exhibits long-memory effects and other
`stylized facts' common to financial time series \citep{Bollerslev1990, Bollersev1994}. Moreover, the application of general time series analysis 
techniques, such as autoregressive and moving average models, has demonstrated positive predictive power for forecasting asset mean returns
(see for example \citep{Box1970, Campbell1997, Chan1996, Lemperiere2014}). Most relevant to this work,  \citet{Moskowitz2012} demonstrate that the
forward returns of many global futures markets can be reasonably forecasted by the market's average return over a 1 to 12 month period. 
The so-called `time series momentum' factor is therefore the observation that markets generally exhibit a small, but positive autocorrelation in returns,
and is the dominant explanatory factor for Commodity Trading Advisors (CTAs) and systematic hedge fund performance \citep{Bollen2021, Boos2023}. 
Of equal importance, the slope and/or shape of the futures term structure contains information pertaining to commodity inventory levels and hedging activity,
and has been shown to positively predict forward returns \citep{Gorton2012, Szy2014}.  \citet{Koijen2018} demonstrate that the convenience yield, 
or market `carry', can be estimated from the spot market and futures term structure and produce predictions that generalize across commodity and
financial markets alike. To date, the time series momentum and carry factors remain some of the strongest, most durable, and most widely studied factors 
in the global factor literature (see for example \citep{Asness2013, Asness2015, Baltas2012, Baltas2017, Hallerbach2017, He2015, Hong2015, Kim2016, Koijen2017, Marshall2017}).
While this paper focuses primarily on the time series momentum factor, the general framework and modelling approach naturally extend to other factors such as carry, volatility, 
and skewness.

Traditionally, the implementation of factor strategies often assumes a simple response function between factor features and forward expected returns. For example,
the time series momentum factor, frequently defined as the past 12-month return, has been widely studied in the context of equities \citep{Jegadeesh1993},
mutual funds \citep{Carhart1997} and futures markets \citep{Moskowitz2012}. The expected return is typically modelled as a linear function of momentum, with a slope 
coefficient set to 1 and no intercept term. This imposes the assumption that expected returns are zero when the factor score is zero, and increase linearly as
the factor score increases. While straightforward to implement, this simplistic linear specification imposes strong assumptions about the relationship between past returns
(feature values) and forward expected returns. 

Statistical learning methods, on the other hand, are able to capture nonlinearities and interactions that are often overlooked by traditional linear factor models.
Rather than assuming a specific linear functional form, machine learning methods provide a data-driven approach to uncover the optimal mapping from feature values
to forward expected returns. With the rise of advanced statistical methods and data science, much of the recent financial literature on asset pricing focuses on the 
use of machine learning methods to extract price signals from noisy financial data \citep{Brogaard2018, Chen2021, Siegel2001, Kim2003,  Kyriakou2019, Ma2021}.
Notably, \citet{Liew2017} explore the use of neural networks, random forest and support vector machines to forecast the sign of liquid ETF returns. 
They consider price, volume and seasonality features and demonstrate positive predictive power over time horizons ranging from 1 to 3 months. 
Related, \citet{Ta2018} explore the use of linear regression and support vector machines for forecasting stock returns.  More recently, \citet{Blitz2023} survey
a variety of machine learning techniques including linear regression, elastic net, gradient boosting trees and neural networks for forecasting US stock returns. 
They demonstrate that, in recent years, a naive machine learning implementation results in zero net performance. Significantly positive net returns, however, are 
achievable by forecasting over longer time horizons and applying efficient portfolio construction techniques.

Closely related is the work of \citet{Carver2023}, who presents a binning procedure for mapping time series momentum features to expected volatility-adjusted returns. 
To attenuate the potential for model overfitting, the author recommends fitting a single response function aggregated across multiple markets, and, consistent with the
factor literature, does not recommend fitting at the individual market level. Related, \citet{Waldow2021} explore the use of time series momentum features in conjunction 
with logistic regression, random forest and gradient boosting classifiers to forecast the cross-section of futures market returns. Similarly, \citet{Schmid2021} uses 
time series momentum features and  provides a parametric model for forecasting futures markets. The author demonstrates that, in contrast to the factor literature, market 
expected returns can be modelled as a cubic polynomial of the observed trend strength, with model parameters that generalize across markets. Most relevant to this paper is the work of \citet{MoskowitzNonlinear} who document a persistent nonlinear relationship between time series momentum features and forward 
volatility-adjusted returns. The authors explore kernel regression and neural networks, and demonstrate that nonlinear models fit across
all markets within a given sector can result in improved out-of-sample performance in comparison to the traditional linear model.

The remainder of the paper is outlined as follows. In Section \ref{sec:method} we follow the work of \citet{MoskowitzNonlinear} and present a framework for fitting regression
models to forecast forward volatility-adjusted market returns.  We limit our exploration to simple linear regression and non-parametric kernel regression,
acknowledging that alternative techniques such as local regression (LOESS), neural networks, random forest and gradient boosting regression are equally valid.
In Section \ref{sec:exp} we present motivating examples, and in Section \ref{sec:hist} we perform a walk-forward analysis across $50$ global futures markets, spanning over $30$ years of market data. In contrast to the 
factor literature, where a single model is fit to generalize across a group of markets, we explore fitting prediction models at the individual market level. 
In general, we confirm the findings from \citet{Schmid2021}, \citet{Carver2023} and \citet{MoskowitzNonlinear} and demonstrate empirically that market 
forward returns are a nonlinear function of the time series momentum feature values. Moreover, when fit in aggregate across all markets, the response 
functions converge on the previously discovered linear and nonlinear model definitions \citep{Moskowitz2012, MoskowitzNonlinear}. However, we show that when models are fit at the individual market
level, they demonstrate a unique nonlinear response that varies across markets in both the strength and character of the signal. The proposed framework therefore enables each feature-market pair to express its unique prediction 
behaviour. This is in direct contrast to the traditional factor literature, which advocates for fitting a unified single model that generalizes across all markets within a group 
and therefore borrows statistical robustness by fitting over substantially larger sample sizes \citep{Baltas2022, Koijen2018, Moskowitz2012, MoskowitzNonlinear}.
Empirical results demonstrate that fitting nonlinear models to individual markets is an equally robust and often overlooked modelling approach. For the time series momentum
factor explored in this paper, we demonstrate that, in aggregate, fitting to individual markets results in improved and differentiated out-of-sample performance
at the margin. From a statistical standpoint, we find the total effect size to be small, with market mean returns that are statistically equivalent to those
produced by both the linear and nonlinear global factor models. This is largely explained by the observation that most markets exhibit a linear forward 
response when feature values are within the normal range. Nonlinear effects therefore occur primarily at extreme feature values, which are observed
much less frequently. Statistical analysis of realized market mean returns at feature extremes provides some evidence that the market-specific nonlinear response 
functions represent genuine market signal and are not purely an artifact of sampling error. Taken together, we are encouraged by the robustness of 
the proposed modelling approach as it enables data-driven factor exploration in settings that require a more fine-grained individual market
perspective.

\section{Methodology} \label{sec:method}
We follow the work of \citet{MoskowitzNonlinear} and cast the market return forecast as an empirical conditional estimation problem. 
We denote the price of the market, $k$, at time $t$ as $\toi{\bp_k}$. The associated return from period $t-1$ to $t$ is therefore 
$\toi{\br_k} =\toi{\bp_k}/\toi[t-1]{\bp_k} - 1.$ Given that market volatility is known to be highly non-stationary \citep{Starica2004}, we choose to work with volatility-adjusted
returns. Specifically, we let the estimated volatility of market $k$ at time $t$ be $\toi{\hat{\bsigma}_k}$. We therefore define the response variable,
$\bY  = \{\toi{\by}\}_{t=1}^T$, as the forward volatility-adjusted average market return, where 
\begin{equation}\label{eq:response_0}
\toi{\by_k} = \frac{1}{\sqrt{m}}\sum_{i=0}^{m-1} \toi[t+i+l]{\br_k}/\toi[t+i]{\hat{\bsigma}_k},
\end{equation}
with forecast horizon, $m$, and lag time, $l$, between signal generation and trade execution. The associated feature variables, $\bX =  \{\toi{\bx}\}_{t=1}^T$,
consist of time series momentum feature values
\begin{equation}\label{eq:feature}
\toi{\bx_k} = \frac{1}{\sqrt{n}}\sum_{i=0}^{n-1} \toi[t - i]{\br_k}/\toi[t]{\bsigma_k},
\end{equation} 
where $\toi[t]{\bsigma_k}$ denotes the realized market volatility over the lookback $n$. Note that both feature values and forward response are standardized 
to have approximately unit variance.

We follow \citet{MoskowitzNonlinear} and enforce both symmetry and zero mean by augmenting each training observation $(  \toi{\bx}, \toi{\by} )$ with its negation
$(  -\toi{\bx}, -\toi{\by} )$. Therefore, given the augmented training dataset ${D = \{(\toi{\tilde{\bx}}, \toi{\tilde{\by}}) \}_{t=1}^T}$, where 
$\toi{\tilde{\bx}} = [\toi{\bx}, -\toi{\bx}]$ and $\toi{\tilde{\by}}=[\toi{\by}, -\toi{\by}]$, we seek to estimate a 
univariate prediction model $f \colon \mathbb{R} \times \mathbb{R}^{d_\theta } \to \mathbb{R}$ such that: 
\begin{equation}\label{eq:expectation}
\toi{\hat{\by}_k} =  f(\toi{\bx_k}, \btheta) = \mathbb{E}_D[\bY \mid \bX = \toi{\bx_k}].
\end{equation}

In this paper, we consider two statistical models: simple linear regression and Nadaraya-Watson kernel regression \citep{Nad1964, Watson1964}. As we will see 
in Section \ref{sec:exp}, under the above assumptions, a simple linear regression model fit globally to all markets recovers the traditional time series momentum factor 
definition. Conversely, the kernel regression model produces locally weighted conditional expected values for each feature value, $\toi{\bx_k}$, and is therefore capable of 
modelling nonlinear response functions. In the following subsections we provide a brief description of the models and their associated parameterization.

\subsection{Linear regression}
We consider a simple univariate linear regression model which models the response function as a multiple of the feature variable:
\begin{equation}
\toi{\by_k} =  \beta \toi{\bx_k} + \toi{\epsilon},
\end{equation}
with noise $\toi{\epsilon} \sim \mathcal{N}(0, \sigma^2)$.  In least-squares regression, the objective is to determine the coefficient, $\hat{\beta}$,
that minimizes the least-squares loss:
\begin{equation} \label{eq:beta_hat}
\begin{split}
    \hat{\beta} = \argmin_{ \beta } \sum_{t=1}^T (\toi{\by_k} - \beta \toi{\bx_k})^2,
\end{split}
\end{equation}
with least-squares estimate $\toi{\hat{\by}_k} = \hat{\beta} \toi{\bx_k} $ and optimal coefficient  
$\hat{\beta}  = \sum_{t=1}^T\toi{\bx_k}\toi{\by_k}/ \sum_{t=1}^T(\toi{\bx_k})^2$.

\subsection{Kernel regression}
The Nadaraya-Watson kernel regression \citep{Nad1964, Watson1964} is a non-parametric method that models the response function, $\toi{\by_k}$, as a 
convex combination of response values, $\{  \toi{\by_k} \}_{t=1}^T$, in the neighbourhood of $\toi{\bx_k}$. We define a non-negative weight in the neighbourhood of
$\toi{\bx_k}$ as:
\begin{equation} \label{eq:kernel_weight}
\bomega(\toi[j]{\bx_k} - \toi{\bx_k}) = \frac{\bkappa_b( \toi[j]{\bx_k} - \toi{\bx_k}) } { \sum_{i=1}^T \bkappa_b( \toi[i]{\bx_k} - \toi{\bx_k}) },
\end{equation}
where $\bkappa_b \colon \mathbb{R}  \to \mathbb{R}$  is a kernel with bandwidth parameter $b > 0$. The final regression estimate is therefore:
\begin{equation} \label{eq:kernel_est}
\toi{\hat{\by}_k} = \sum_{j=1}^T \bomega(\toi[j]{\bx_k} - \toi{\bx_k}) \toi[j]{\by_k}.
\end{equation}

The kernel function, $\bkappa_b( \toi[j]{\bx_k} - \toi{\bx_k})$, assigns a non-negative weight to all observations, $\{  \toi{\bx_k} \}_{t=1}^T$, that decreases
with the distance to the observation $\toi{\bx_k}$, whereas the bandwidth controls the span of the local neighbourhood. There are many choices of kernel function, such as 
uniform,  $\bkappa_b(\bx) = \mathbb{I}(|| \frac{\bx}{b} || \leq 1)$, Epanechnikov, $\bkappa_b(\bx) = (1 - ||\frac{\bx}{ b}||^2) \mathbb{I}(|| \frac{\bx}{ b} || \leq 1)$ 
and Gaussian, $\bkappa_b(\bx) = \frac{1}{\sqrt{2\pi}} e^{ -0.5 ||\frac{\bx}{b}||^2}$. The choice of kernel function and bandwidth is important as it determines the
degree of nonlinearity in the model. A small bandwidth will result in a model that closely fits the data, potentially capturing fine-grained patterns 
but also being more susceptible to noise.  A large bandwidth, on the other hand, will produce a smoother model that may overlook subtle patterns.

\section{Experimental Results} \label{sec:exp}
\noindent \textbf{Experiment data:} We consider an asset universe of $50$ global futures markets, described in Table \ref{table:universe}.
 Daily price data is provided by Commodity Systems Inc and is available as early as January $1950$ and extends through June $2026$. 
 Each futures market is rolled on a calendar basis in order to remain invested in the nearest most liquid available contract. 
\newline

\noindent \textbf{Feature variables:} As described in Section \ref{sec:method} we define the time series momentum feature as the volatility-adjusted average return 
observed over lookbacks $n \in \{21, 126, 252\}$, corresponding approximately to $1$, $6$ and $12$ month momentum:
\begin{equation}\label{eq:x_feature}
\toi{\bx_k} = \frac{1}{\sqrt{n}}\sum_{i=0}^{n-1} \toi[t - i]{\br_k}/\toi[t]{\bsigma_k},
\end{equation} 
where $\toi[t]{\bsigma_k}$ denotes the realized market volatility over the lookback horizon $n$. Feature values are winsorized at a value of $\pm 10$ and are further 
binned into $256$ equally spaced bins in order to reduce complexity and the propensity for overfitting.
\newline

\noindent  \textbf{Response variables:} Similarly, we define the response variable for each market as the forward volatility-adjusted average return
\begin{equation}\label{eq:response}
\toi{\by_k} = \frac{1}{\sqrt{m}}\sum_{i=0}^{m-1} \toi[t+i+l]{\br_k}/\toi[t+i]{\hat{\bsigma}_k},
\end{equation}
with lag time $l=2$, executing the trade signal at the closing price the day after the feature calculation, and $m \in \{1, 5, 21\}$, corresponding to
daily, weekly and monthly forecast horizons. Recall that the volatility of market $k$ at time  $t$ is denoted as $\toi{\hat{\bsigma}_k}$, and is estimated using 
a $40$-day exponential moving average.
\newline

\noindent \textbf{Model specification:}
For each market and forecast horizon, models are fit on one feature at a time (1-dimensional). Kernel regression models use a Gaussian kernel with bandwidth of $0.30$. 
In general we find that a Gaussian kernel with a bandwidth in the range of $0.20$--$0.50$ produces consistent results. Finally, each model is trained using a 
bootstrap aggregation (bagging) of the in-sample training data in order to attenuate the impact of outliers on the model fit \citep{Breiman1996}. Bagging is performed using
an average of $100$ models, each fit on $10\%$ of the training data.  We compare $4$ models:
\begin{enumerate}
\item \textbf{Global Linear:} a univariate linear OLS regression model fit globally across all markets.
\item \textbf{Global Nonlinear:} a univariate kernel regression model fit globally across all markets. 
\item \textbf{Sector Nonlinear:} a univariate kernel regression model fit across all markets within a sector. 
\item \textbf{Market Nonlinear:} a univariate kernel regression model fit locally to each individual market. 
\end{enumerate}

\newpage

\subsection{Motivating examples} \label{sec:examples}
\begin{figure}[H]
  \centering
  \begin{subfigure}[b]{0.45\linewidth}
    \includegraphics[width=\linewidth ]{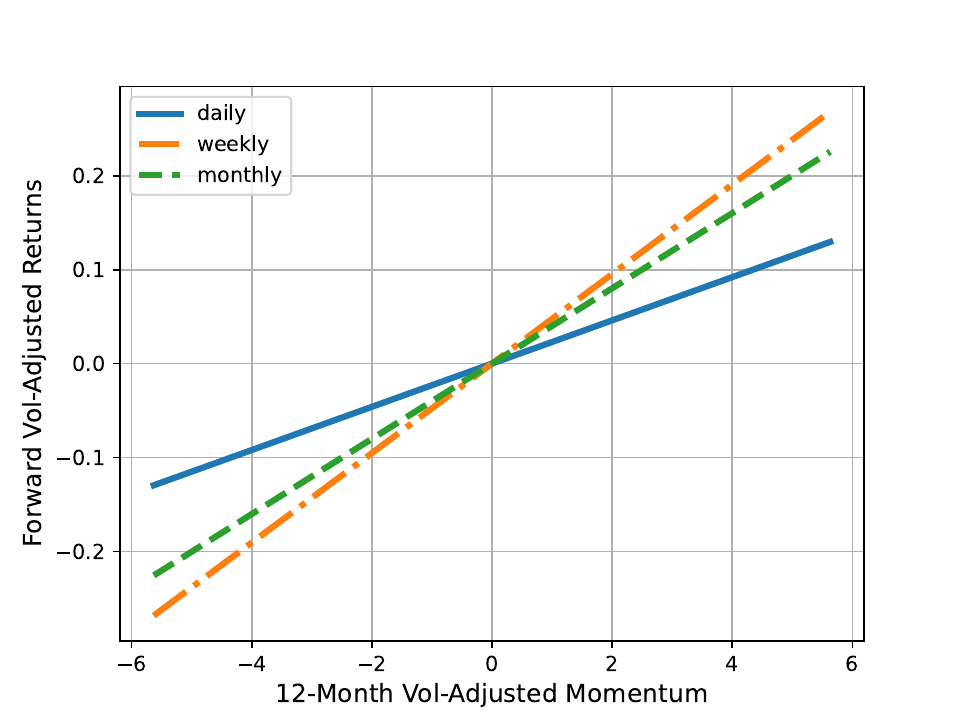}
    \caption{Global Linear Model}
    \label{fig:factor_linear}
  \end{subfigure}
  \begin{subfigure}[b]{0.45\linewidth}
    \includegraphics[width=\linewidth ]{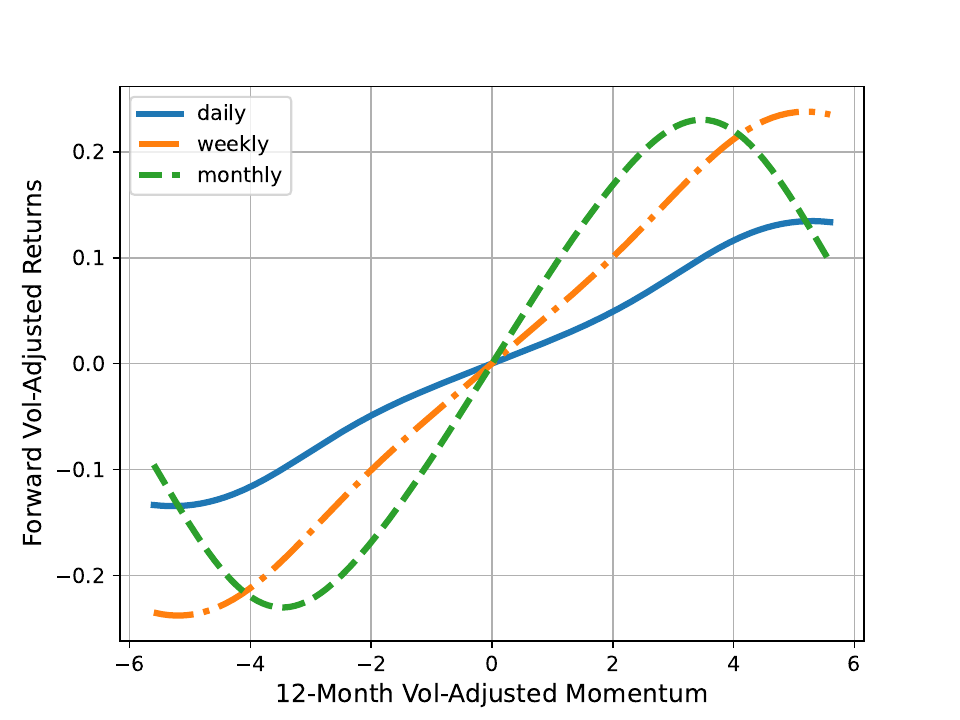}
    \caption{Global Nonlinear Model}
    \label{fig:factor_nonlinear}
  \end{subfigure}
  
  \caption{Estimated forward vol-adjusted returns as a function of the $12$-month time series momentum factor for Global Linear and Global Nonlinear models.
  Models are fit using all available training data from January $1950$ through June $2026$. }
  \label{fig:factor}
\end{figure}

We begin with some examples to help illustrate the idiosyncratic behaviour of the nonlinear models in comparison to the global linear and nonlinear factor definitions.
Note that models shown below are fit using all available training data from January $1950$ through June $2026$.
Figure \ref{fig:factor} plots the estimated response curve as a function of the $12$-month time series momentum factor. Recall that the Global Linear and Global Nonlinear
models are fit using all market data. As expected, the empirical linear model perfectly recovers the traditional time series momentum factor definition
where historical average returns positively predict forward expected returns \citep{Moskowitz2012}. We observe that the magnitude of the slope varies slightly across forecast
horizons;  weekly forecast horizons exhibit the strongest positive slope and in contrast daily horizons exhibit the weakest positive slope. The response function of the 
Global Nonlinear model is provided in Figure \ref{fig:factor_nonlinear}. First, we observe that for all forecast horizons, the Global Nonlinear model 
exhibits a positive and monotonic relationship for feature values in the range $-2 \leq \toi{\bx} \leq 2$ (hereafter referred to as the `normal range').  
This is a commonly observed effect for most model variants and helps to explain why nonlinear models correlate strongly with their linear counterpart: 
the range $-2 \leq \toi{\bx} \leq 2$ spans over $95\%$ of the time series momentum distribution and therefore for most feature values the linear and
nonlinear models will produce similar forecasts. However, for daily and weekly forecasts, the nonlinear model exhibits a modest sigmoidal relationship with marginally
decreasing response values as feature values become strongly positive $(\toi{\bx} > 3)$ or strongly negative $(\toi{\bx} < -3)$. For monthly forecasts,
the response relationship begins to flatten as feature values approach $\pm 3$ and then begins to invert as feature values exceed the 
extreme level $|\toi{\bx}| \geq 3$. 

\begin{figure}[H]
  \centering
  \begin{subfigure}[b]{0.43\linewidth}
    \includegraphics[width=\linewidth ]{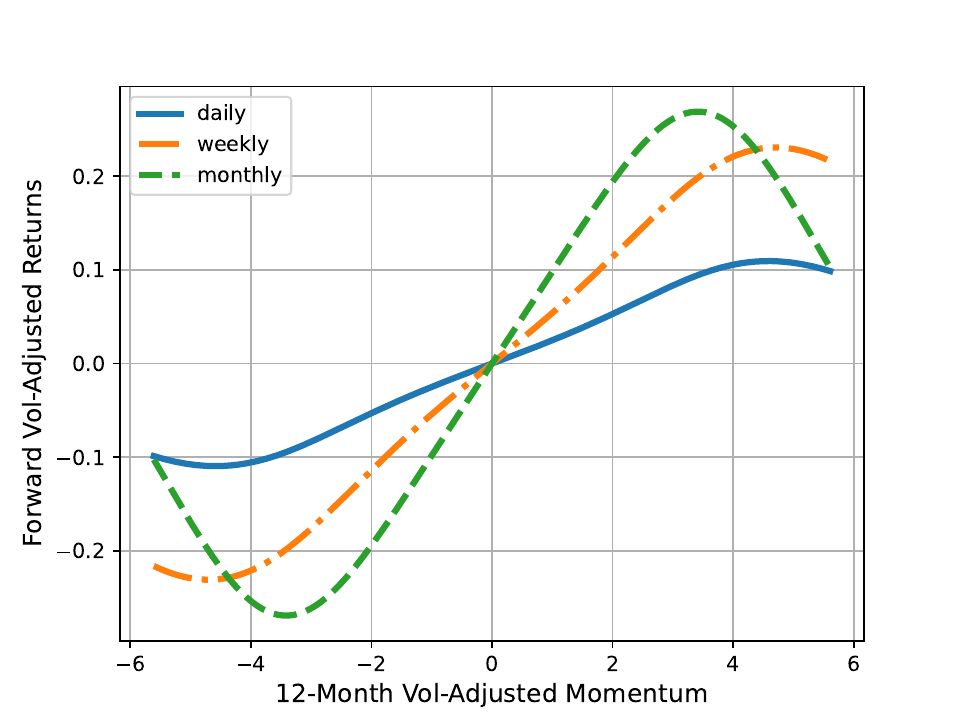}
    \caption{Nonlinear Model: Commodity Sectors}
    \label{fig:nonlinear_commodity}
  \end{subfigure}
  \begin{subfigure}[b]{0.43\linewidth}
    \includegraphics[width=\linewidth ]{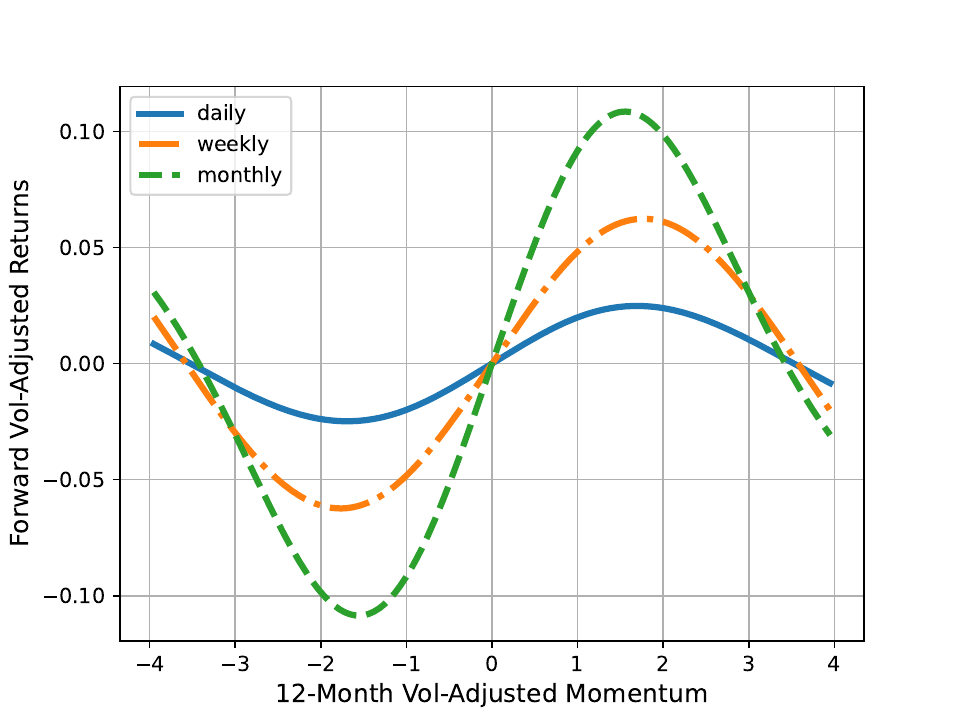}
    \caption{Nonlinear Model: Financial Sectors}
    \label{fig:nonlinear_financial}
  \end{subfigure}

  \caption{Estimated forward vol-adjusted returns as a function of the $12$-month time series momentum factor for Nonlinear models fit across commodity sectors
   (energy, metal, grain, soft) and  financial sectors (bond, currency, equity). Models are fit using all available training data from January $1950$ through June $2026$.}
  \label{fig:asset_class}
\end{figure}

\begin{figure}[H]
  \centering
  \begin{subfigure}[b]{0.43\linewidth}
    \includegraphics[width=\linewidth ]{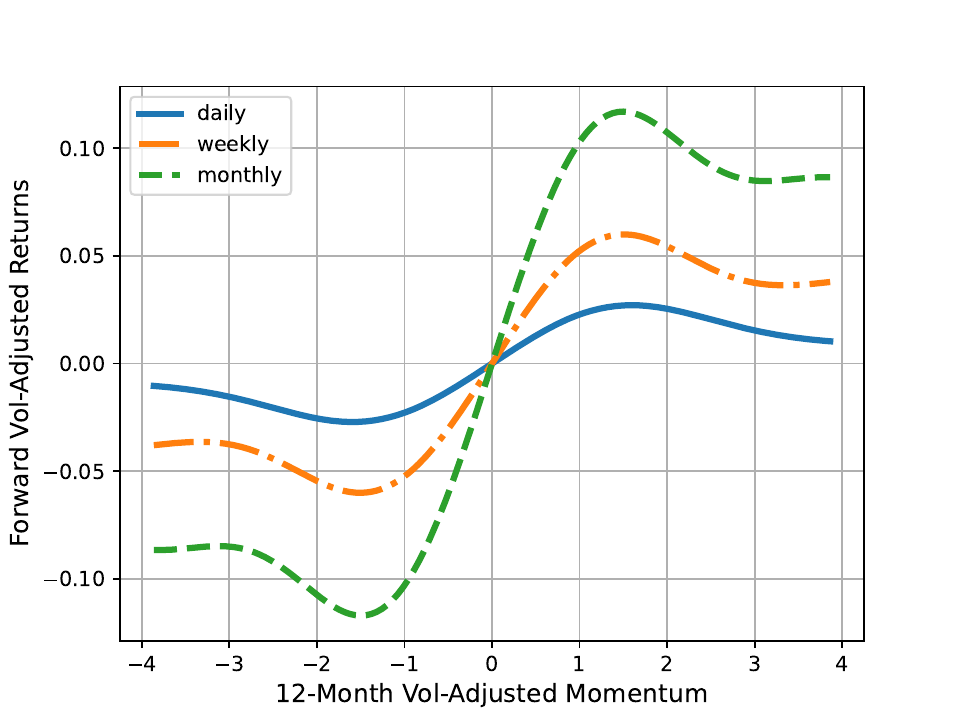}
    \caption{Nonlinear Model: Bond Sector}
    \label{fig:nonlinear_bond}
  \end{subfigure}
  \begin{subfigure}[b]{0.43\linewidth}
    \includegraphics[width=\linewidth ]{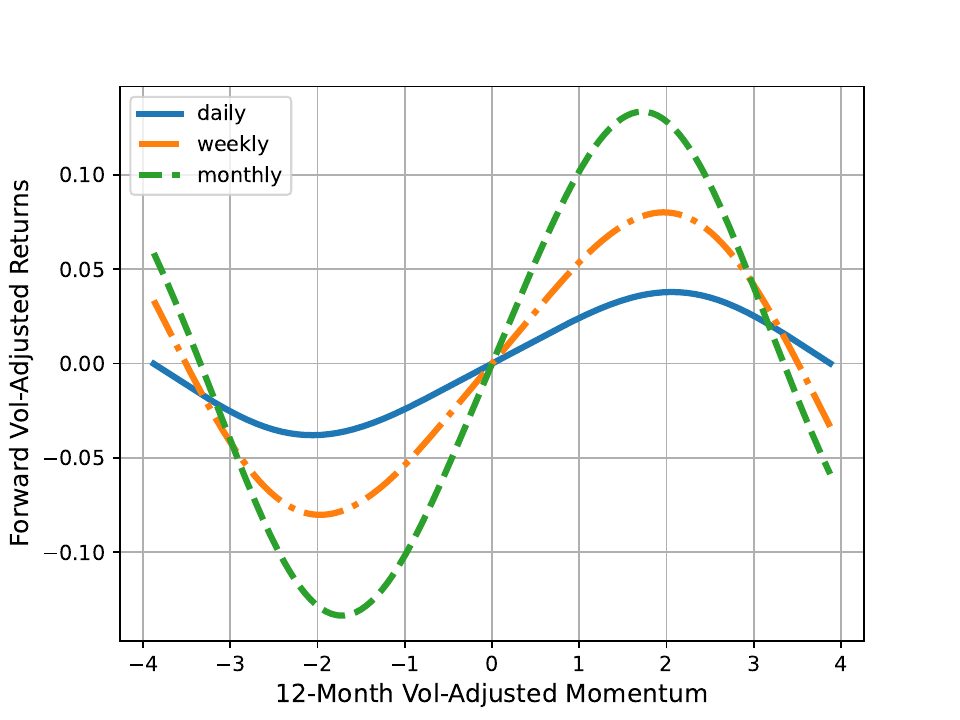}
    \caption{Nonlinear Model: Currency Sector}
    \label{fig:nonlinear_currency}
  \end{subfigure}
  \begin{subfigure}[b]{0.43\linewidth}
    \includegraphics[width=\linewidth ]{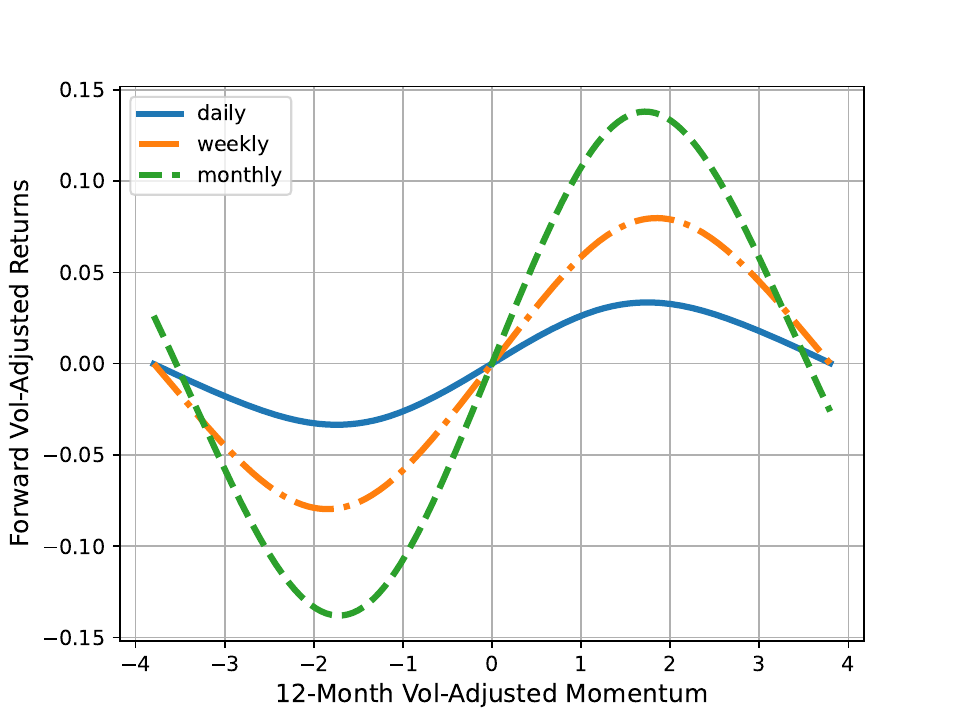}
    \caption{Nonlinear Model: Energy Sector}
    \label{fig:nonlinear_energy}
  \end{subfigure}
  \begin{subfigure}[b]{0.43\linewidth}
    \includegraphics[width=\linewidth ]{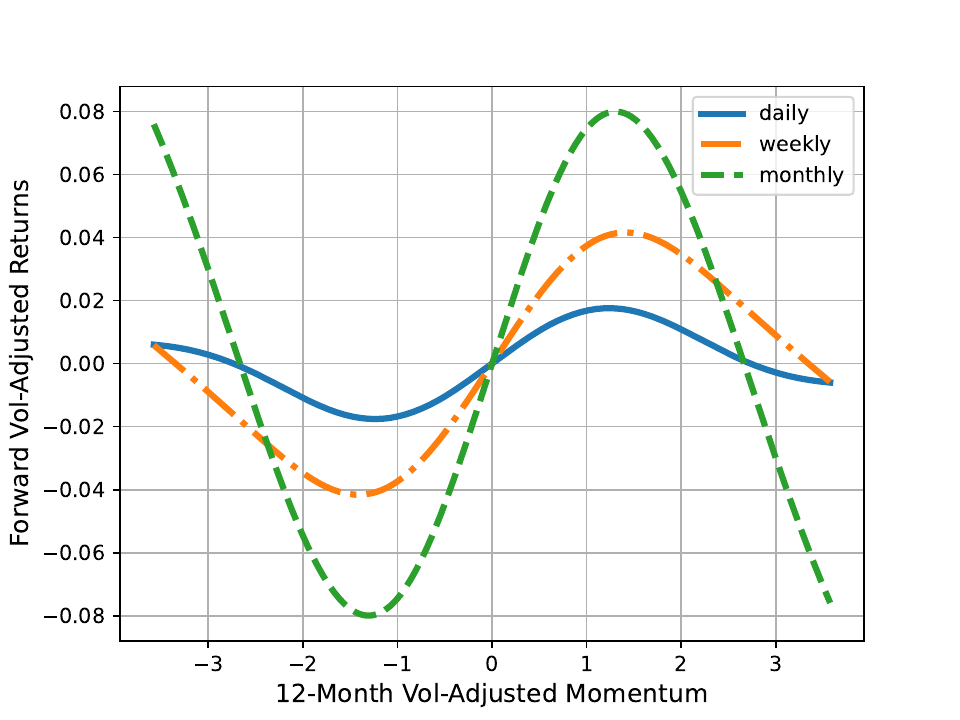}
    \caption{Nonlinear Model: Equity Sector}
    \label{fig:nonlinear_equity}
  \end{subfigure}
  \begin{subfigure}[b]{0.43\linewidth}
    \includegraphics[width=\linewidth ]{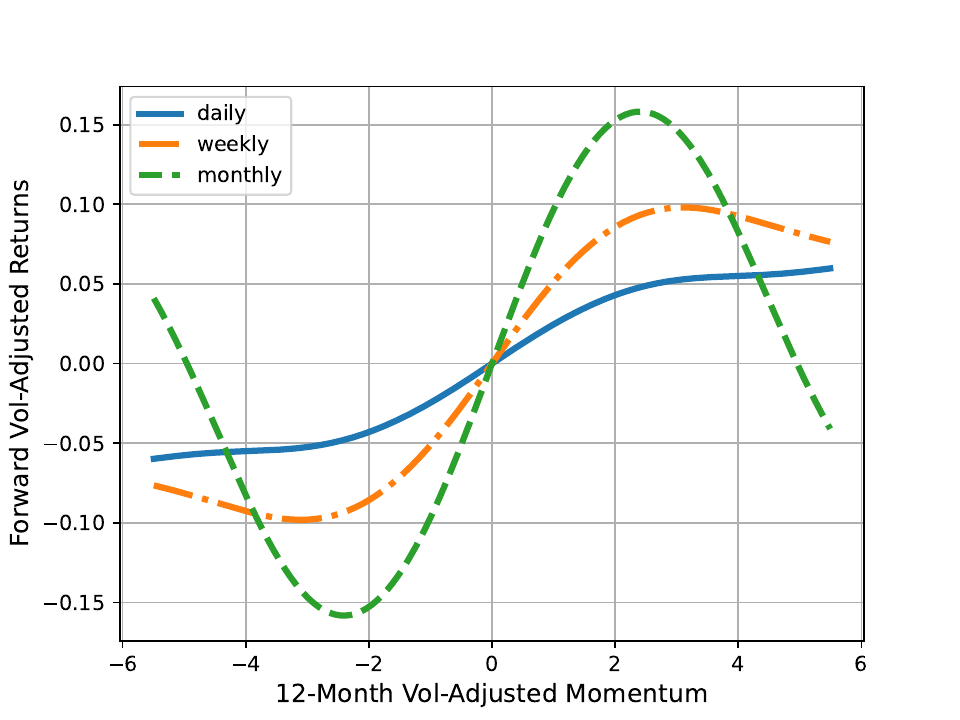}
    \caption{Nonlinear Model: Metal Sector}
    \label{fig:nonlinear_metal}
  \end{subfigure}
  \begin{subfigure}[b]{0.43\linewidth}
    \includegraphics[width=\linewidth ]{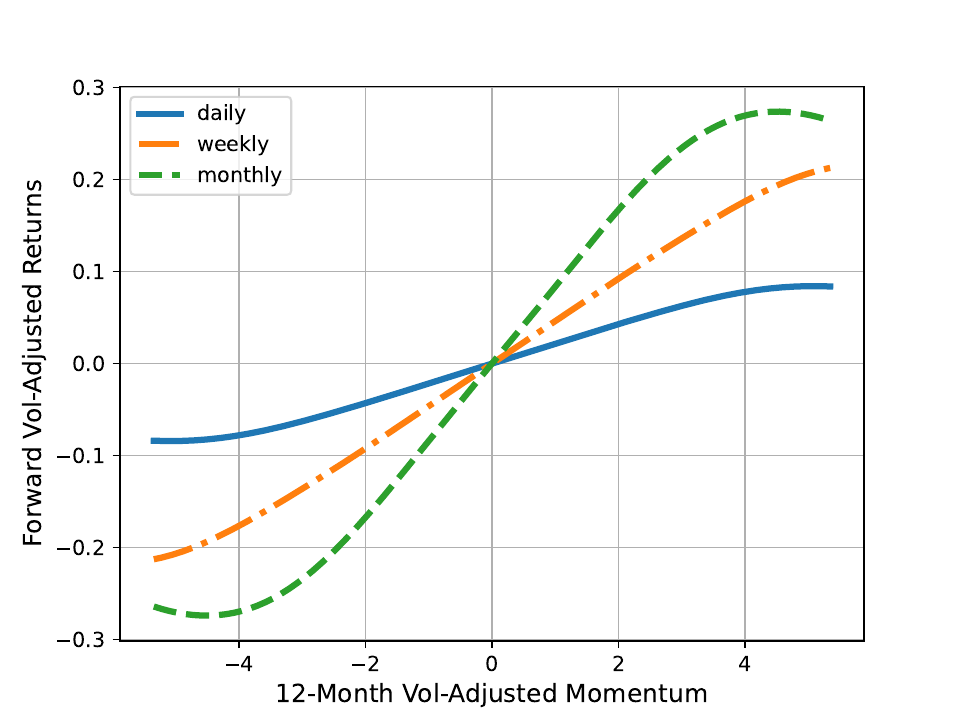}
    \caption{Nonlinear Model: Grain Sector}
    \label{fig:nonlinear_grain}
  \end{subfigure}
  
  \caption{Estimated forward vol-adjusted returns as a function of the $12$-month time series momentum factor for Sector Nonlinear models.
  Models are fit using all available training data from January $1950$ through June $2026$. }
  \label{fig:sector}
\end{figure}

\newpage
Figure \ref{fig:sector} plots the estimated response curve as a function of the $12$-month time series momentum factor for the Sector Nonlinear models. Note that 
the majority of Sector Nonlinear models exhibit traditional time series momentum factor behaviour across most forecast horizons when 
feature values are in the normal range. However, there are notable differences in the response function strength and character across various sectors, particularly at extreme
feature values.  For example, the metal and grain sectors exhibit predominantly linear time series momentum behaviour across the entire feature range. 
In contrast, currency and equity sectors display strong mean-reversion characteristics across all forecast horizons when feature values exceed $|\toi{\bx}| \approx 3$. 
Moreover, grain predictions range between $\pm 0.30$, indicating a strong predictive response to the time series momentum feature, whereas bond predictions
are notably weaker and range between $\pm 0.10$. From an empirical perspective, it would be easy to dismiss these observed differences as mere sampling error around a global time series momentum model. 
Alternatively, a causal explanation for the differences across sectors may arise from the composition and behaviour of the market participants within 
each sector. Hedgers, speculators and financial actors may respond to changes in prices with different objectives and time horizons. For example, in 
Figure \ref{fig:asset_class} we observe that commodity markets produce consistent trend-like behaviour, even at price extremes, whereas financial markets mean-revert.
Indeed, in commodity markets, speculative trading and hedging, in combination with unique supply-demand dynamics, may reinforce one another, resulting in sustained trends at
price extremes. Conversely, in financial markets, policy-driven rebalancing may trade against extended prices, resulting in a negative feedback that manifests as 
mean-reversion. 

\begin{figure}[H]
  \centering
  \begin{subfigure}[b]{0.43\linewidth}
    \includegraphics[width=\linewidth ]{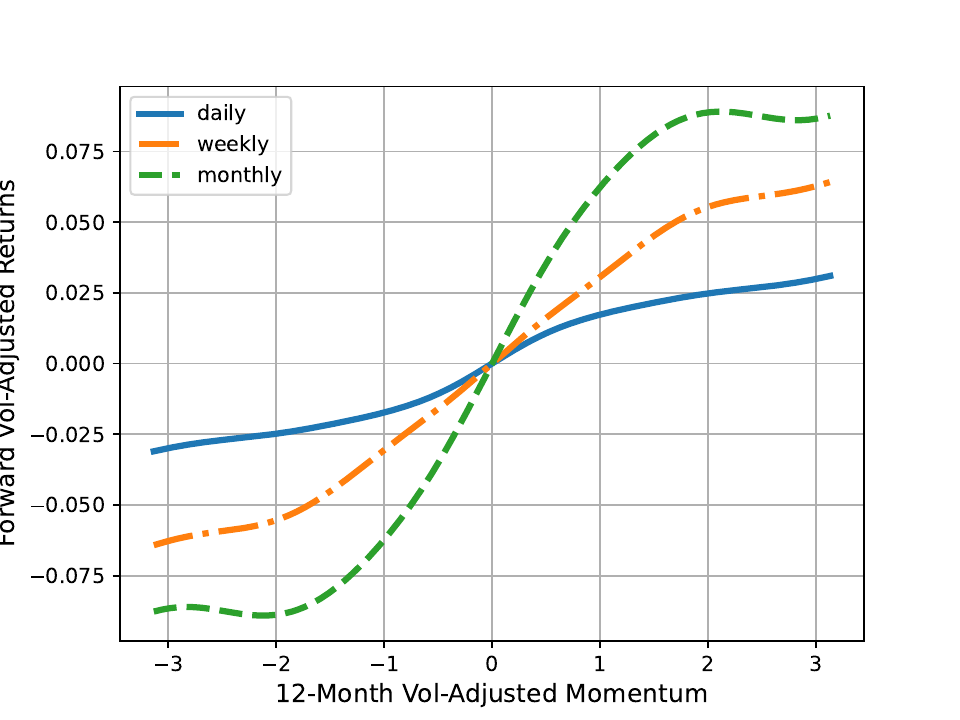}
    \caption{Nonlinear Model: Natural Gas Market}
    \label{fig:nonlinear_CL}
  \end{subfigure}
  \begin{subfigure}[b]{0.43\linewidth}
    \includegraphics[width=\linewidth ]{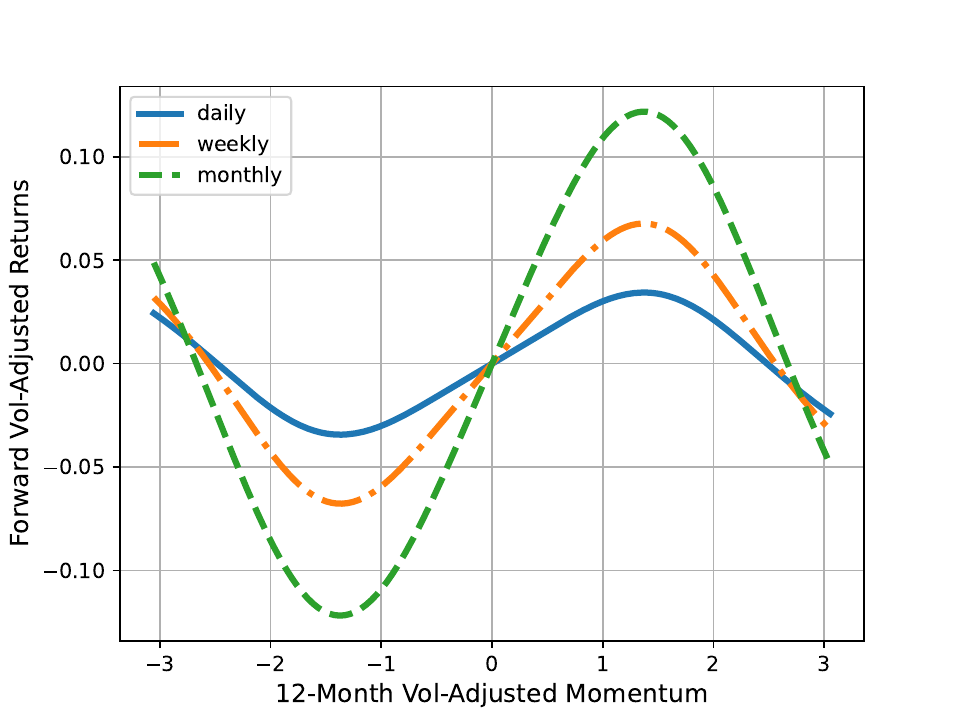}
    \caption{Nonlinear Model: Heating Oil Market}
    \label{fig:nonlinear_RBOB}
  \end{subfigure}
  \begin{subfigure}[b]{0.43\linewidth}
    \includegraphics[width=\linewidth ]{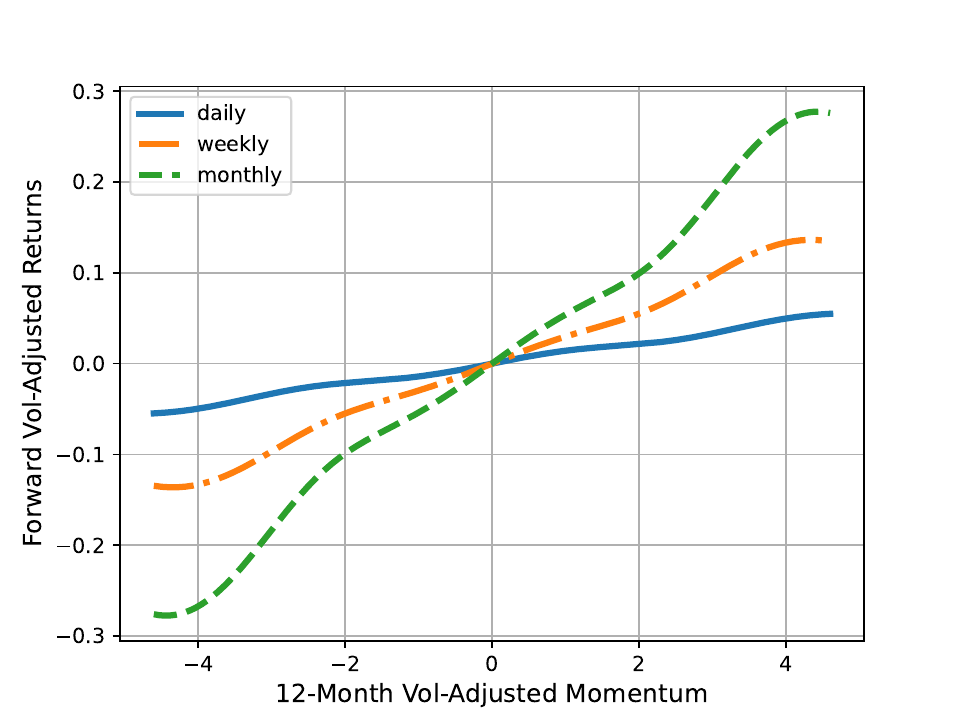}
    \caption{Nonlinear Model: Coffee Market}
    \label{fig:nonlinear_KC}
  \end{subfigure}
  \begin{subfigure}[b]{0.43\linewidth}
    \includegraphics[width=\linewidth ]{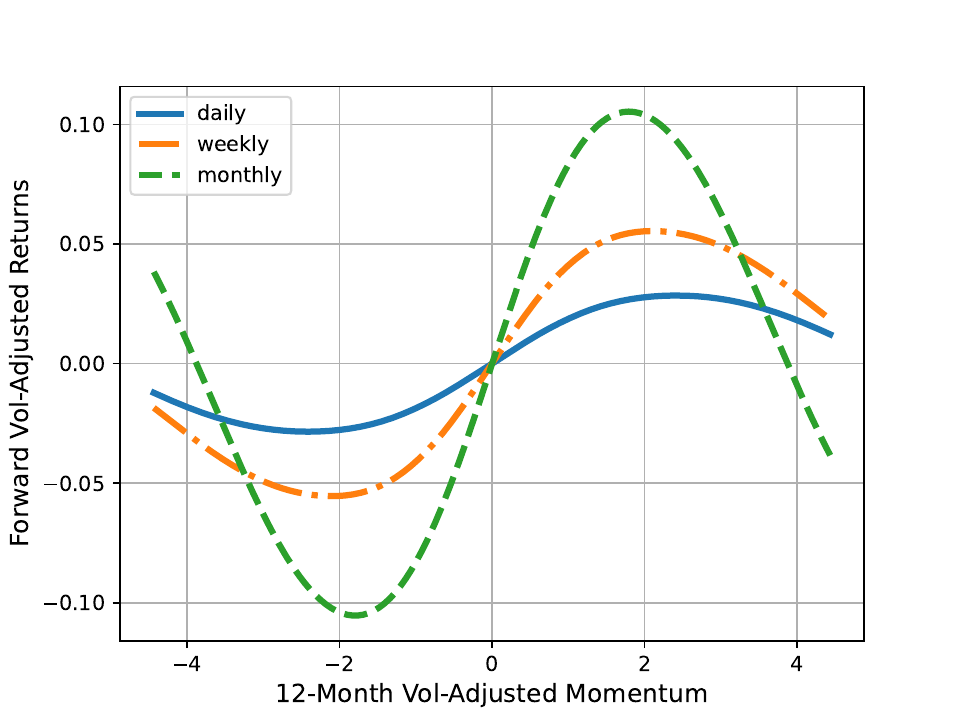}
    \caption{Nonlinear Model: Cotton Market}
    \label{fig:nonlinear_W}
  \end{subfigure}
  
  \caption{Estimated forward vol-adjusted returns as a function of the $12$-month time series momentum factor for Nonlinear models fit at the individual market level.
  Models are fit using all available training data from January $1950$ through June $2026$. }
  \label{fig:market}
\end{figure}

\newpage
The observation that response curves vary on a per sector basis motivates the empirical question: do individual markets within a sector display idiosyncratic
forward return dynamics in response to their observed historical performance? In general we find that for the $12$-month time series momentum feature,
financial sectors tend to display consistent response functions at the individual market level. As illustrated in \ref{sec:appendix}, this does not necessarily
hold true for all time series momentum features. In contrast, as depicted in Figure \ref{fig:market}, commodity markets tend to exhibit more idiosyncratic response functions. 
For example, within the energy sector, markets like Brent, crude and natural gas tend to exhibit traditional trend behaviour across the entire $12$-month time series
momentum range. In contrast, refined products such as gasoil, gasoline and heating oil display mean-reverting characteristics at price extremes.
Similarly, in agriculture, markets such as coffee and cocoa exhibit more linear time series momentum response functions, whereas cotton markets exhibit 
strong mean-reversion characteristics at the extremes. Moreover, the strength of the prediction response also varies across markets. For example, 
both natural gas and coffee exhibit a fairly linear response character but with very different intensities: the monthly response predictions in
natural gas are generally weak, ranging between $\pm 0.08$, whereas the monthly response predictions in coffee are approximately $4$ times stronger, 
and range between $\pm 0.30$.  A full catalogue of all feature-market pair models is presented in \ref{sec:appendix}.

It is difficult to determine with certainty whether the observed differences in response function strength and character are a result of sampling error, or indicate genuine 
differences in market response dynamics. We hypothesize that the differences in commodity markets' supply and demand dynamics, as well as inventory availability,
product substitution effects, and sensitivity to seasonality may serve as reasonable candidate explanations for the differing trend and mean-reversion behaviour observed 
across individual markets within a given sector.

\section{Historical Analysis} \label{sec:hist}
Fitting nonlinear models at the individual market level allows for the expression of idiosyncratic market dynamics, but may be vulnerable to sampling error as a result of
small sample sizes. In contrast, fitting a single model across all markets garners robustness from increased sample sizes, but fails to capture 
unique market response characteristics. In this section we test the validity of a framework that fits models at the market level.
Historical model performance is evaluated using the following experiment design: for each market, feature and
forecast horizon,  models are fit using all available data prior to January $1$, $1995$. We then apply walk-forward training and testing: each model is refit
every year using all available data and model predictions are applied to the next year's out-of-sample segment. To ensure no training data overlap, we 
apply a $22$-day embargo period between model training and next-year predictions. The testing period is between January $1$, $1995$ and 
June $30$, $2026$. Portfolios are formed by averaging the predictions across forecast horizons. The predictions for weekly and monthly forecast horizons are 
smoothed using a simple $5$-day and $21$-day moving average, respectively. Individual market positions are formed by dividing the average market prediction by 
the $40$-day market volatility estimate. Portfolios are rebalanced on a daily basis, assuming formation at the market close, and
trade execution at the next-day market closing price. Results are scaled to a volatility of $20\%$ and are net of slippage and 
commission transaction costs, described in more detail in Table \ref{table:universe}.

\subsection{1-Month Time Series Momentum}
\begin{figure}[ht]
  \centering
  \begin{subfigure}[b]{0.49\linewidth}
    \includegraphics[trim={0.5cm 1.cm 1cm 1.cm}, clip, width=\linewidth ]{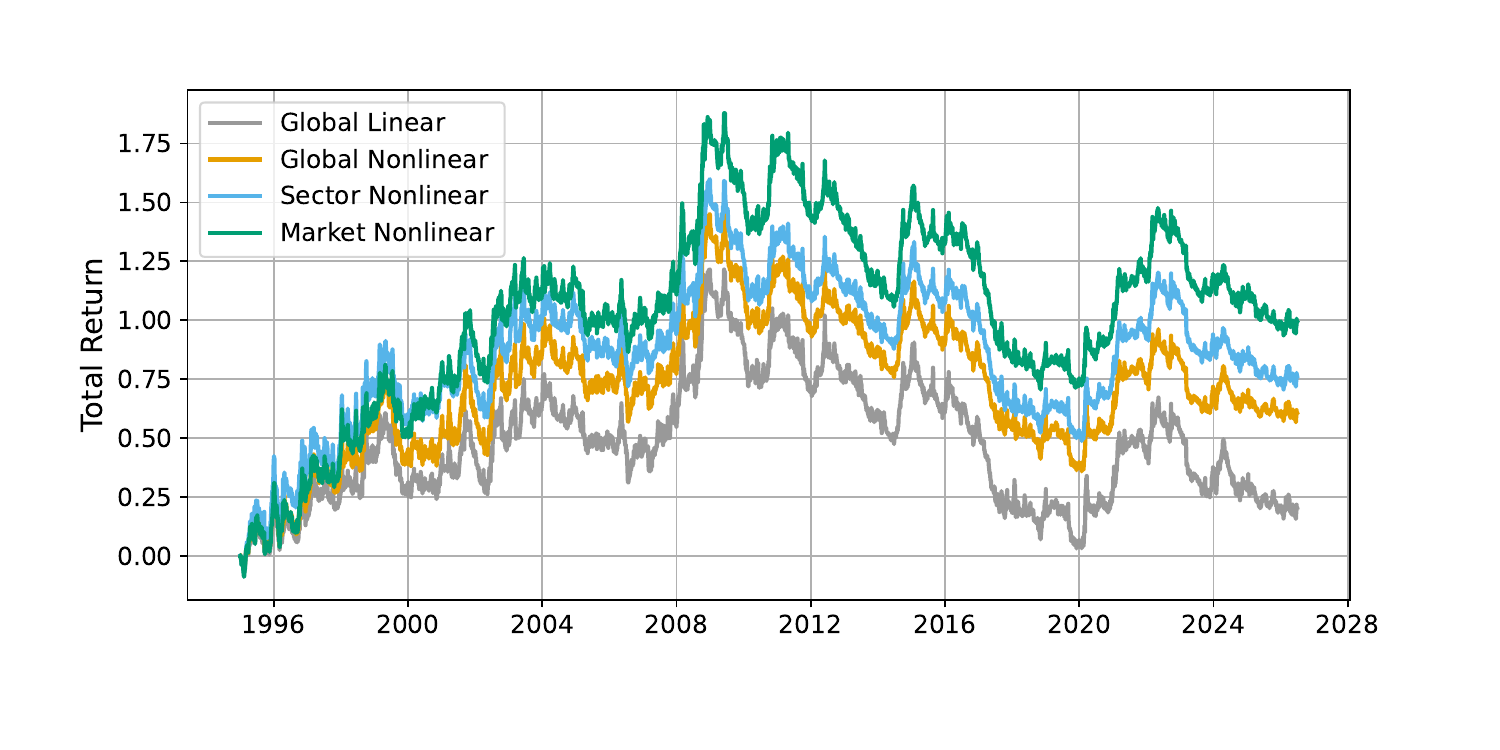}
    \caption{Net out-of-sample total return.}
    \label{fig:momo_1_month_eq}
  \end{subfigure}
  \begin{subfigure}[b]{0.49\linewidth}
    \includegraphics[width=\linewidth ]{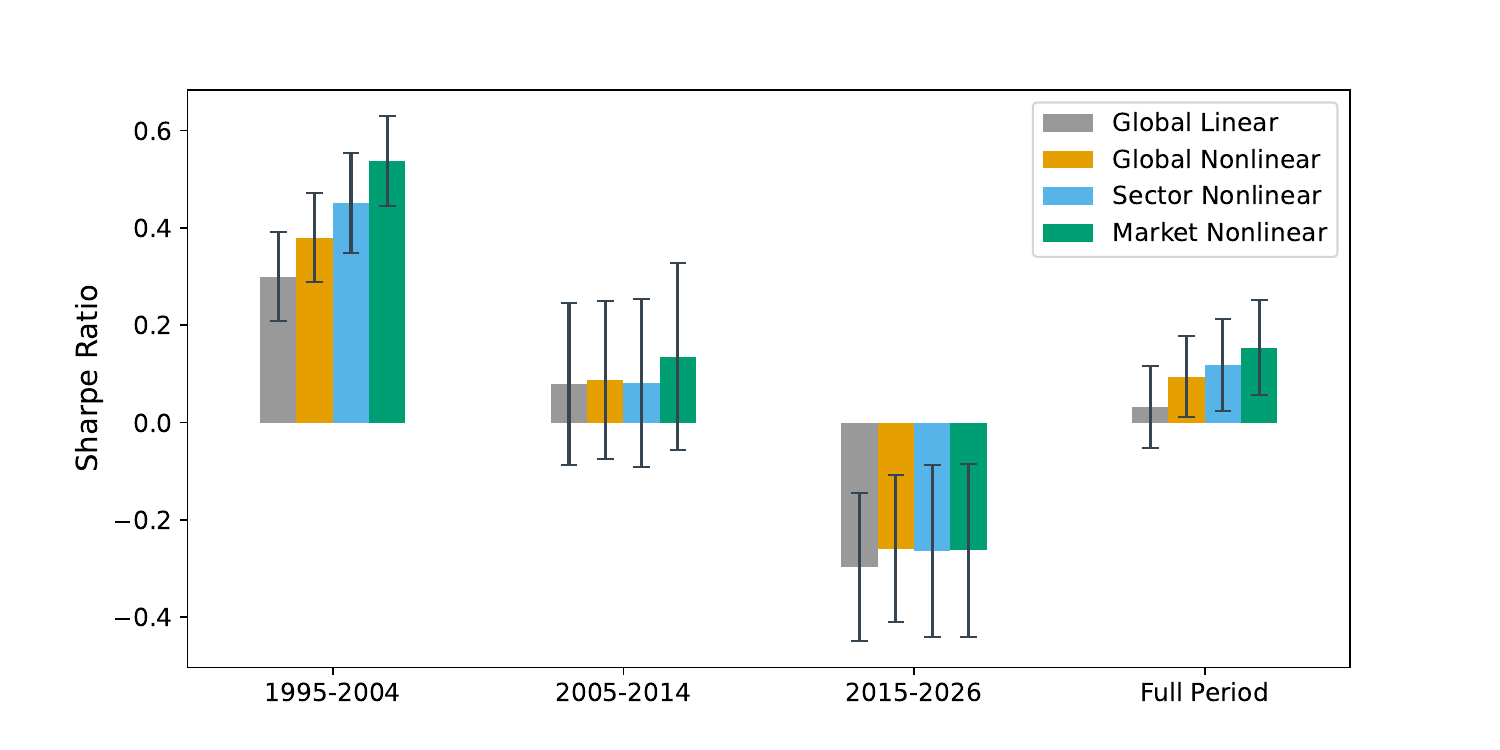}
    \caption{Net out-of-sample Sharpe ratio.}
    \label{fig:momo_1_month_sharpe}
  \end{subfigure}

  \caption{Out-of-sample equity growth and Sharpe ratio for models fit using the $1$-month time series momentum feature. Results are 
  net of estimated slippage and commission transaction costs. The out-of-sample period is January $1995$ through June $2026$.}
  \label{fig:momo_1_month}
\end{figure}

Figure \ref{fig:momo_1_month} reports the out-of-sample equity growth and Sharpe ratio for models fit using the $1$-month time series momentum feature.
We observe that the Global Linear model exhibits a near-zero Sharpe ratio over the entire test horizon. In contrast, the nonlinear models produce Sharpe ratios in 
the $0.10$--$0.20$ range. Moreover, for all models, the Sharpe ratios in the $1990$s and early $2000$s are significantly higher than Sharpe ratios exhibited over 
the most recent decade; a common theme observed across all testing. Importantly, we observe that over every evaluation decade, the models fit on individual markets exhibit
the highest Sharpe ratio. Notably, between $1995$ and $2004$, the Market Nonlinear model yields a Sharpe ratio of $0.57$, whereas the Global Nonlinear model and 
Sector Nonlinear model yield Sharpe ratios of $0.38$ and $0.44$, respectively.

Table \ref{table:momo_1_month} reports the annualized market mean returns, averaged across individual sectors and the total portfolio, for
the $1$-month time series momentum models. All results are net of transaction costs. The values in parentheses report the t-statistic from 
a paired t-test of the null hypothesis that the average market return  of the stated model is equal to the average market return produced by 
the Market Nonlinear model. First, we note that, in aggregate, the average market return of the Market Nonlinear model is greater than 
the average market return of all other models. Moreover, the effect is monotonic: models specified with finer-grained precision result in 
higher average market returns. However, the effect size is small and inconsistent across market sectors; 
for example in currencies and metals, the Market Nonlinear model produces the lowest mean return. Statistically, we are unable to reject
the null hypothesis that all models produce equivalent market mean returns.

\begin{table}[H]
\centering
\small

\begin{tabular}{lllll}
\toprule
 & Global Linear & Global Nonlinear & Sector Nonlinear & Market Nonlinear \\
\midrule
Bond & 0.001878 (1.1686)  & 0.002145 (0.8479)  & 0.002438 (0.2126)  & 0.002505 (0.0)  \\
Currency & 0.001223 (-0.3144)  & 0.001524 (-0.8204)  & 0.00128 (-0.5772)  & 0.001016 (0.0)  \\
Energy & 0.002668 (1.3615)  & 0.002748 (1.2117)  & 0.003231 (0.296)  & 0.003362 (0.0)  \\
Equity & -0.000858 (0.4027)  & -0.000576 (0.0578)  & -0.000302 (-0.5874)  & -0.00053 (0.0)  \\
Grain & 0.000973 (1.0539)  & 0.001152 (0.9229)  & 0.001135 (0.9026)  & 0.002200 (0.0)  \\
Metal & -0.002341* (-2.2542)  & -0.002025 (-1.7555)  & -0.002618 (-1.8833)  & -0.003488 (0.0)  \\
Soft & -0.003394 (1.0236)  & -0.003087 (0.7901)  & -0.002747 (0.5927)  & -0.001956 (0.0)  \\
\midrule
Portfolio & 0.000126 (1.4425)  & 0.000374 (0.7042)  & 0.00047 (0.4881)  & 0.000611 (0.0)  \\
\bottomrule
\end{tabular}

\caption{Annualized market mean returns, net of transaction costs, and reported at the individual sector and total portfolio level
for models fit using the $1$-month time series momentum feature.
Values in parentheses denote the t-statistic from a paired t-test of the null hypothesis that the average market return 
of the stated model is equal to the average market return produced by the Market Nonlinear model. Statistical significance
at the $0.10$, $0.05$ and $0.01$ level is denoted by $*$, $**$ and $***$, respectively. 
The analysis period is January $1995$ through June $2026$. }
\label{table:momo_1_month}
\end{table}

\subsection{6-Month Time Series Momentum}
\begin{figure}[ht]
  \centering
  \begin{subfigure}[b]{0.49\linewidth}
    \includegraphics[trim={0.5cm 1.cm 1cm 1.cm}, clip, width=\linewidth ]{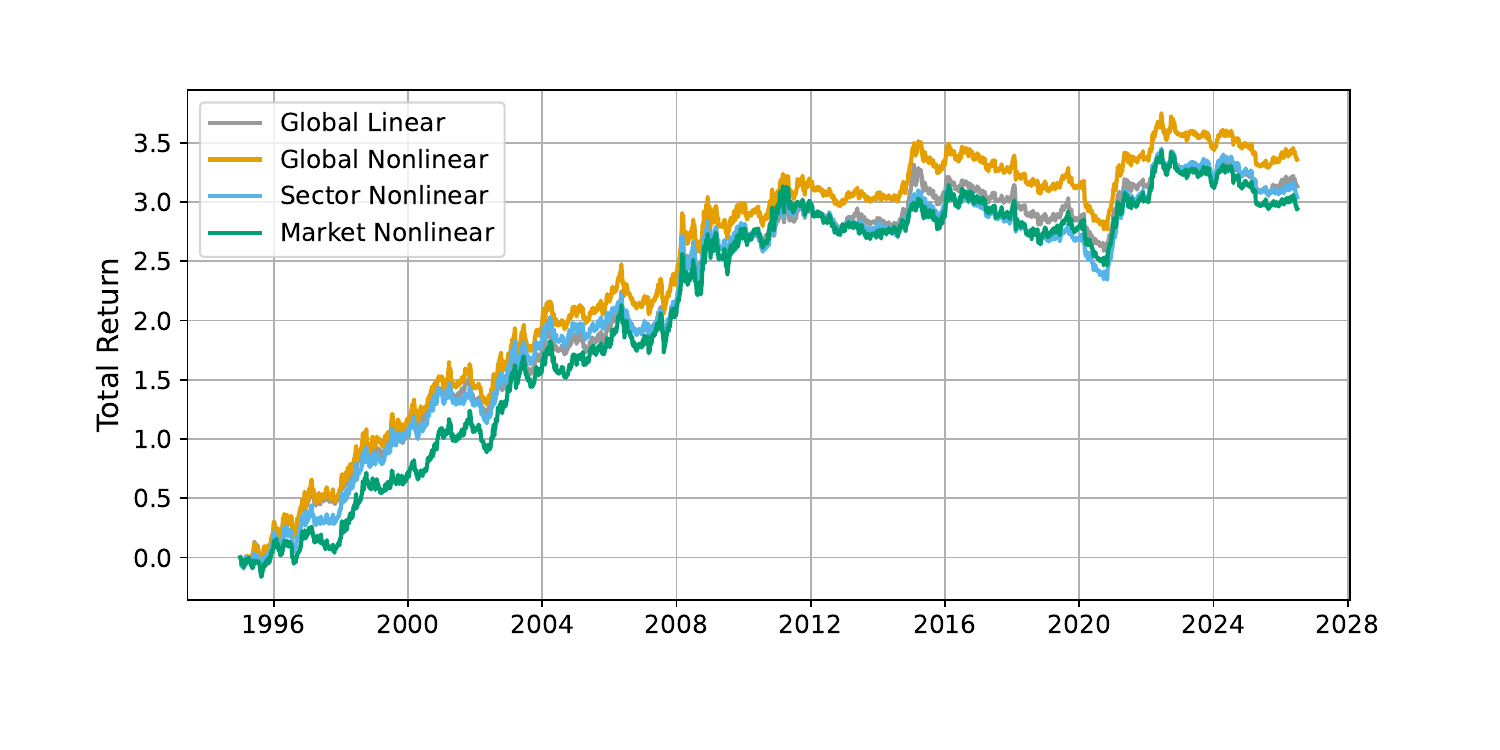}
    \caption{Net out-of-sample total return.}
    \label{fig:momo_6_month_eq}
  \end{subfigure}
  \begin{subfigure}[b]{0.49\linewidth}
    \includegraphics[width=\linewidth ]{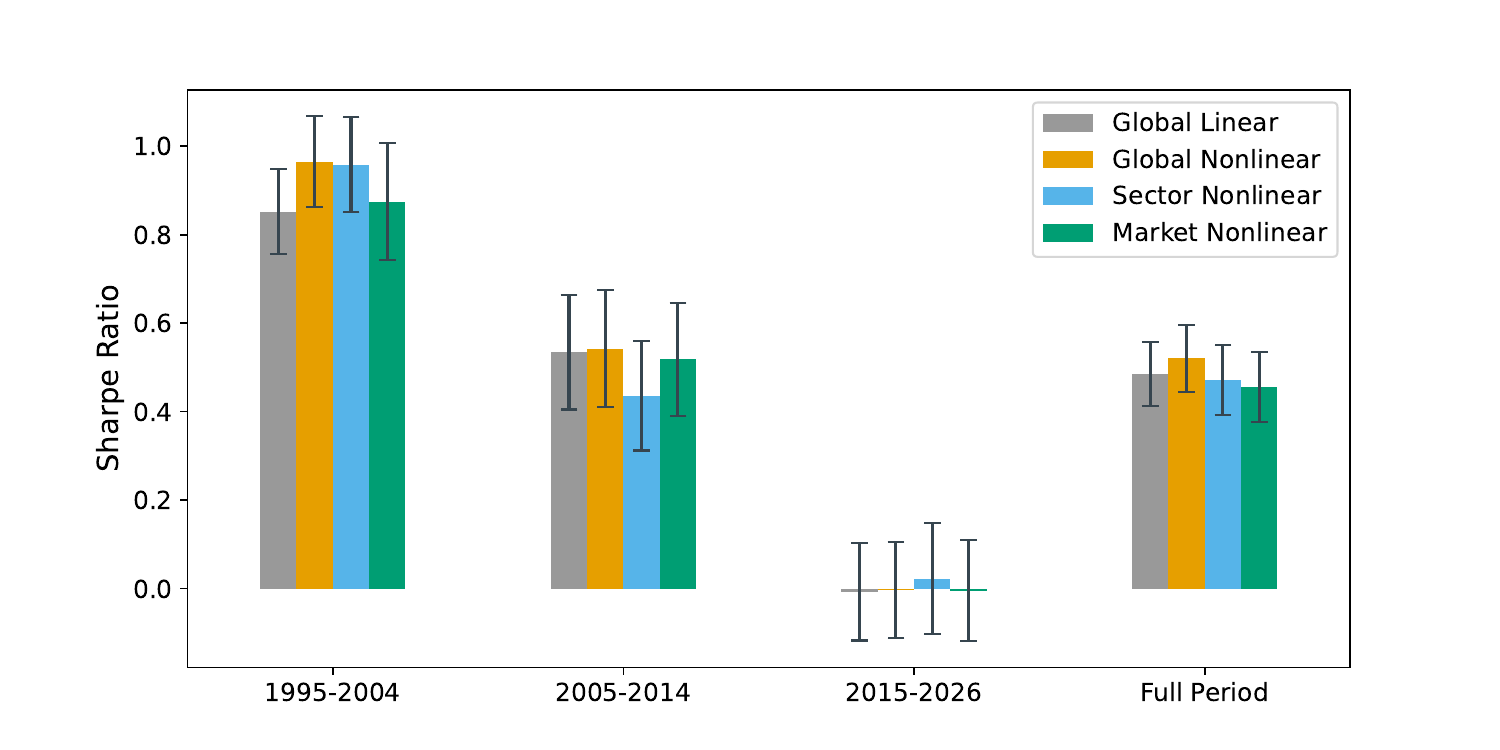}
    \caption{Net out-of-sample Sharpe ratio.}
    \label{fig:momo_6_month_sharpe}
  \end{subfigure}

  \caption{Out-of-sample equity growth and Sharpe ratio for models fit using the $6$-month time series momentum feature. Results are 
  net of estimated slippage and commission transaction costs. The out-of-sample period is January $1995$ through June $2026$.}
  \label{fig:momo_6_month}
\end{figure}

Figure \ref{fig:momo_6_month} reports the out-of-sample equity growth and Sharpe ratio for models fit using the $6$-month time series momentum feature.
In contrast to the $1$-month feature, here we observe no advantage to fitting nonlinear models at the individual market level. In fact, over the full test
horizon, the Market Nonlinear model yields the lowest (though statistically equal) Sharpe ratio of $0.47$, in comparison to a Sharpe ratio of $0.48$ exhibited by the Global
Linear model and the Sector Nonlinear model and a modestly higher Sharpe ratio of $0.52$ exhibited by the Global Nonlinear model. Consistent with the findings above,
we observe a decline in Sharpe ratios over the most recent decade, with all models producing net Sharpe ratios near zero.

Table \ref{table:momo_6_month} reports the annualized market mean returns, averaged across individual sectors and the total portfolio, for
the $6$-month time series momentum models. We observe that, in aggregate, the average market return of the Market Nonlinear model is less than 
the average market return of all other models. Again, however, we note that the effect is inconsistent across market sectors, 
and, with the exception of a handful of sector-level comparisons (bond, energy, equity and soft), all other tests, including at the portfolio level, are statistically insignificant.
In totality we fail to reject the null hypothesis that all models produce equivalent market mean returns.

\begin{table}[H]
\centering
\small

\begin{tabular}{lllll}
\toprule
 & Global Linear & Global Nonlinear & Sector Nonlinear & Market Nonlinear \\
\midrule
Bond & 0.002318* (1.9936)  & 0.002705 (1.5295)  & 0.002462 (1.6477)  & 0.003787 (0.0)  \\
Currency & 0.001813 (0.1072)  & 0.002036 (-0.3195)  & 0.002501 (-1.1108)  & 0.001869 (0.0)  \\
Energy & 0.004699* (-2.3217)  & 0.004572* (-2.1228)  & 0.004026 (-1.5147)  & 0.002688 (0.0)  \\
Equity & 0.001945 (-1.7546)  & 0.002302** (-2.4285)  & 0.001342 (-0.5598)  & 0.001019 (0.0)  \\
Grain & 0.000131 (-0.3668)  & -0.000118 (0.0304)  & 0.000065 (-0.2437)  & -0.000100 (0.0)  \\
Metal & 0.002537 (1.0097)  & 0.002667 (0.8632)  & 0.001529 (1.4464)  & 0.003261 (0.0)  \\
Soft & 0.000835 (-0.1781)  & 0.001017 (-0.6312)  & 0.001538* (-2.0017)  & 0.000782 (0.0)  \\
\midrule
Portfolio & 0.001928 (-0.4454)  & 0.002066 (-0.9972)  & 0.001871 (-0.2091)  & 0.00181 (0.0)  \\
\bottomrule
\end{tabular}

\caption{Annualized market mean returns, net of transaction costs, and reported at the individual sector and total portfolio level
for models fit using the $6$-month time series momentum feature.
Values in parentheses denote the t-statistic from a paired t-test of the null hypothesis that the average market return 
of the stated model is equal to the average market return produced by the Market Nonlinear model. Statistical significance
at the $0.10$, $0.05$ and $0.01$ level is denoted by $*$, $**$ and $***$, respectively. 
The analysis period is January $1995$ through June $2026$. }
\label{table:momo_6_month}
\end{table}

\subsection{12-Month Time Series Momentum}
\begin{figure}[ht]
  \centering
  \begin{subfigure}[b]{0.49\linewidth}
    \includegraphics[trim={0.5cm 1.cm 1cm 1.cm}, clip, width=\linewidth ]{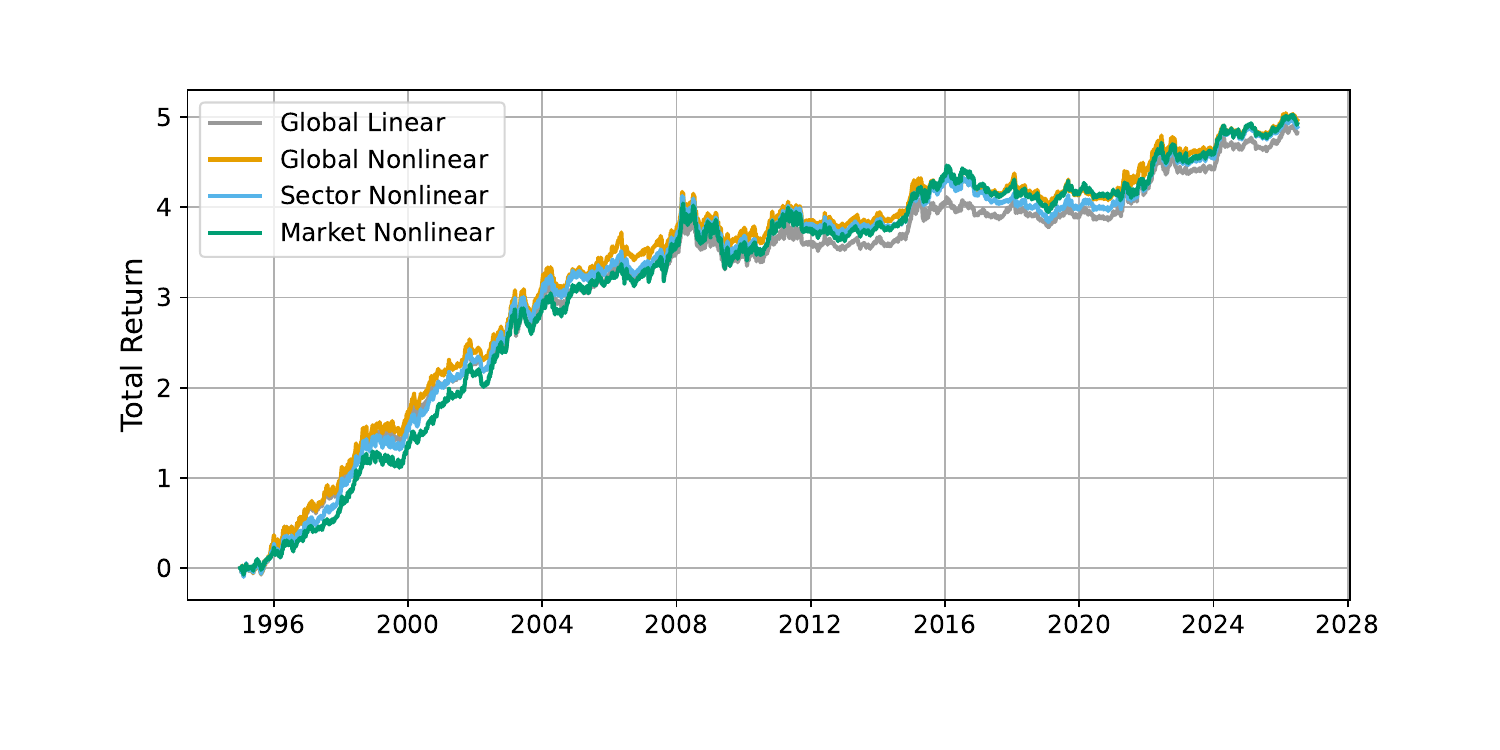}
    \caption{Net out-of-sample total return.}
    \label{fig:momo_12_month_eq}
  \end{subfigure}
  \begin{subfigure}[b]{0.49\linewidth}
    \includegraphics[width=\linewidth ]{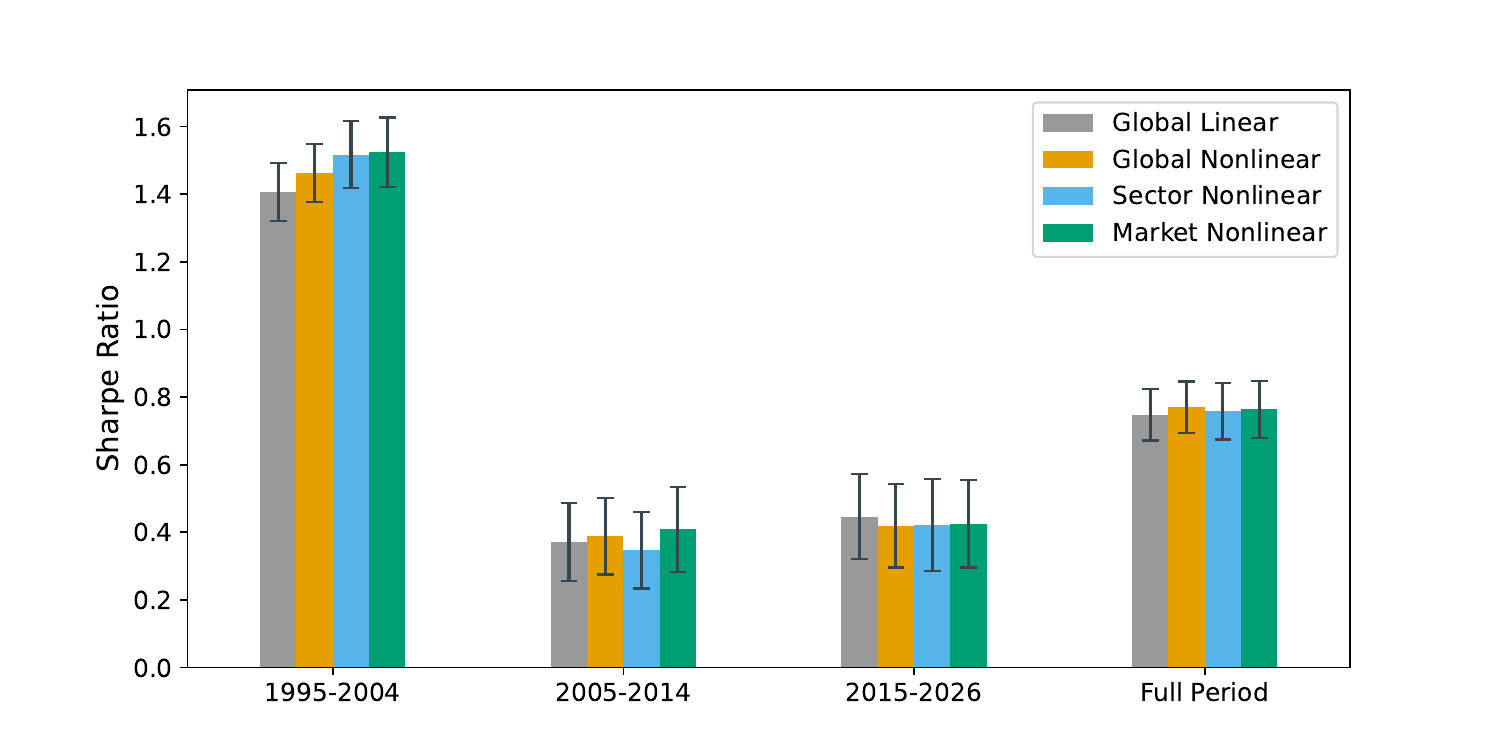}
    \caption{Net out-of-sample Sharpe ratio.}
    \label{fig:momo_12_month_sharpe}
  \end{subfigure}

   \caption{Out-of-sample equity growth and Sharpe ratio for models fit using the $12$-month time series momentum feature. Results are 
  net of estimated slippage and commission transaction costs. The out-of-sample period is January $1995$ through June $2026$.}
  \label{fig:momo_12_month}
\end{figure}

Figure \ref{fig:momo_12_month} reports the out-of-sample equity growth and Sharpe ratio for models fit using the $12$-month time series momentum feature.
Consistent with the $1$-month results, we observe that the Market Nonlinear models produce a marginally higher Sharpe ratio over the $1995$--$2004$ and 
$2005$--$2014$ periods. Moreover, the Market Nonlinear model yields the highest Sharpe ratio over the entire test horizon, while simultaneously producing 
strategy characteristics that are highly correlated with the Global Linear and Global Nonlinear time series momentum models.
In contrast, however, the Global Linear model exhibits the best performance over the most recent decade, a result that is generally
consistent with \citet{MoskowitzNonlinear}.

Table \ref{table:momo_12_month} reports the annualized market mean returns and paired t-statistic for
the $12$-month time series momentum models. Once again, we observe a statistically insignificant difference in market mean returns and
are therefore unable to reject the null hypothesis that, in aggregate, each model produces statistically equivalent out-of-sample
market performance. In comparison to the $1$-month and $6$-month models, we find that the $12$-month time series momentum models produce
the smallest portfolio-level t-statistics, indicating a strong statistical equivalence across each model. As illustrated above,
and highlighted in more detail in \ref{sec:appendix}, this is expected and is a direct result of the dominant linear relationship displayed by 
most models when the feature values are in the normal range. 

\begin{table}[H]
\centering
\small

\begin{tabular}{lllll}
\toprule
 & Global Linear & Global Nonlinear & Sector Nonlinear & Market Nonlinear \\
\midrule
Bond & 0.004084 (1.4606)  & 0.004271 (1.2825)  & 0.004675 (1.0093)  & 0.005712 (0.0)  \\
Currency & 0.002738 (-0.2369)  & 0.002827 (-0.3735)  & 0.003024 (-1.0348)  & 0.002569 (0.0)  \\
Energy & 0.005087 (-0.3636)  & 0.005064 (-0.3243)  & 0.005079 (-0.3532)  & 0.004858 (0.0)  \\
Equity & 0.003207 (-1.2975)  & 0.003418 (-1.6922)  & 0.001985 (1.1041)  & 0.00237 (0.0)  \\
Grain & 0.001349 (-1.5203)  & 0.001371 (-1.7279)  & 0.001394 (-1.6932)  & 0.000652 (0.0)  \\
Metal & 0.00276 (0.7572)  & 0.002915 (0.6798)  & 0.002873 (0.8600)  & 0.003319 (0.0)  \\
Soft & 0.001856 (0.6432)  & 0.001806 (0.7003)  & 0.002402 (-0.3800)  & 0.002182 (0.0)  \\
\midrule
Portfolio & 0.002967 (0.2176)  & 0.003054 (-0.0893)  & 0.003006 (0.0989)  & 0.003029 (0.0)  \\
\bottomrule
\end{tabular}

\caption{Annualized market mean returns, net of transaction costs, and reported at the individual sector and total portfolio level
for models fit using the $12$-month time series momentum feature.
Values in parentheses denote the t-statistic from a paired t-test of the null hypothesis that the average market return 
of the stated model is equal to the average market return produced by the Market Nonlinear model. Statistical significance
at the $0.10$, $0.05$ and $0.01$ level is denoted by $*$, $**$ and $***$, respectively. 
The analysis period is January $1995$ through June $2026$. }
\label{table:momo_12_month}
\end{table}

\subsection{Average Time Series Momentum}
\begin{figure}[H]
  \centering
  \begin{subfigure}[b]{0.49\linewidth}
    \includegraphics[trim={0.5cm 1.cm 1cm 1.cm}, clip, width=\linewidth ]{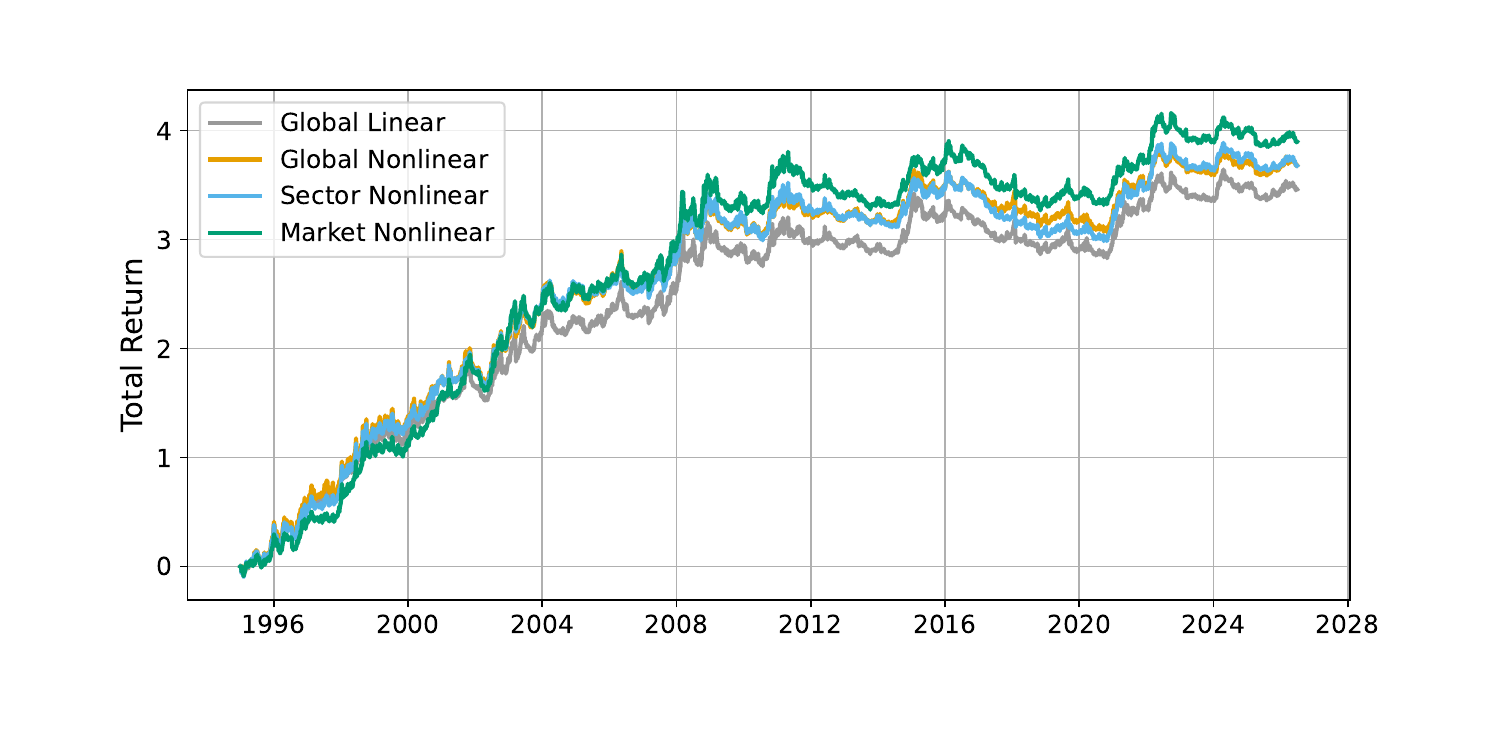}
    \caption{Net out-of-sample total return.}
    \label{fig:average_eq}
  \end{subfigure}
  \begin{subfigure}[b]{0.49\linewidth}
    \includegraphics[width=\linewidth ]{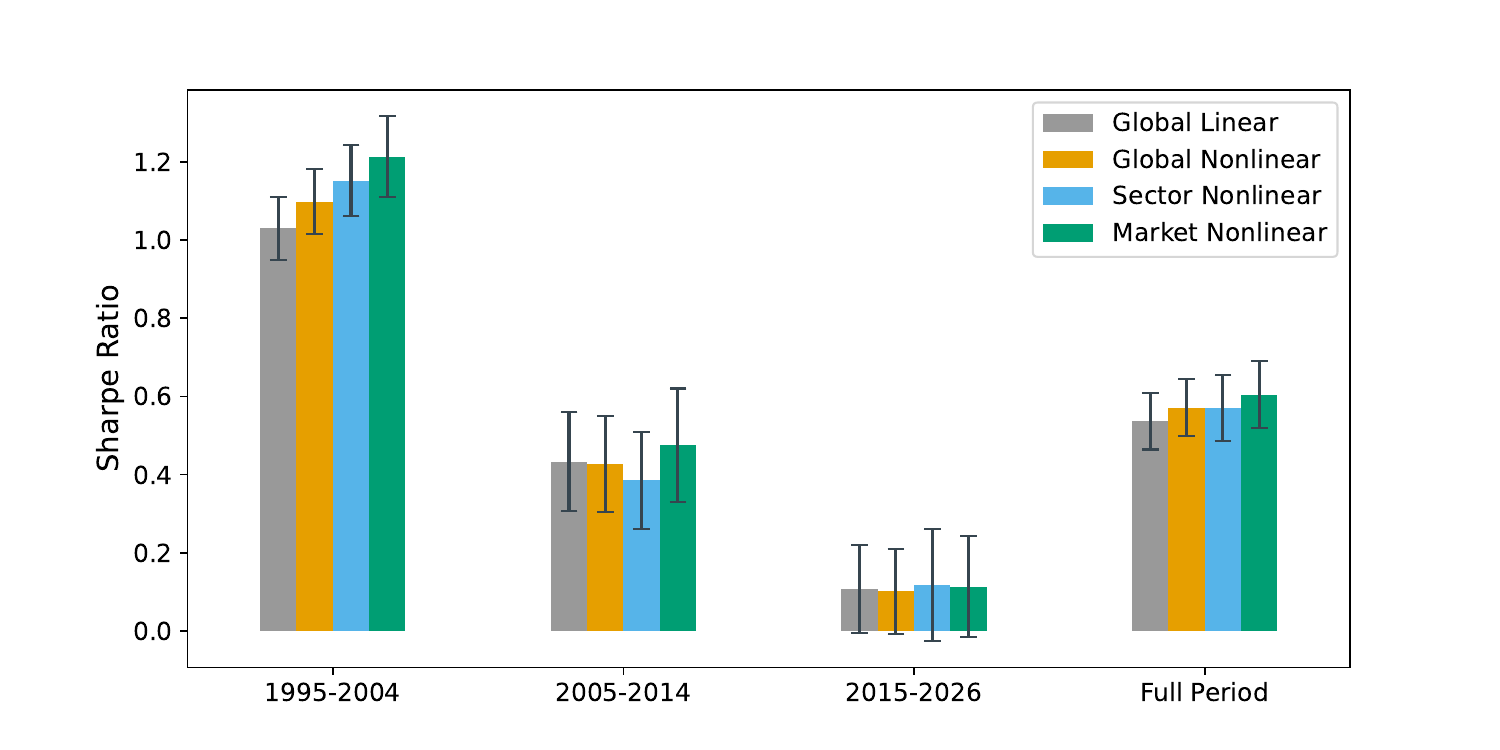}
    \caption{Net out-of-sample Sharpe ratio.}
    \label{fig:average_sharpe}
  \end{subfigure}

  \caption{Out-of-sample equity growth and Sharpe ratio for portfolios formed using a simple average of the $1$, $6$ and $12$-month time series momentum models.
  Results are net of estimated slippage and commission transaction costs. The out-of-sample period is January $1995$ through June $2026$.}
  \label{fig:average}
\end{figure}

Next, Figure \ref{fig:average} reports the out-of-sample equity growth and Sharpe ratio for models fit using an average of the $1$, $6$ and $12$-month 
time series momentum features. The average results provide clear positive evidence that fitting nonlinear models at the individual market level 
is a robust and often overlooked modelling alternative that can result in improved out-of-sample performance. We observe that over every evaluation decade, 
the Market Nonlinear model produces net Sharpe ratios that are no worse than the Global Linear and Nonlinear alternatives, and in fact over the majority of
evaluation horizons produces a marginal Sharpe ratio improvement. With the exception of the $1995$-$2004$ decade, the relationship between model precision 
and out-of-sample Sharpe ratio is not monotonic. Indeed, we observe that in the $2005$-$2014$ period, fitting models at the sector level actually results in a slight 
decrease in net Sharpe ratio in comparison to the Global Linear and Global Nonlinear models. When averaged across all features and evaluated over the full period, 
the Global Linear model produces a net Sharpe ratio of $0.54$, whereas the Global Nonlinear and Sector Nonlinear models produce net Sharpe ratios of $0.57$.
In contrast, the Market Nonlinear model produces a net Sharpe ratio of $0.60$ when evaluated over the full 
test period. The Sharpe ratio error bars in Figure \ref{fig:average_sharpe} indicate that this improvement is not statistically significant; 
but nonetheless provides evidence in support of a framework that fits models at the individual market level as opposed to the traditional 
global factor approach. Lastly, we note that all models exhibit strong positive correlation to the Global Linear model, with correlations ranging from $0.95$ in the
case of the Market Nonlinear model, to $0.99$ for the Global Nonlinear model. This realization is consistent with the observation that most markets exhibit 
traditional linear time series momentum behaviour in the normal feature range: $-2 \leq \toi{\bx} \leq 2$. The benefit of nonlinear models, therefore, occurs 
when feature values are at extreme levels, which occurs with a much lower frequency.

In Table \ref{table:average_ext_2} and Table \ref{table:average_ext_3} we test the hypothesis that the market-specific nonlinear response 
functions, predominantly observed outside of the feature normal range, represent genuine market signal and are not purely an artifact of sampling error.
Specifically, Table \ref{table:average_ext_2} reports the annualized market mean returns formed using a simple average of the $1$, $6$ and $12$-month time series momentum 
out-of-sample results, evaluated for all observations where the feature values fall immediately outside the normal range $|\toi{\bx}| \geq 2$. Similarly, Table
\ref{table:average_ext_3} reports the same results evaluated for all observations where the feature values are at extreme levels: $|\toi{\bx}| \geq 3$.
The conclusions are mixed: Table \ref{table:average_ext_2} reports positive paired t-statistics on average, indicating that the Market Nonlinear models
produce positive excess mean market returns for feature values $|\toi{\bx}| \geq 2$. The effect size, however, is not statistically significant and 
the results are inconsistent across sectors. Conversely, Table \ref{table:average_ext_3} demonstrates that the Market Nonlinear model produces a higher mean return 
than the Global Linear and Global Nonlinear models that is statistically significant at the $1\%$ level, and significant at the 
$10\%$ level in comparison to the Sector Nonlinear models. These results provide evidence that the unique market response characteristics, 
observed when feature values are at extreme levels $(|\toi{\bx}| \geq 3)$, are not purely an artifact of random sampling error. 
From a practitioner's perspective, however, we should interpret these results with some reservation. Importantly, for all models, 
we observe a net negative portfolio return when feature values are at extreme levels that is generally consistent across all sectors. 
Therefore, from a practical standpoint there is evidence to suggest that, in expectation, it would be more favourable to 
completely exit the market position when the time series momentum values rise to extreme levels.

\begin{table}[H]
\centering
\small

\begin{tabular}{lllll}
\toprule
 & Global Linear & Global Nonlinear & Sector Nonlinear & Market Nonlinear \\
\midrule
Bond & -0.005801 (1.6733)  & -0.004403 (1.4709)  & -0.002216 (1.7925)  & 0.0021 (0.0)  \\
Currency & 0.005317 (-0.6104)  & 0.005721 (-0.92)  & 0.004473 (-0.5044)  & 0.003433 (0.0)  \\
Energy & 0.005988 (0.3261)  & 0.006609 (0.1235)  & 0.00567 (1.0829)  & 0.006856 (0.0)  \\
Equity & -0.002702 (0.1405)  & -0.003124 (0.2398)  & -0.002325 (0.108)  & -0.002039 (0.0)  \\
Grain & 0.004380 (-0.2133)  & 0.004465 (-0.2483)  & 0.004471 (-0.2187)  & 0.003456 (0.0)  \\
Metal & 0.012087 (-0.8134)  & 0.011653 (-0.8843)  & 0.008563 (-0.3112)  & 0.007592 (0.0)  \\
Soft & 0.010439 (0.606)  & 0.009475 (0.7883)  & 0.015235 (-0.3009)  & 0.013503 (0.0)  \\
\midrule
Portfolio & 0.003580 (0.6105)  & 0.003710 (0.5729)  & 0.004374 (0.165)  & 0.004585 (0.0)  \\
\bottomrule
\end{tabular}

\caption{Annualized market mean returns, net of transaction costs, and reported at the individual sector and total portfolio level,
formed using a simple average of the $1$, $6$ and $12$-month time series momentum results and evaluated for all observations outside the normal range $|\toi{\bx}| \geq 2$.
Values in parentheses denote the t-statistic from a paired t-test of the null hypothesis that the average market return 
of the stated model is equal to the average market return produced by the Market Nonlinear model. Statistical significance
at the $0.10$, $0.05$ and $0.01$ level is denoted by $*$, $**$ and $***$, respectively. 
The analysis period is January $1995$ through June $2026$. }
\label{table:average_ext_2}
\end{table}

\begin{table}[H]
\centering
\small

\begin{tabular}{lllll}
\toprule
 & Global Linear & Global Nonlinear & Sector Nonlinear & Market Nonlinear \\
\midrule
Bond & -0.097848 (1.802)  & -0.076609 (1.6412)  & -0.016211 (1.3119)  & 0.045998 (0.0)  \\
Currency & -0.169342 (1.3222)  & -0.157799 (1.3237)  & -0.076206 (-0.7331)  & -0.087565 (0.0)  \\
Energy & -0.108877 (1.854)  & -0.083548 (1.4831)  & -0.050413 (0.8084)  & -0.028639 (0.0)  \\
Equity & -0.070390 (1.1868)  & -0.052659 (1.1321)  & -0.020043 (0.4898)  & -0.015157 (0.0)  \\
Grain & -0.018701 (0.9167)  & -0.018596 (1.1518)  & -0.017806 (0.8176)  & -0.006307 (0.0)  \\
Metal & -0.082016 (1.071)  & -0.066332 (1.1982)  & -0.026293 (1.1765)  & -0.002886 (0.0)  \\
Soft & -0.055297* (2.3163)  & -0.040832 (1.1254)  & -0.032737 (0.4349)  & -0.026286 (0.0)  \\
\midrule
Portfolio & -0.086013*** (3.4612)  & -0.071282*** (3.1611)  & -0.034104* (1.7523)  & -0.01749 (0.0)  \\
\bottomrule
\end{tabular}

\caption{Annualized market mean returns, net of transaction costs, and reported at the individual sector and total portfolio level,
formed using a simple average of the $1$, $6$ and $12$-month time series momentum results and evaluated for all observations where feature values are at extreme levels $|\toi{\bx}| \geq 3$.
Values in parentheses denote the t-statistic from a paired t-test of the null hypothesis that the average market return 
of the stated model is equal to the average market return produced by the Market Nonlinear model. Statistical significance
at the $0.10$, $0.05$ and $0.01$ level is denoted by $*$, $**$ and $***$, respectively. 
The analysis period is January $1995$ through June $2026$. }
\label{table:average_ext_3}
\end{table}

Finally, Table \ref{table:average} reports the annualized market mean returns and paired t-statistic for the 
aggregate portfolios using a simple average of the $1$, $6$ and $12$-month time series momentum features. 
In contrast to previous results, here we observe that, for bond and soft sectors, the Market Nonlinear model produces a higher mean return 
than the Global Linear model that is statistically significant at the $10\%$ level. This result, however, is not consistent
across all other sectors; notably in currency, energy and equity sectors the Market Nonlinear model produces marginally smaller mean returns
in comparison to all other models. Moreover, we observe a statistically insignificant difference in market mean returns when 
evaluated across all markets in the portfolio. Therefore, when  aggregated 
across multiple features and forecast horizons, a time series momentum portfolio formed by fitting nonlinear models at the individual market level
produces average market returns that are statistically equivalent to those produced by Global Linear, Global Nonlinear and Sector Nonlinear models.  
Taken together, it is difficult to state with statistical confidence that the  unique response characteristics displayed by individual feature-market 
pairs are genuine signal, above sampling error. However, we are encouraged by the robustness of the proposed modelling approach;
fitting models at the individual market level is as valid as the grouped factor approach and therefore 
enables data-driven factor exploration in settings that may require a more fine-grained individual market perspective.

\begin{table}[H]
\centering
\small

\begin{tabular}{lllll}
\toprule
 & Global Linear & Global Nonlinear & Sector Nonlinear & Market Nonlinear \\
\midrule
Bond & 0.003185* (2.3106)  & 0.003511 (1.87)  & 0.00374 (1.5091)  & 0.004774 (0.0)  \\
Currency & 0.00222 (-0.0875)  & 0.002459 (-0.4865)  & 0.002658 (-0.9568)  & 0.002169 (0.0)  \\
Energy & 0.004789 (-0.8858)  & 0.004767 (-0.7789)  & 0.004818 (-0.8475)  & 0.004337 (0.0)  \\
Equity & 0.001651 (-0.9948)  & 0.00198 (-1.6475)  & 0.001181 (-0.1498)  & 0.001137 (0.0)  \\
Grain & 0.000943 (0.2113)  & 0.000926 (0.2364)  & 0.001014 (0.109)  & 0.001094 (0.0)  \\
Metal & 0.001136 (0.229)  & 0.001369 (-0.7076)  & 0.000697 (1.1373)  & 0.00123 (0.0)  \\
Soft & -0.00027* (2.1819)  & -0.000101 (1.7939)  & 0.000466 (-0.2268)  & 0.000401 (0.0)  \\
\midrule
Portfolio & 0.00193 (1.0332)  & 0.002115 (0.2327)  & 0.002088 (0.3794)  & 0.002167 (0.0)  \\
\bottomrule
\end{tabular}

\caption{Annualized market mean returns, net of transaction costs, and reported at the individual sector and total portfolio level,
formed using a simple average of the $1$, $6$ and $12$-month time series momentum model results.
Values in parentheses denote the t-statistic from a paired t-test of the null hypothesis that the average market return 
of the stated model is equal to the average market return produced by the Market Nonlinear model. Statistical significance
at the $0.10$, $0.05$ and $0.01$ level is denoted by $*$, $**$ and $***$, respectively. 
The analysis period is January $1995$ through June $2026$. }
\label{table:average}
\end{table}


\begin{figure}[h]
  \centering
  \includegraphics[width=\linewidth, height=0.40\linewidth ]{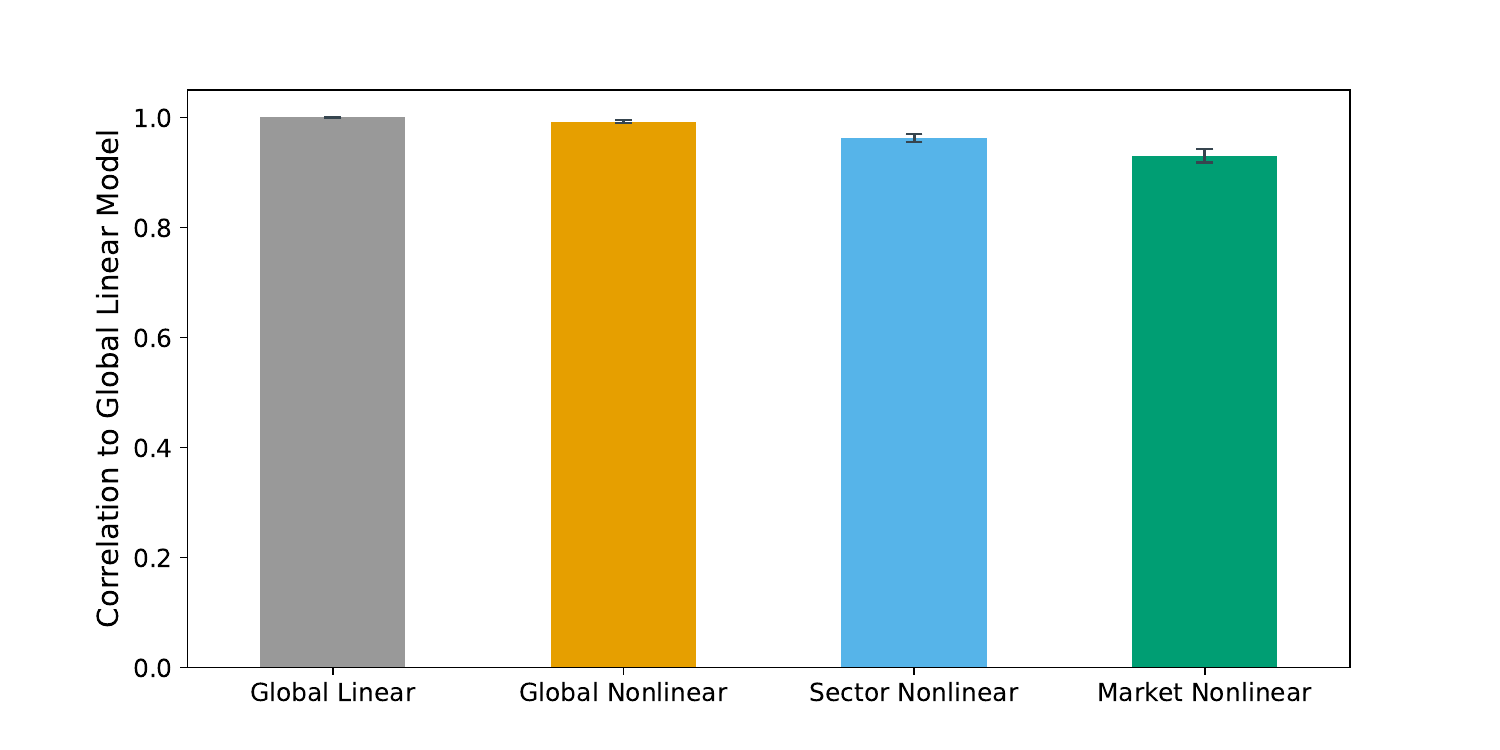}
  
  \caption{Average out-of-sample correlation to the Global Linear model, evaluated across $1$, $6$ and $12$-month time series momentum features.
   The out-of-sample period is January $1995$ through June $2026$.}
  \label{fig:correlation}
\end{figure}

\newpage
\section{Conclusion}
We extended the work of \citet{MoskowitzNonlinear} and demonstrated that models fit at the individual market level exhibit idiosyncratic forward return characteristics
that are distinct from the global linear and nonlinear time series momentum factor models. Nonlinear models were fit using non-parametric kernel regression
and provide a framework that enables each feature-market pair to express its potentially unique prediction behaviour. A comprehensive catalogue of model response curves
is provided in \ref{sec:appendix} and helps to illustrate that, even within sectors, individual markets can display heterogeneous forward return behaviour.
We provided two primary empirical findings. First, financial markets, such as bonds, currencies and equities, tend to display stronger mean-reversion characteristics 
when the time series momentum factor is at extreme levels. In contrast, commodity markets, such as energies, metals, grains and softs, typically exhibit more 
traditional linear time series momentum effects, on average and across the entire feature range. Second, for commodity markets, we also observed differences in
the prediction strength and response behaviour at the individual market level. We hypothesize that the tendency for markets to display 
sustained trends or mean reversion at the extremes may be governed by both economic feedback mechanisms and the participants specific to each market.

Empirical results demonstrated that, despite substantially smaller sample sizes, fitting nonlinear models to individual markets is a robust alternative to the
traditional global modelling approach that preserves the characteristics of the time series momentum strategy with a high degree of correlation.
Historical walk-forward simulations, spanning over $30$ years of market data, indicated that, despite increased complexity, 
models fit at the individual market level generalized to the out-of-sample period. Portfolio-level results exhibited positive out-of-sample net Sharpe ratios
and, on average, demonstrated that fitting models at the individual market level has the potential to provide differentiated and improved out-of-sample performance
at the margin. In general, all nonlinear models were found to be strongly positively correlated with the Global Linear factor model, 
indicating that they build upon traditional factor intuitions, while also allowing for the expression of idiosyncratic nonlinear response characteristics.

Statistical tests of paired differences in mean returns across markets provided mixed results. When evaluated over the entire feature range, 
we demonstrated that models fit at the individual market level produce average market returns that are statistically equivalent to those 
generated by the grouped linear and nonlinear models. We observed a statistically significant increase in market mean returns when out-of-sample
results were isolated to periods of extreme feature values. However, under such circumstances, all models displayed negative expected portfolio 
returns; therefore, a higher-expected-value decision would be to completely exit the market position when feature values are at extreme levels.
In totality, we are unable to claim with statistical confidence that the unique response functions displayed by individual markets
represent genuine signal and therefore recommend further testing and analysis. 

This paper serves as an initial proof-of-concept showcasing the potential of machine learning techniques for advancing factor-based investment modelling
at the individual market level. There are many promising future extensions and areas for exploration. Indeed, fitting models at the individual market
level is generally robust and enables factor discovery that requires market-level analysis, such as  
seasonality \citep{Sorensen2002, Merc2023} or Commitments of Traders (COT) data \citep{Boos2023}.
Machine learning techniques may also be well suited to discovering conditional multivariate relationships that exist globally or at 
the individual market level. Recent advances in deep learning and causal inference also 
offer avenues for uncovering subtle factor relationships amidst market noise \citep{Anis2021, Fischer2018, Wang2018}. The results 
suggest that machine learning can empirically derive factor relationships directly from market data, with the potential for flexibility
and specificity exceeding that of the traditional factor modelling approach. This framework establishes a starting point for the continued exploration of 
machine learning's benefits for data-driven factor discovery in market forecasting.

\appendix
\setcounter{table}{0}
\renewcommand{\thetable}{A\arabic{table}}
\setcounter{figure}{0}                       
\renewcommand\thefigure{A\arabic{figure}}

\section{Time Series Momentum Models} \label{sec:appendix}
\subsection{Global Linear Models}
\begin{figure}[H]
  \centering
  \begin{subfigure}[b]{0.30\linewidth}
    \includegraphics[width=\linewidth ]{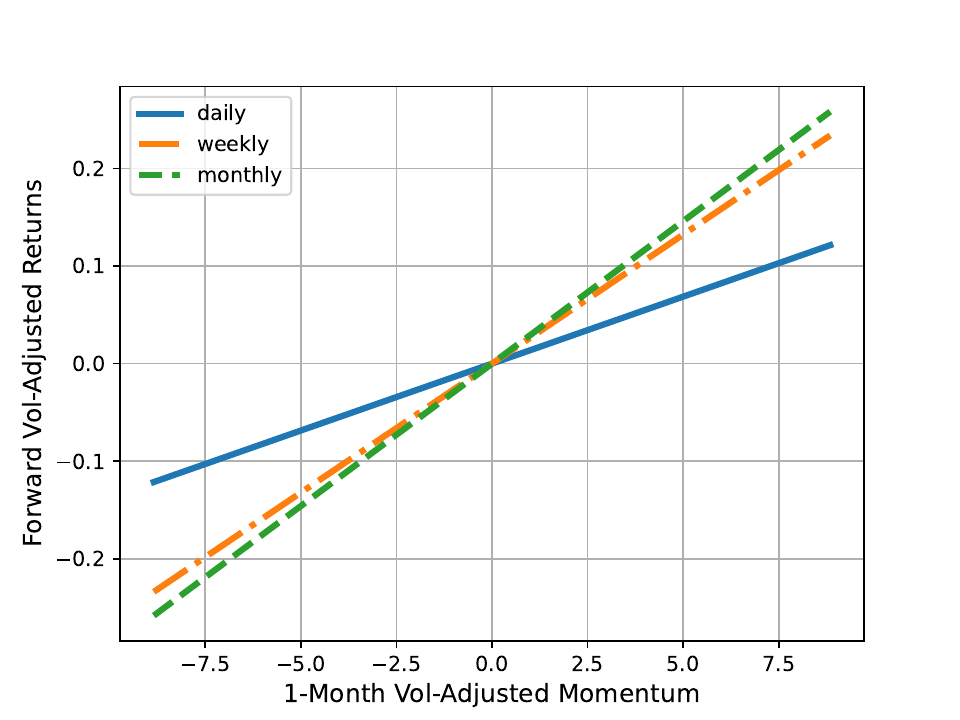}
    \caption{$1$-Month Momentum}
  \end{subfigure}
  \begin{subfigure}[b]{0.30\linewidth}
    \includegraphics[width=\linewidth ]{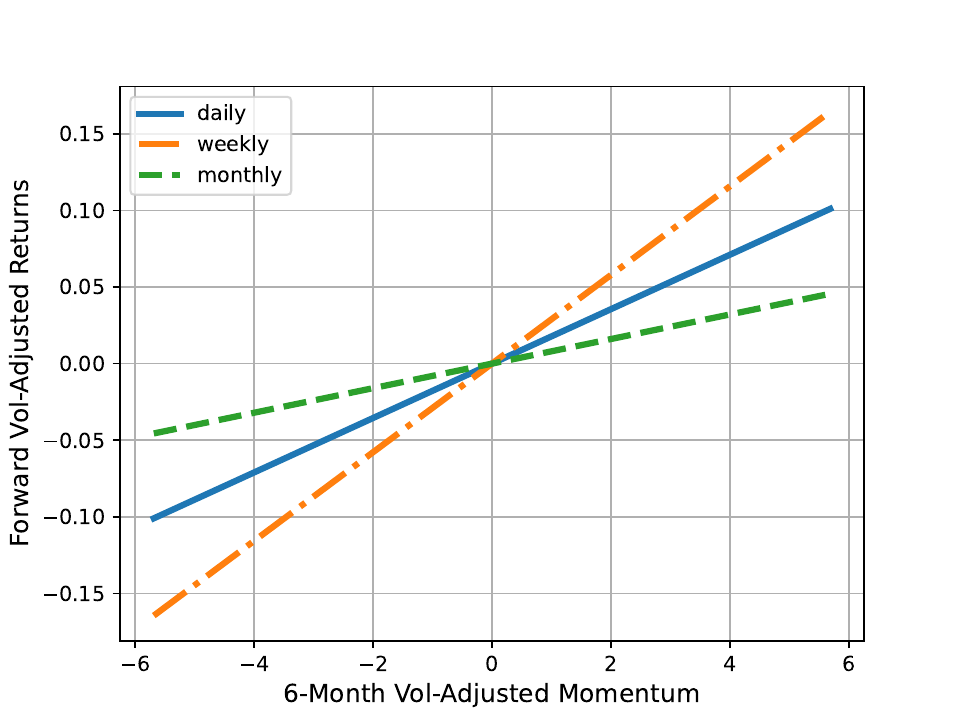}
    \caption{$6$-Month Momentum}
  \end{subfigure}
  \begin{subfigure}[b]{0.30\linewidth}
    \includegraphics[width=\linewidth ]{figs/model__linear_momo_12_month.pdf}
    \caption{$12$-Month Momentum}
  \end{subfigure}
  
  \caption{Estimated forward vol-adjusted returns as a function of the $1$, $6$ and $12$-month time series momentum factor for Global Linear models.
  Models are fit using all available training data from January $1950$ through June $2026$. }
\end{figure}

\subsection{Global Nonlinear Models}
\begin{figure}[H]
  \centering
  \begin{subfigure}[b]{0.30\linewidth}
    \includegraphics[width=\linewidth ]{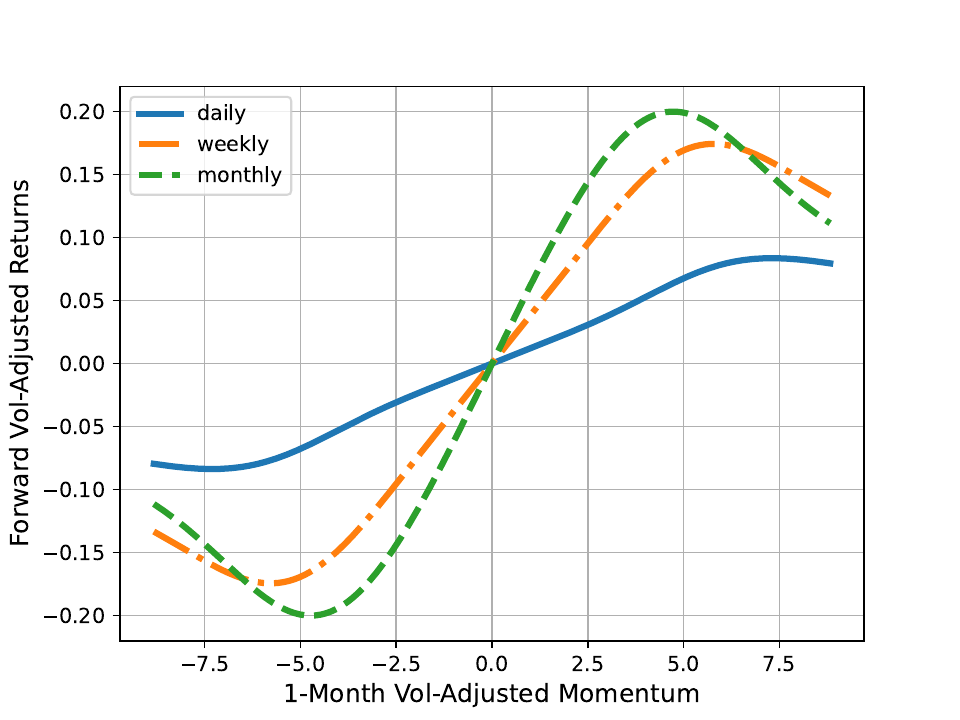}
    \caption{$1$-Month Momentum}
  \end{subfigure}
  \begin{subfigure}[b]{0.30\linewidth}
    \includegraphics[width=\linewidth ]{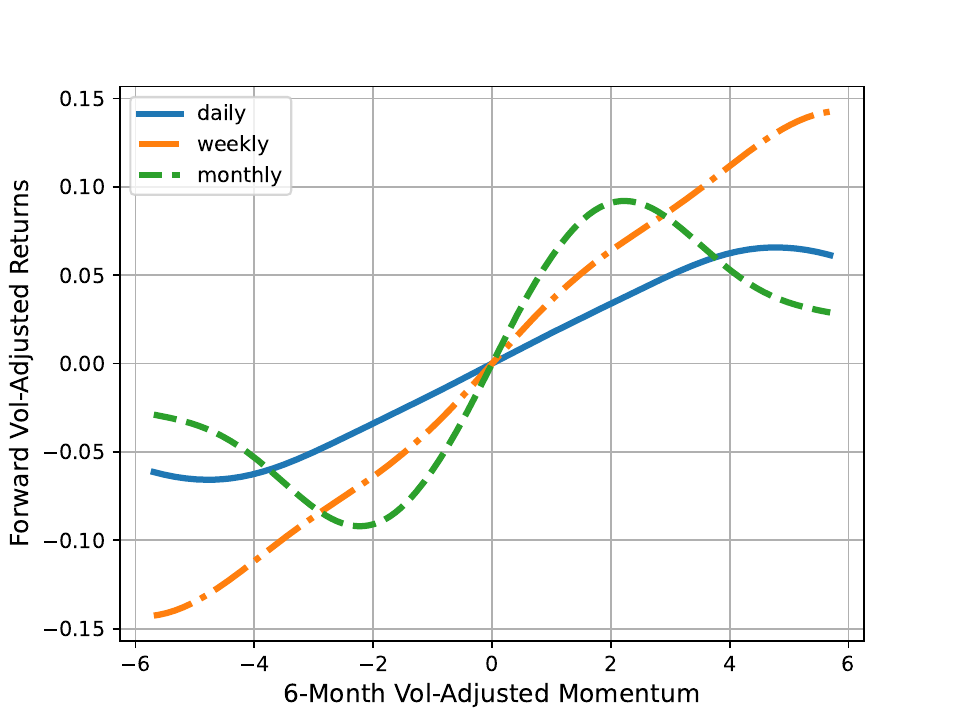}
    \caption{$6$-Month Momentum}
  \end{subfigure}
  \begin{subfigure}[b]{0.30\linewidth}
    \includegraphics[width=\linewidth ]{figs/model__nonlinear_momo_12_month.pdf}
    \caption{$12$-Month Momentum}
  \end{subfigure}
  
  \caption{Estimated forward vol-adjusted returns as a function of the $1$, $6$ and $12$-month time series momentum factor for Global Nonlinear models.
  Models are fit using all available training data from January $1950$ through June $2026$. }
\end{figure}

\subsection{Bond Sector Nonlinear Models}
\begin{figure}[H]
  \centering
  \begin{subfigure}[b]{0.30\linewidth}
    \includegraphics[width=\linewidth ]{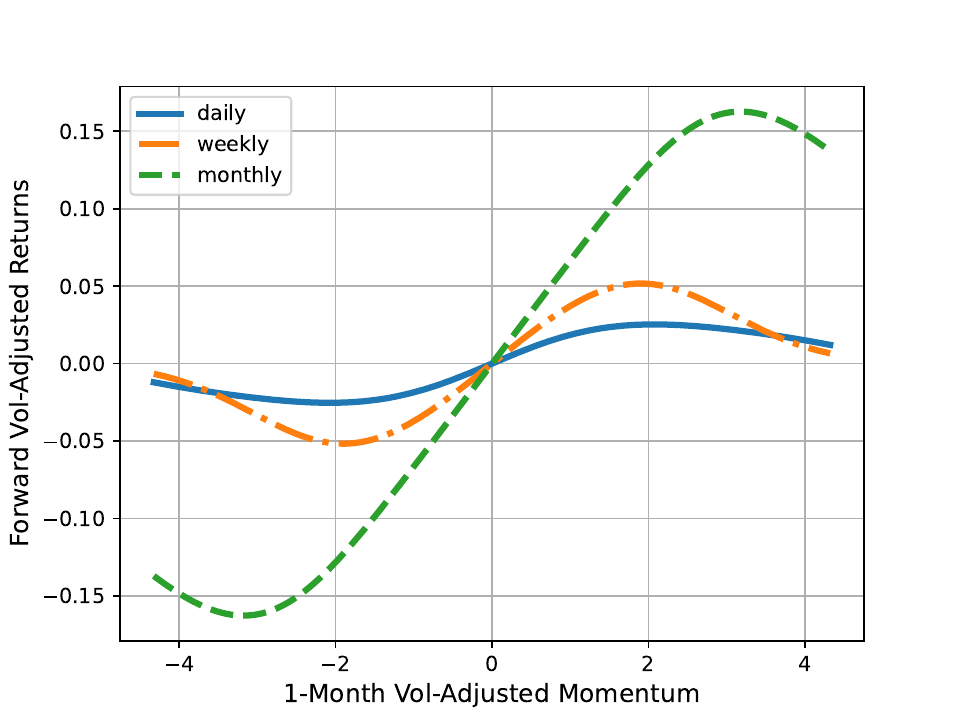}
    \caption{$1$-Month Momentum}
  \end{subfigure}
  \begin{subfigure}[b]{0.30\linewidth}
    \includegraphics[width=\linewidth ]{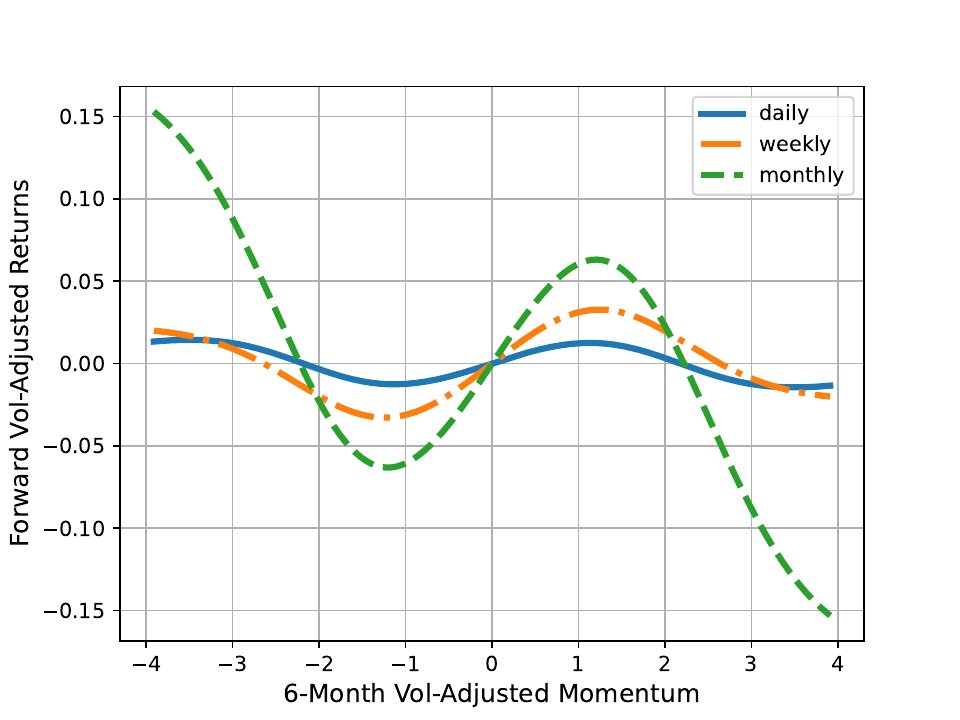}
    \caption{$6$-Month Momentum}
  \end{subfigure}
  \begin{subfigure}[b]{0.30\linewidth}
    \includegraphics[width=\linewidth ]{figs/model_bond_momo_12_month.pdf}
    \caption{$12$-Month Momentum}
  \end{subfigure}
  
  \caption{Estimated forward vol-adjusted returns as a function of the $1$, $6$ and $12$-month time series momentum factor for Bond Sector Nonlinear models.
  Models are fit using all available training data from January $1950$ through June $2026$. }
\end{figure}

\noindent \textbf{Canadian 10 Year Bond}
\begin{figure}[H]
  \centering
  \begin{subfigure}[b]{0.30\linewidth}
    \includegraphics[width=\linewidth ]{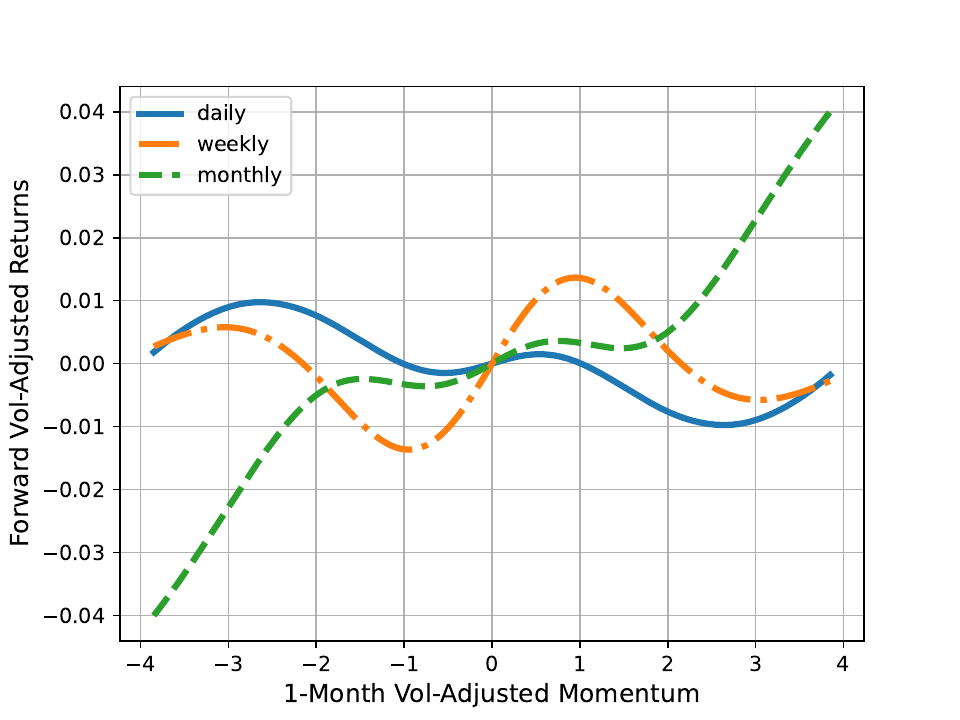}
    \caption{$1$-Month Momentum}
  \end{subfigure}
  \begin{subfigure}[b]{0.30\linewidth}
    \includegraphics[width=\linewidth ]{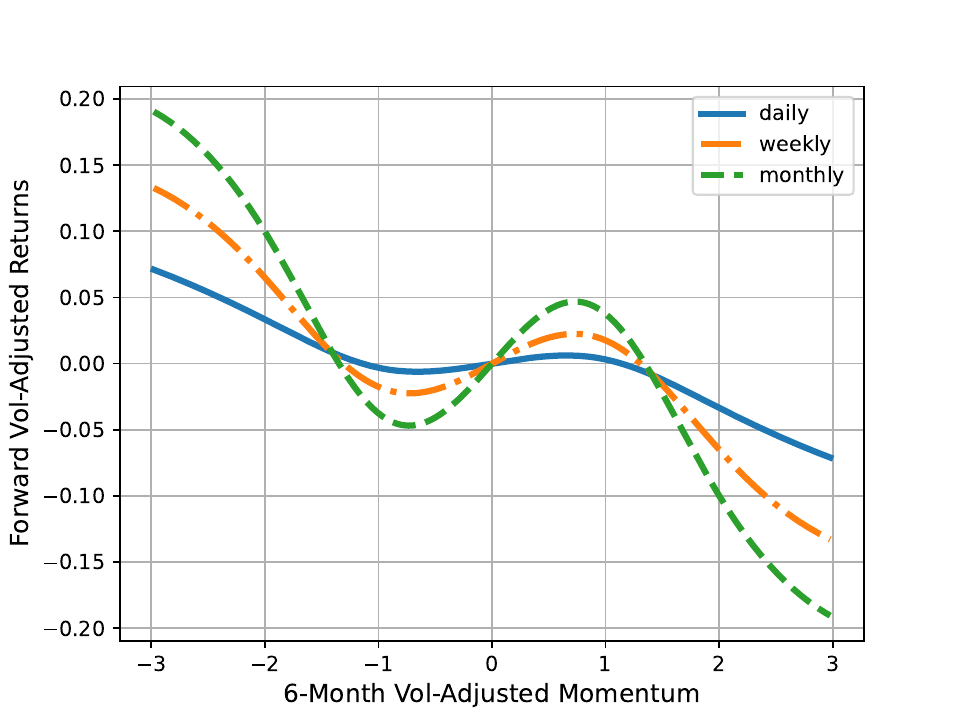}
    \caption{$6$-Month Momentum}
  \end{subfigure}
  \begin{subfigure}[b]{0.30\linewidth}
    \includegraphics[width=\linewidth ]{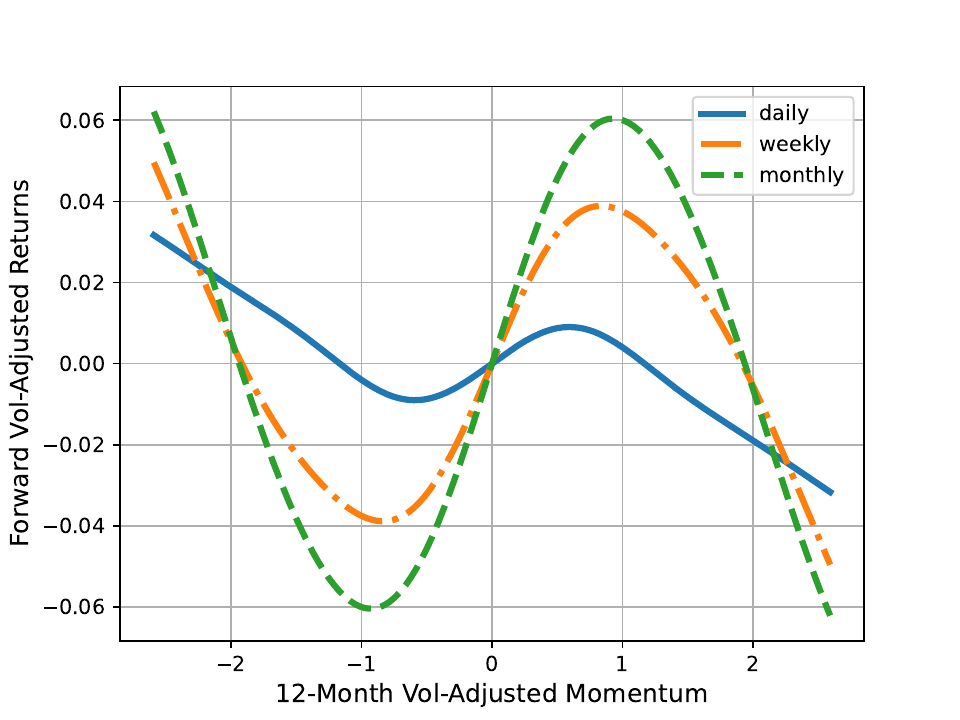}
    \caption{$12$-Month Momentum}
  \end{subfigure}
  
  \caption{Estimated forward vol-adjusted returns as a function of the $1$, $6$ and $12$-month time series momentum factor for Canadian 10 Year Bond Nonlinear models.
  Models are fit using all available training data from January $1950$ through June $2026$. }
\end{figure}

\noindent \textbf{Gilts}
\begin{figure}[H]
  \centering
  \begin{subfigure}[b]{0.30\linewidth}
    \includegraphics[width=\linewidth ]{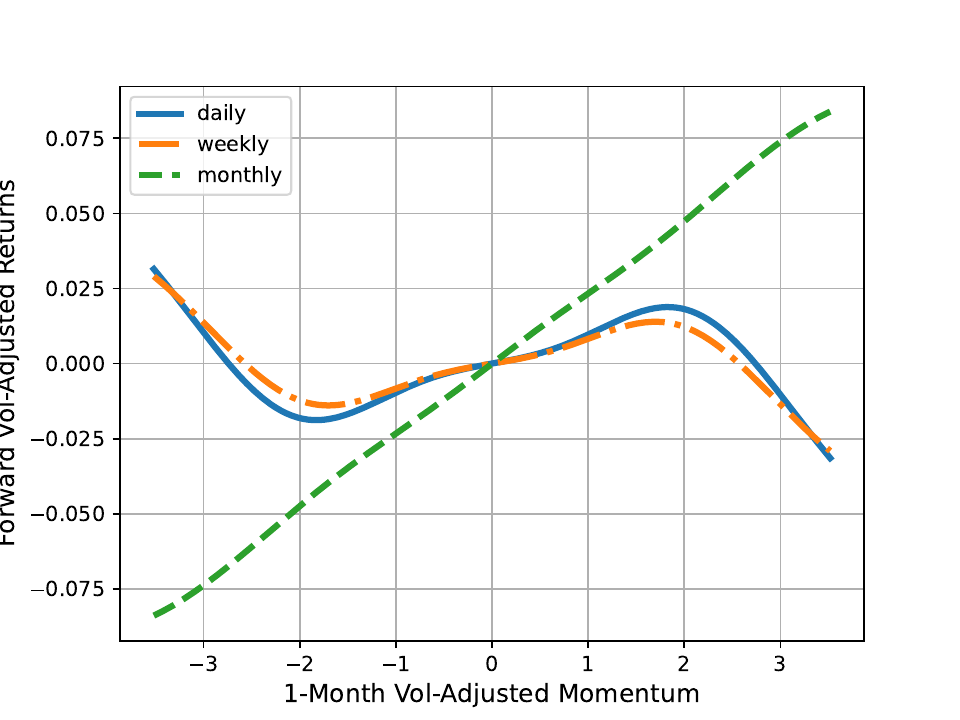}
    \caption{$1$-Month Momentum}
  \end{subfigure}
  \begin{subfigure}[b]{0.30\linewidth}
    \includegraphics[width=\linewidth ]{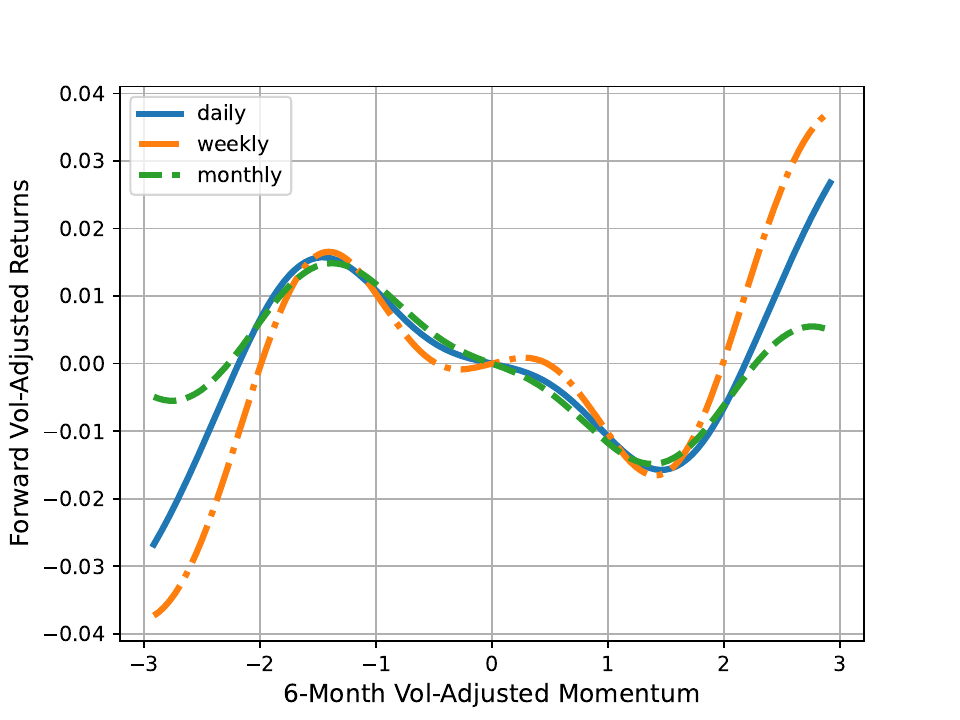}
    \caption{$6$-Month Momentum}
  \end{subfigure}
  \begin{subfigure}[b]{0.30\linewidth}
    \includegraphics[width=\linewidth ]{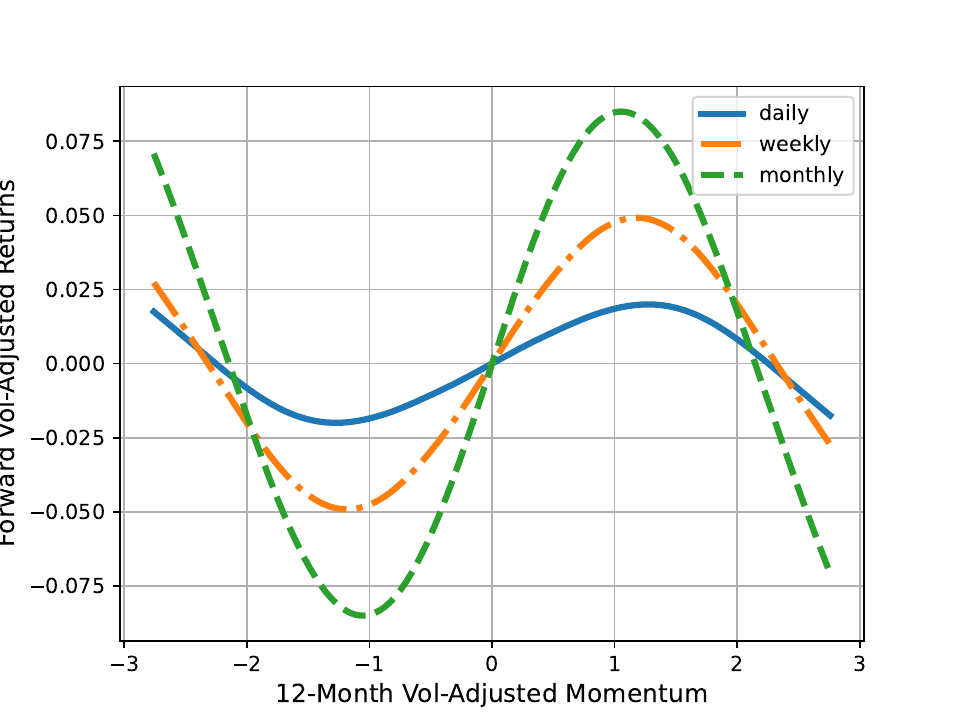}
    \caption{$12$-Month Momentum}
  \end{subfigure}
  
  \caption{Estimated forward vol-adjusted returns as a function of the $1$, $6$ and $12$-month time series momentum factor for Gilts Nonlinear models.
  Models are fit using all available training data from January $1950$ through June $2026$. }
\end{figure}

\noindent \textbf{Japan 10 Year Bond}
\begin{figure}[H]
  \centering
  \begin{subfigure}[b]{0.30\linewidth}
    \includegraphics[width=\linewidth ]{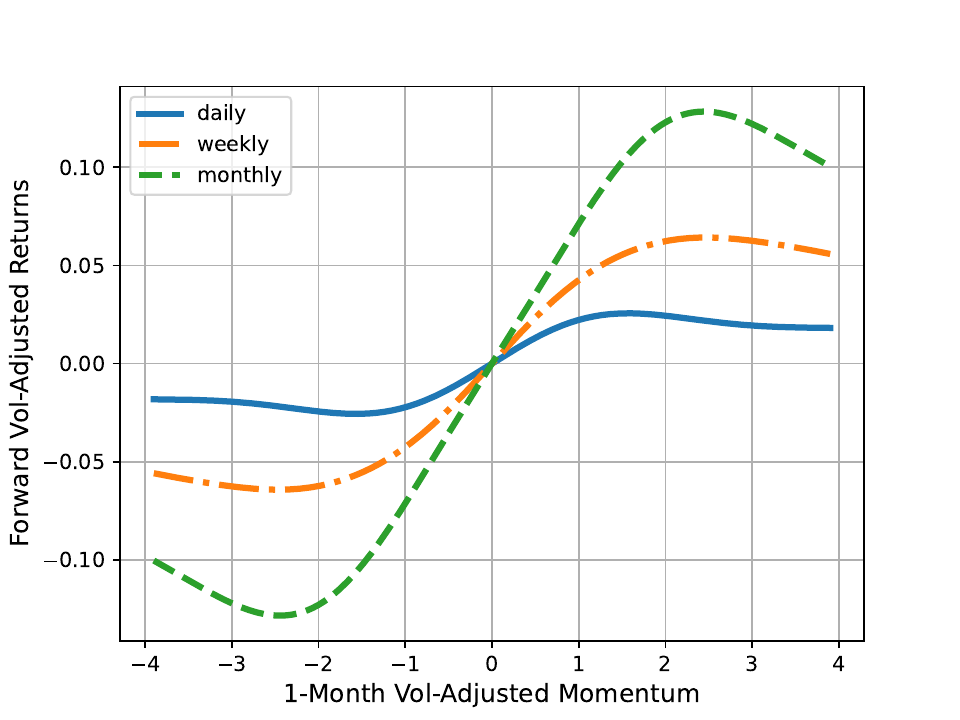}
    \caption{$1$-Month Momentum}
  \end{subfigure}
  \begin{subfigure}[b]{0.30\linewidth}
    \includegraphics[width=\linewidth ]{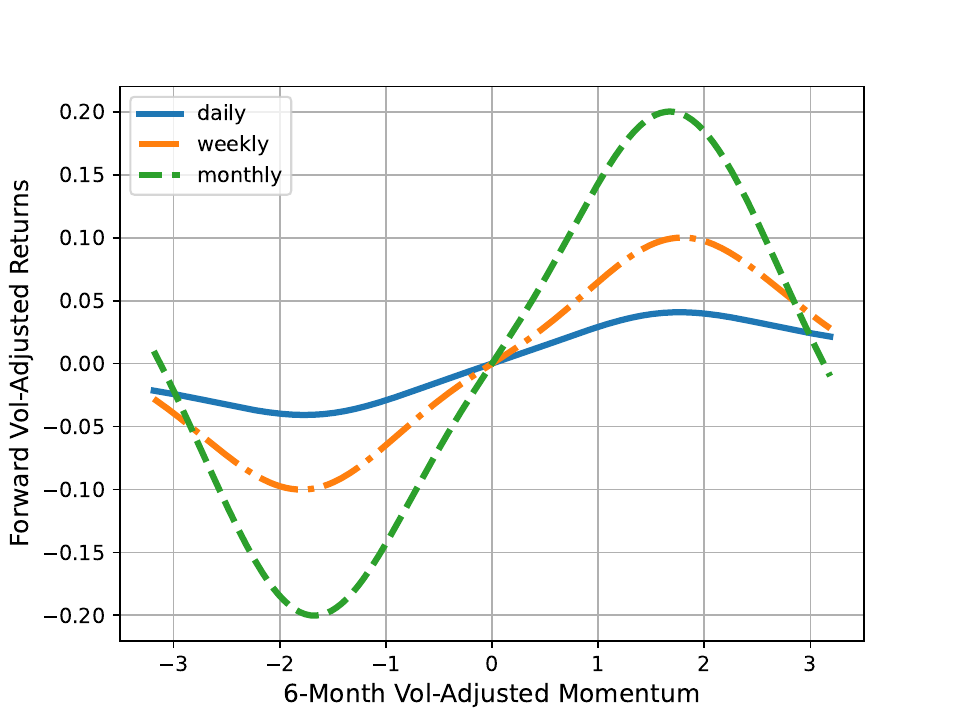}
    \caption{$6$-Month Momentum}
  \end{subfigure}
  \begin{subfigure}[b]{0.30\linewidth}
    \includegraphics[width=\linewidth ]{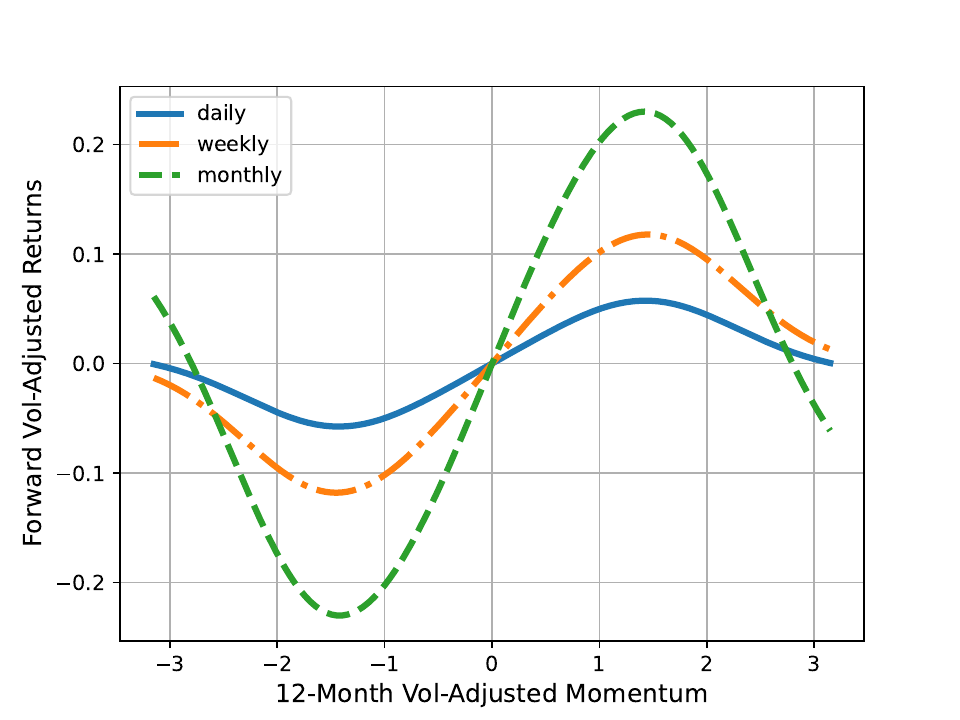}
    \caption{$12$-Month Momentum}
  \end{subfigure}
  
  \caption{Estimated forward vol-adjusted returns as a function of the $1$, $6$ and $12$-month time series momentum factor for Japan 10 Year Nonlinear models.
  Models are fit using all available training data from January $1950$ through June $2026$. }
\end{figure}

\noindent \textbf{Bobl}
\begin{figure}[H]
  \centering
  \begin{subfigure}[b]{0.30\linewidth}
    \includegraphics[width=\linewidth ]{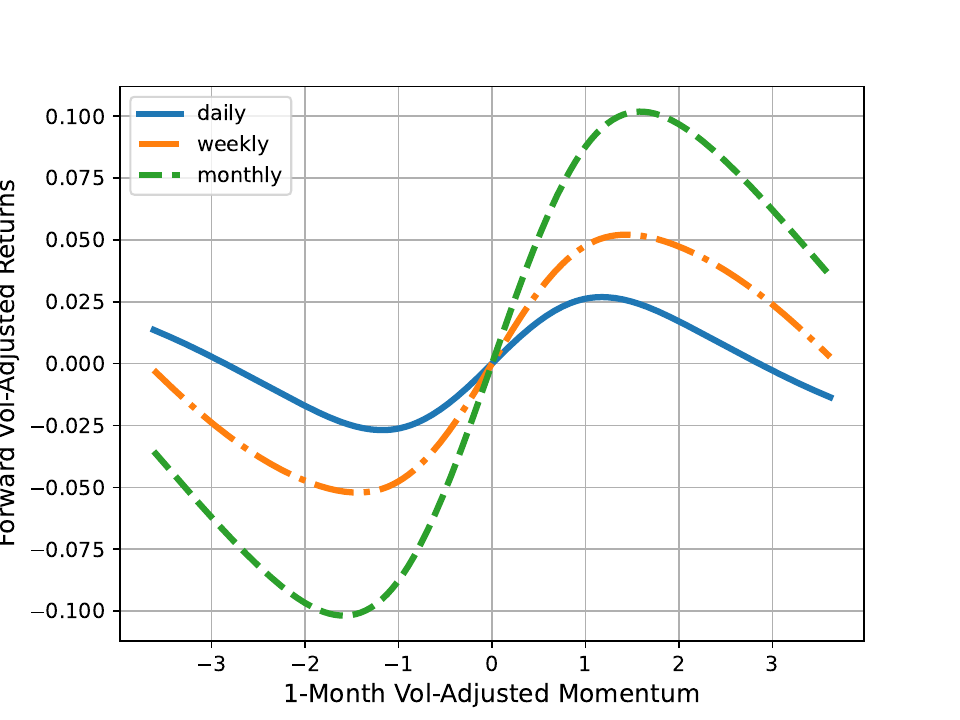}
    \caption{$1$-Month Momentum}
  \end{subfigure}
  \begin{subfigure}[b]{0.30\linewidth}
    \includegraphics[width=\linewidth ]{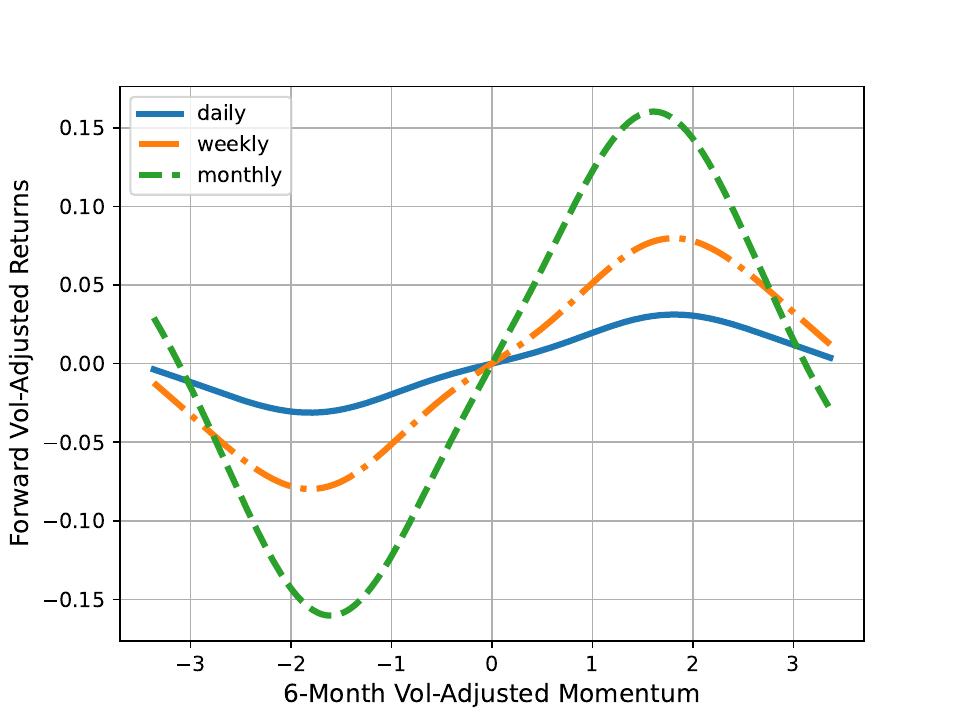}
    \caption{$6$-Month Momentum}
  \end{subfigure}
  \begin{subfigure}[b]{0.30\linewidth}
    \includegraphics[width=\linewidth ]{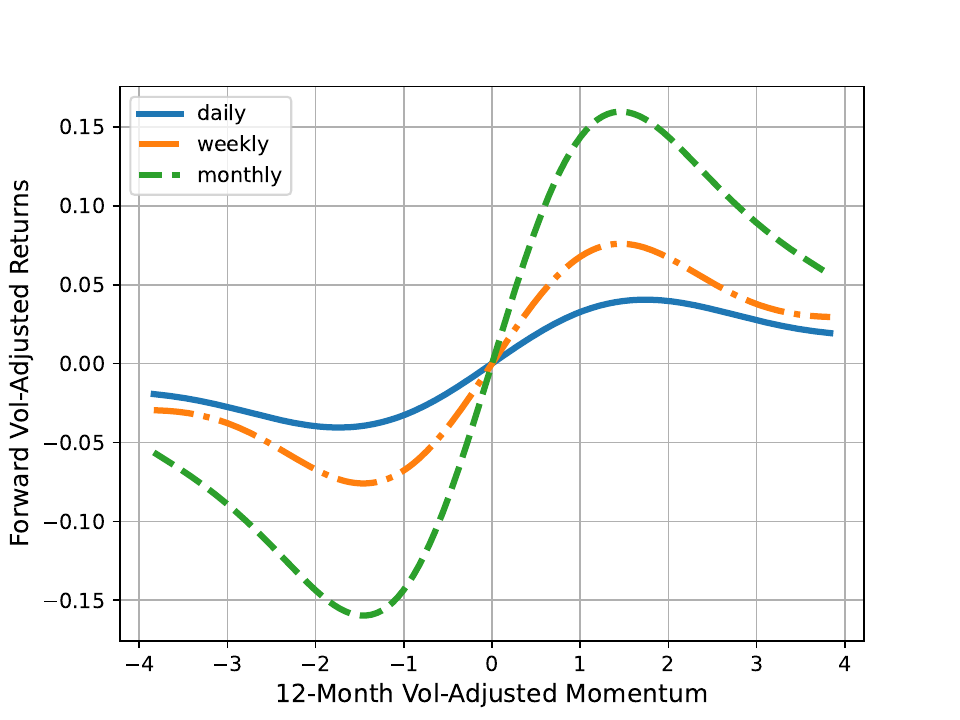}
    \caption{$12$-Month Momentum}
  \end{subfigure}
  
  \caption{Estimated forward vol-adjusted returns as a function of the $1$, $6$ and $12$-month time series momentum factor for Bobl Nonlinear models.
  Models are fit using all available training data from January $1950$ through June $2026$. }
\end{figure}

\noindent \textbf{Bund}
\begin{figure}[H]
  \centering
  \begin{subfigure}[b]{0.30\linewidth}
    \includegraphics[width=\linewidth ]{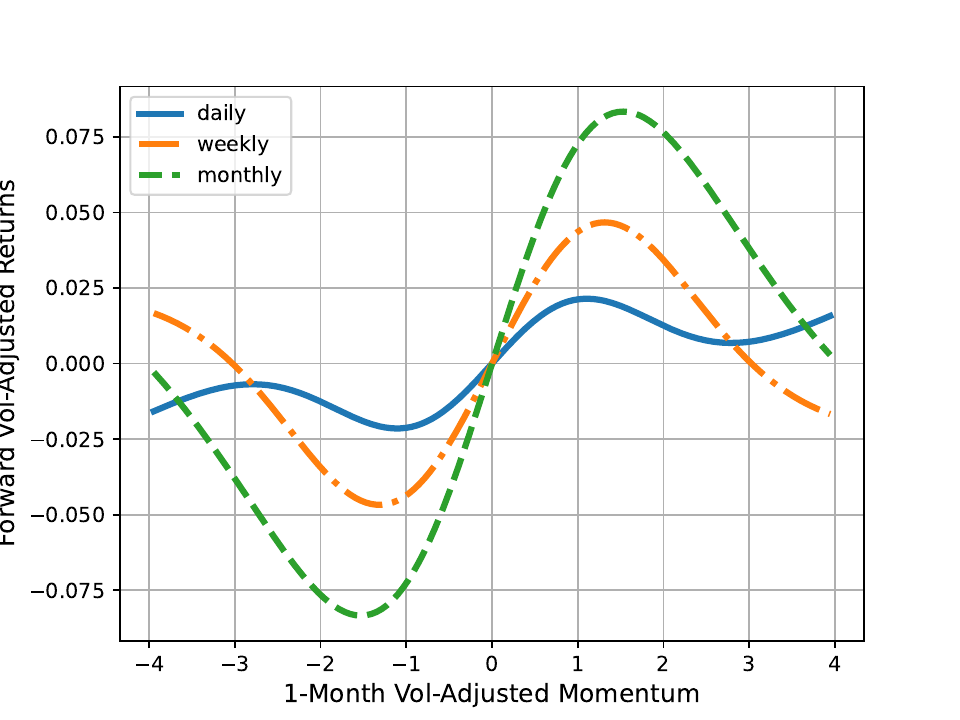}
    \caption{$1$-Month Momentum}
  \end{subfigure}
  \begin{subfigure}[b]{0.30\linewidth}
    \includegraphics[width=\linewidth ]{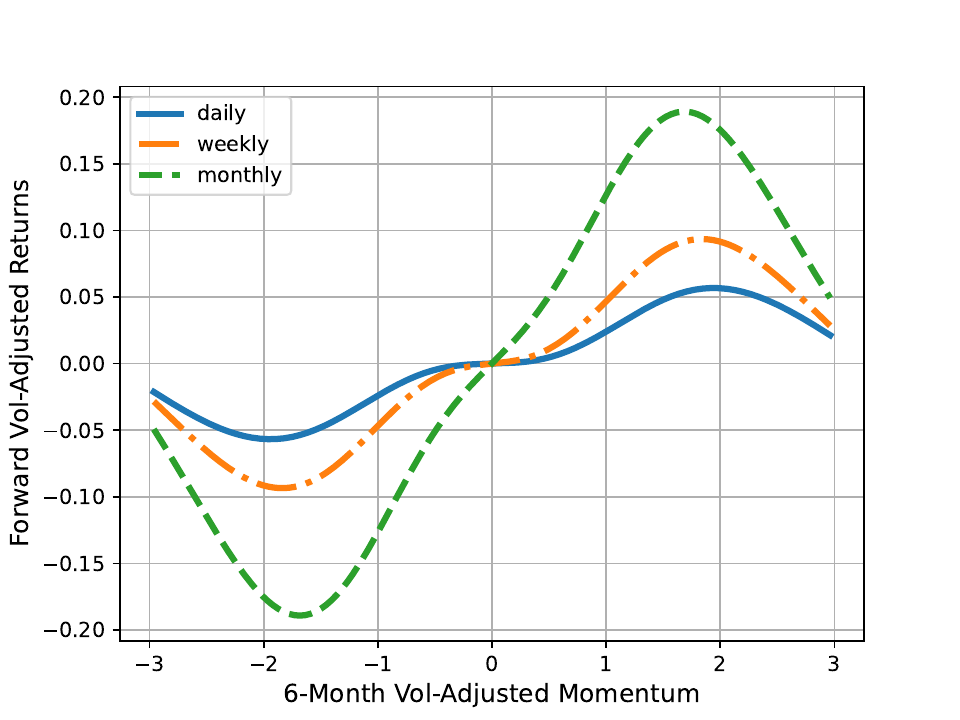}
    \caption{$6$-Month Momentum}
  \end{subfigure}
  \begin{subfigure}[b]{0.30\linewidth}
    \includegraphics[width=\linewidth ]{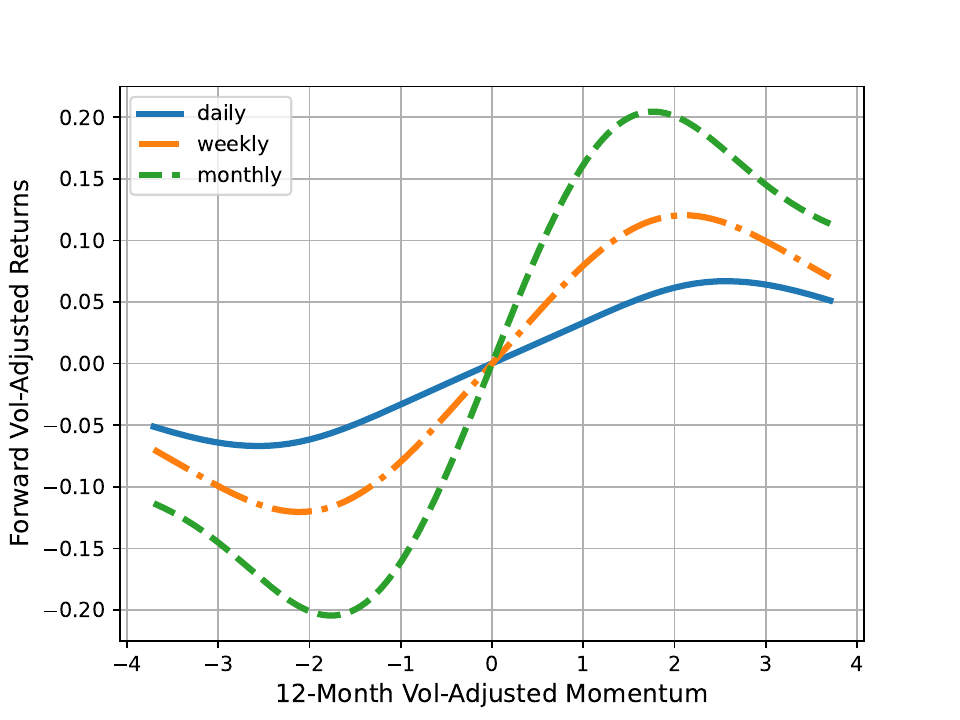}
    \caption{$12$-Month Momentum}
  \end{subfigure}
  
  \caption{Estimated forward vol-adjusted returns as a function of the $1$, $6$ and $12$-month time series momentum factor for Bund Nonlinear models.
  Models are fit using all available training data from January $1950$ through June $2026$. }
\end{figure}

\noindent \textbf{US 5 Year Bond}
\begin{figure}[H]
  \centering
  \begin{subfigure}[b]{0.30\linewidth}
    \includegraphics[width=\linewidth ]{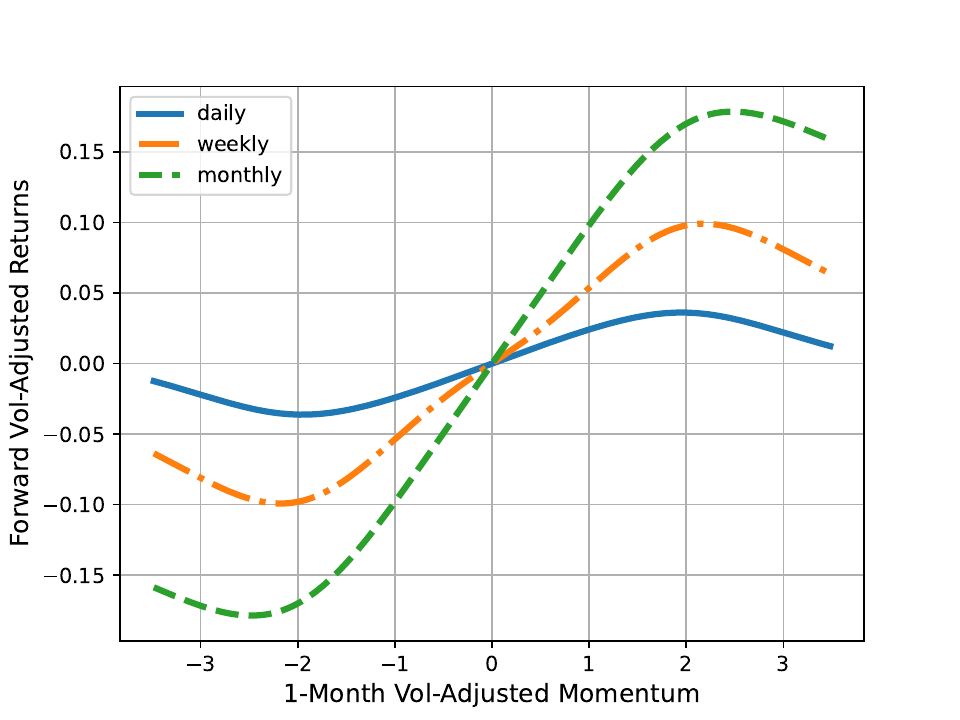}
    \caption{$1$-Month Momentum}
  \end{subfigure}
  \begin{subfigure}[b]{0.30\linewidth}
    \includegraphics[width=\linewidth ]{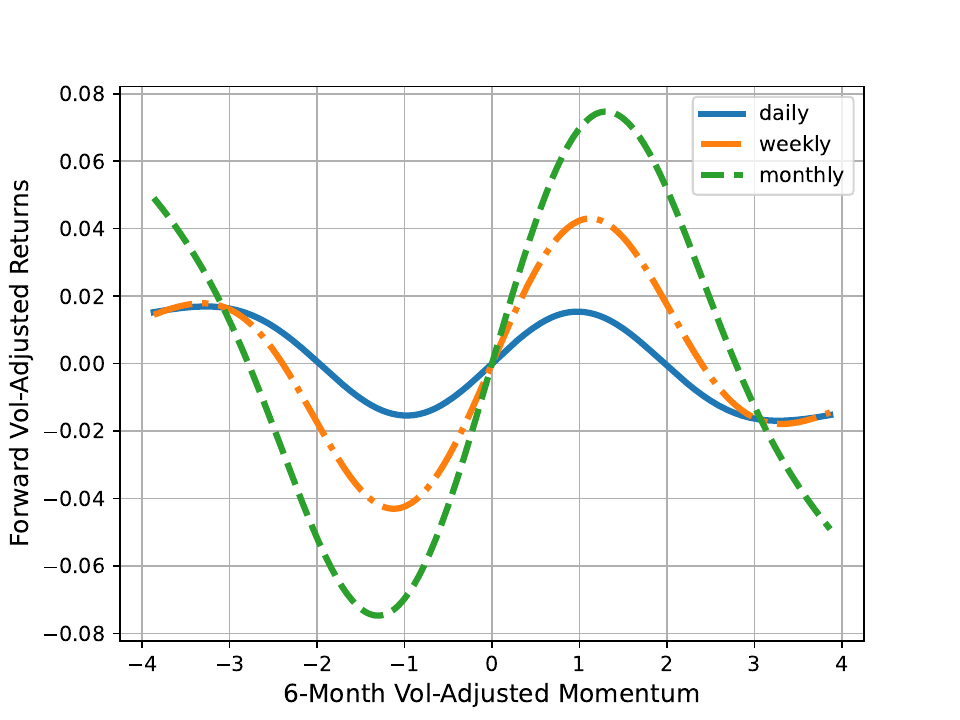}
    \caption{$6$-Month Momentum}
  \end{subfigure}
  \begin{subfigure}[b]{0.30\linewidth}
    \includegraphics[width=\linewidth ]{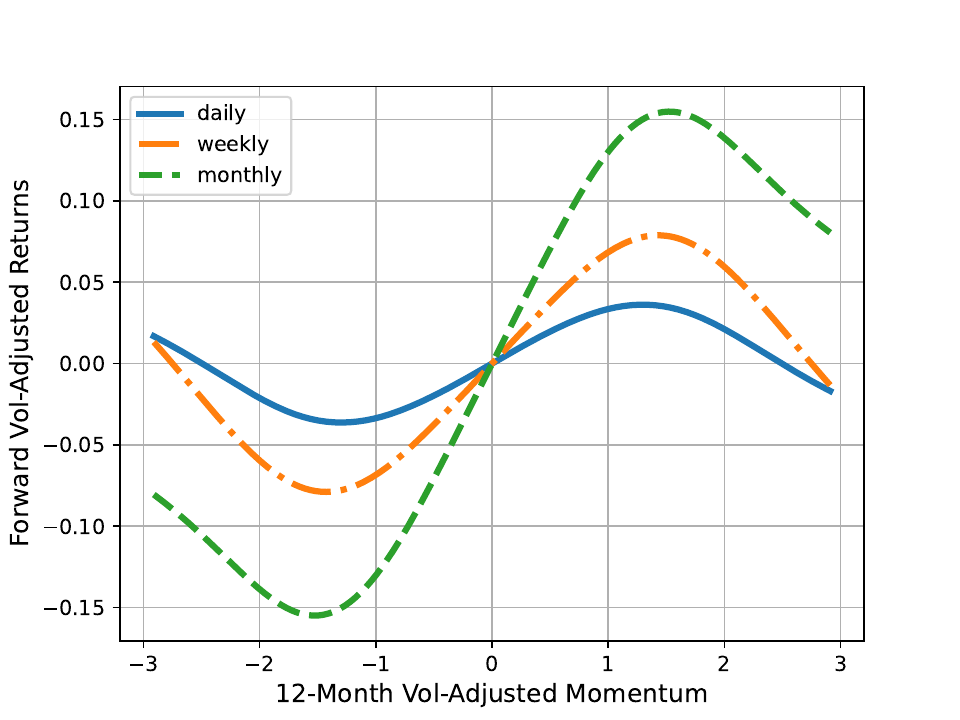}
    \caption{$12$-Month Momentum}
  \end{subfigure}
  
  \caption{Estimated forward vol-adjusted returns as a function of the $1$, $6$ and $12$-month time series momentum factor for US 5 Year Bond Nonlinear models.
  Models are fit using all available training data from January $1950$ through June $2026$. }
\end{figure}

\noindent \textbf{US 10 Year Bond}
\begin{figure}[H]
  \centering
  \begin{subfigure}[b]{0.30\linewidth}
    \includegraphics[width=\linewidth ]{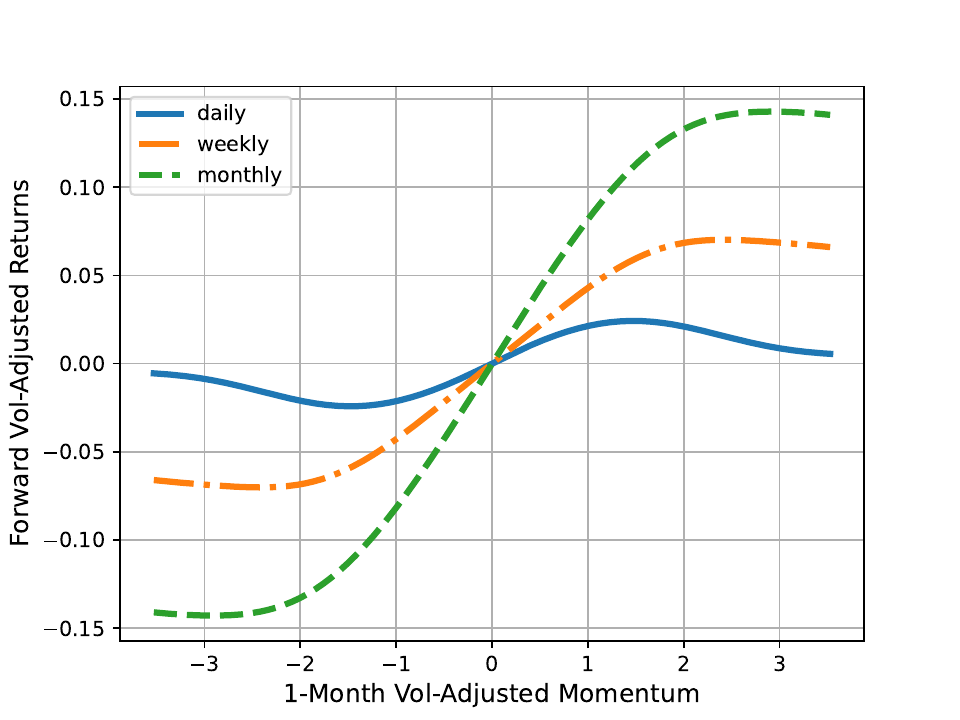}
    \caption{$1$-Month Momentum}
  \end{subfigure}
  \begin{subfigure}[b]{0.30\linewidth}
    \includegraphics[width=\linewidth ]{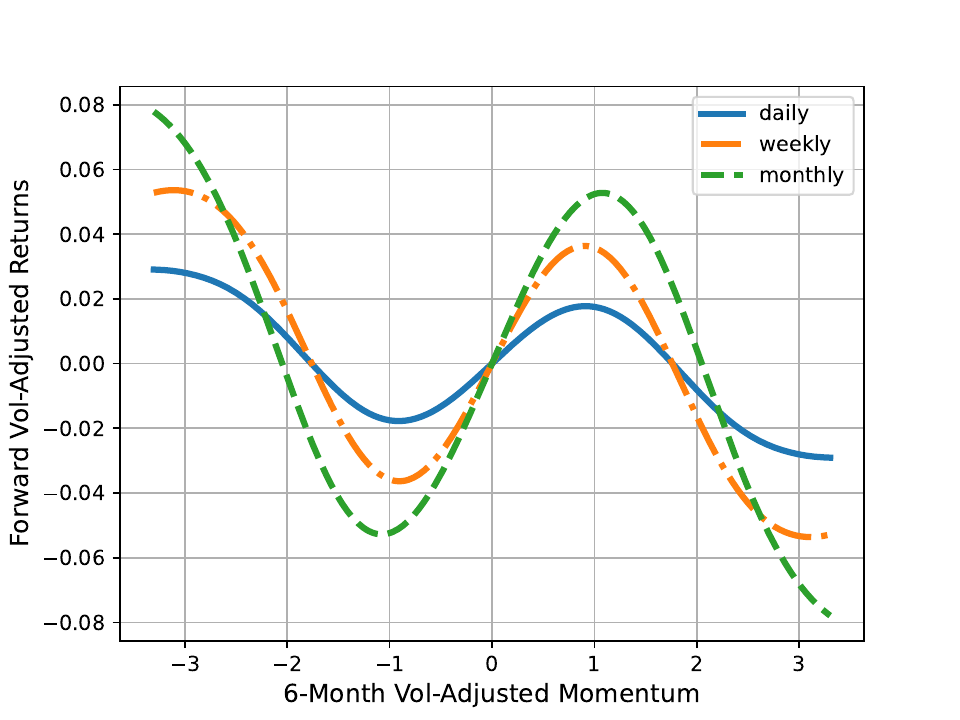}
    \caption{$6$-Month Momentum}
  \end{subfigure}
  \begin{subfigure}[b]{0.30\linewidth}
    \includegraphics[width=\linewidth ]{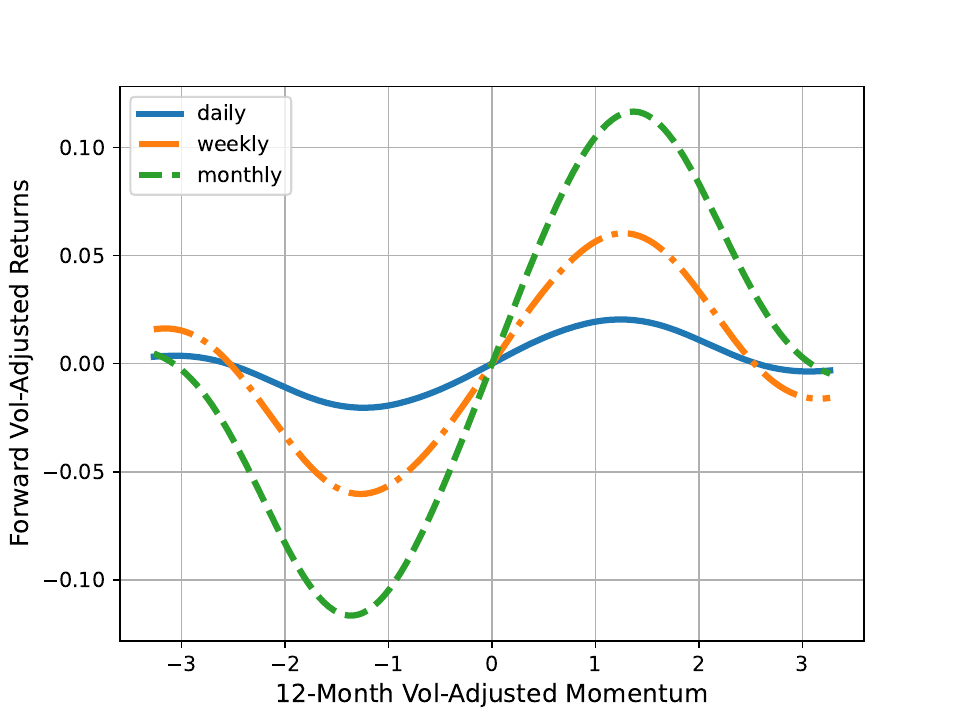}
    \caption{$12$-Month Momentum}
  \end{subfigure}
  
  \caption{Estimated forward vol-adjusted returns as a function of the $1$, $6$ and $12$-month time series momentum factor for US 10 Year Nonlinear models.
  Models are fit using all available training data from January $1950$ through June $2026$. }
\end{figure}

\noindent \textbf{US 30 Year Bond}
\begin{figure}[H]
  \centering
  \begin{subfigure}[b]{0.30\linewidth}
    \includegraphics[width=\linewidth ]{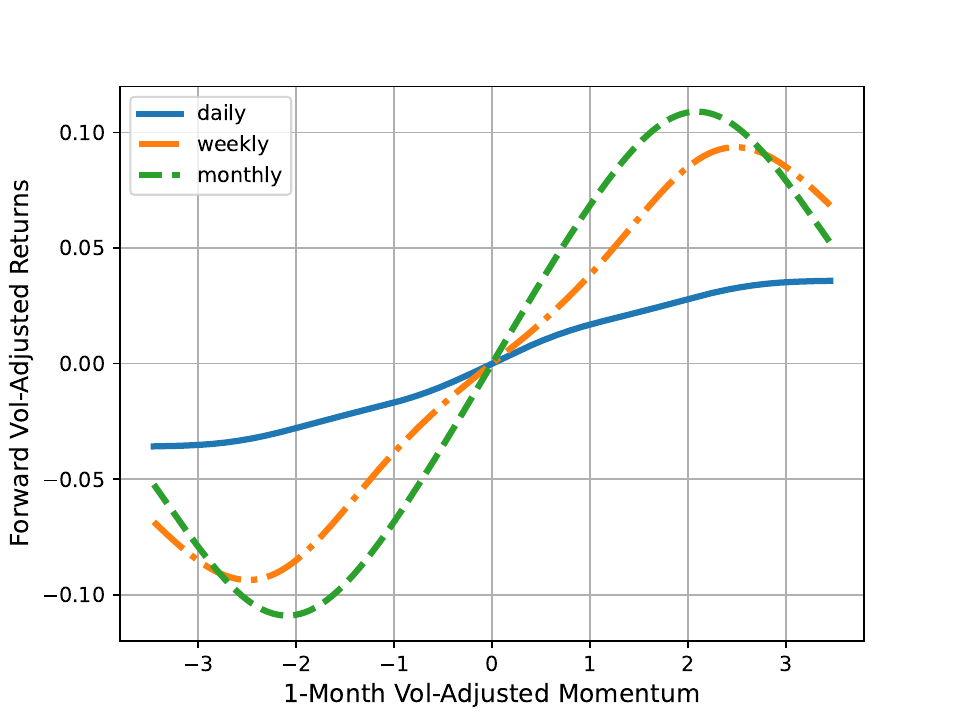}
    \caption{$1$-Month Momentum}
  \end{subfigure}
  \begin{subfigure}[b]{0.30\linewidth}
    \includegraphics[width=\linewidth ]{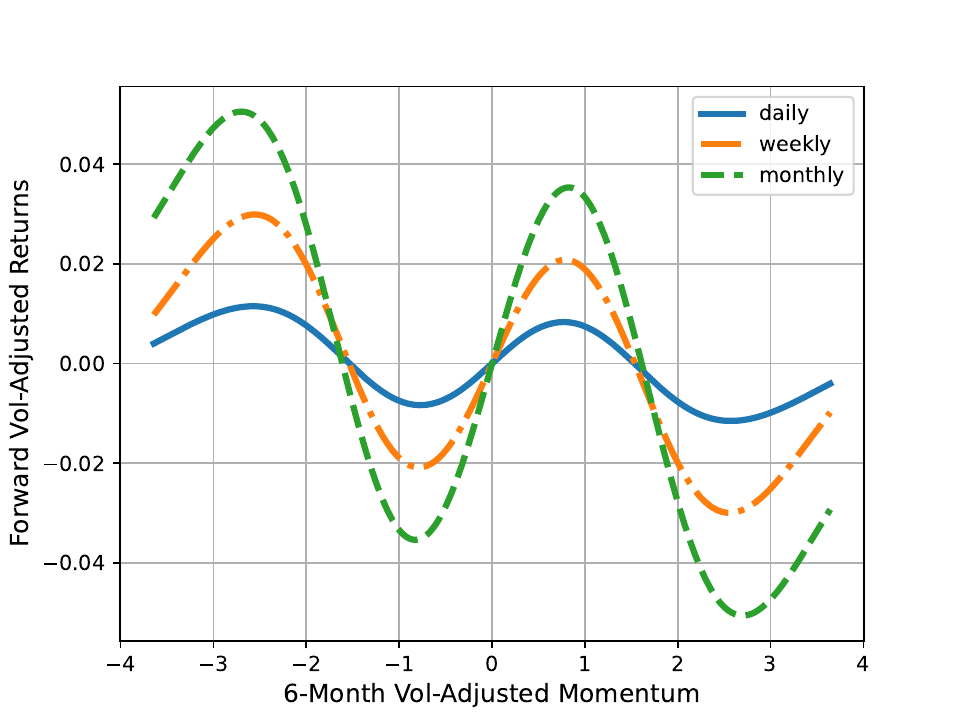}
    \caption{$6$-Month Momentum}
  \end{subfigure}
  \begin{subfigure}[b]{0.30\linewidth}
    \includegraphics[width=\linewidth ]{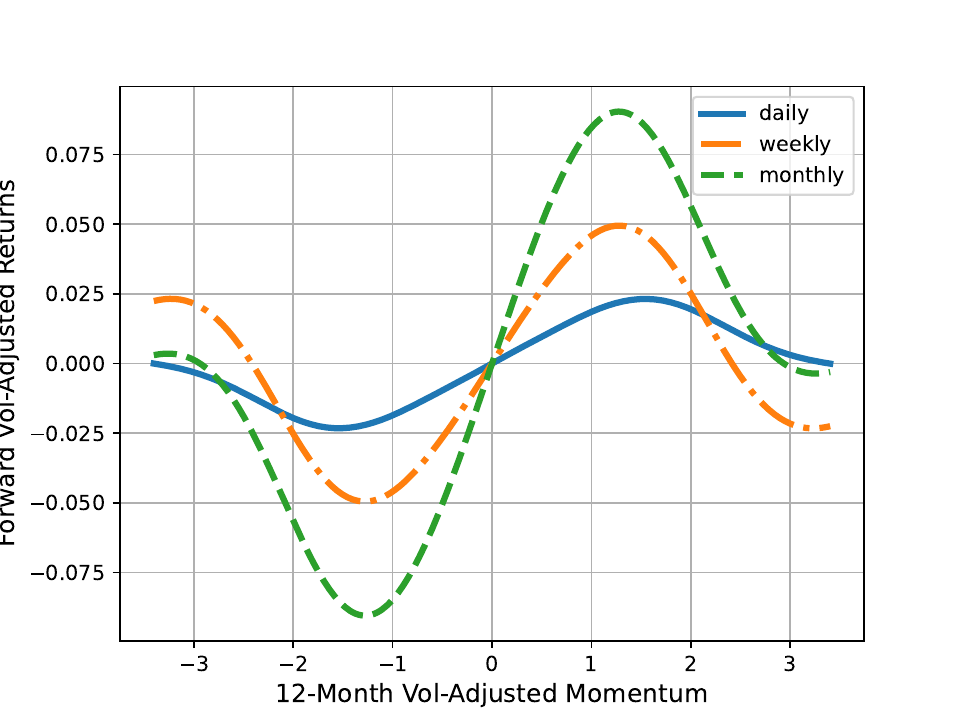}
    \caption{$12$-Month Momentum}
  \end{subfigure}
  
  \caption{Estimated forward vol-adjusted returns as a function of the $1$, $6$ and $12$-month time series momentum factor for US 30 Year Nonlinear models.
  Models are fit using all available training data from January $1950$ through June $2026$. }
\end{figure}

\subsection{Currency Sector Nonlinear Models}
\begin{figure}[H]
  \centering
  \begin{subfigure}[b]{0.30\linewidth}
    \includegraphics[width=\linewidth ]{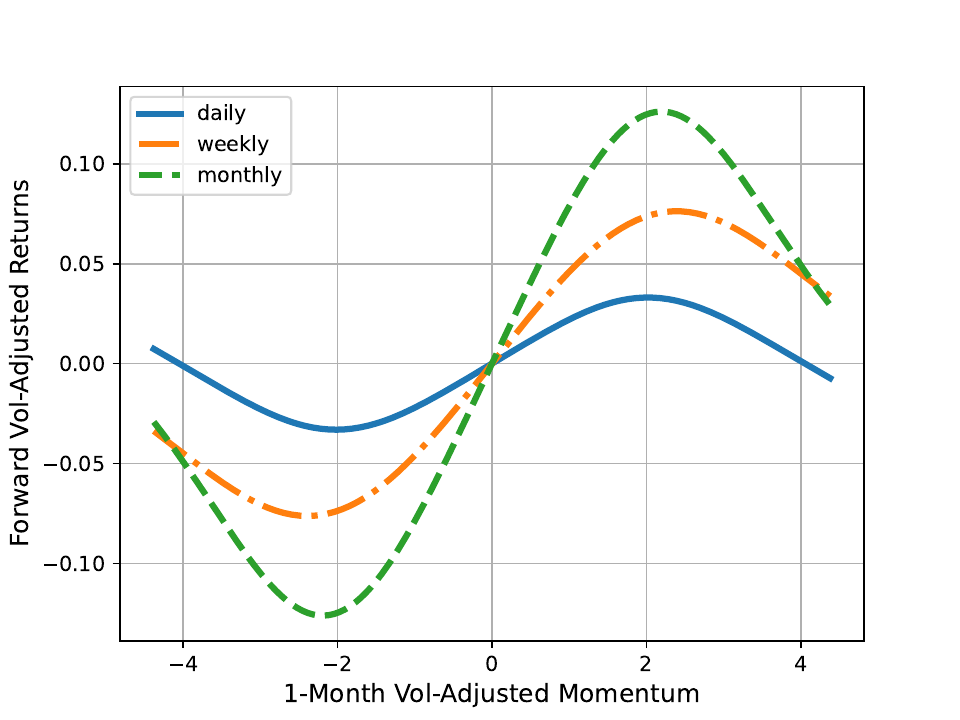}
    \caption{$1$-Month Momentum}
  \end{subfigure}
  \begin{subfigure}[b]{0.30\linewidth}
    \includegraphics[width=\linewidth ]{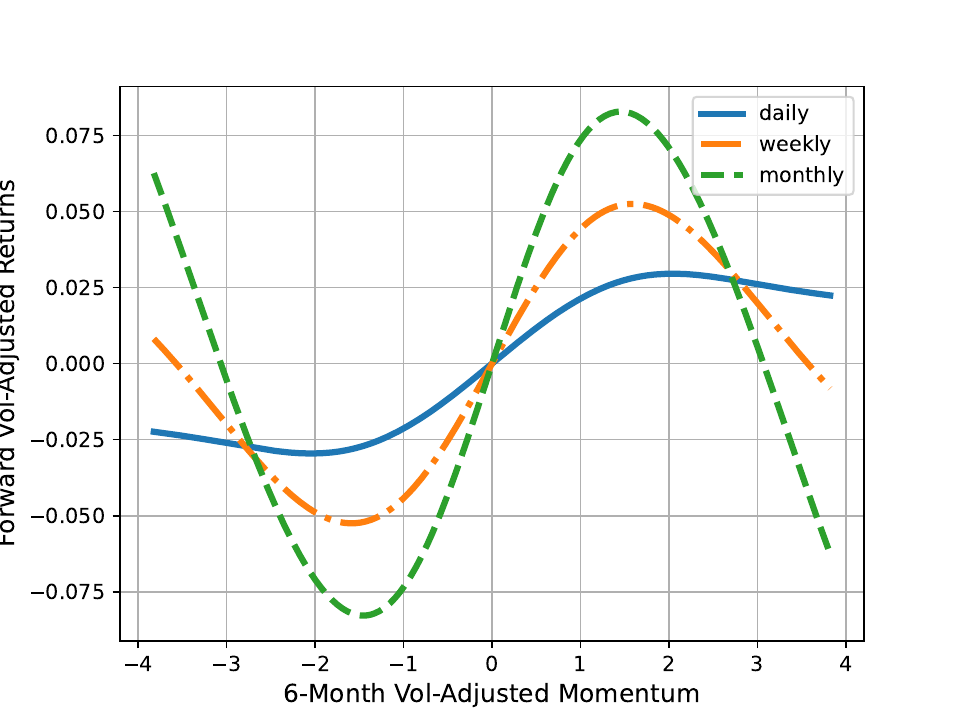}
    \caption{$6$-Month Momentum}
  \end{subfigure}
  \begin{subfigure}[b]{0.30\linewidth}
    \includegraphics[width=\linewidth ]{figs/model_currency_momo_12_month.pdf}
    \caption{$12$-Month Momentum}
  \end{subfigure}
  
  \caption{Estimated forward vol-adjusted returns as a function of the $1$, $6$ and $12$-month time series momentum factor for Currency Sector Nonlinear models.
  Models are fit using all available training data from January $1950$ through June $2026$. }
\end{figure}

\noindent \textbf{Australian Dollar}
\begin{figure}[H]
  \centering
  \begin{subfigure}[b]{0.30\linewidth}
    \includegraphics[width=\linewidth ]{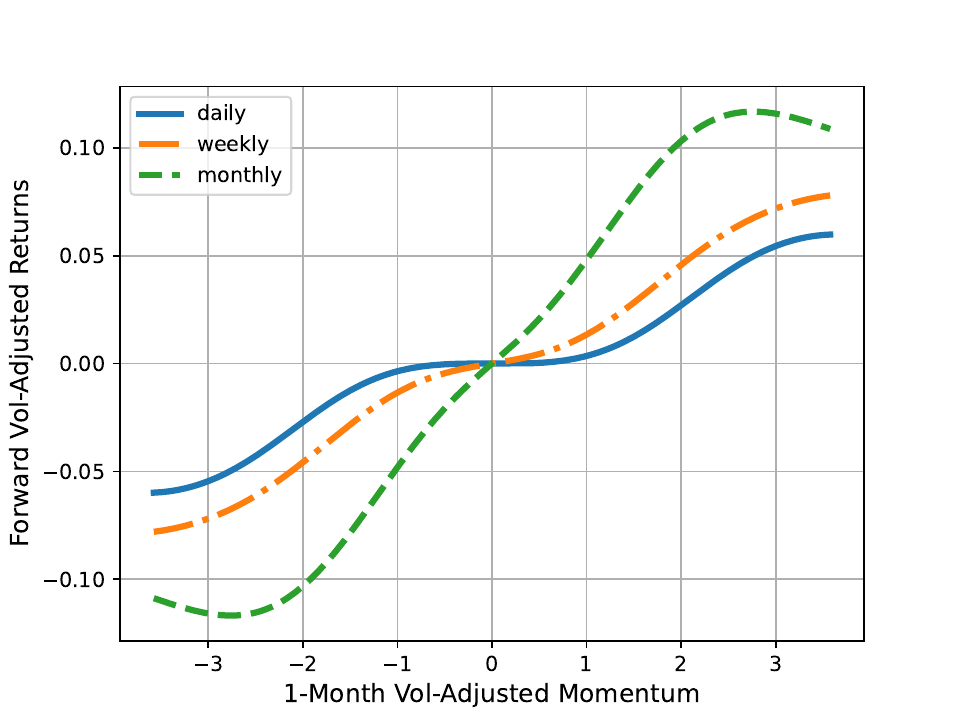}
    \caption{$1$-Month Momentum}
  \end{subfigure}
  \begin{subfigure}[b]{0.30\linewidth}
    \includegraphics[width=\linewidth ]{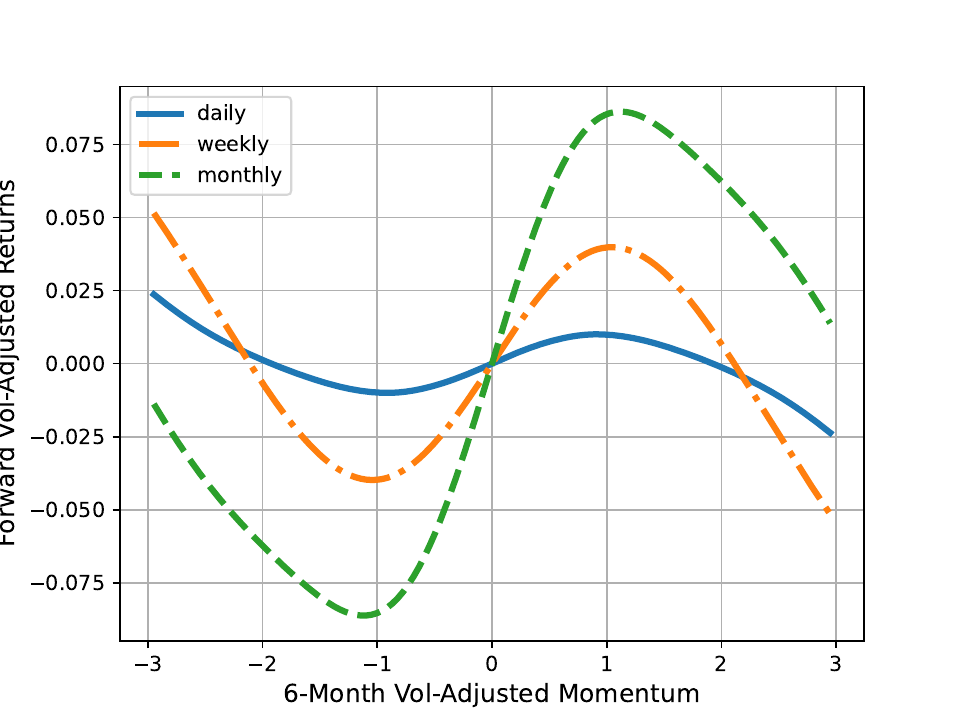}
    \caption{$6$-Month Momentum}
  \end{subfigure}
  \begin{subfigure}[b]{0.30\linewidth}
    \includegraphics[width=\linewidth ]{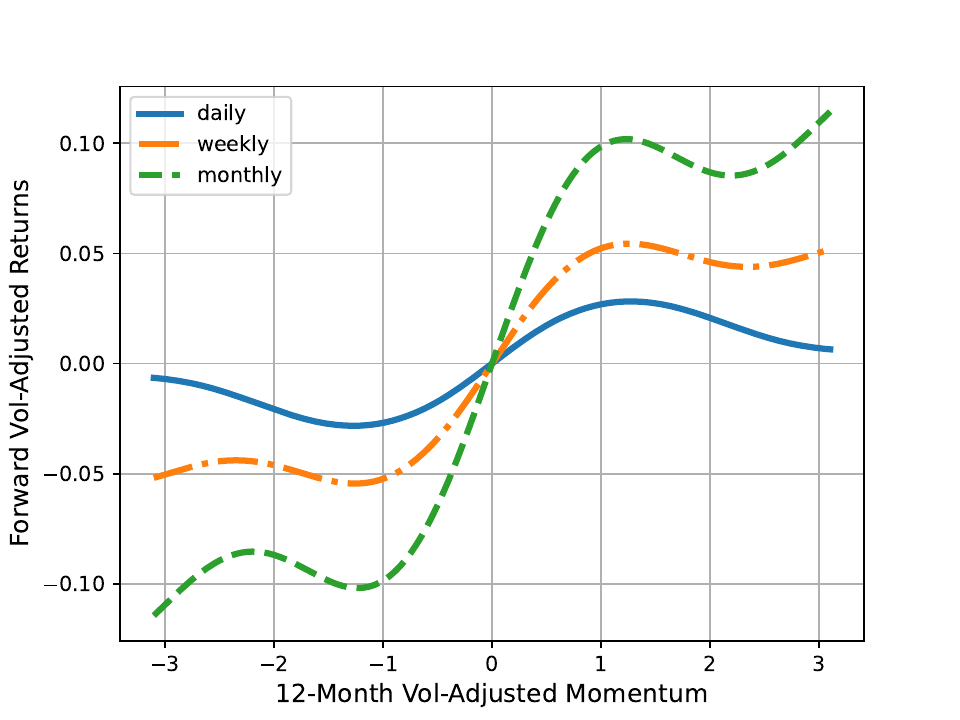}
    \caption{$12$-Month Momentum}
  \end{subfigure}
  
  \caption{Estimated forward vol-adjusted returns as a function of the $1$, $6$ and $12$-month time series momentum factor for Australian Dollar Nonlinear models.
  Models are fit using all available training data from January $1950$ through June $2026$. }
\end{figure}

\noindent \textbf{British Pound}
\begin{figure}[H]
  \centering
  \begin{subfigure}[b]{0.30\linewidth}
    \includegraphics[width=\linewidth ]{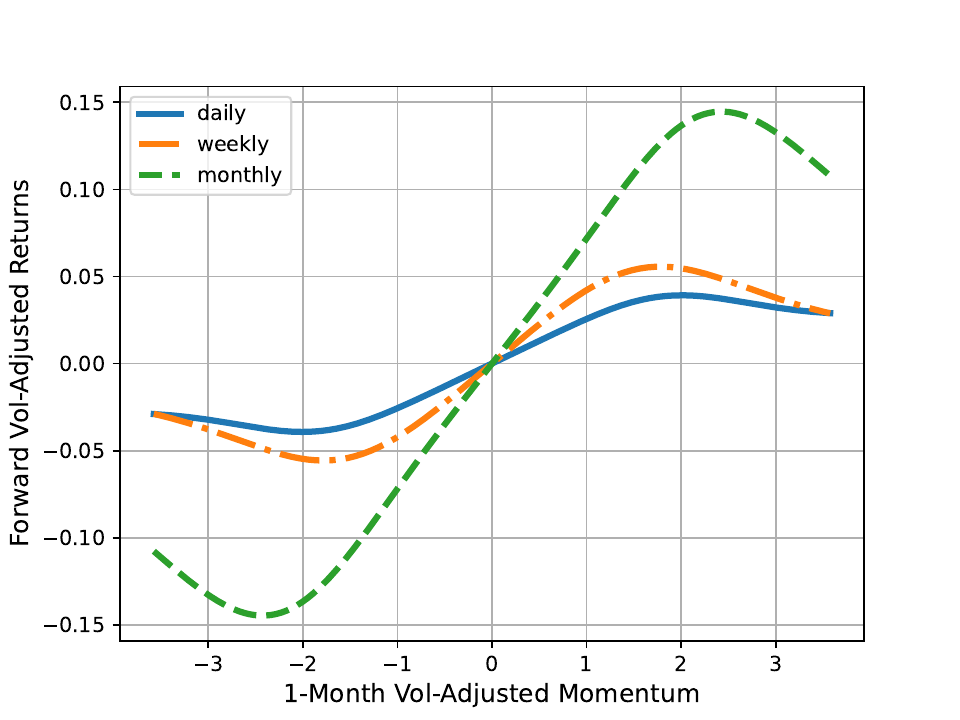}
    \caption{$1$-Month Momentum}
  \end{subfigure}
  \begin{subfigure}[b]{0.30\linewidth}
    \includegraphics[width=\linewidth ]{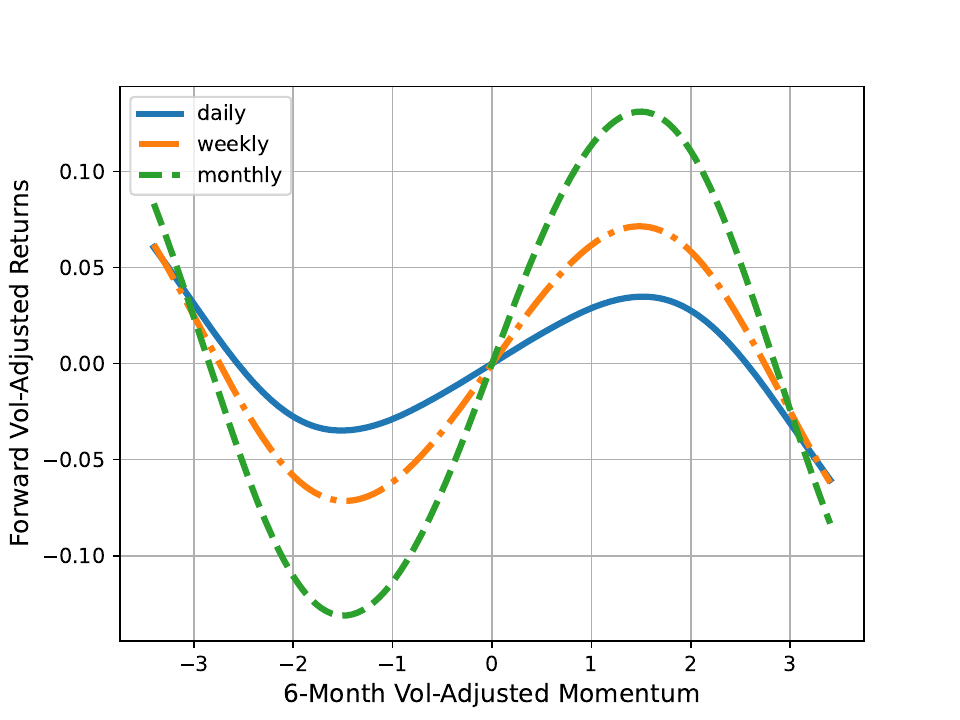}
    \caption{$6$-Month Momentum}
  \end{subfigure}
  \begin{subfigure}[b]{0.30\linewidth}
    \includegraphics[width=\linewidth ]{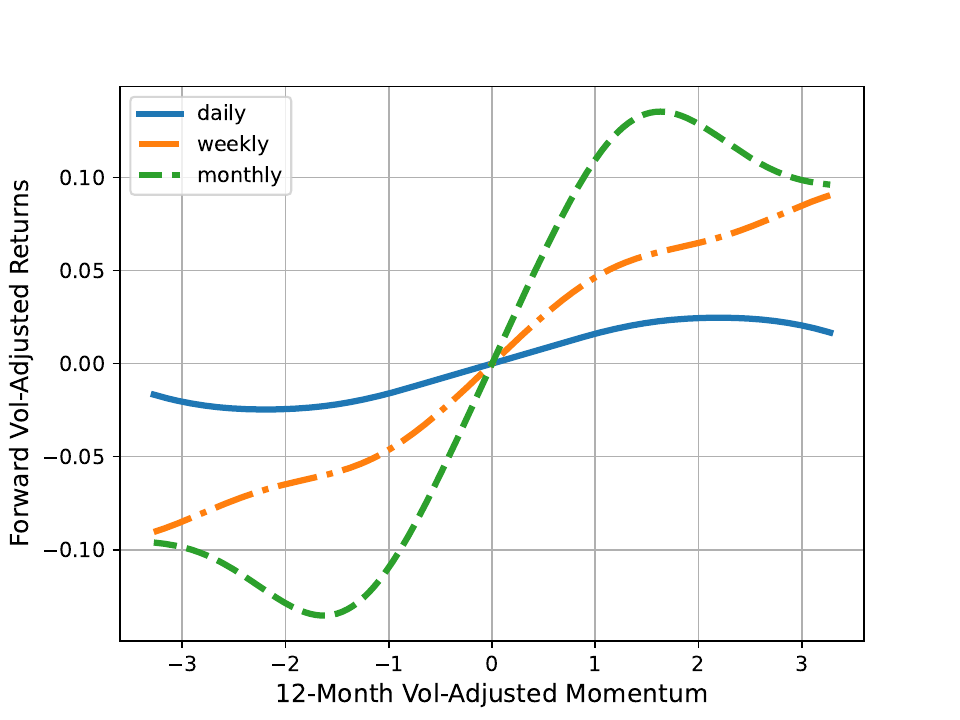}
    \caption{$12$-Month Momentum}
  \end{subfigure}
  
  \caption{Estimated forward vol-adjusted returns as a function of the $1$, $6$ and $12$-month time series momentum factor for British Pound Nonlinear models.
  Models are fit using all available training data from January $1950$ through June $2026$. }
\end{figure}

\noindent \textbf{Canadian Dollar}
\begin{figure}[H]
  \centering
  \begin{subfigure}[b]{0.30\linewidth}
    \includegraphics[width=\linewidth ]{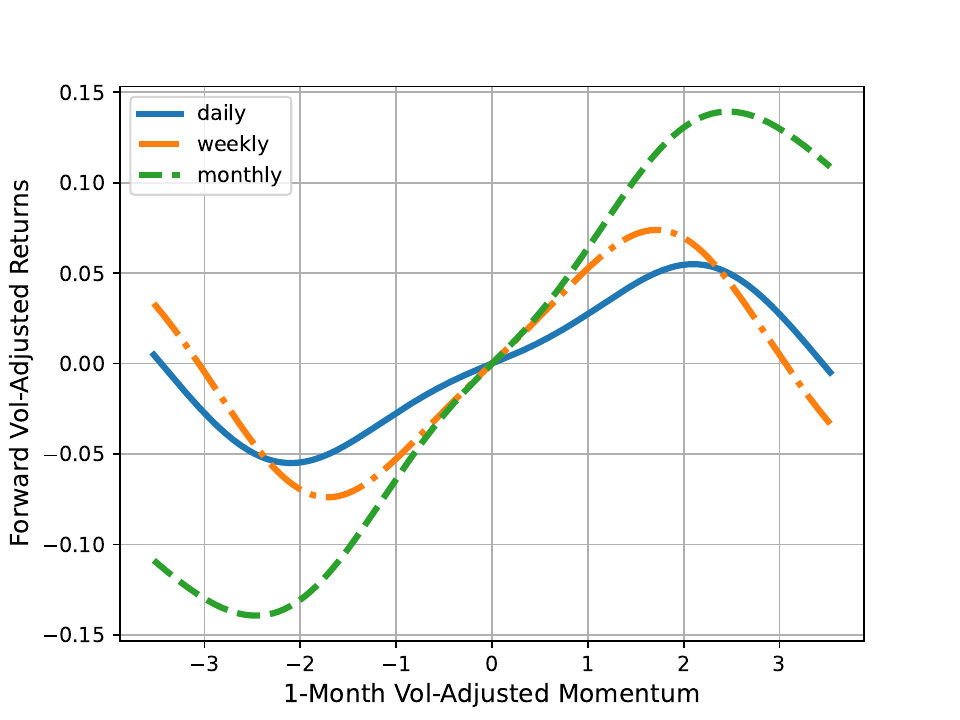}
    \caption{$1$-Month Momentum}
  \end{subfigure}
  \begin{subfigure}[b]{0.30\linewidth}
    \includegraphics[width=\linewidth ]{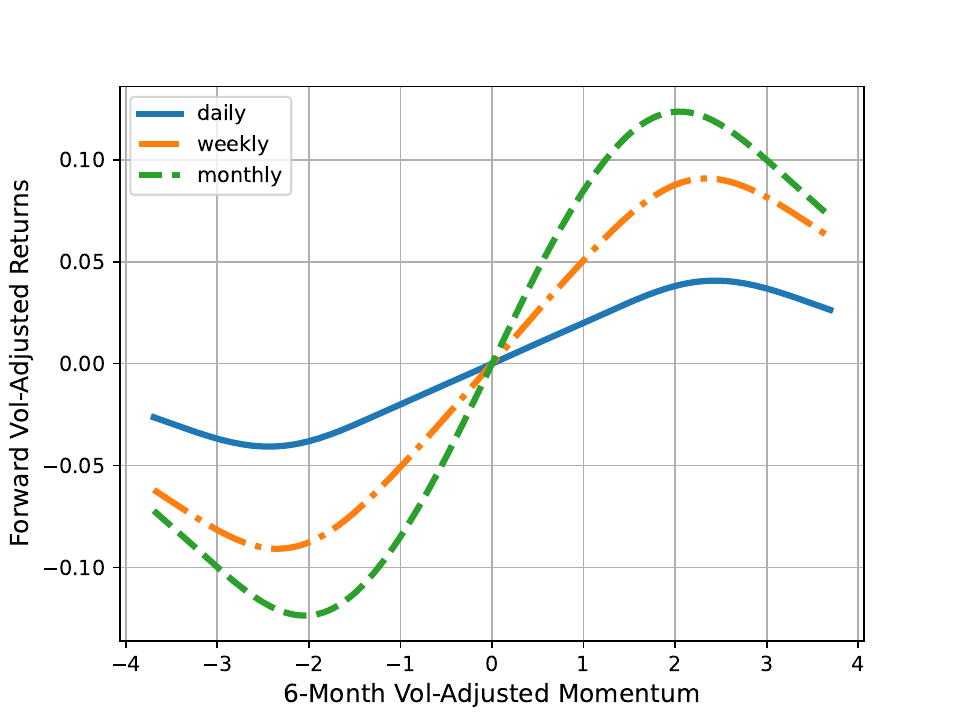}
    \caption{$6$-Month Momentum}
  \end{subfigure}
  \begin{subfigure}[b]{0.30\linewidth}
    \includegraphics[width=\linewidth ]{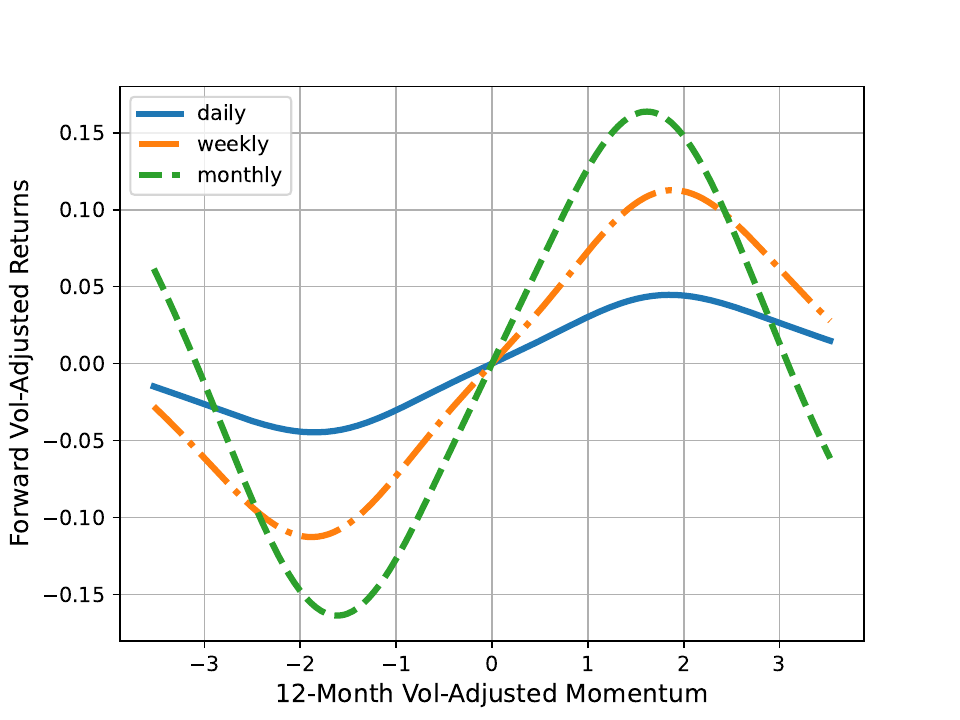}
    \caption{$12$-Month Momentum}
  \end{subfigure}
  
  \caption{Estimated forward vol-adjusted returns as a function of the $1$, $6$ and $12$-month time series momentum factor for Canadian Dollar Nonlinear models.
  Models are fit using all available training data from January $1950$ through June $2026$. }
\end{figure}

\noindent \textbf{US Dollar}
\begin{figure}[H]
  \centering
  \begin{subfigure}[b]{0.30\linewidth}
    \includegraphics[width=\linewidth ]{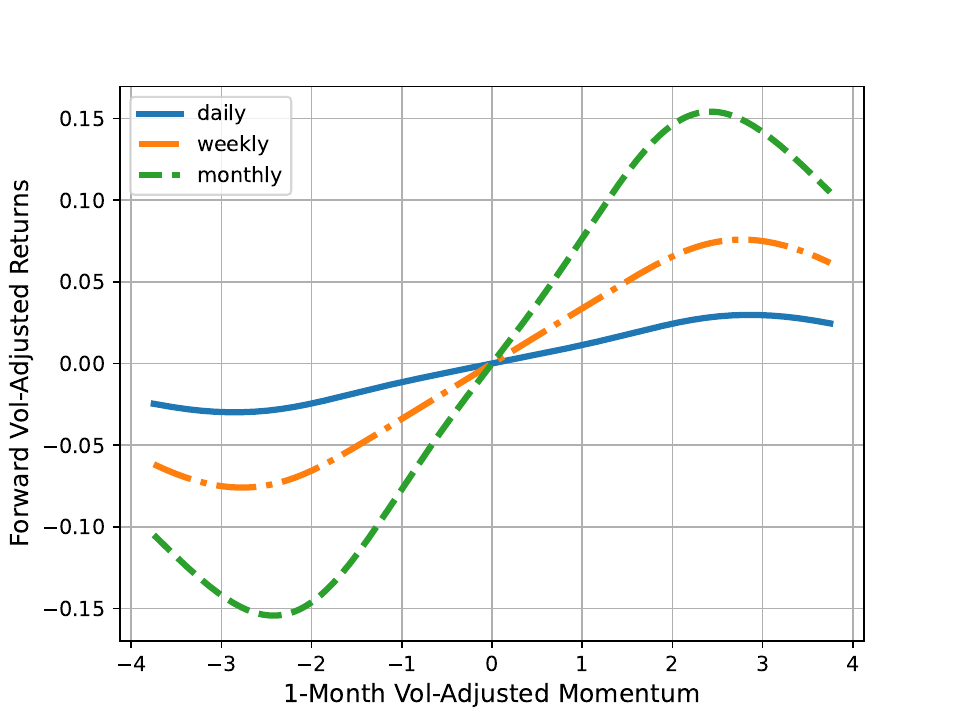}
    \caption{$1$-Month Momentum}
  \end{subfigure}
  \begin{subfigure}[b]{0.30\linewidth}
    \includegraphics[width=\linewidth ]{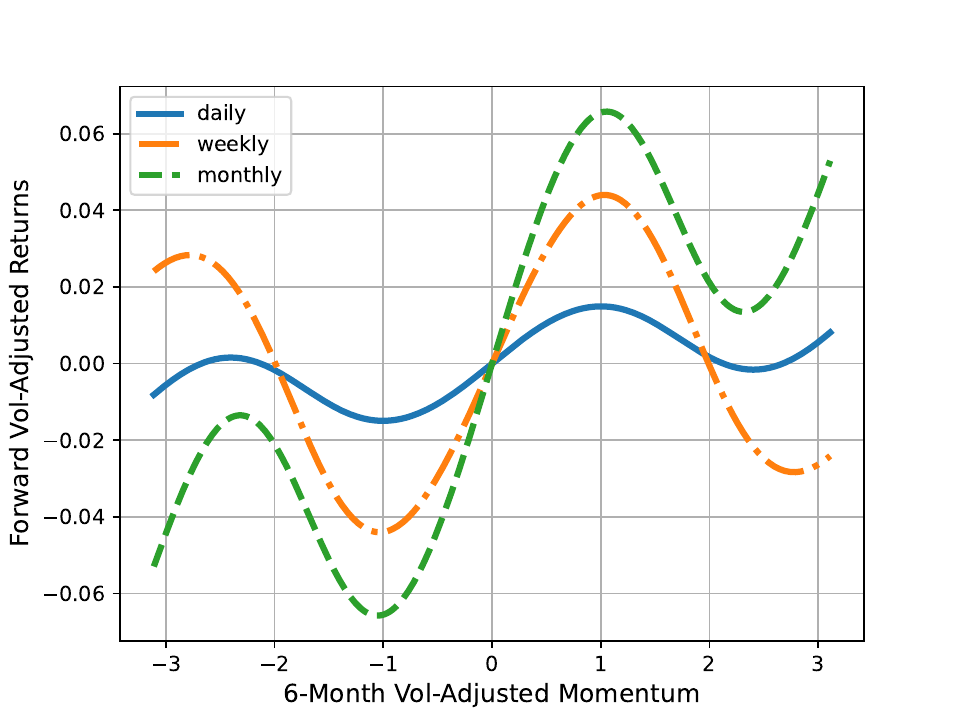}
    \caption{$6$-Month Momentum}
  \end{subfigure}
  \begin{subfigure}[b]{0.30\linewidth}
    \includegraphics[width=\linewidth ]{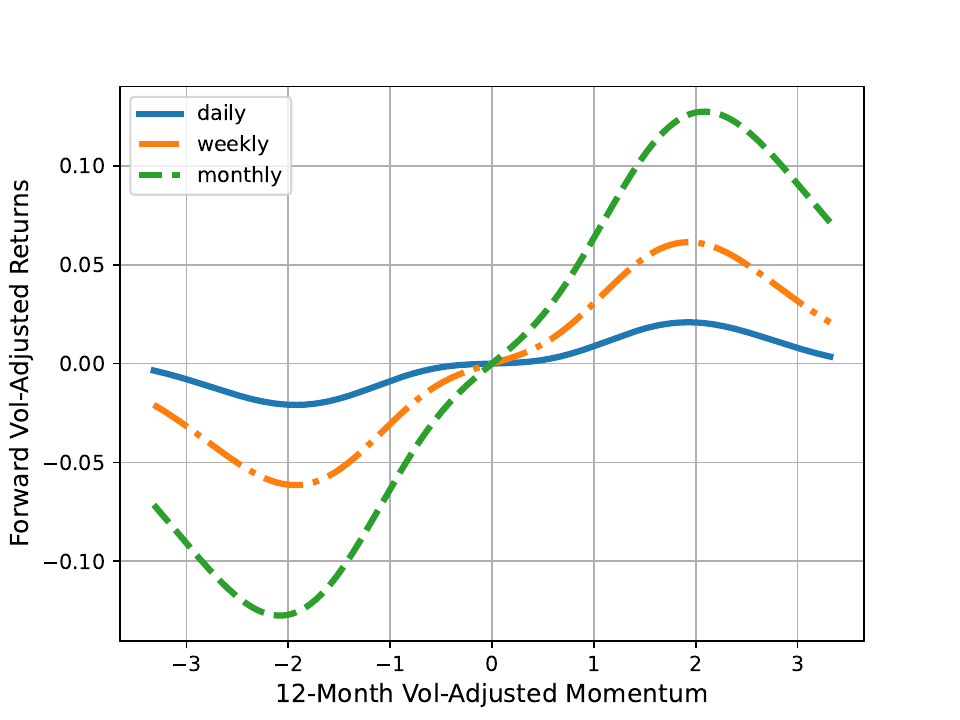}
    \caption{$12$-Month Momentum}
  \end{subfigure}
  
  \caption{Estimated forward vol-adjusted returns as a function of the $1$, $6$ and $12$-month time series momentum factor for US Dollar Nonlinear models.
  Models are fit using all available training data from January $1950$ through June $2026$. }
\end{figure}

\noindent \textbf{Euro}
\begin{figure}[H]
  \centering
  \begin{subfigure}[b]{0.30\linewidth}
    \includegraphics[width=\linewidth ]{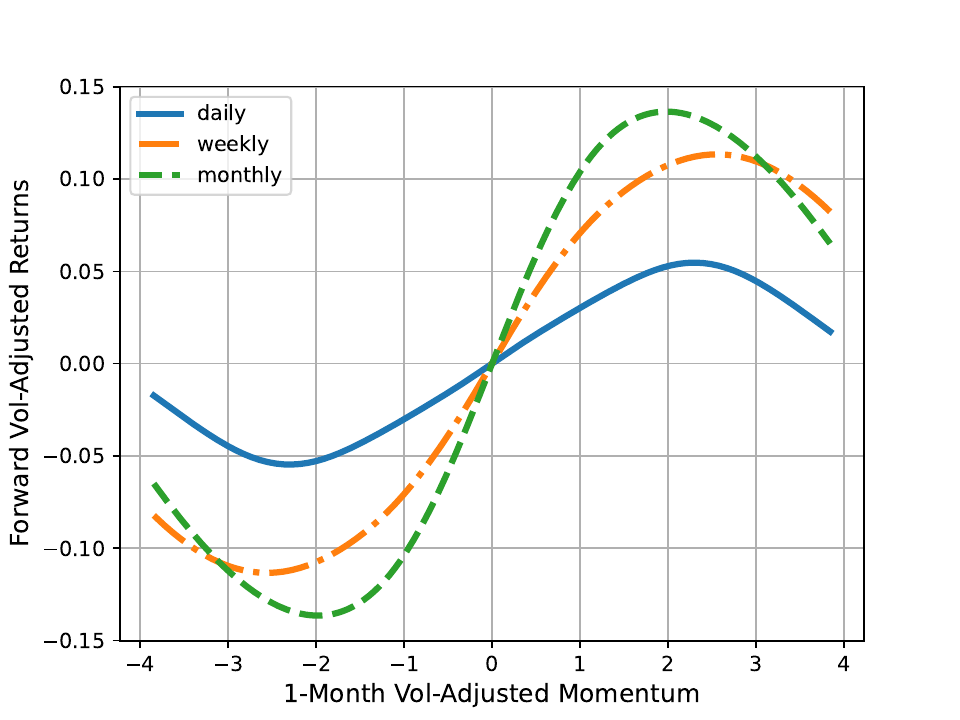}
    \caption{$1$-Month Momentum}
  \end{subfigure}
  \begin{subfigure}[b]{0.30\linewidth}
    \includegraphics[width=\linewidth ]{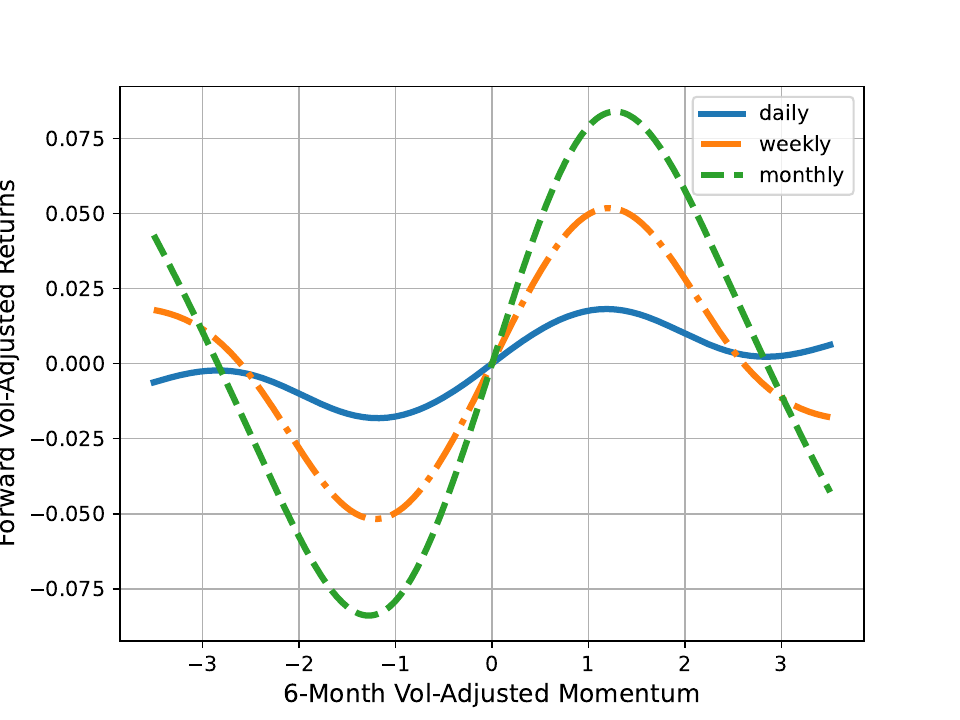}
    \caption{$6$-Month Momentum}
  \end{subfigure}
  \begin{subfigure}[b]{0.30\linewidth}
    \includegraphics[width=\linewidth ]{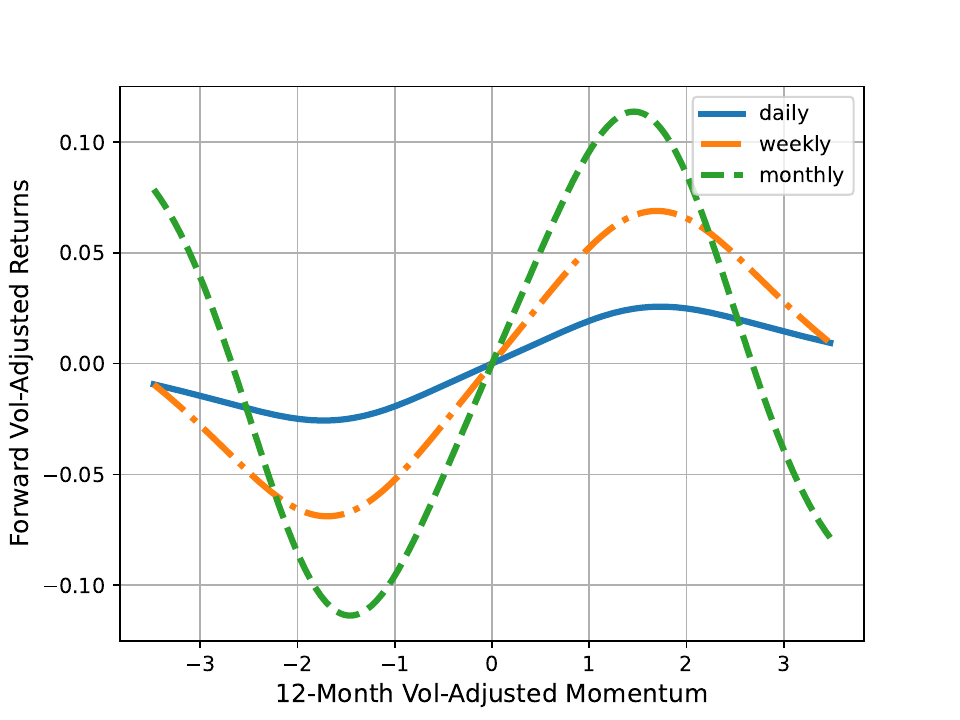}
    \caption{$12$-Month Momentum}
  \end{subfigure}
  
  \caption{Estimated forward vol-adjusted returns as a function of the $1$, $6$ and $12$-month time series momentum factor for Euro Nonlinear models.
  Models are fit using all available training data from January $1950$ through June $2026$. }
\end{figure}

\noindent \textbf{Japanese Yen}
\begin{figure}[H]
  \centering
  \begin{subfigure}[b]{0.30\linewidth}
    \includegraphics[width=\linewidth ]{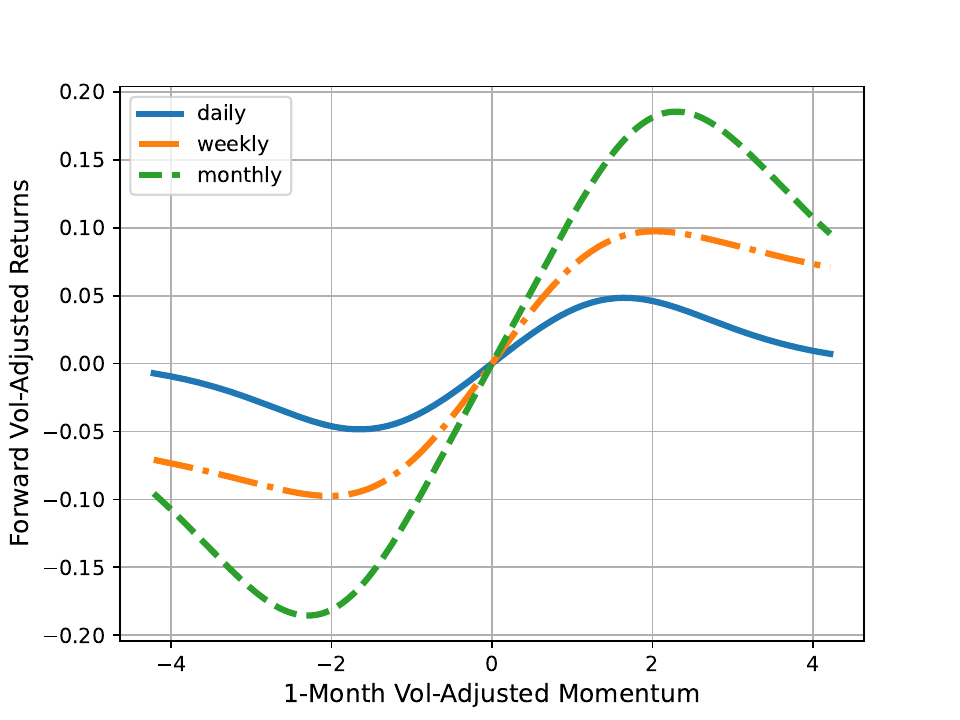}
    \caption{$1$-Month Momentum}
  \end{subfigure}
  \begin{subfigure}[b]{0.30\linewidth}
    \includegraphics[width=\linewidth ]{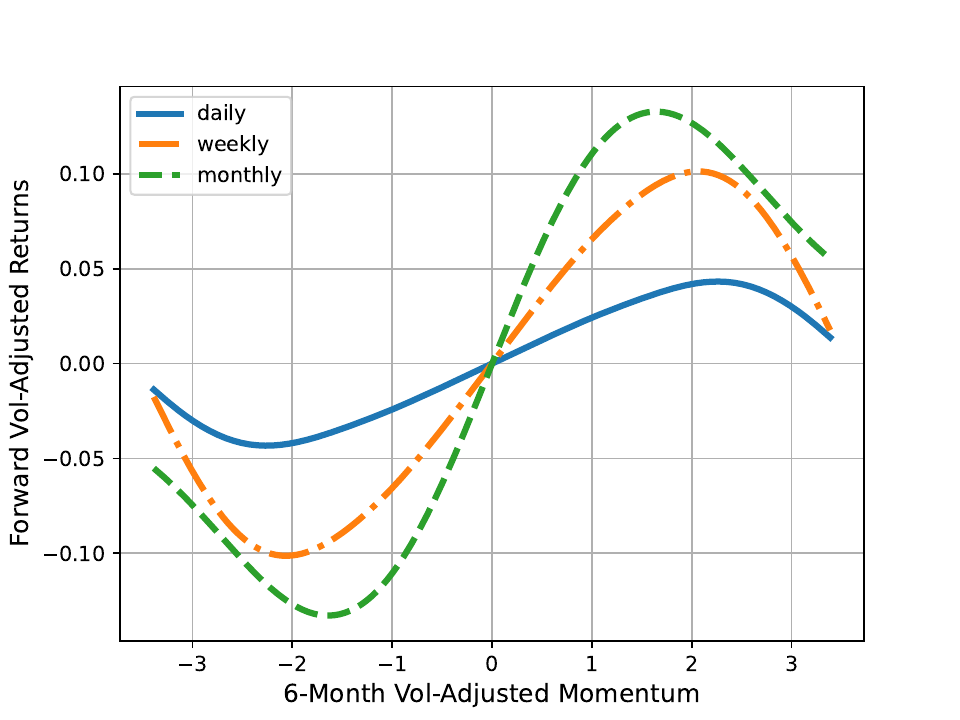}
    \caption{$6$-Month Momentum}
  \end{subfigure}
  \begin{subfigure}[b]{0.30\linewidth}
    \includegraphics[width=\linewidth ]{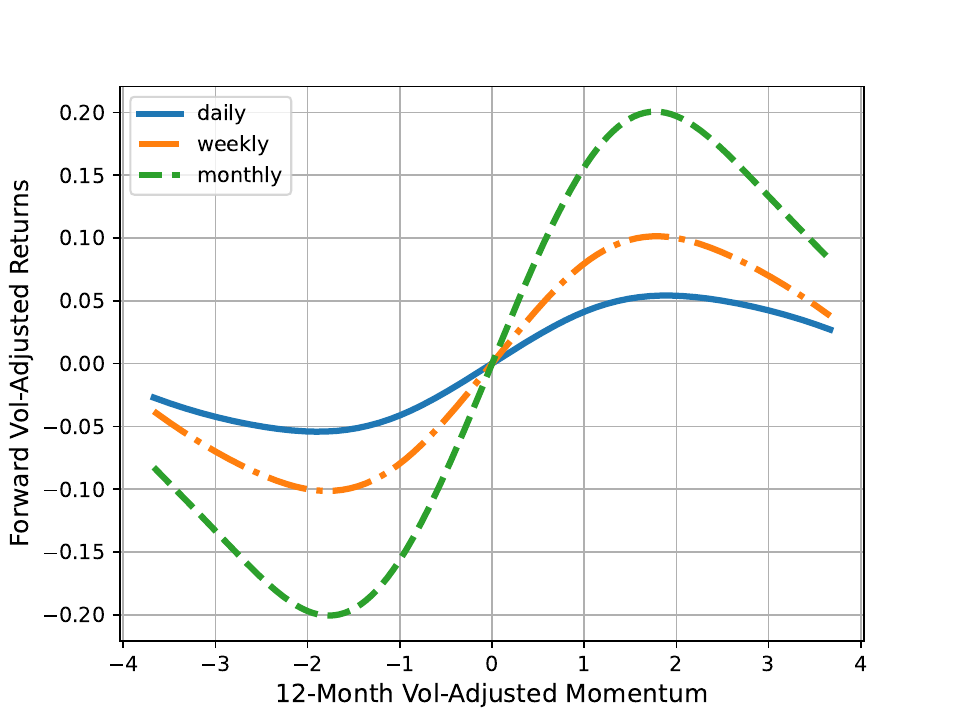}
    \caption{$12$-Month Momentum}
  \end{subfigure}
  
  \caption{Estimated forward vol-adjusted returns as a function of the $1$, $6$ and $12$-month time series momentum factor for Japanese Yen Nonlinear models.
  Models are fit using all available training data from January $1950$ through June $2026$. }
\end{figure}

\noindent \textbf{New Zealand Dollar}
\begin{figure}[H]
  \centering
  \begin{subfigure}[b]{0.30\linewidth}
    \includegraphics[width=\linewidth ]{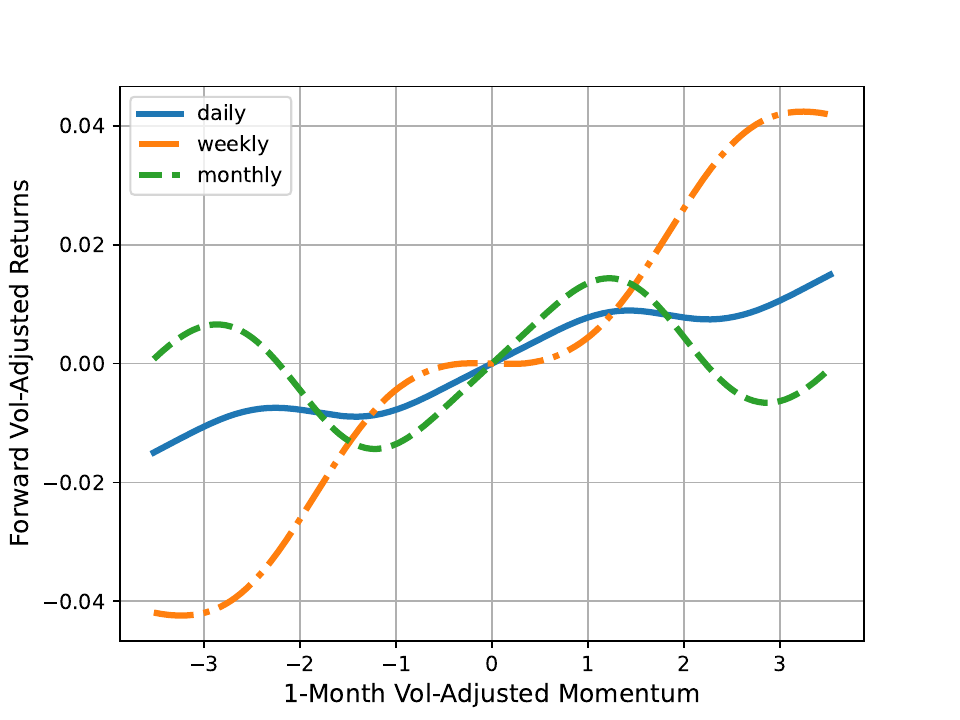}
    \caption{$1$-Month Momentum}
  \end{subfigure}
  \begin{subfigure}[b]{0.30\linewidth}
    \includegraphics[width=\linewidth ]{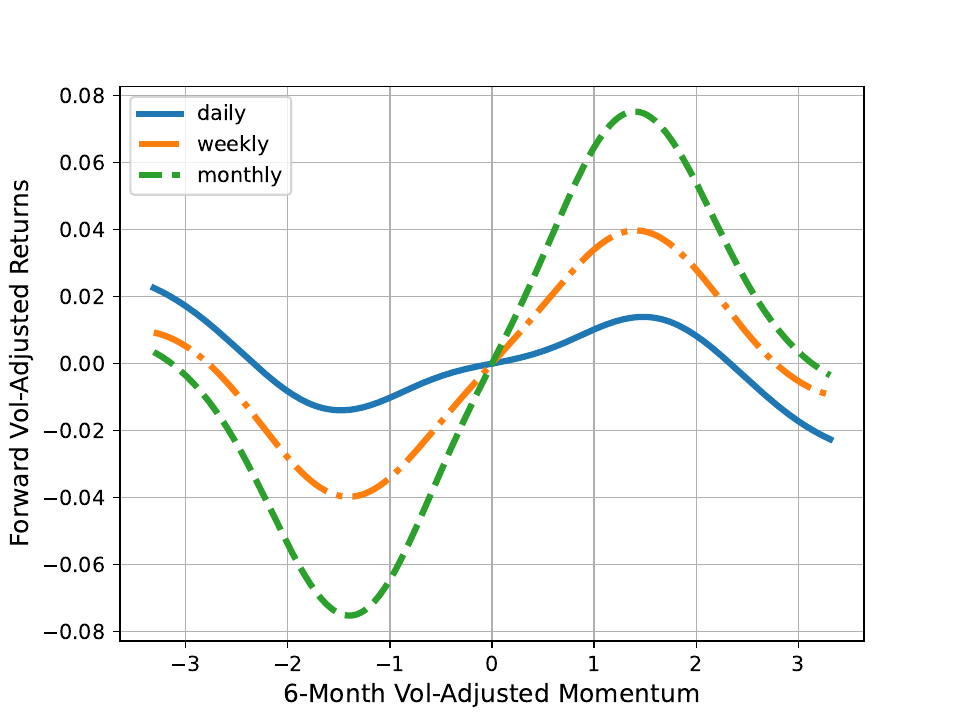}
    \caption{$6$-Month Momentum}
  \end{subfigure}
  \begin{subfigure}[b]{0.30\linewidth}
    \includegraphics[width=\linewidth ]{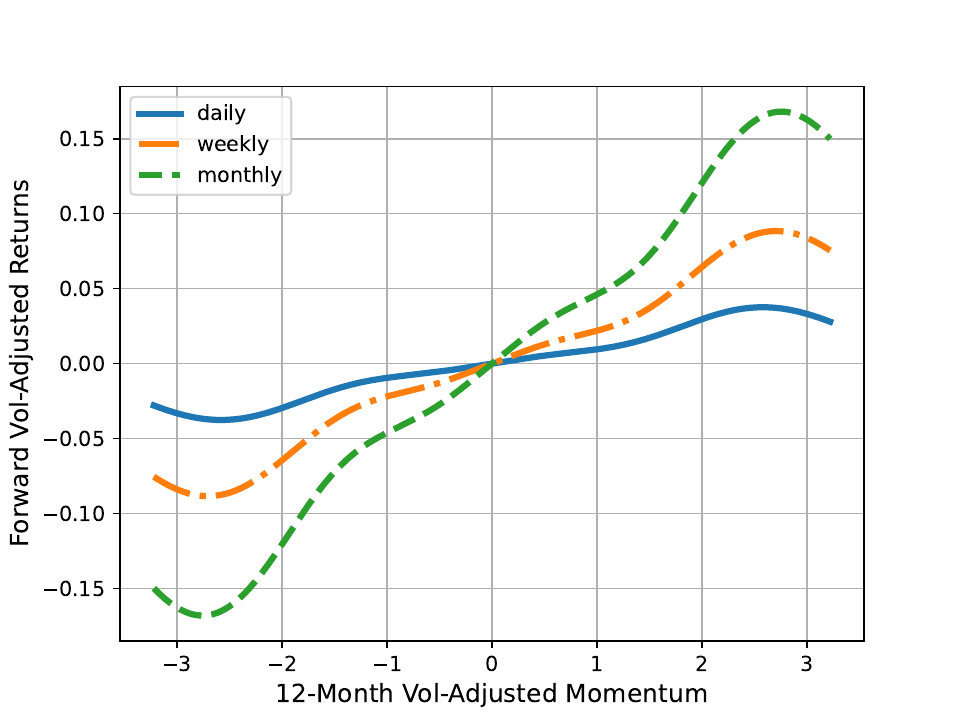}
    \caption{$12$-Month Momentum}
  \end{subfigure}
  
  \caption{Estimated forward vol-adjusted returns as a function of the $1$, $6$ and $12$-month time series momentum factor for New Zealand Dollar Nonlinear models.
  Models are fit using all available training data from January $1950$ through June $2026$. }
\end{figure}

\noindent \textbf{Swiss Franc}
\begin{figure}[H]
  \centering
  \begin{subfigure}[b]{0.30\linewidth}
    \includegraphics[width=\linewidth ]{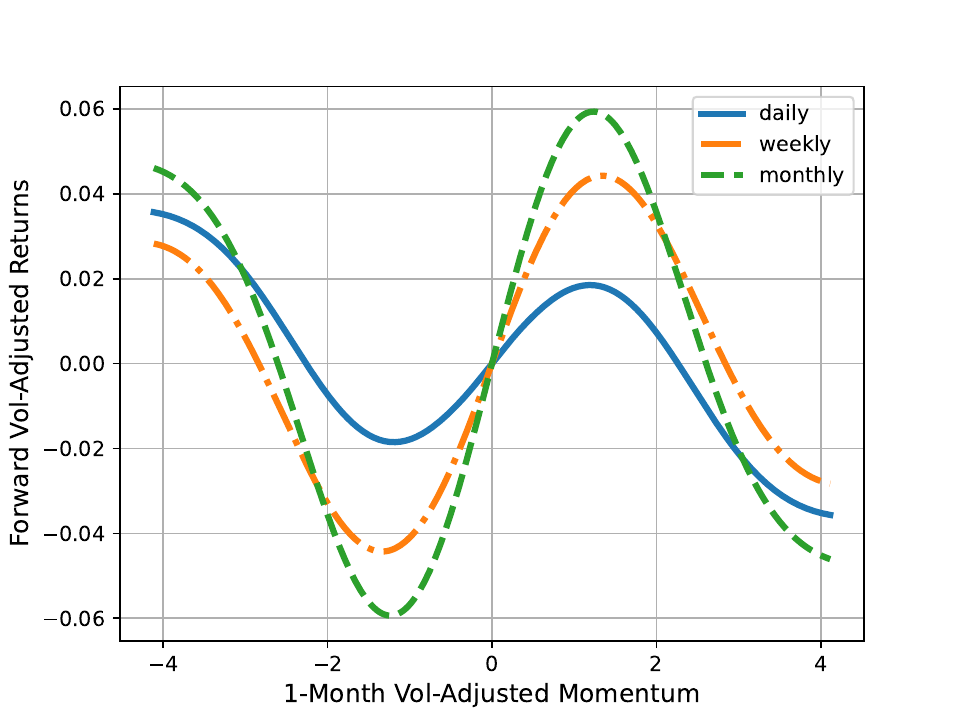}
    \caption{$1$-Month Momentum}
  \end{subfigure}
  \begin{subfigure}[b]{0.30\linewidth}
    \includegraphics[width=\linewidth ]{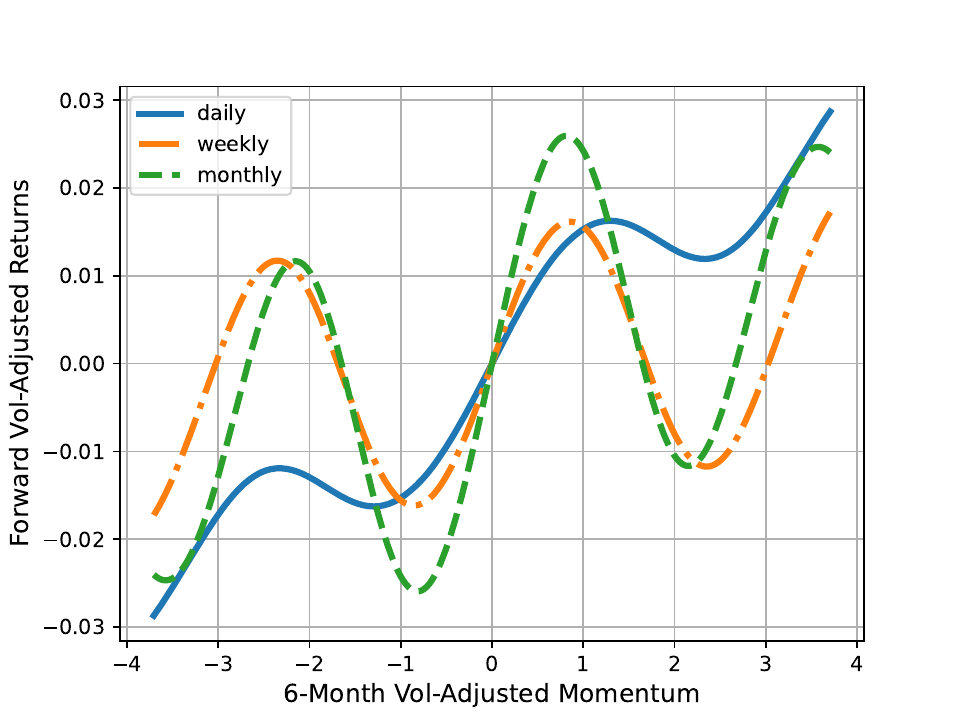}
    \caption{$6$-Month Momentum}
  \end{subfigure}
  \begin{subfigure}[b]{0.30\linewidth}
    \includegraphics[width=\linewidth ]{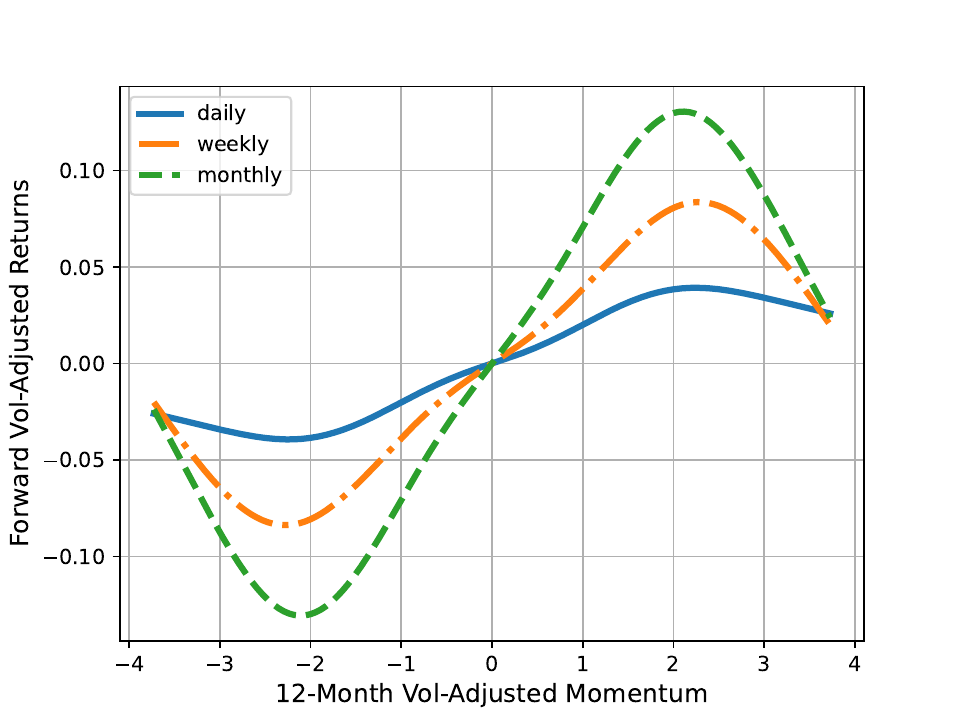}
    \caption{$12$-Month Momentum}
  \end{subfigure}
  
  \caption{Estimated forward vol-adjusted returns as a function of the $1$, $6$ and $12$-month time series momentum factor for Swiss Franc Nonlinear models.
  Models are fit using all available training data from January $1950$ through June $2026$. }
\end{figure}

\subsection{Energy Sector Nonlinear Models}
\begin{figure}[H]
  \centering
  \begin{subfigure}[b]{0.30\linewidth}
    \includegraphics[width=\linewidth ]{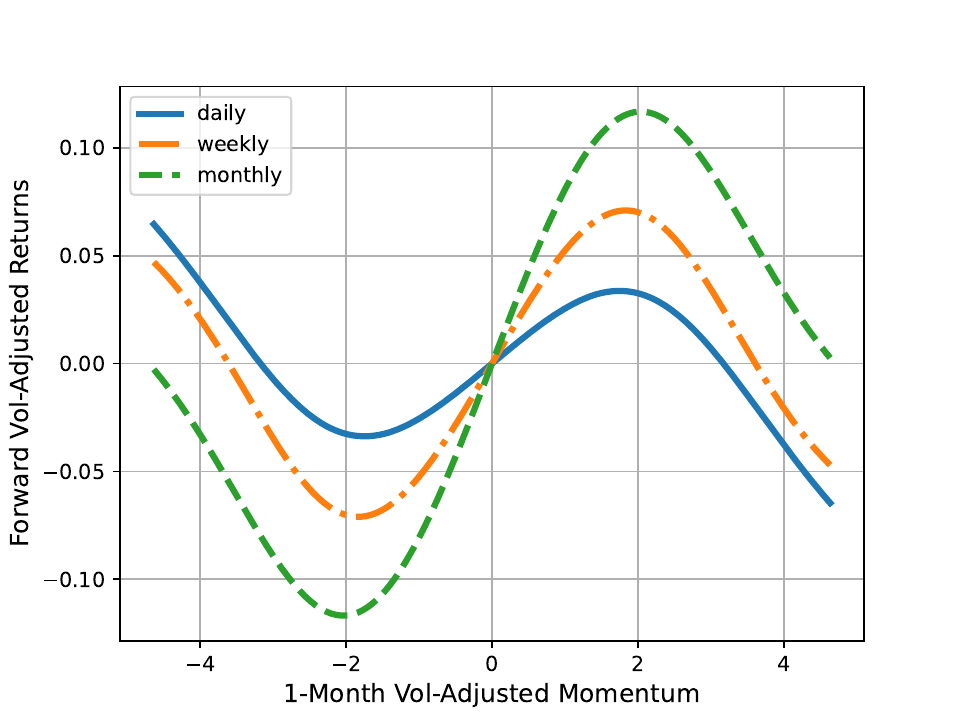}
    \caption{$1$-Month Momentum}
  \end{subfigure}
  \begin{subfigure}[b]{0.30\linewidth}
    \includegraphics[width=\linewidth ]{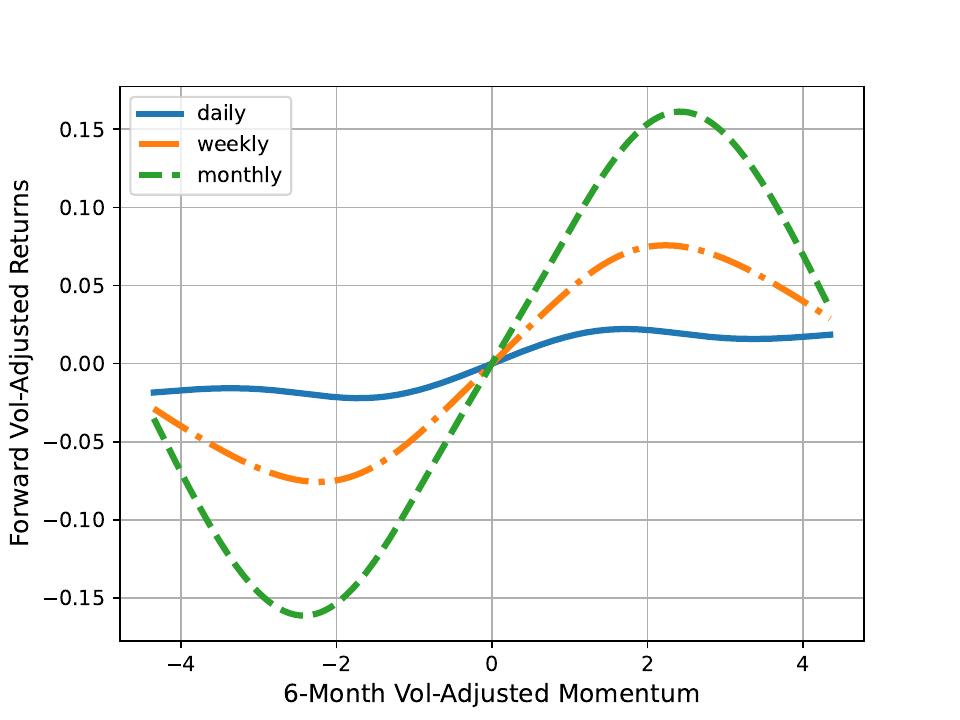}
    \caption{$6$-Month Momentum}
  \end{subfigure}
  \begin{subfigure}[b]{0.30\linewidth}
    \includegraphics[width=\linewidth ]{figs/model_energy_momo_12_month.pdf}
    \caption{$12$-Month Momentum}
  \end{subfigure}
  
  \caption{Estimated forward vol-adjusted returns as a function of the $1$, $6$ and $12$-month time series momentum factor for Energy Sector Nonlinear models.
  Models are fit using all available training data from January $1950$ through June $2026$. }
\end{figure}

\noindent \textbf{Crude Oil}
\begin{figure}[H]
  \centering
  \begin{subfigure}[b]{0.30\linewidth}
    \includegraphics[width=\linewidth ]{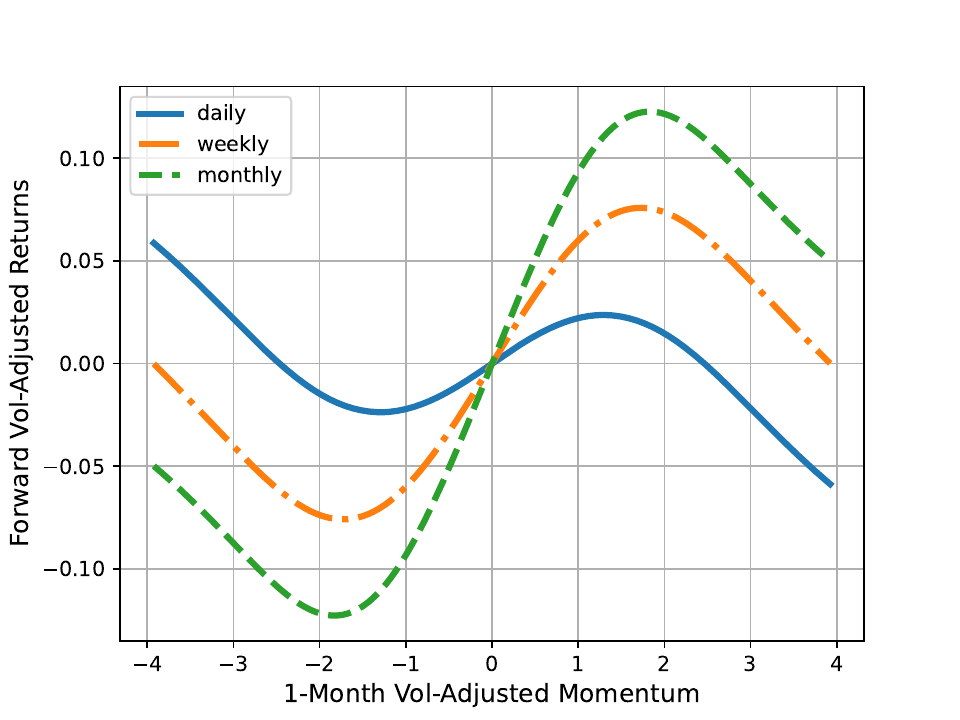}
    \caption{$1$-Month Momentum}
  \end{subfigure}
  \begin{subfigure}[b]{0.30\linewidth}
    \includegraphics[width=\linewidth ]{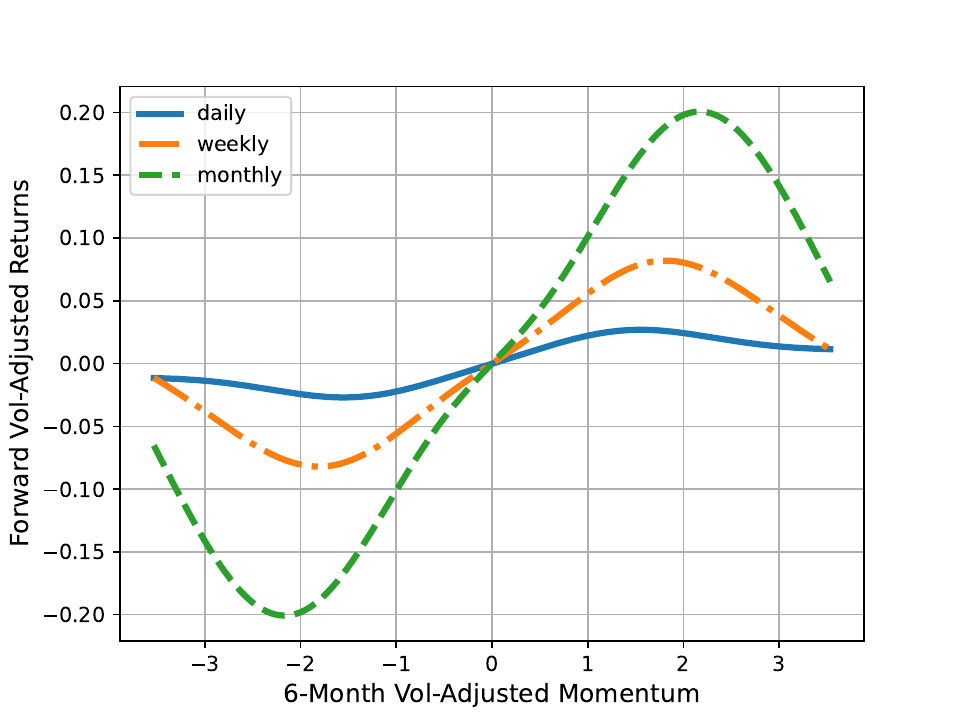}
    \caption{$6$-Month Momentum}
  \end{subfigure}
  \begin{subfigure}[b]{0.30\linewidth}
    \includegraphics[width=\linewidth ]{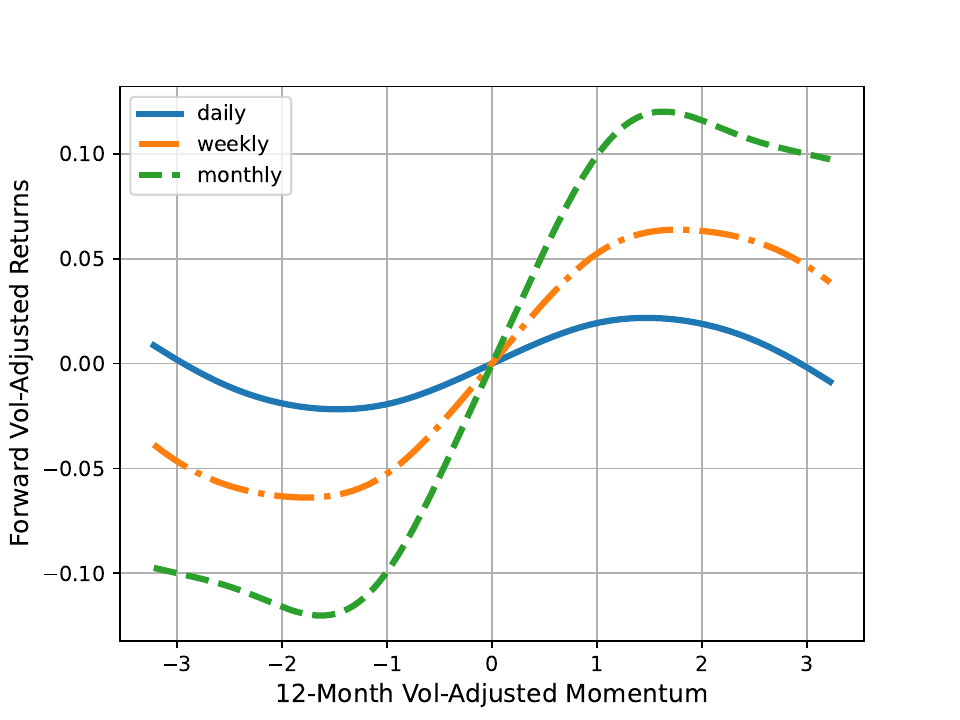}
    \caption{$12$-Month Momentum}
  \end{subfigure}
  
  \caption{Estimated forward vol-adjusted returns as a function of the $1$, $6$ and $12$-month time series momentum factor for Crude Oil Nonlinear models.
  Models are fit using all available training data from January $1950$ through June $2026$. }
\end{figure}

\noindent \textbf{Brent Crude}
\begin{figure}[H]
  \centering
  \begin{subfigure}[b]{0.30\linewidth}
    \includegraphics[width=\linewidth ]{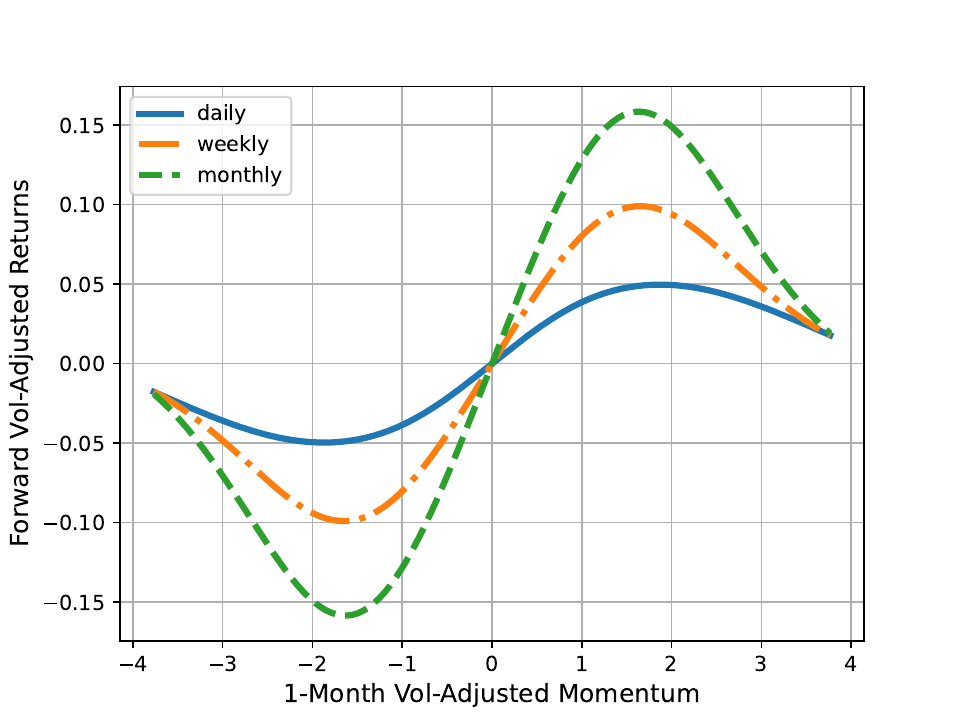}
    \caption{$1$-Month Momentum}
  \end{subfigure}
  \begin{subfigure}[b]{0.30\linewidth}
    \includegraphics[width=\linewidth ]{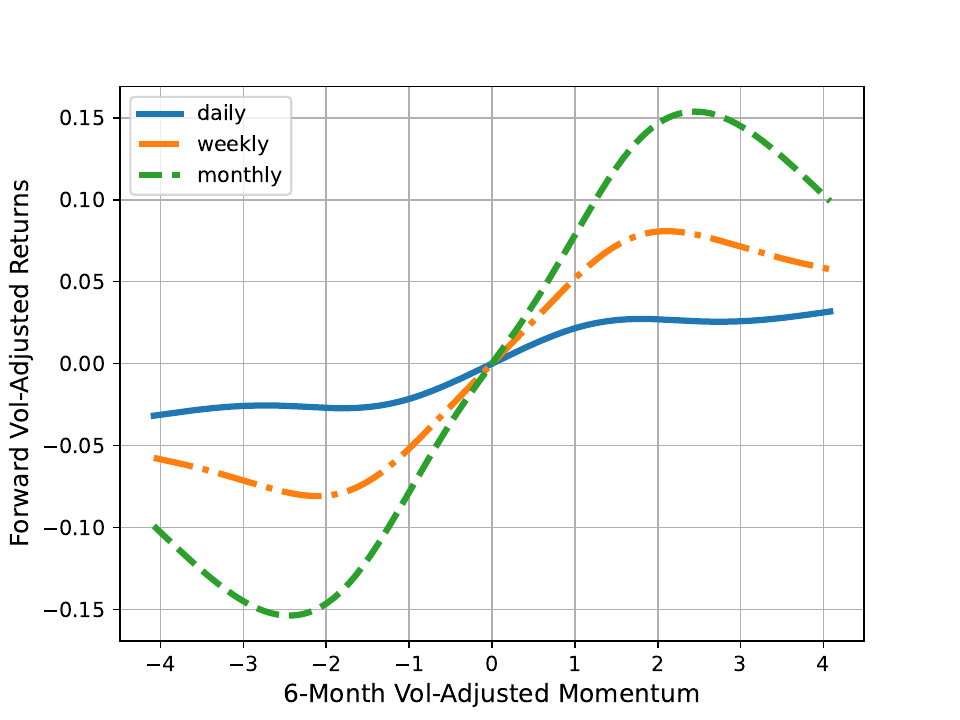}
    \caption{$6$-Month Momentum}
  \end{subfigure}
  \begin{subfigure}[b]{0.30\linewidth}
    \includegraphics[width=\linewidth ]{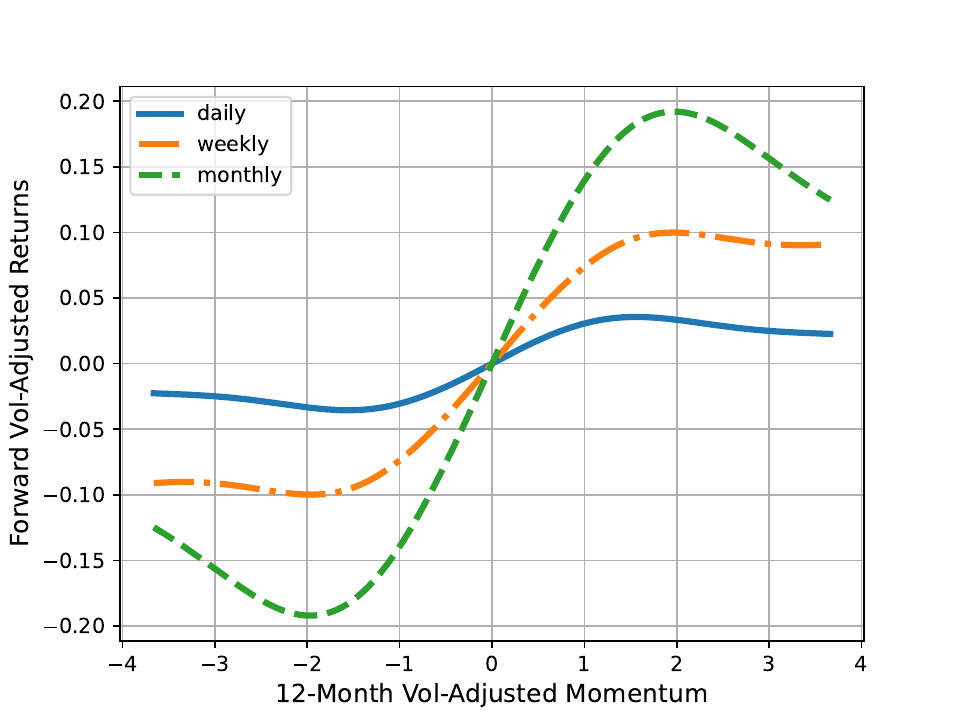}
    \caption{$12$-Month Momentum}
  \end{subfigure}
  
  \caption{Estimated forward vol-adjusted returns as a function of the $1$, $6$ and $12$-month time series momentum factor for Brent Crude Nonlinear models.
  Models are fit using all available training data from January $1950$ through June $2026$. }
\end{figure}

\noindent \textbf{Heating Oil}
\begin{figure}[H]
  \centering
  \begin{subfigure}[b]{0.30\linewidth}
    \includegraphics[width=\linewidth ]{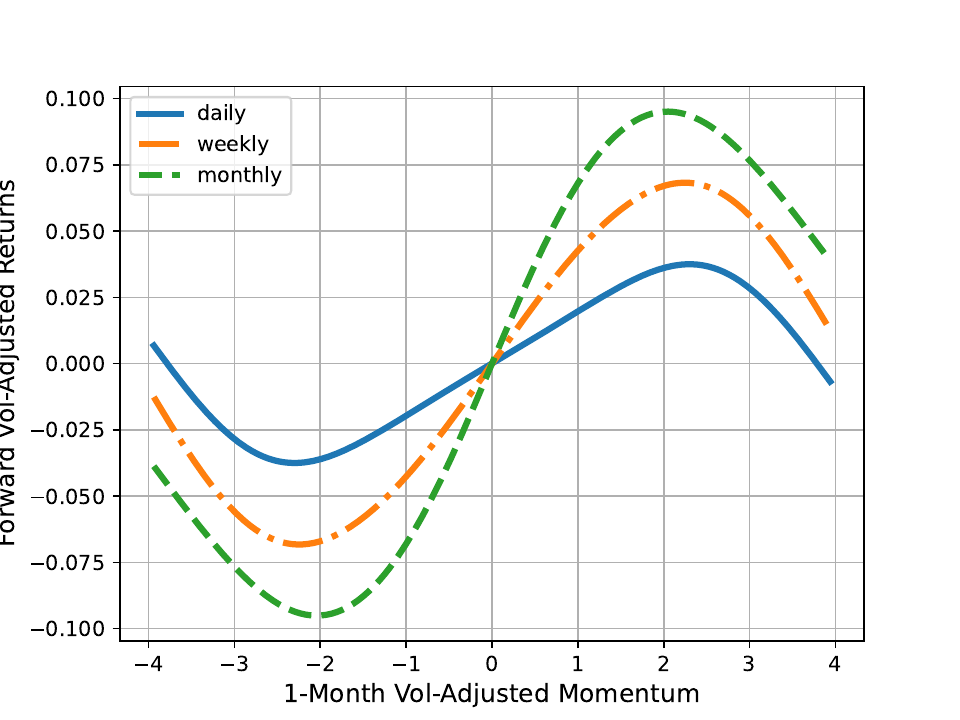}
    \caption{$1$-Month Momentum}
  \end{subfigure}
  \begin{subfigure}[b]{0.30\linewidth}
    \includegraphics[width=\linewidth ]{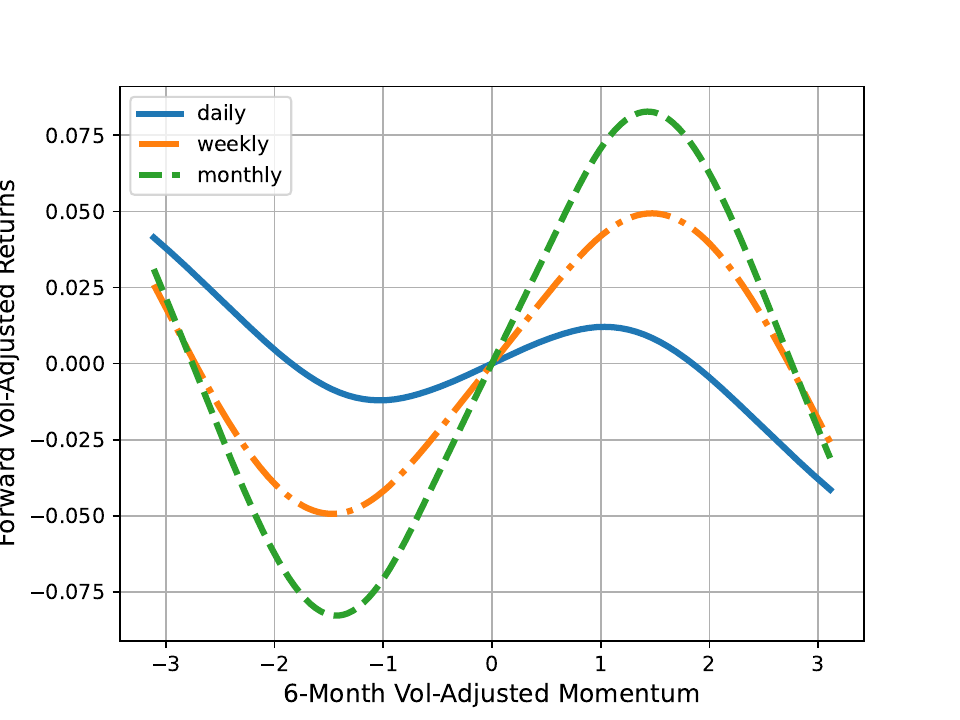}
    \caption{$6$-Month Momentum}
  \end{subfigure}
  \begin{subfigure}[b]{0.30\linewidth}
    \includegraphics[width=\linewidth ]{figs/model_energy_HO_momo_12_month.pdf}
    \caption{$12$-Month Momentum}
  \end{subfigure}
  
  \caption{Estimated forward vol-adjusted returns as a function of the $1$, $6$ and $12$-month time series momentum factor for Heating Oil Nonlinear models.
  Models are fit using all available training data from January $1950$ through June $2026$. }
\end{figure}

\noindent \textbf{Natural Gas}
\begin{figure}[H]
  \centering
  \begin{subfigure}[b]{0.30\linewidth}
    \includegraphics[width=\linewidth ]{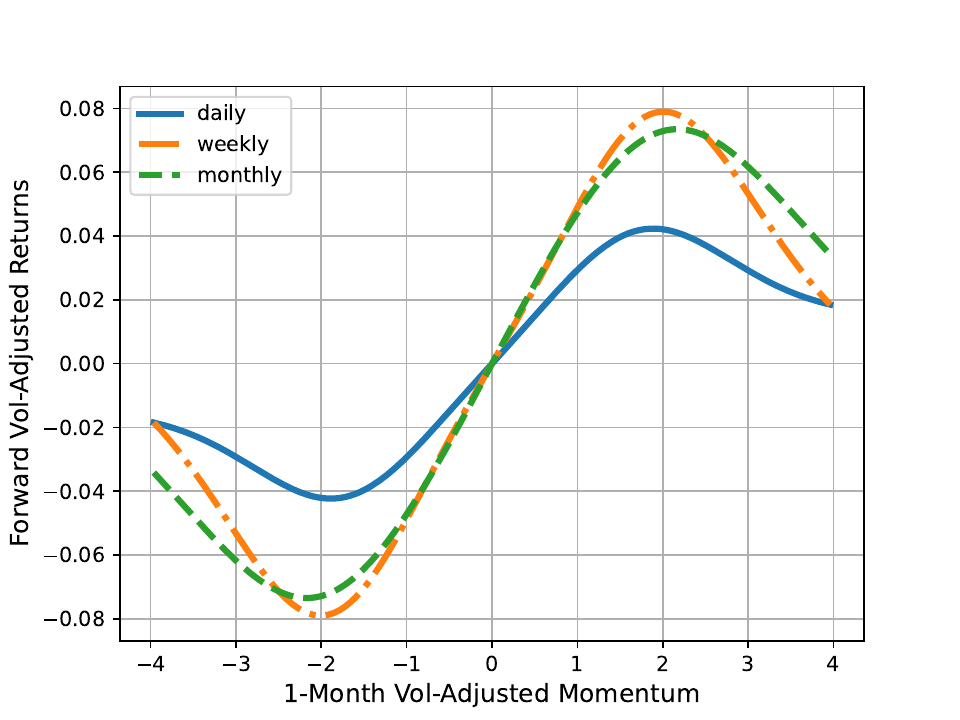}
    \caption{$1$-Month Momentum}
  \end{subfigure}
  \begin{subfigure}[b]{0.30\linewidth}
    \includegraphics[width=\linewidth ]{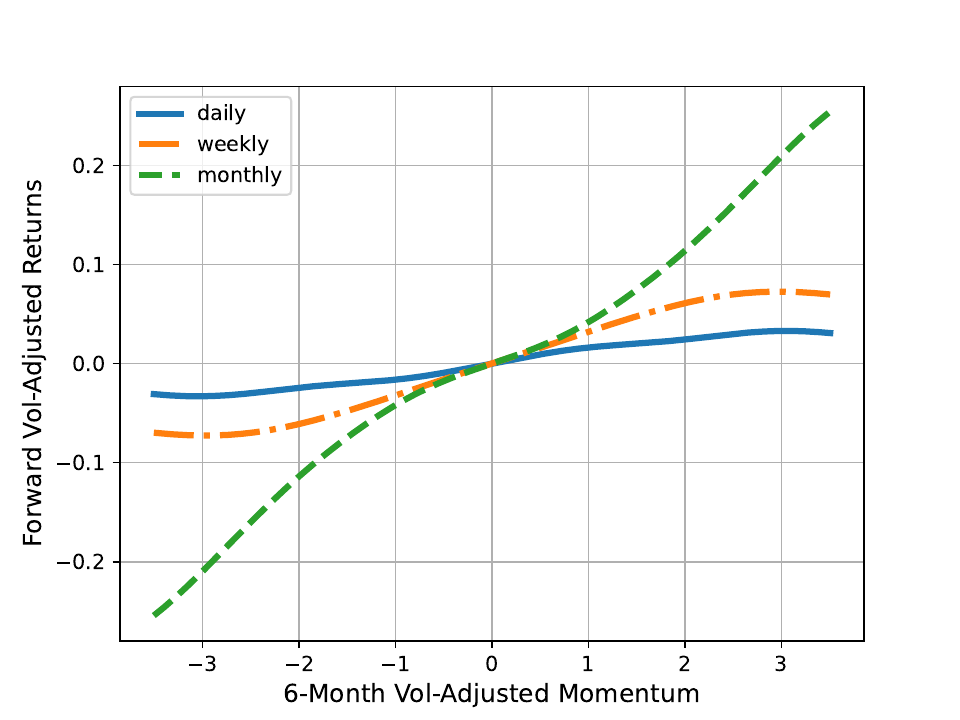}
    \caption{$6$-Month Momentum}
  \end{subfigure}
  \begin{subfigure}[b]{0.30\linewidth}
    \includegraphics[width=\linewidth ]{figs/model_energy_NG_momo_12_month.pdf}
    \caption{$12$-Month Momentum}
  \end{subfigure}
  
  \caption{Estimated forward vol-adjusted returns as a function of the $1$, $6$ and $12$-month time series momentum factor for Natural Gas Nonlinear models.
  Models are fit using all available training data from January $1950$ through June $2026$. }
\end{figure}

\noindent \textbf{Gasoil}
\begin{figure}[H]
  \centering
  \begin{subfigure}[b]{0.30\linewidth}
    \includegraphics[width=\linewidth ]{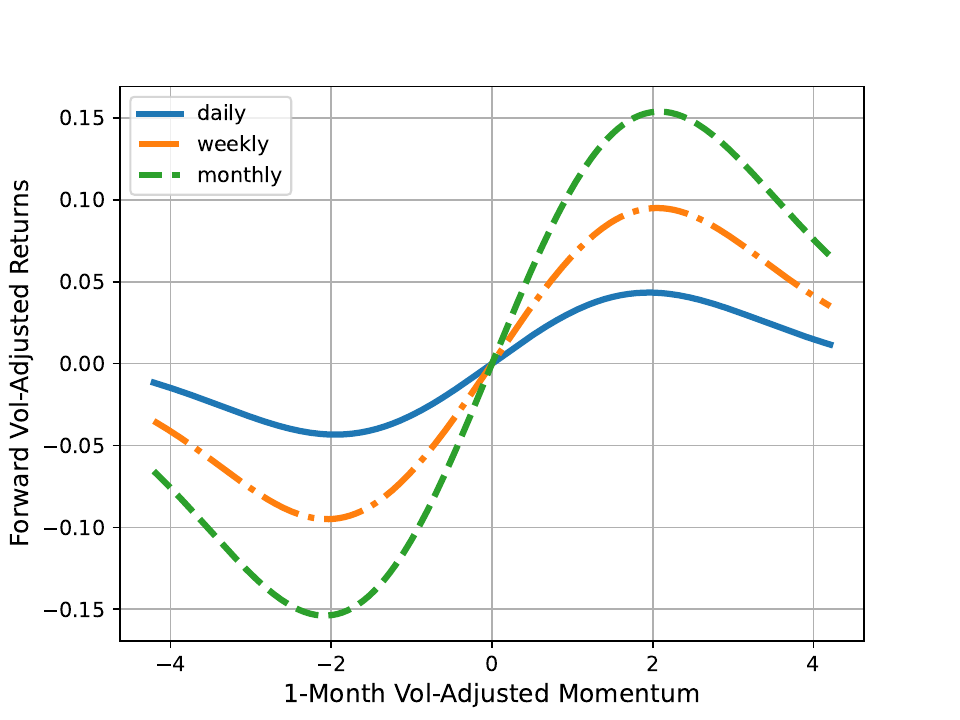}
    \caption{$1$-Month Momentum}
  \end{subfigure}
  \begin{subfigure}[b]{0.30\linewidth}
    \includegraphics[width=\linewidth ]{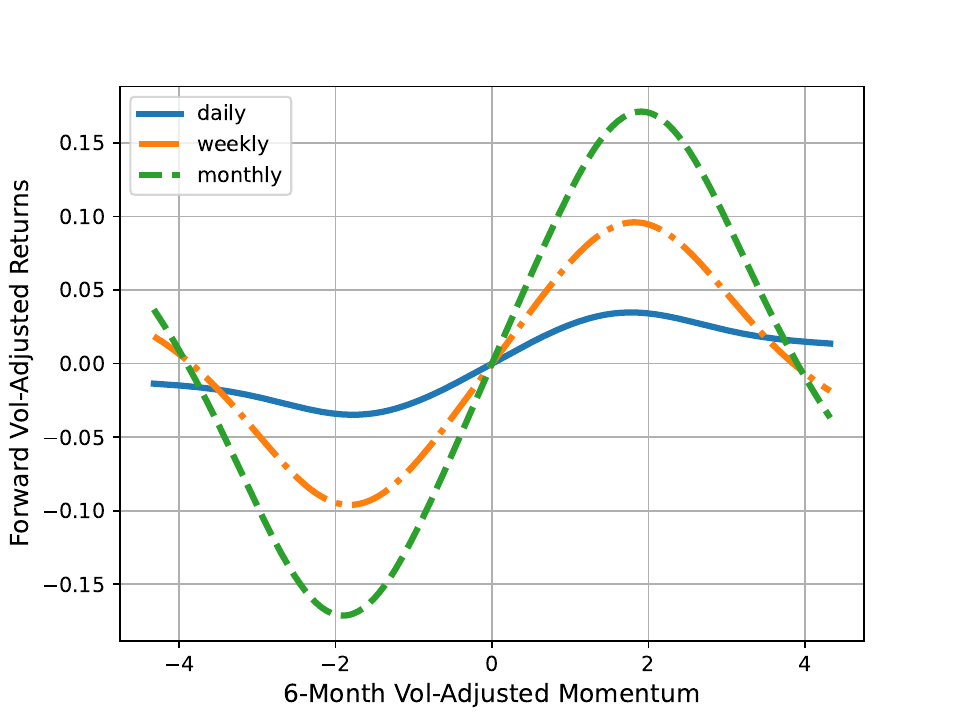}
    \caption{$6$-Month Momentum}
  \end{subfigure}
  \begin{subfigure}[b]{0.30\linewidth}
    \includegraphics[width=\linewidth ]{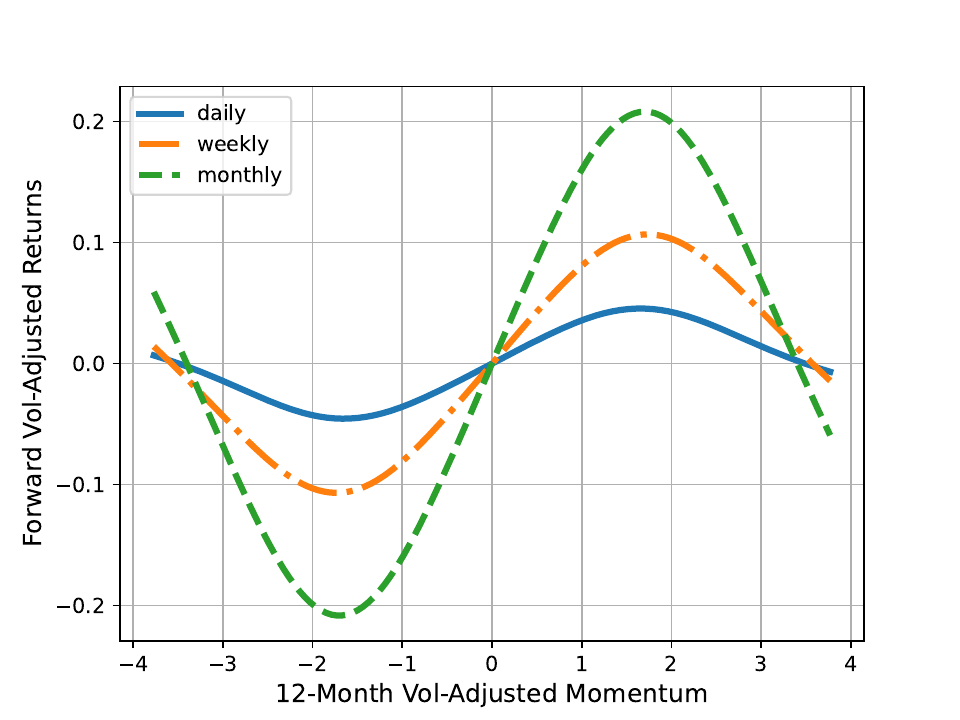}
    \caption{$12$-Month Momentum}
  \end{subfigure}
  
  \caption{Estimated forward vol-adjusted returns as a function of the $1$, $6$ and $12$-month time series momentum factor for Gasoil Nonlinear models.
  Models are fit using all available training data from January $1950$ through June $2026$. }
\end{figure}

\noindent \textbf{Gasoline}
\begin{figure}[H]
  \centering
  \begin{subfigure}[b]{0.30\linewidth}
    \includegraphics[width=\linewidth ]{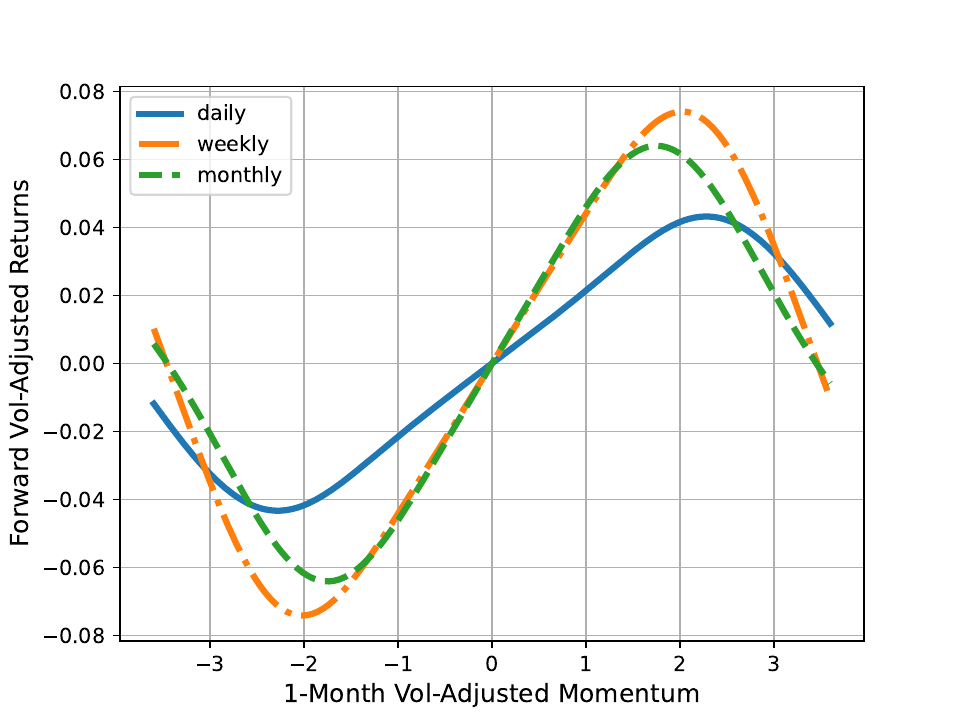}
    \caption{$1$-Month Momentum}
  \end{subfigure}
  \begin{subfigure}[b]{0.30\linewidth}
    \includegraphics[width=\linewidth ]{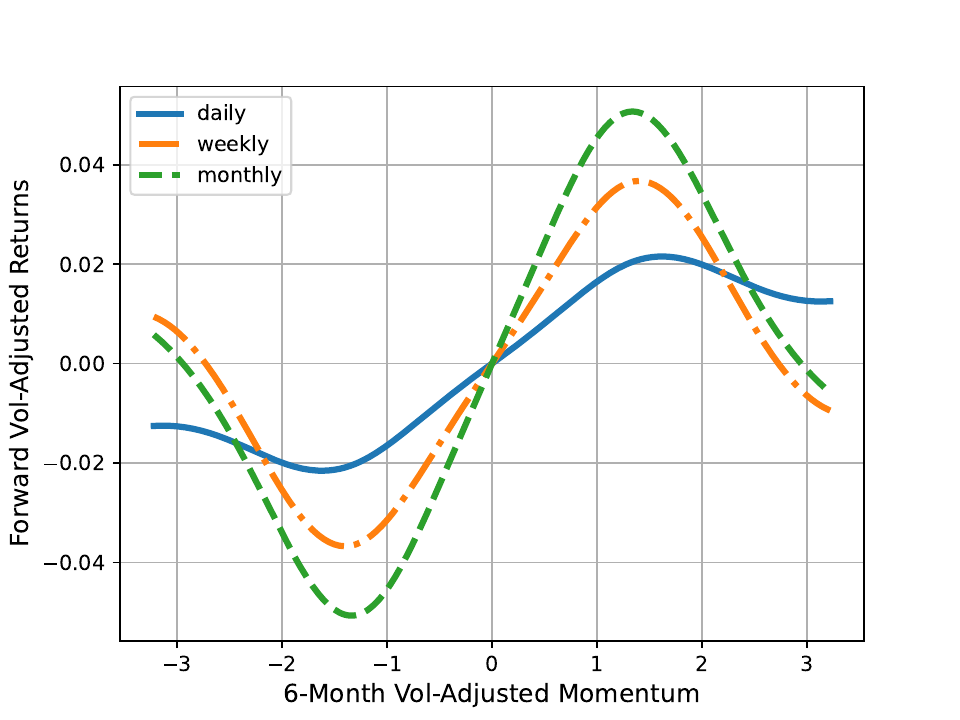}
    \caption{$6$-Month Momentum}
  \end{subfigure}
  \begin{subfigure}[b]{0.30\linewidth}
    \includegraphics[width=\linewidth ]{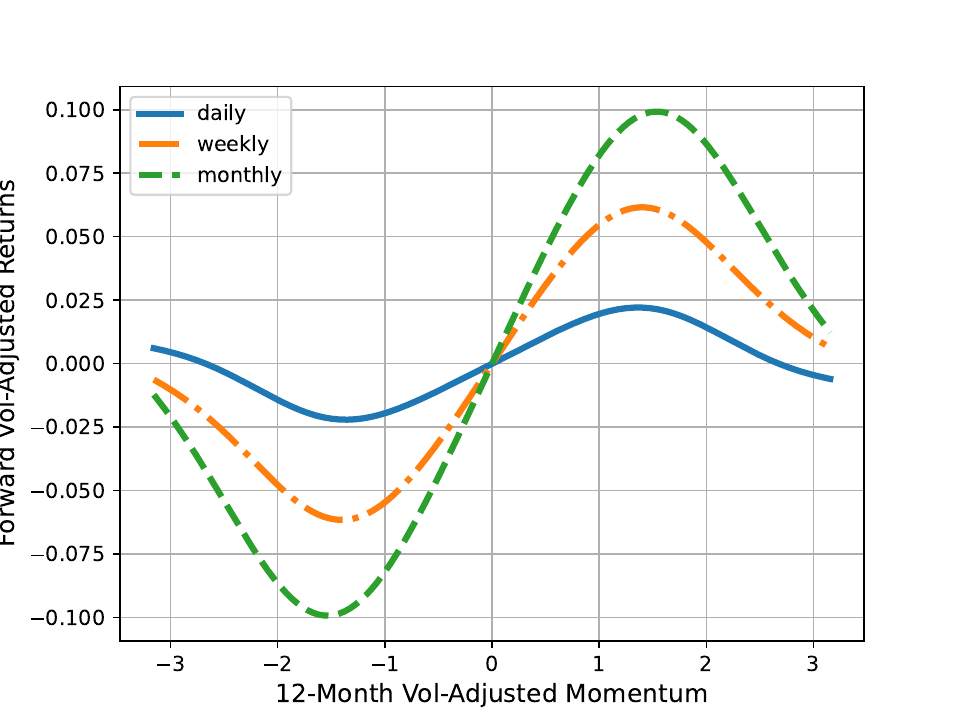}
    \caption{$12$-Month Momentum}
  \end{subfigure}
  
  \caption{Estimated forward vol-adjusted returns as a function of the $1$, $6$ and $12$-month time series momentum factor for Gasoline Nonlinear models.
  Models are fit using all available training data from January $1950$ through June $2026$. }
\end{figure}

\subsection{Equity Sector Nonlinear Models}
\begin{figure}[H]
  \centering
  \begin{subfigure}[b]{0.30\linewidth}
    \includegraphics[width=\linewidth ]{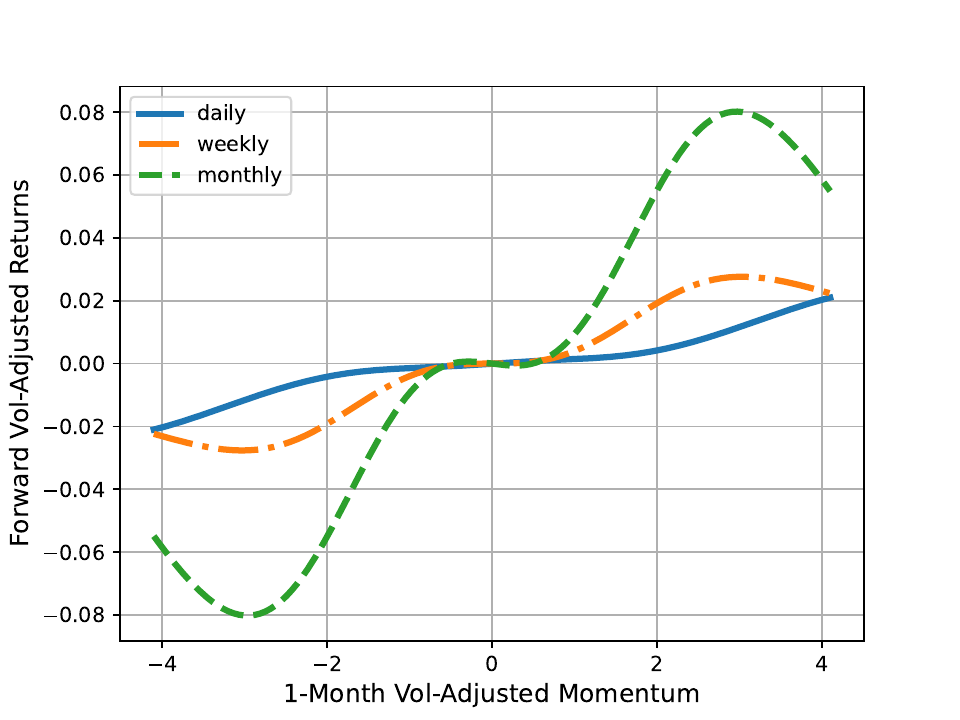}
    \caption{$1$-Month Momentum}
  \end{subfigure}
  \begin{subfigure}[b]{0.30\linewidth}
    \includegraphics[width=\linewidth ]{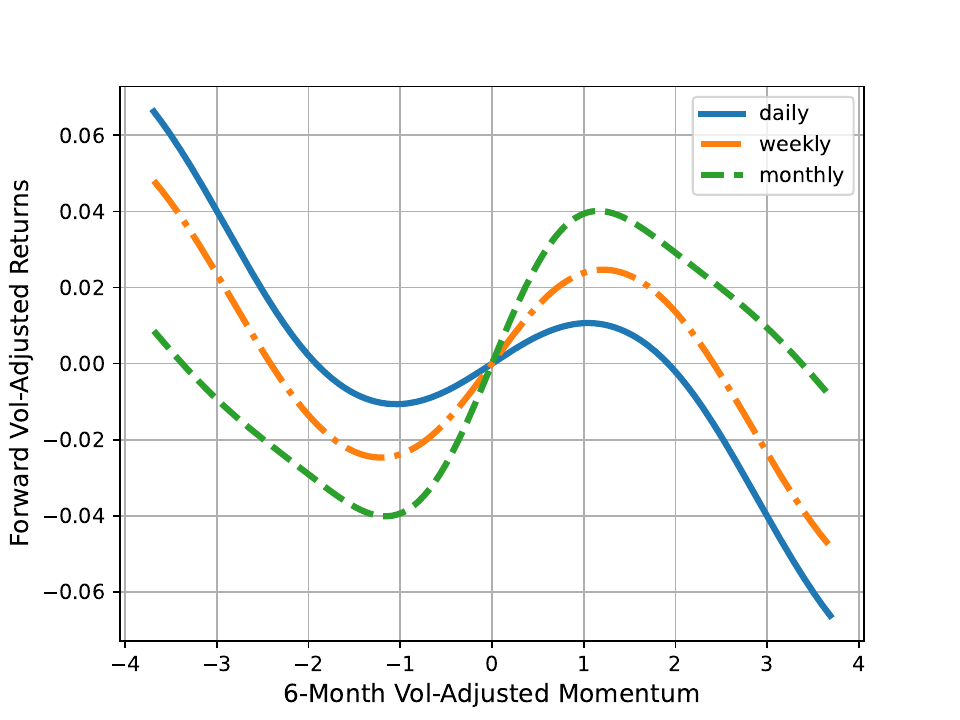}
    \caption{$6$-Month Momentum}
  \end{subfigure}
  \begin{subfigure}[b]{0.30\linewidth}
    \includegraphics[width=\linewidth ]{figs/model_equity_momo_12_month.pdf}
    \caption{$12$-Month Momentum}
  \end{subfigure}
  
  \caption{Estimated forward vol-adjusted returns as a function of the $1$, $6$ and $12$-month time series momentum factor for Equity Sector Nonlinear models.
  Models are fit using all available training data from January $1950$ through June $2026$. }
\end{figure}

\noindent \textbf{CAC 40}
\begin{figure}[H]
  \centering
  \begin{subfigure}[b]{0.30\linewidth}
    \includegraphics[width=\linewidth ]{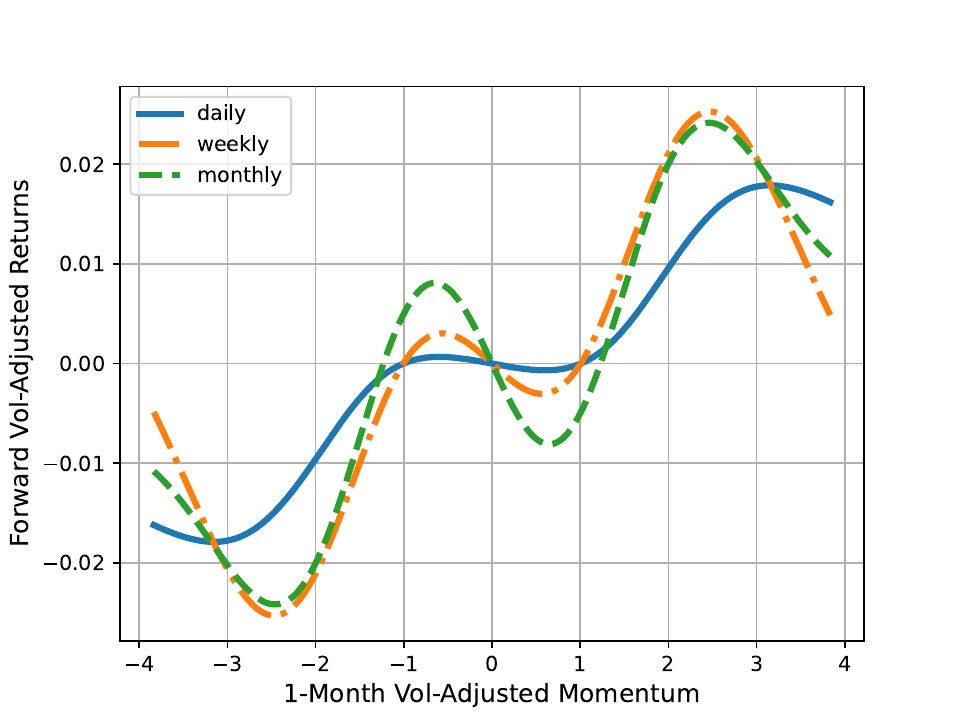}
    \caption{$1$-Month Momentum}
  \end{subfigure}
  \begin{subfigure}[b]{0.30\linewidth}
    \includegraphics[width=\linewidth ]{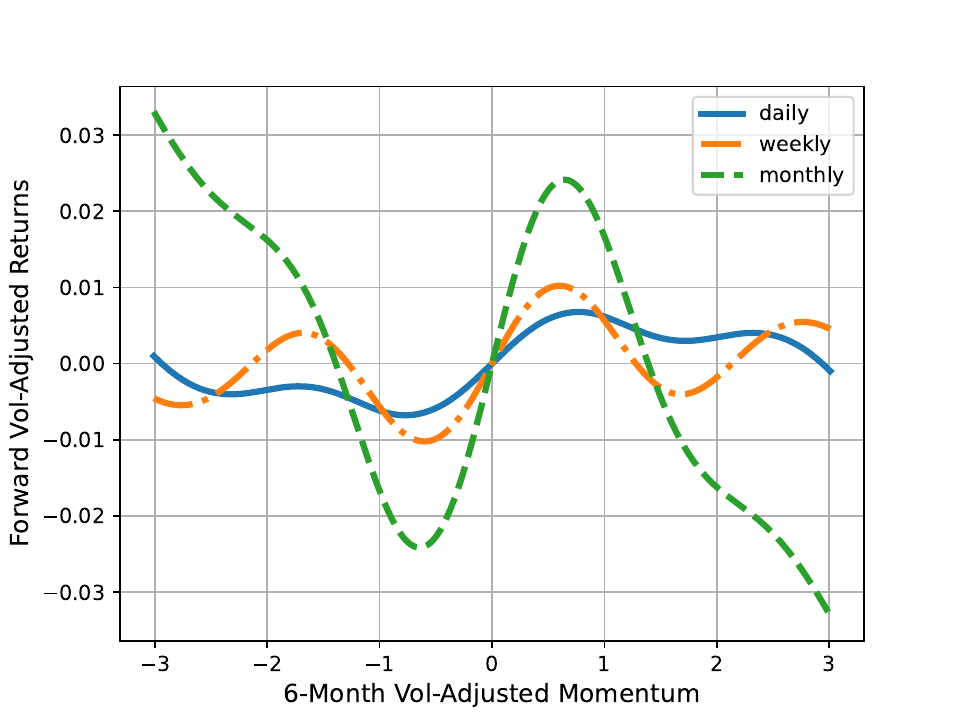}
    \caption{$6$-Month Momentum}
  \end{subfigure}
  \begin{subfigure}[b]{0.30\linewidth}
    \includegraphics[width=\linewidth ]{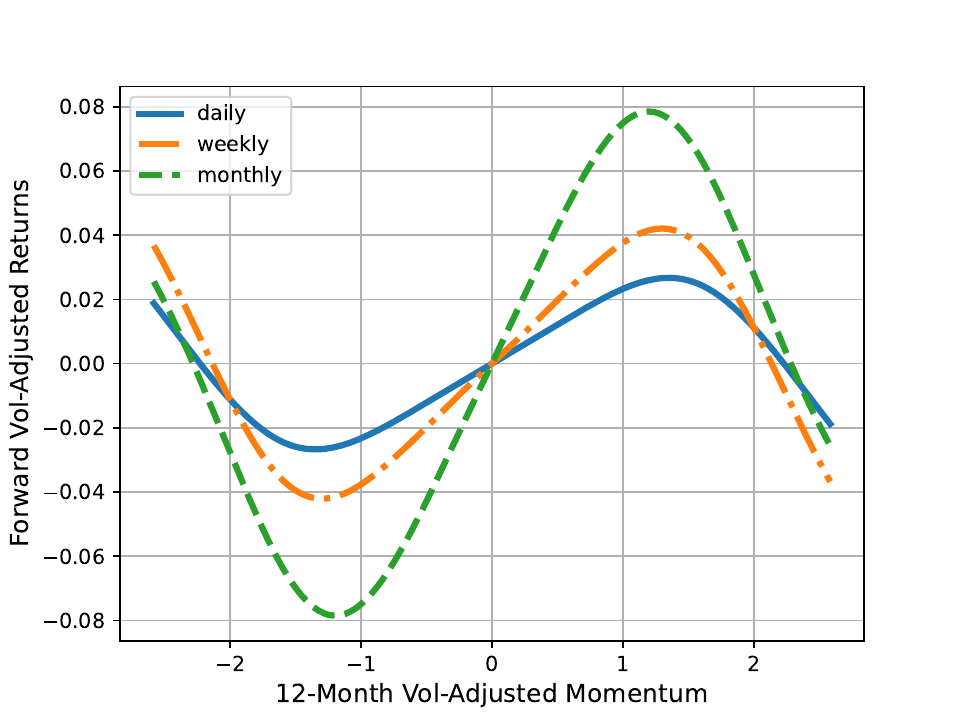}
    \caption{$12$-Month Momentum}
  \end{subfigure}
  
  \caption{Estimated forward vol-adjusted returns as a function of the $1$, $6$ and $12$-month time series momentum factor for CAC 40 Nonlinear models.
  Models are fit using all available training data from January $1950$ through June $2026$. }
\end{figure}

\noindent \textbf{S\&P 500}
\begin{figure}[H]
  \centering
  \begin{subfigure}[b]{0.30\linewidth}
    \includegraphics[width=\linewidth ]{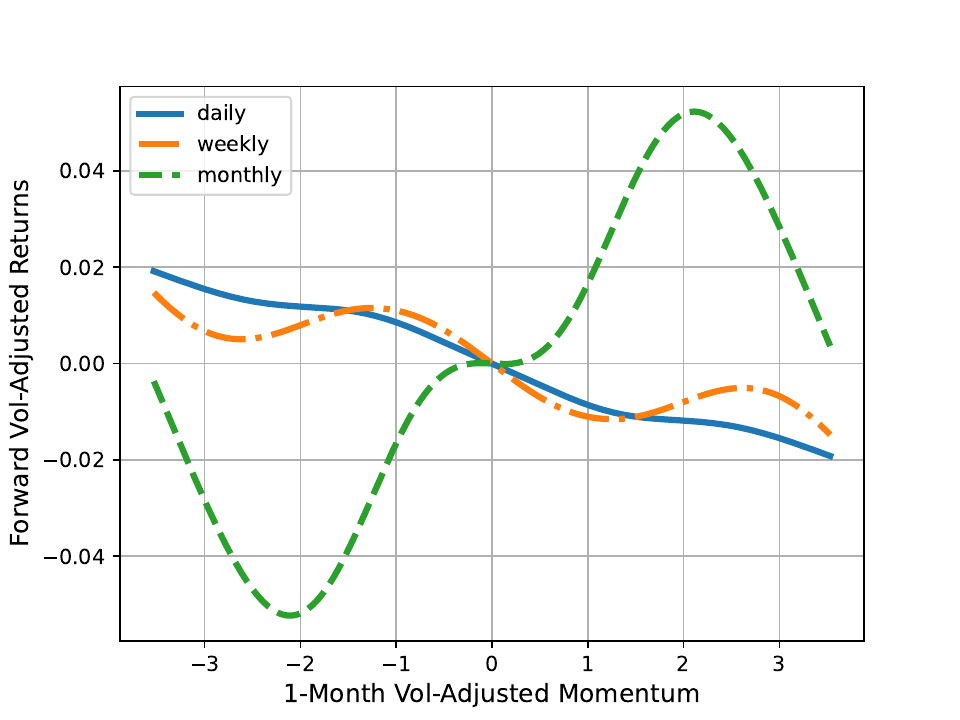}
    \caption{$1$-Month Momentum}
  \end{subfigure}
  \begin{subfigure}[b]{0.30\linewidth}
    \includegraphics[width=\linewidth ]{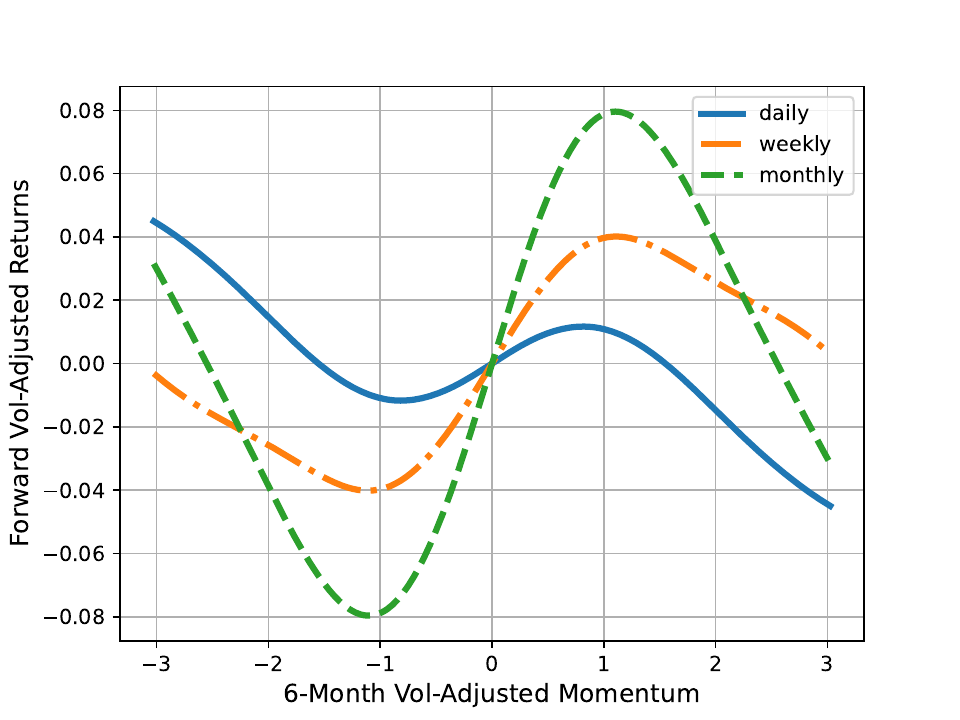}
    \caption{$6$-Month Momentum}
  \end{subfigure}
  \begin{subfigure}[b]{0.30\linewidth}
    \includegraphics[width=\linewidth ]{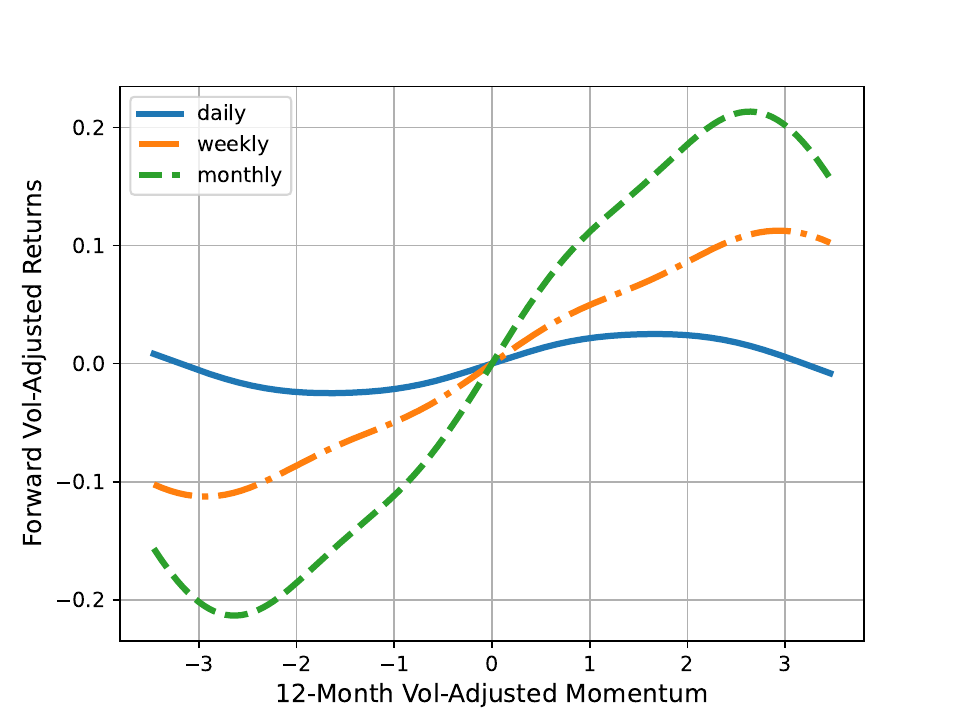}
    \caption{$12$-Month Momentum}
  \end{subfigure}
  
  \caption{Estimated forward vol-adjusted returns as a function of the $1$, $6$ and $12$-month time series momentum factor for S\&P 500 Nonlinear models.
  Models are fit using all available training data from January $1950$ through June $2026$. }
\end{figure}

\noindent \textbf{DAX}
\begin{figure}[H]
  \centering
  \begin{subfigure}[b]{0.30\linewidth}
    \includegraphics[width=\linewidth ]{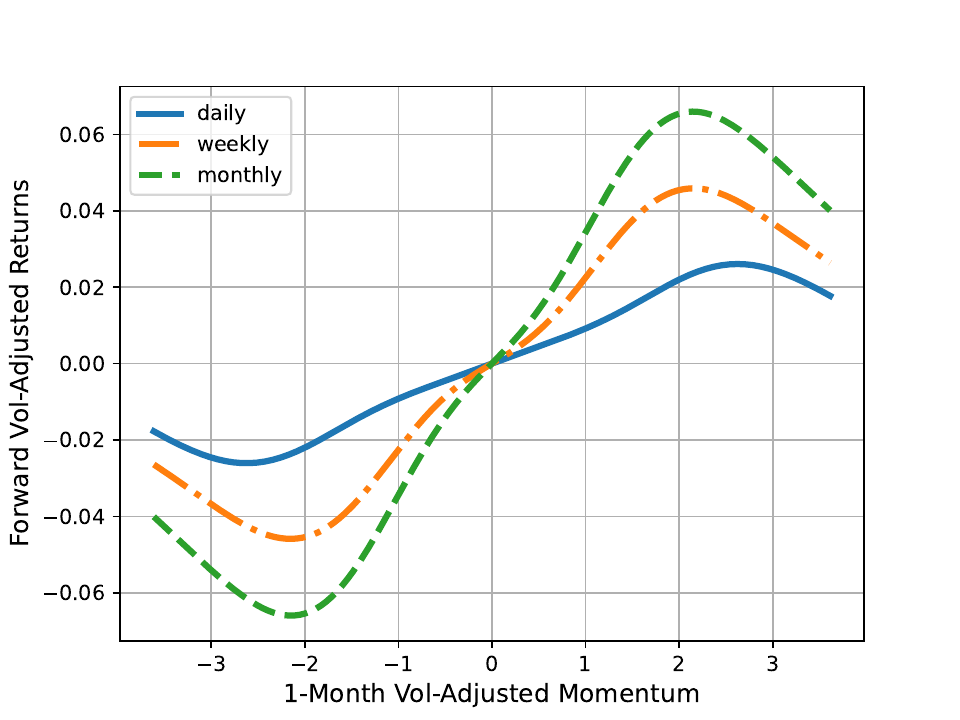}
    \caption{$1$-Month Momentum}
  \end{subfigure}
  \begin{subfigure}[b]{0.30\linewidth}
    \includegraphics[width=\linewidth ]{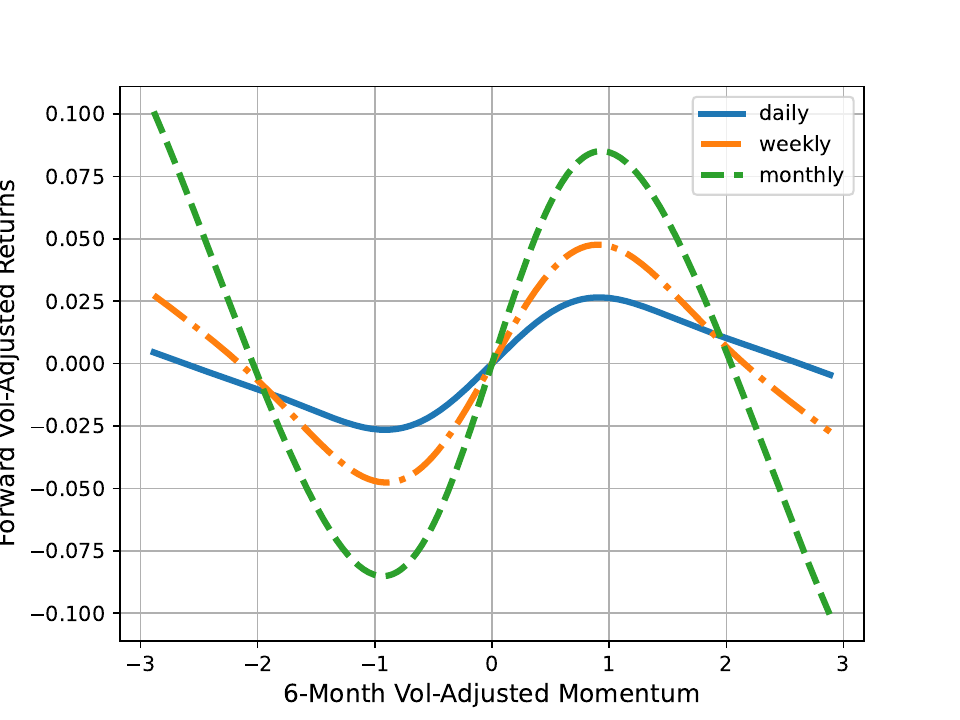}
    \caption{$6$-Month Momentum}
  \end{subfigure}
  \begin{subfigure}[b]{0.30\linewidth}
    \includegraphics[width=\linewidth ]{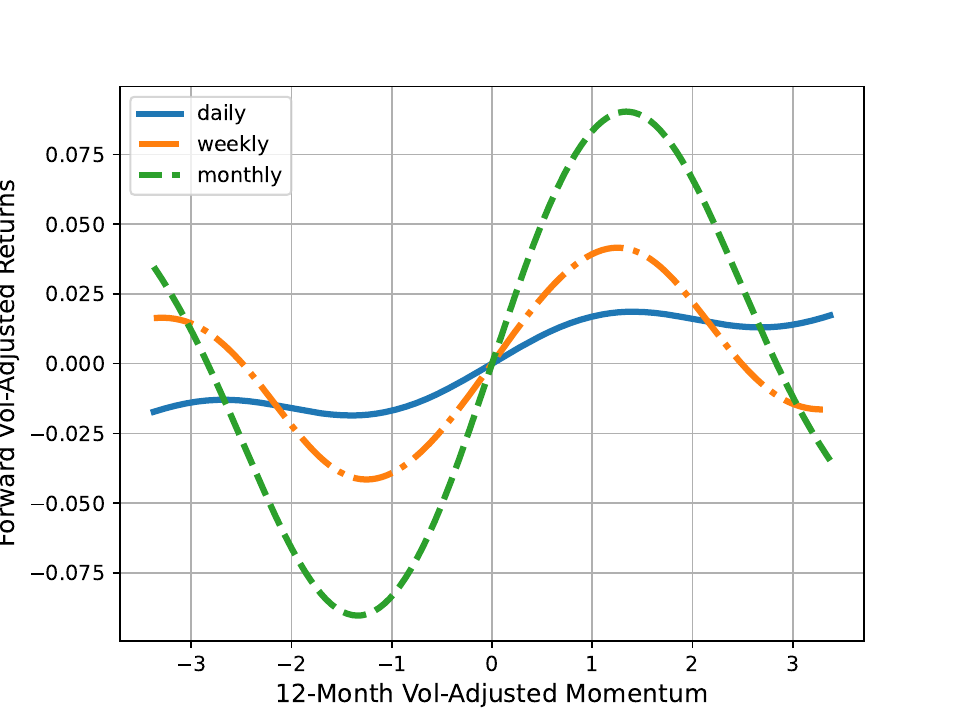}
    \caption{$12$-Month Momentum}
  \end{subfigure}
  
  \caption{Estimated forward vol-adjusted returns as a function of the $1$, $6$ and $12$-month time series momentum factor for DAX Nonlinear models.
  Models are fit using all available training data from January $1950$ through June $2026$. }
\end{figure}

\noindent \textbf{Hang Seng}
\begin{figure}[H]
  \centering
  \begin{subfigure}[b]{0.30\linewidth}
    \includegraphics[width=\linewidth ]{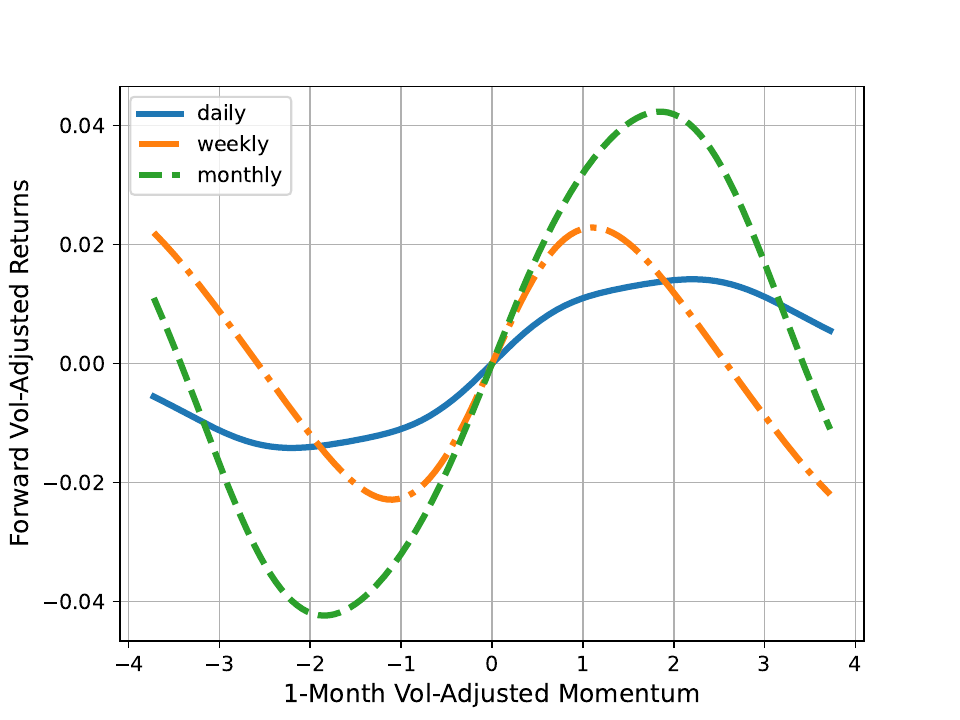}
    \caption{$1$-Month Momentum}
  \end{subfigure}
  \begin{subfigure}[b]{0.30\linewidth}
    \includegraphics[width=\linewidth ]{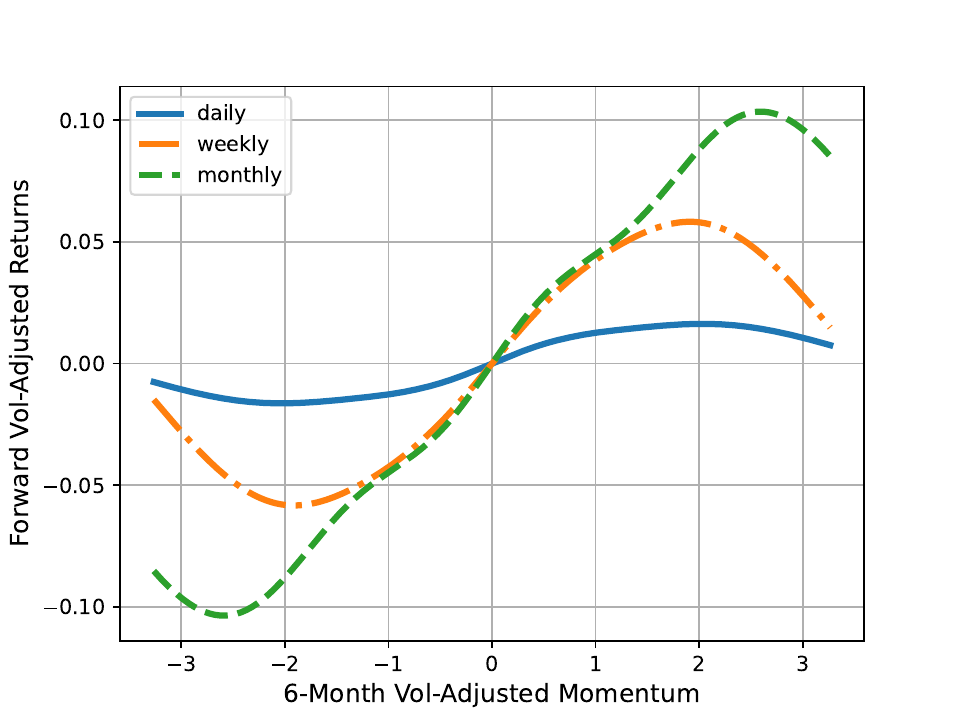}
    \caption{$6$-Month Momentum}
  \end{subfigure}
  \begin{subfigure}[b]{0.30\linewidth}
    \includegraphics[width=\linewidth ]{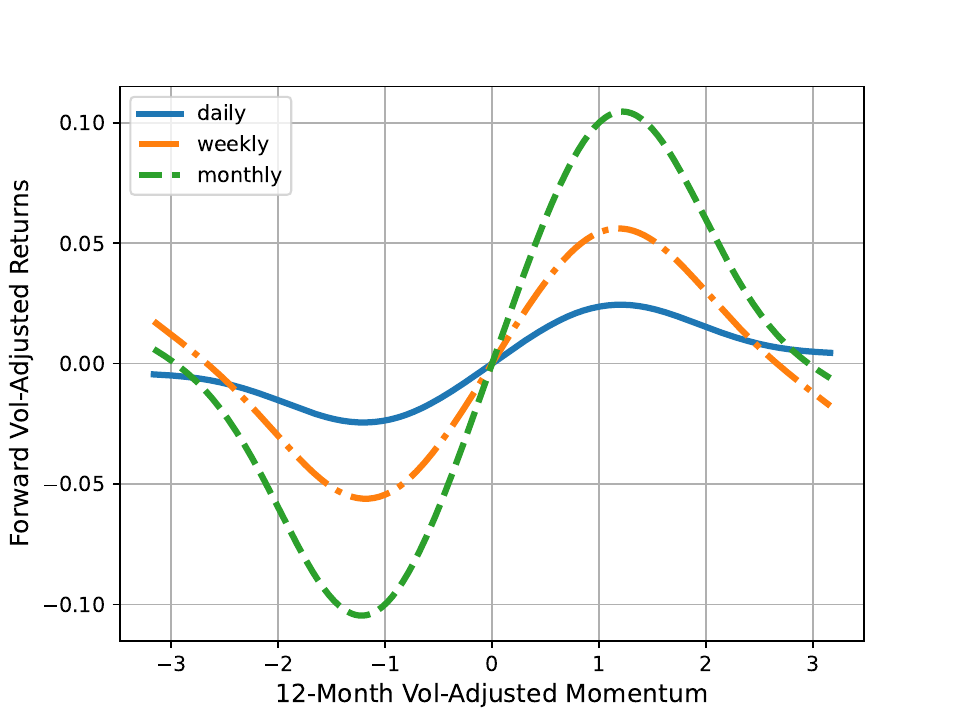}
    \caption{$12$-Month Momentum}
  \end{subfigure}
  
  \caption{Estimated forward vol-adjusted returns as a function of the $1$, $6$ and $12$-month time series momentum factor for Hang Seng Nonlinear models.
  Models are fit using all available training data from January $1950$ through June $2026$. }
\end{figure}

\noindent \textbf{Nikkei}
\begin{figure}[H]
  \centering
  \begin{subfigure}[b]{0.30\linewidth}
    \includegraphics[width=\linewidth ]{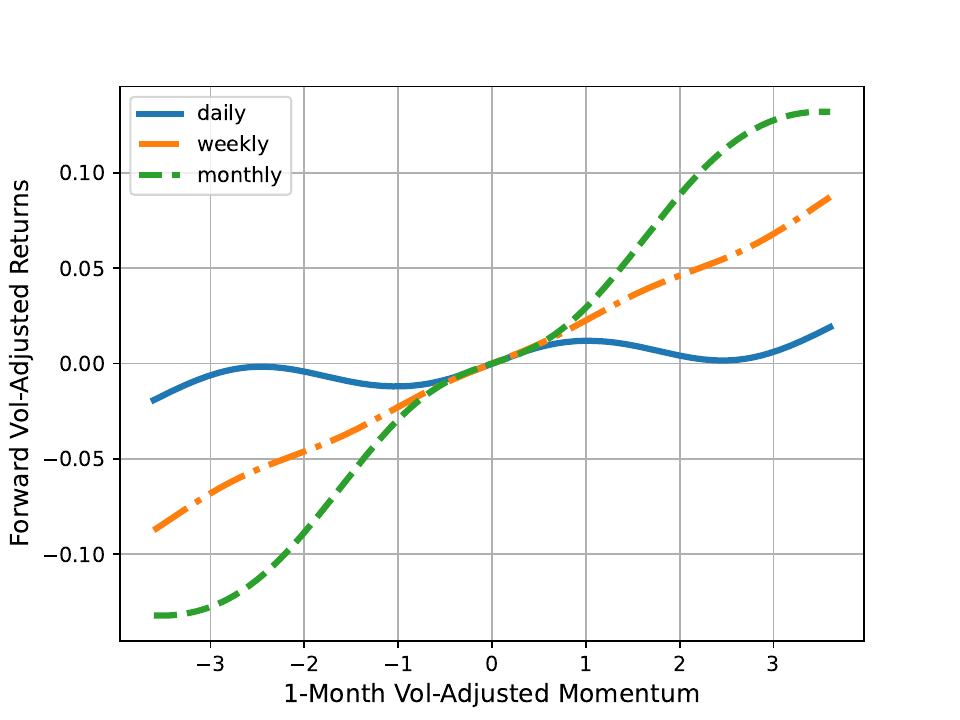}
    \caption{$1$-Month Momentum}
  \end{subfigure}
  \begin{subfigure}[b]{0.30\linewidth}
    \includegraphics[width=\linewidth ]{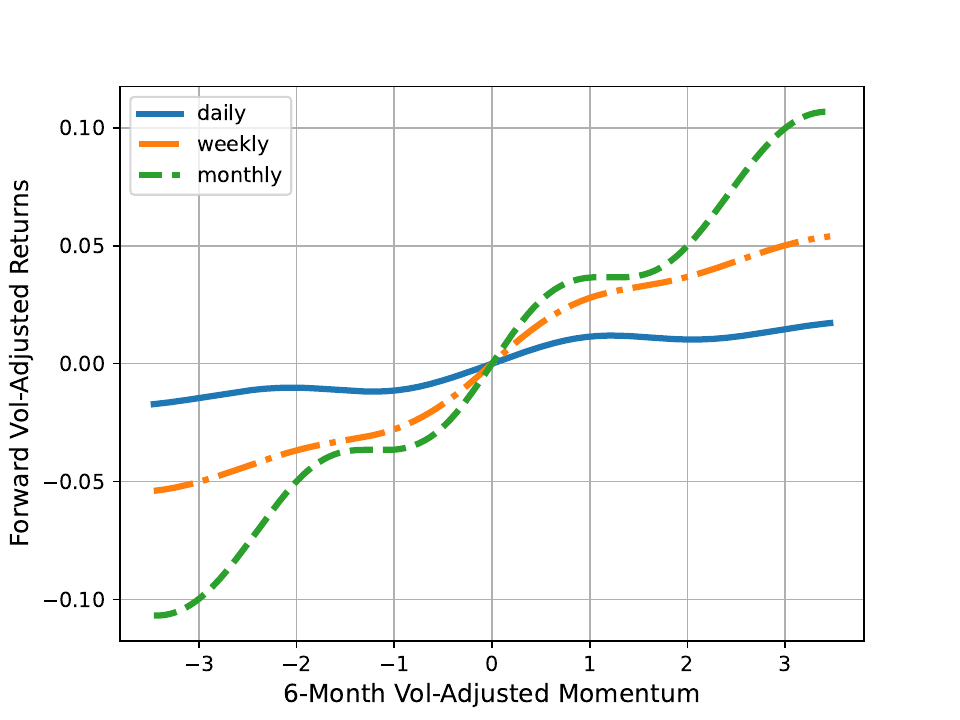}
    \caption{$6$-Month Momentum}
  \end{subfigure}
  \begin{subfigure}[b]{0.30\linewidth}
    \includegraphics[width=\linewidth ]{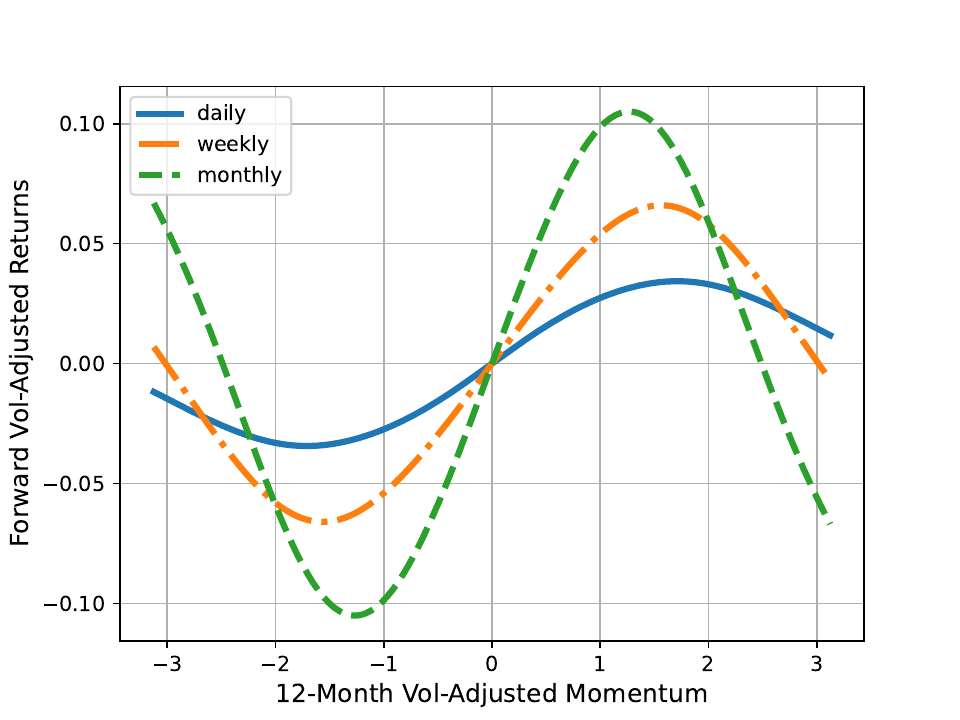}
    \caption{$12$-Month Momentum}
  \end{subfigure}
  
  \caption{Estimated forward vol-adjusted returns as a function of the $1$, $6$ and $12$-month time series momentum factor for Nikkei Nonlinear models.
  Models are fit using all available training data from January $1950$ through June $2026$. }
\end{figure}

\noindent \textbf{TSX 60}
\begin{figure}[H]
  \centering
  \begin{subfigure}[b]{0.30\linewidth}
    \includegraphics[width=\linewidth ]{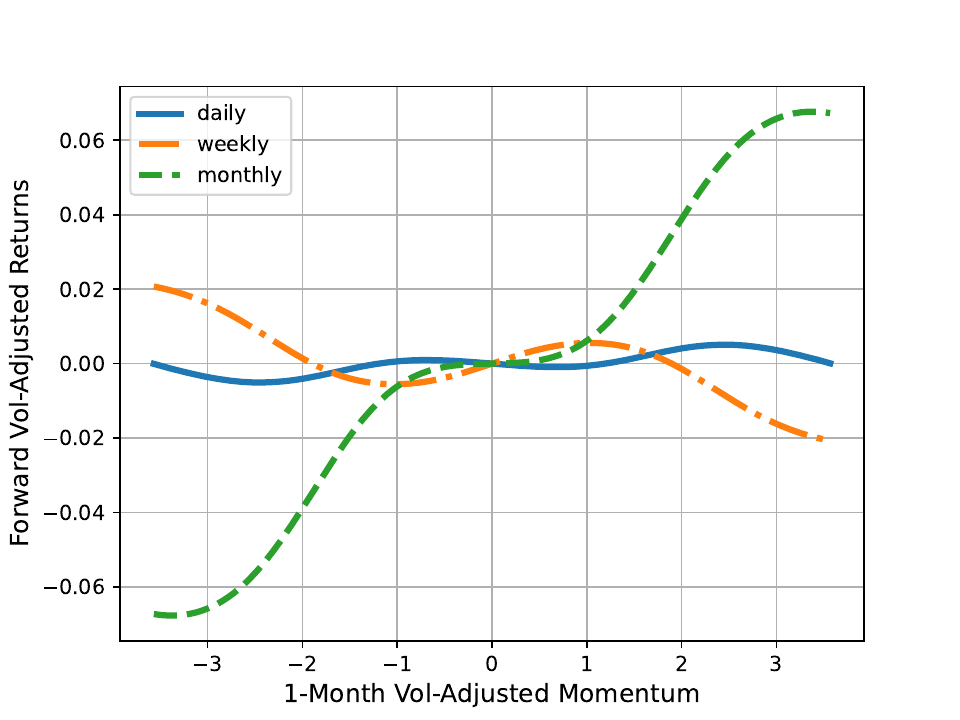}
    \caption{$1$-Month Momentum}
  \end{subfigure}
  \begin{subfigure}[b]{0.30\linewidth}
    \includegraphics[width=\linewidth ]{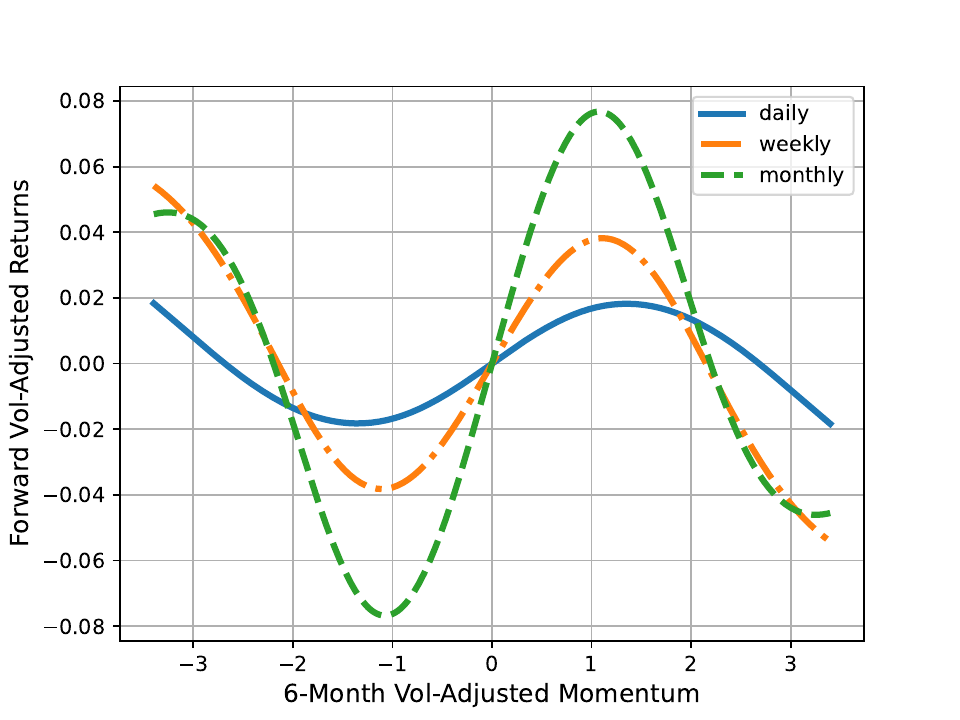}
    \caption{$6$-Month Momentum}
  \end{subfigure}
  \begin{subfigure}[b]{0.30\linewidth}
    \includegraphics[width=\linewidth ]{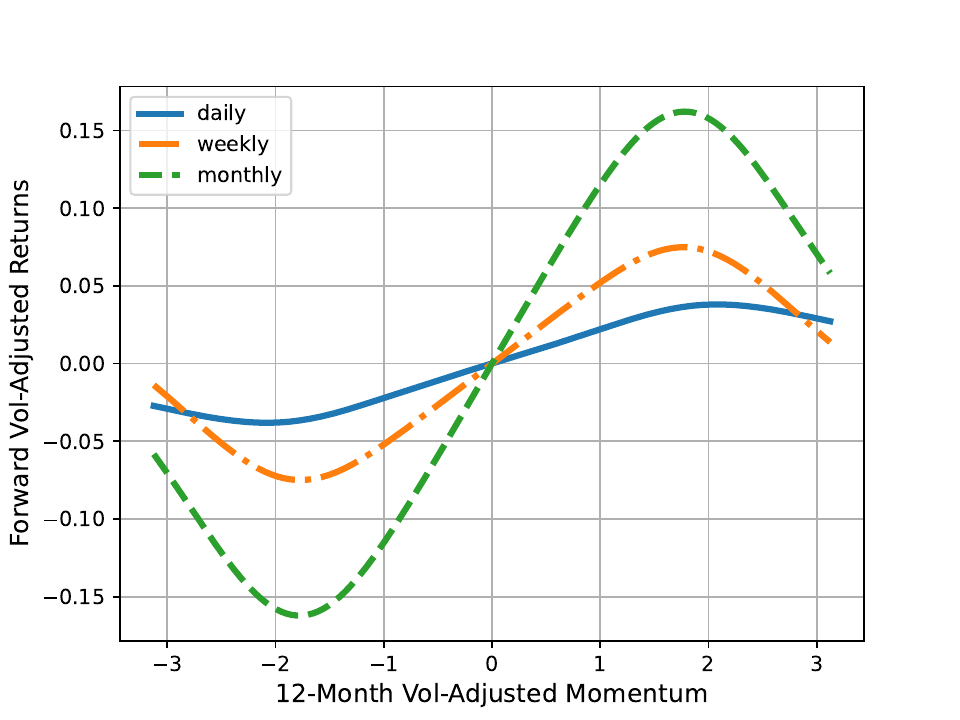}
    \caption{$12$-Month Momentum}
  \end{subfigure}
  
  \caption{Estimated forward vol-adjusted returns as a function of the $1$, $6$ and $12$-month time series momentum factor for TSX 60 Nonlinear models.
  Models are fit using all available training data from January $1950$ through June $2026$. }
\end{figure}

\noindent \textbf{ASX 200}
\begin{figure}[H]
  \centering
  \begin{subfigure}[b]{0.30\linewidth}
    \includegraphics[width=\linewidth ]{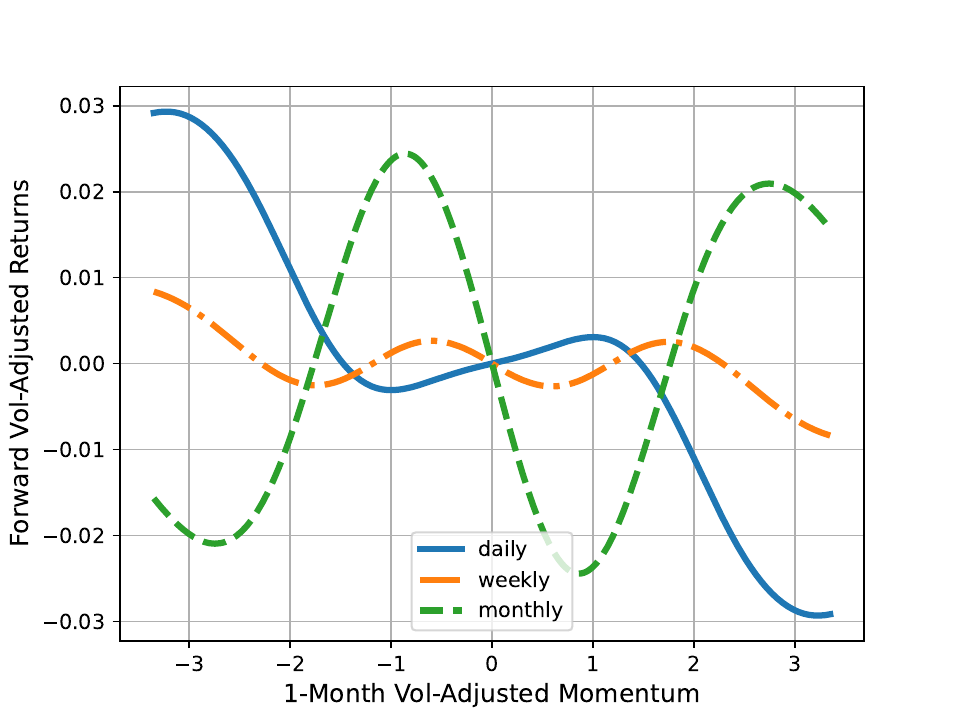}
    \caption{$1$-Month Momentum}
  \end{subfigure}
  \begin{subfigure}[b]{0.30\linewidth}
    \includegraphics[width=\linewidth ]{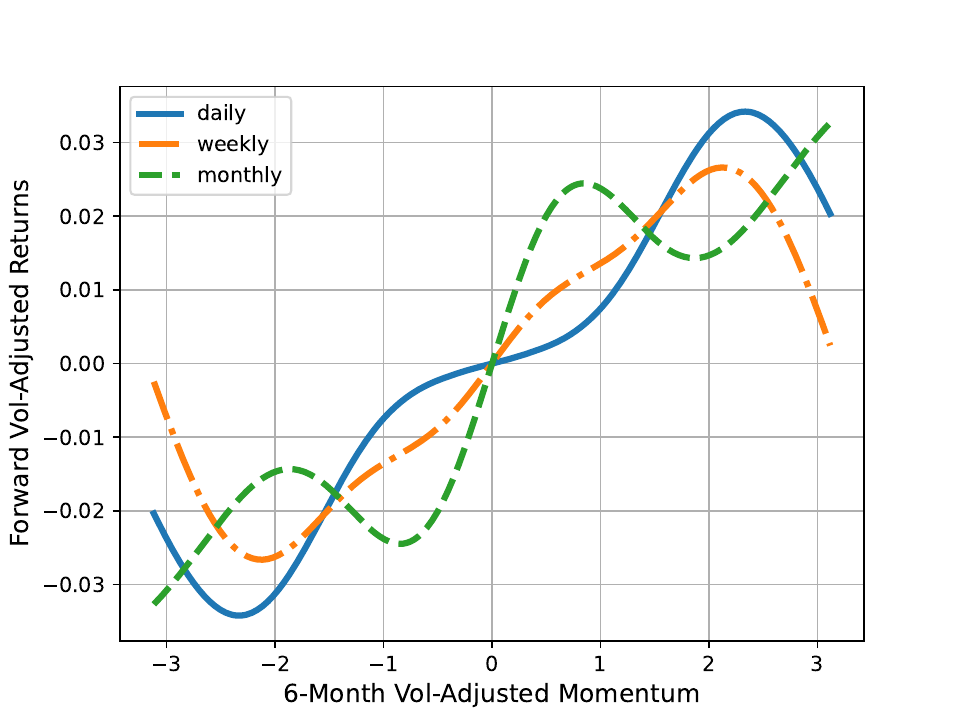}
    \caption{$6$-Month Momentum}
  \end{subfigure}
  \begin{subfigure}[b]{0.30\linewidth}
    \includegraphics[width=\linewidth ]{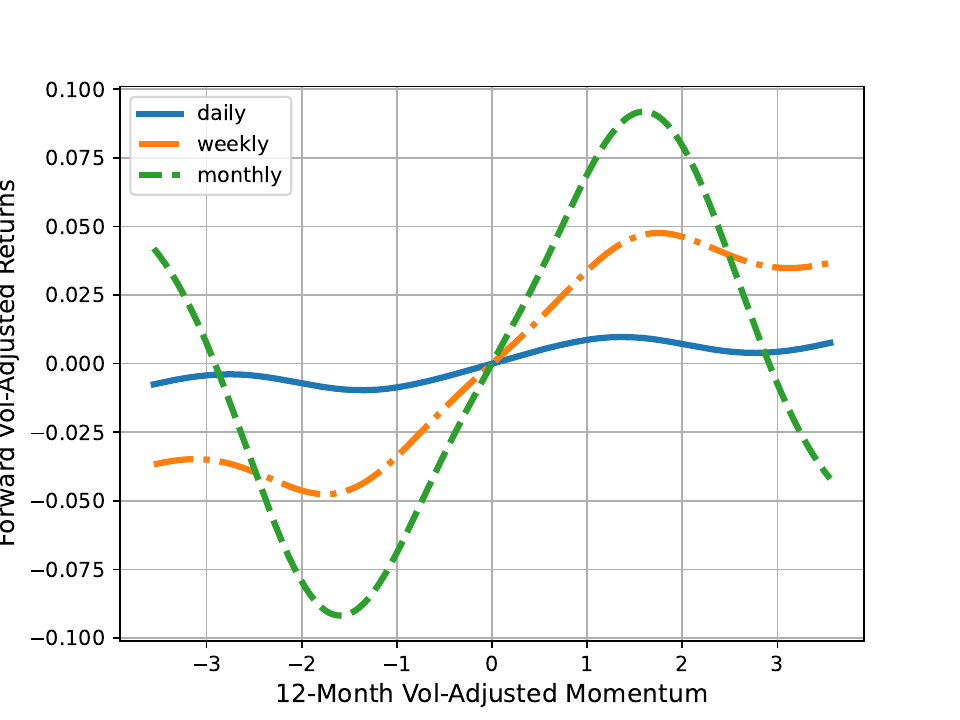}
    \caption{$12$-Month Momentum}
  \end{subfigure}
  
  \caption{Estimated forward vol-adjusted returns as a function of the $1$, $6$ and $12$-month time series momentum factor for ASX 200 Nonlinear models.
  Models are fit using all available training data from January $1950$ through June $2026$. }
\end{figure}

\noindent \textbf{FTSE 100}
\begin{figure}[H]
  \centering
  \begin{subfigure}[b]{0.30\linewidth}
    \includegraphics[width=\linewidth ]{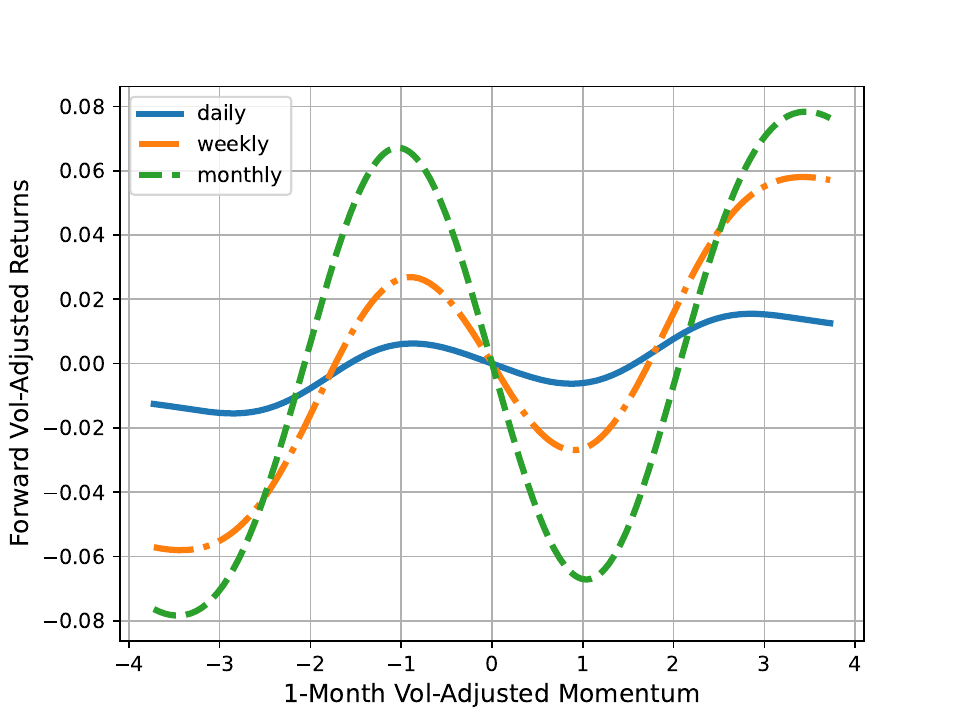}
    \caption{$1$-Month Momentum}
  \end{subfigure}
  \begin{subfigure}[b]{0.30\linewidth}
    \includegraphics[width=\linewidth ]{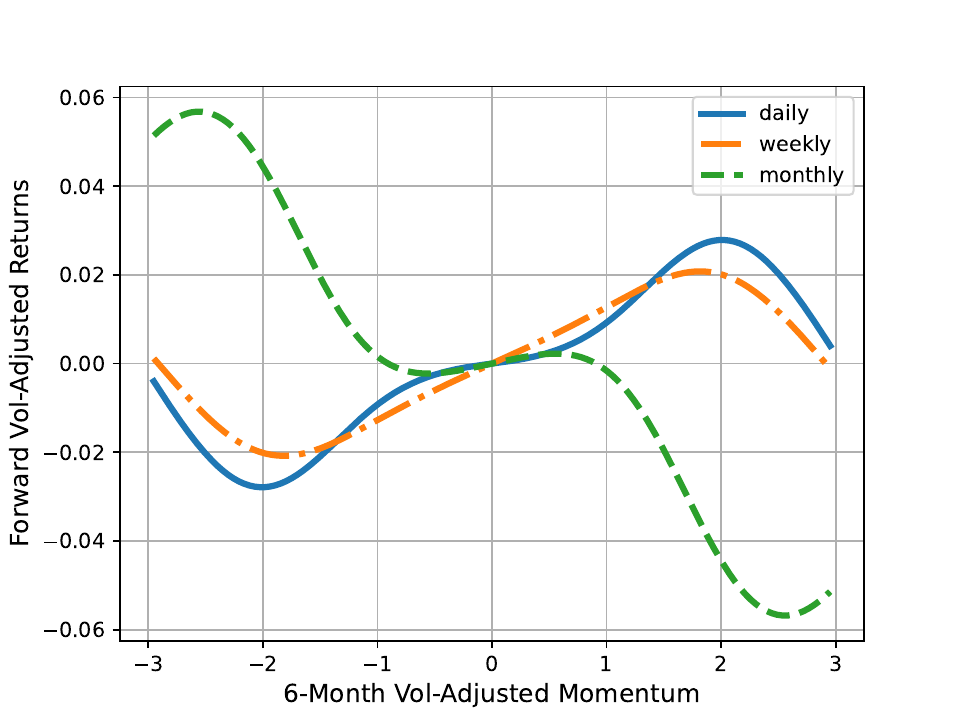}
    \caption{$6$-Month Momentum}
  \end{subfigure}
  \begin{subfigure}[b]{0.30\linewidth}
    \includegraphics[width=\linewidth ]{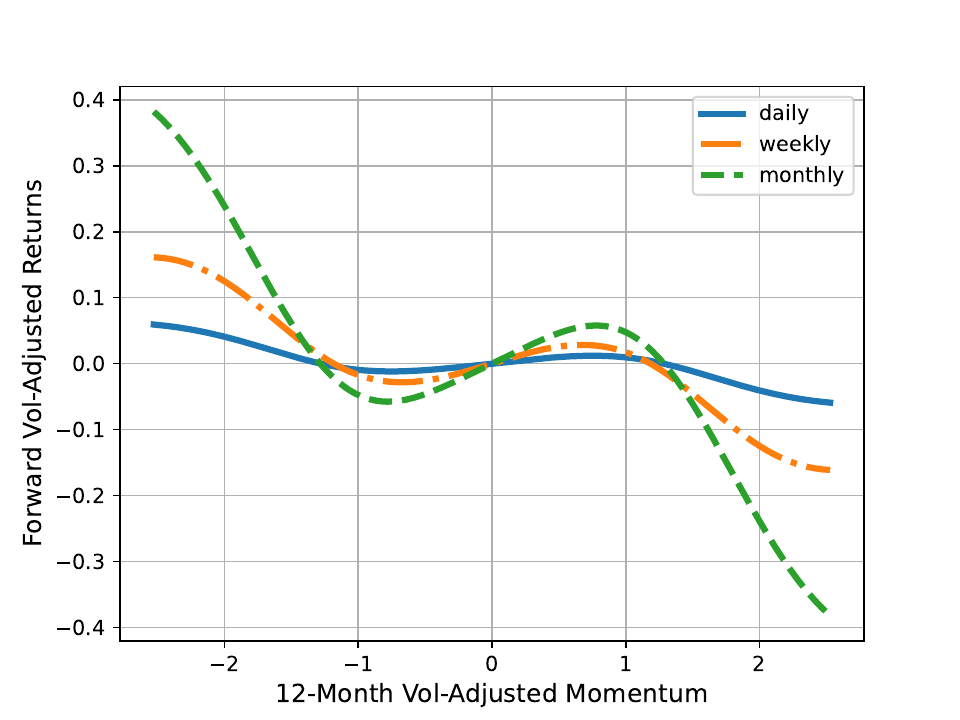}
    \caption{$12$-Month Momentum}
  \end{subfigure}
  
  \caption{Estimated forward vol-adjusted returns as a function of the $1$, $6$ and $12$-month time series momentum factor for FTSE 100 Nonlinear models.
  Models are fit using all available training data from January $1950$ through June $2026$. }
\end{figure}

\subsection{Grain Sector Nonlinear Models}
\begin{figure}[H]
  \centering
  \begin{subfigure}[b]{0.30\linewidth}
    \includegraphics[width=\linewidth ]{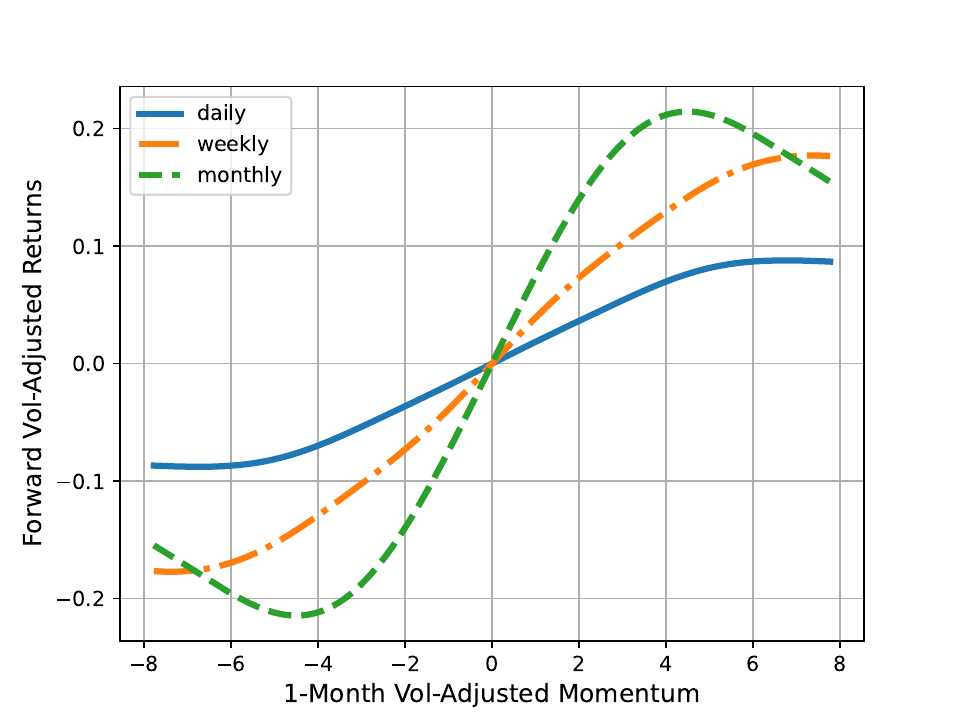}
    \caption{$1$-Month Momentum}
  \end{subfigure}
  \begin{subfigure}[b]{0.30\linewidth}
    \includegraphics[width=\linewidth ]{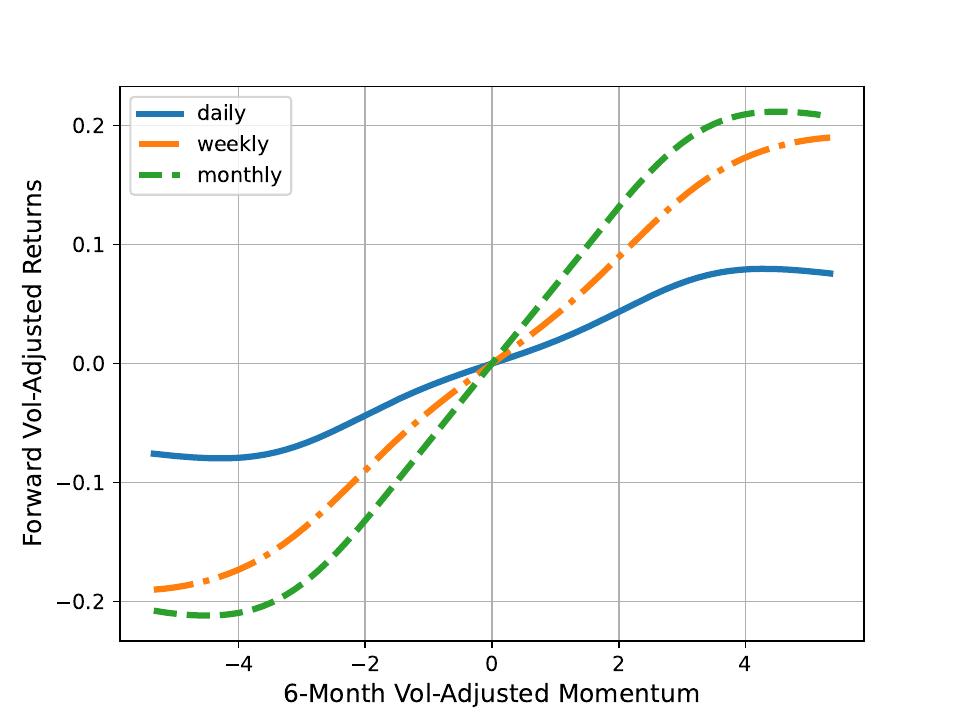}
    \caption{$6$-Month Momentum}
  \end{subfigure}
  \begin{subfigure}[b]{0.30\linewidth}
    \includegraphics[width=\linewidth ]{figs/model_grain_momo_12_month.pdf}
    \caption{$12$-Month Momentum}
  \end{subfigure}
  
  \caption{Estimated forward vol-adjusted returns as a function of the $1$, $6$ and $12$-month time series momentum factor for Grain Sector Nonlinear models.
  Models are fit using all available training data from January $1950$ through June $2026$. }
\end{figure}

\noindent \textbf{Bean Oil}
\begin{figure}[H]
  \centering
  \begin{subfigure}[b]{0.30\linewidth}
    \includegraphics[width=\linewidth ]{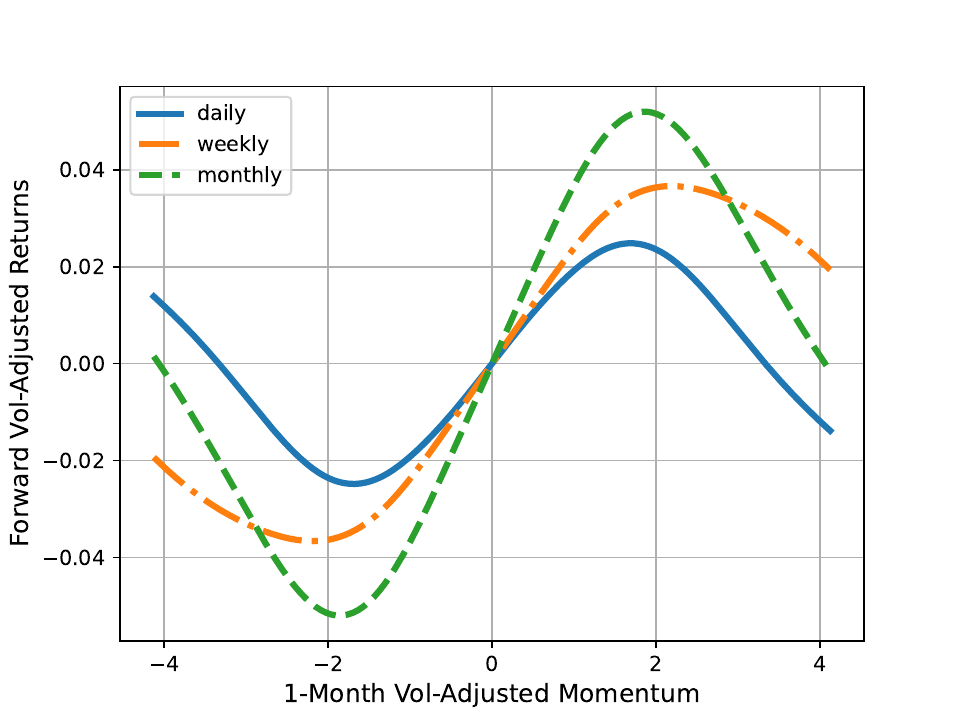}
    \caption{$1$-Month Momentum}
  \end{subfigure}
  \begin{subfigure}[b]{0.30\linewidth}
    \includegraphics[width=\linewidth ]{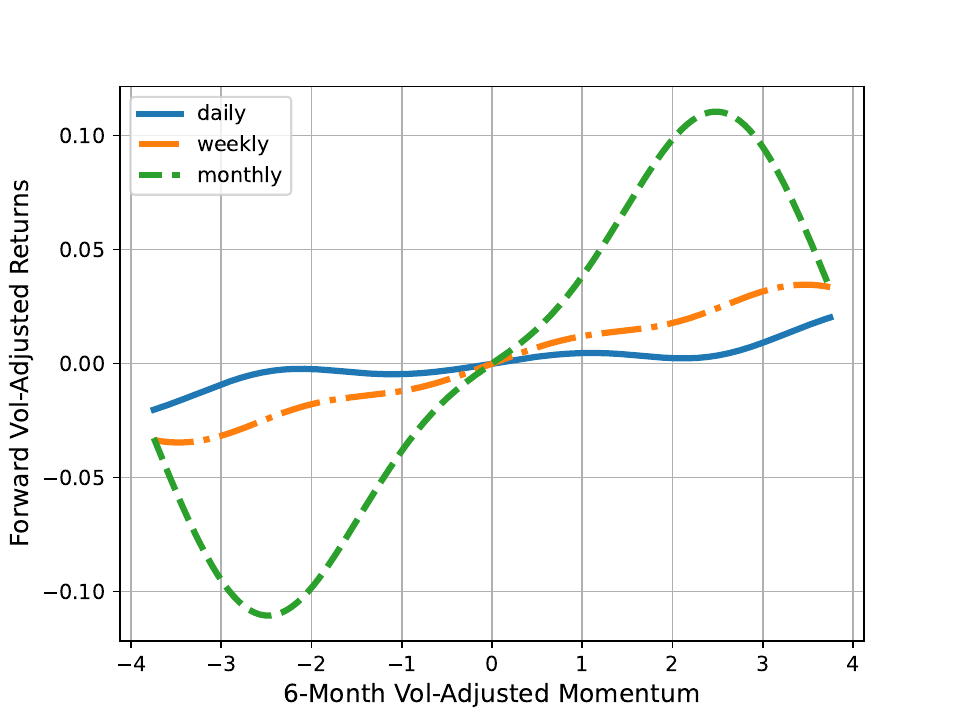}
    \caption{$6$-Month Momentum}
  \end{subfigure}
  \begin{subfigure}[b]{0.30\linewidth}
    \includegraphics[width=\linewidth ]{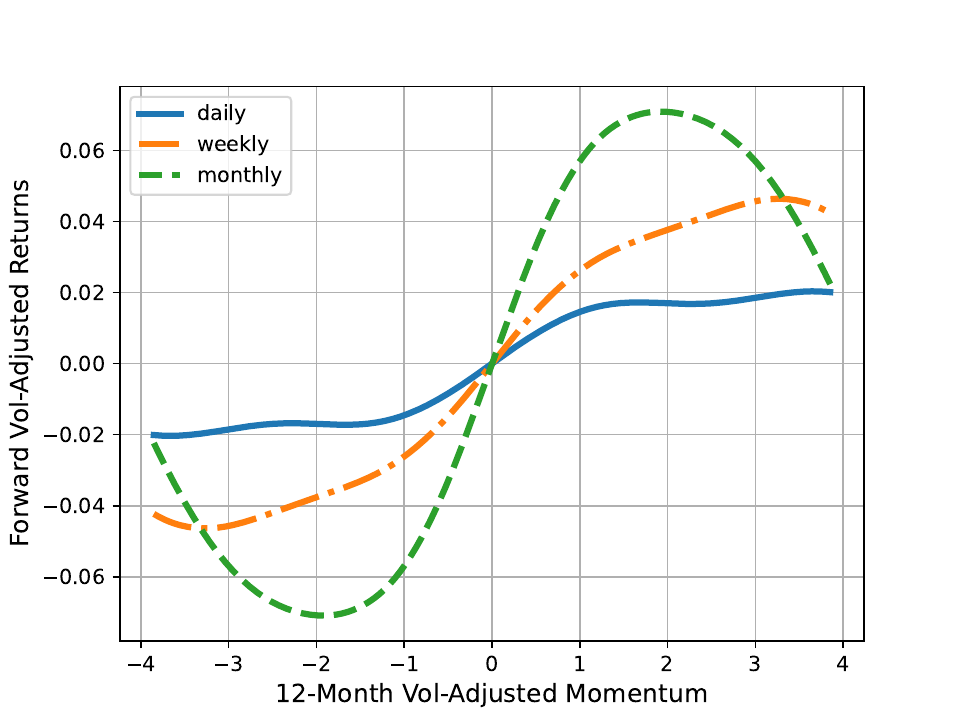}
    \caption{$12$-Month Momentum}
  \end{subfigure}
  
  \caption{Estimated forward vol-adjusted returns as a function of the $1$, $6$ and $12$-month time series momentum factor for Bean Oil Nonlinear models.
  Models are fit using all available training data from January $1950$ through June $2026$. }
\end{figure}

\noindent \textbf{Corn}
\begin{figure}[H]
  \centering
  \begin{subfigure}[b]{0.30\linewidth}
    \includegraphics[width=\linewidth ]{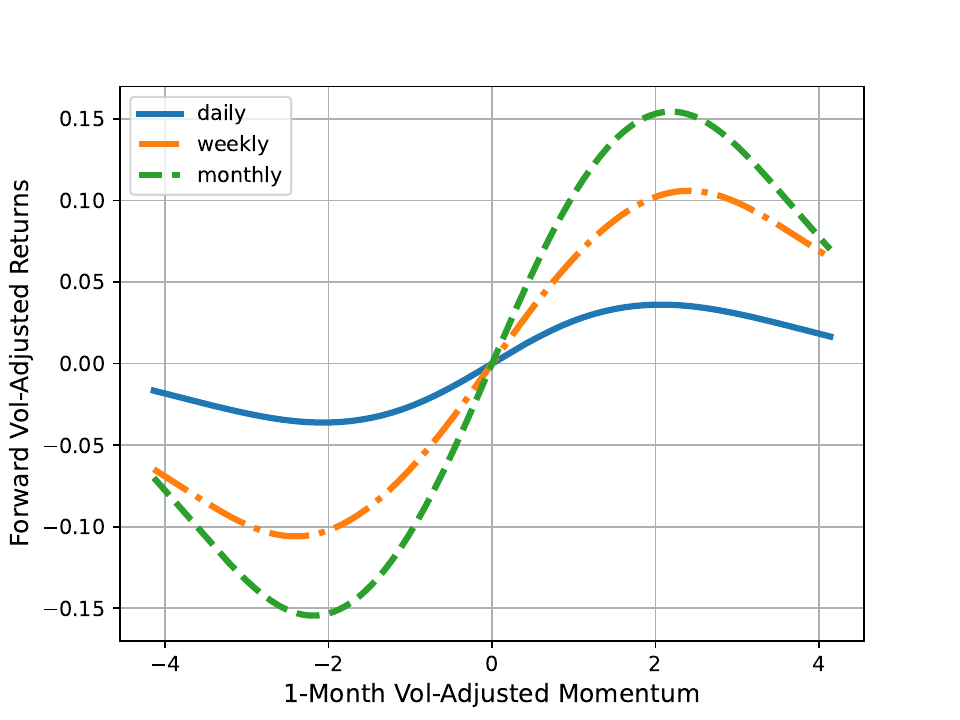}
    \caption{$1$-Month Momentum}
  \end{subfigure}
  \begin{subfigure}[b]{0.30\linewidth}
    \includegraphics[width=\linewidth ]{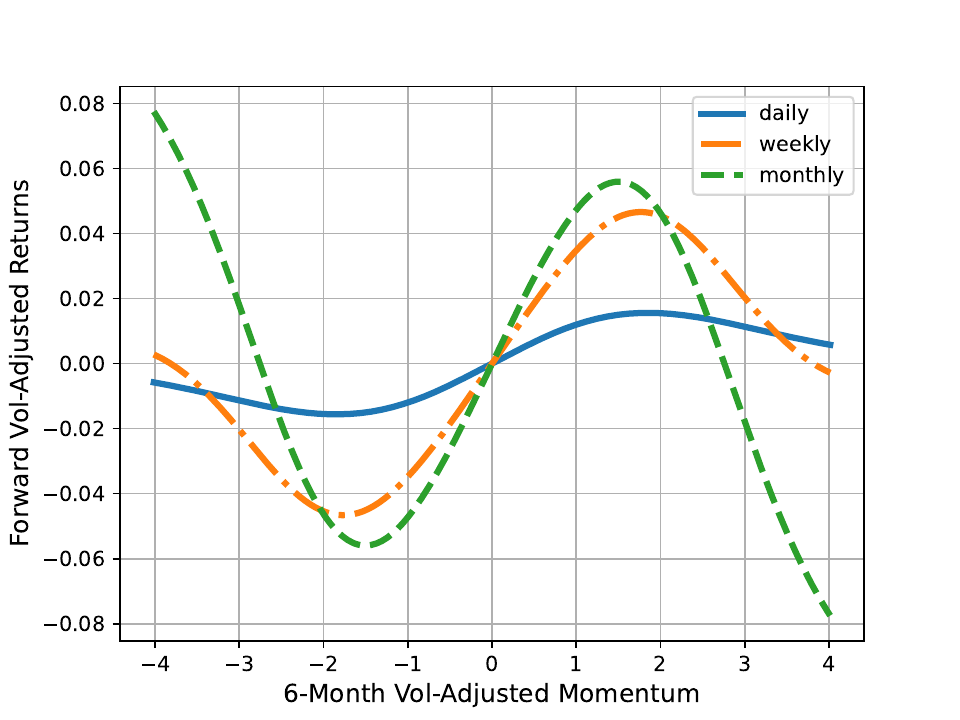}
    \caption{$6$-Month Momentum}
  \end{subfigure}
  \begin{subfigure}[b]{0.30\linewidth}
    \includegraphics[width=\linewidth ]{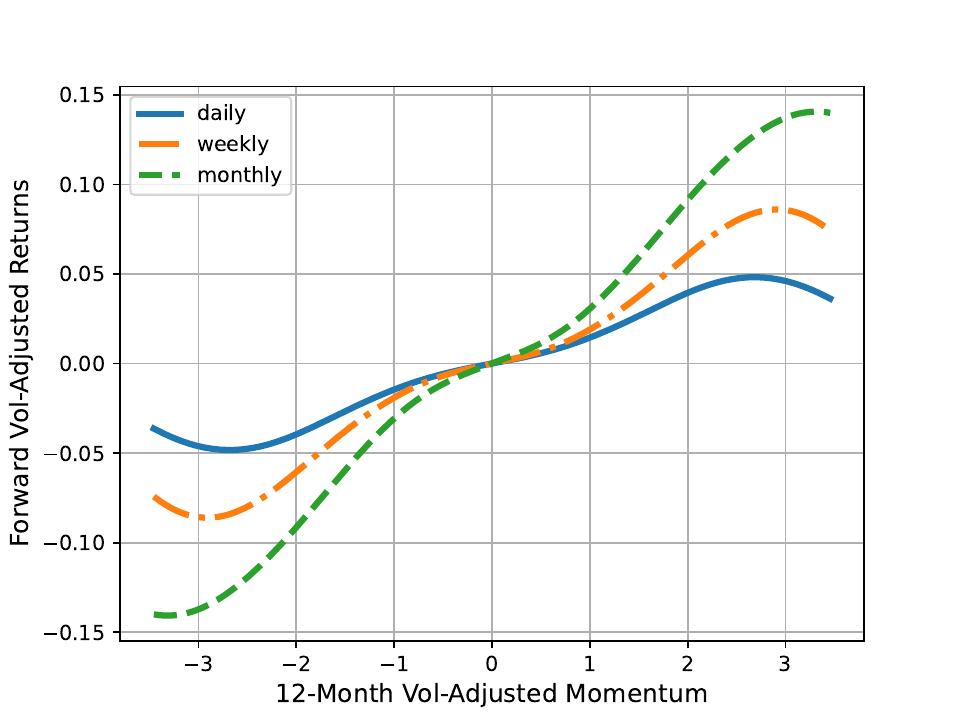}
    \caption{$12$-Month Momentum}
  \end{subfigure}
  
  \caption{Estimated forward vol-adjusted returns as a function of the $1$, $6$ and $12$-month time series momentum factor for Corn Nonlinear models.
  Models are fit using all available training data from January $1950$ through June $2026$. }
\end{figure}

\noindent \textbf{Palm Oil}
\begin{figure}[H]
  \centering
  \begin{subfigure}[b]{0.30\linewidth}
    \includegraphics[width=\linewidth ]{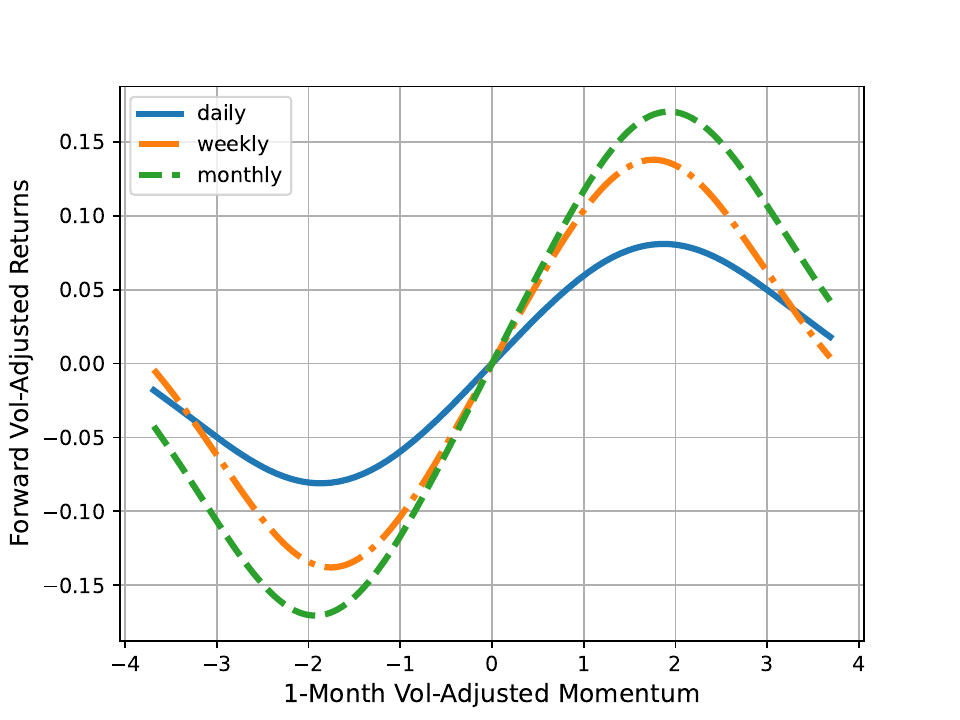}
    \caption{$1$-Month Momentum}
  \end{subfigure}
  \begin{subfigure}[b]{0.30\linewidth}
    \includegraphics[width=\linewidth ]{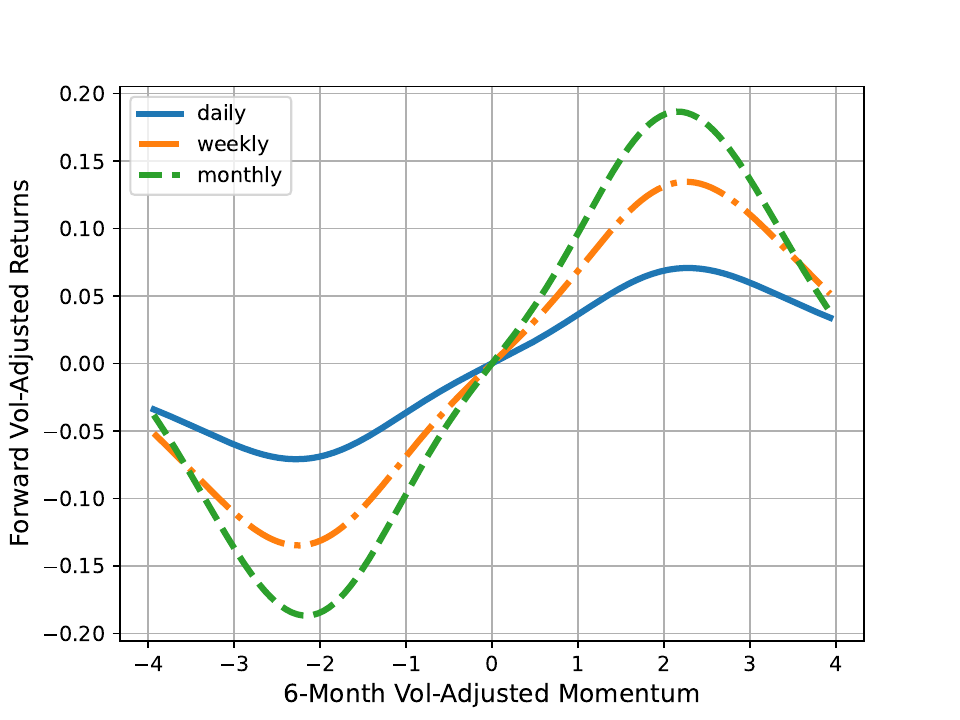}
    \caption{$6$-Month Momentum}
  \end{subfigure}
  \begin{subfigure}[b]{0.30\linewidth}
    \includegraphics[width=\linewidth ]{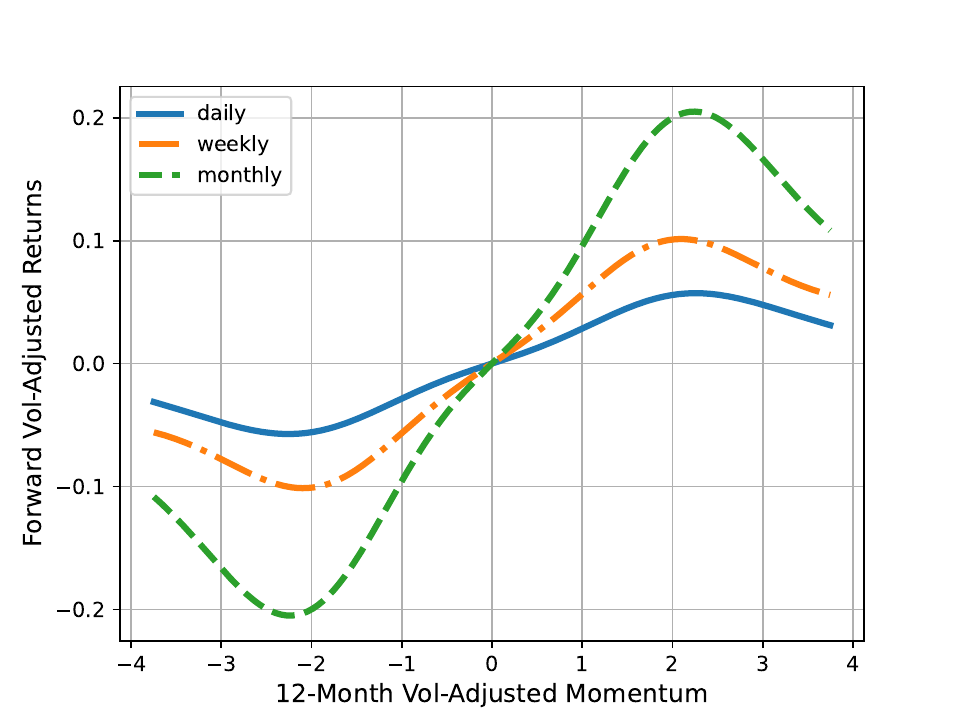}
    \caption{$12$-Month Momentum}
  \end{subfigure}
  
  \caption{Estimated forward vol-adjusted returns as a function of the $1$, $6$ and $12$-month time series momentum factor for Palm Oil Nonlinear models.
  Models are fit using all available training data from January $1950$ through June $2026$. }
\end{figure}

\noindent \textbf{KC Wheat}
\begin{figure}[H]
  \centering
  \begin{subfigure}[b]{0.30\linewidth}
    \includegraphics[width=\linewidth ]{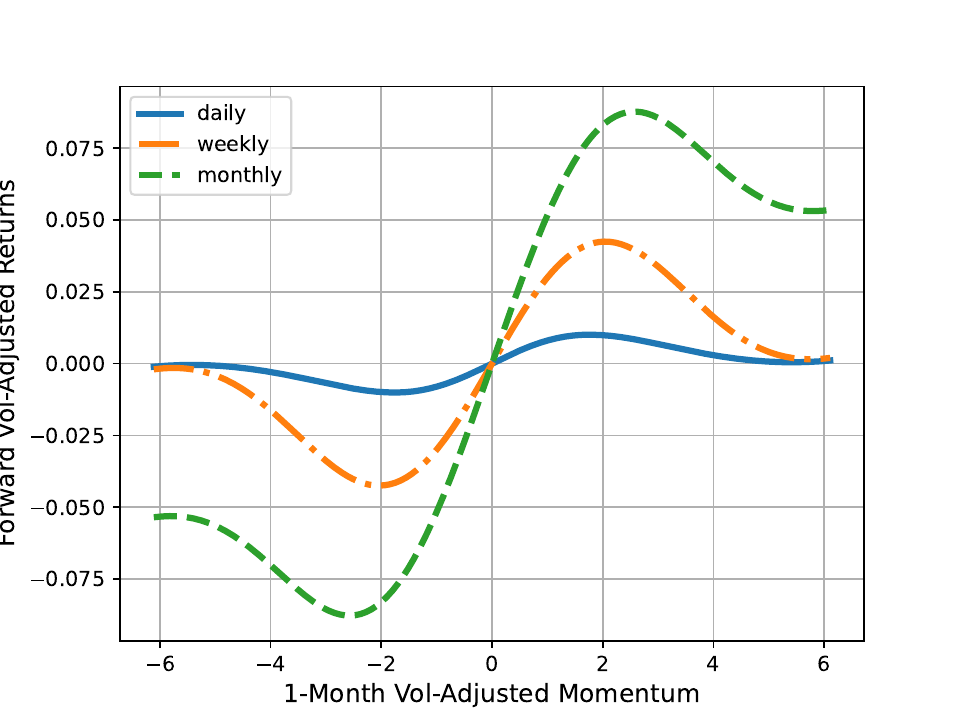}
    \caption{$1$-Month Momentum}
  \end{subfigure}
  \begin{subfigure}[b]{0.30\linewidth}
    \includegraphics[width=\linewidth ]{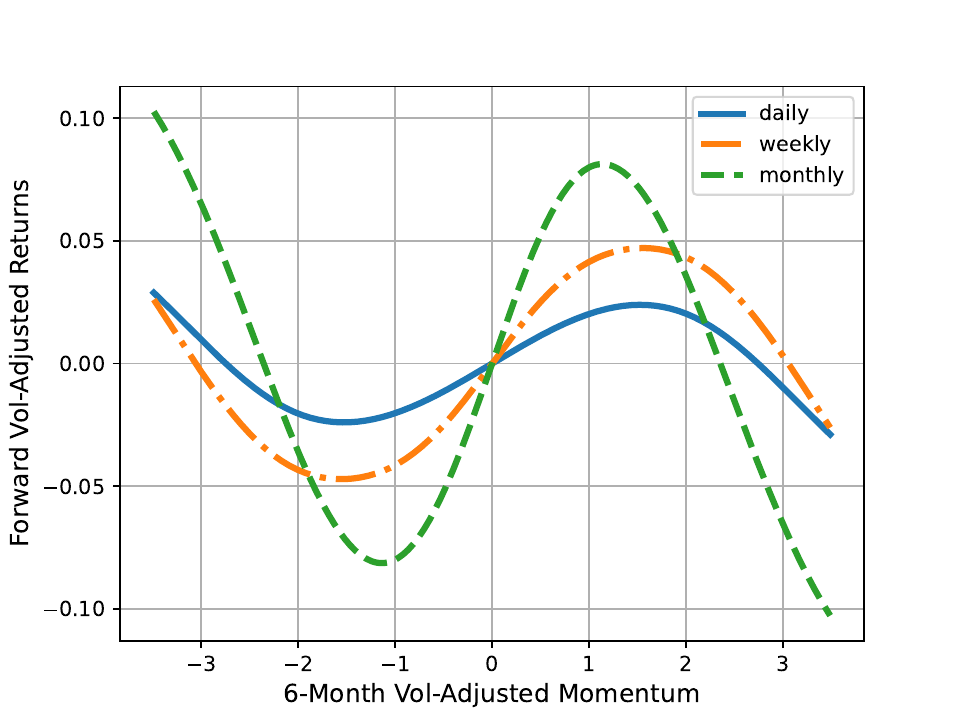}
    \caption{$6$-Month Momentum}
  \end{subfigure}
  \begin{subfigure}[b]{0.30\linewidth}
    \includegraphics[width=\linewidth ]{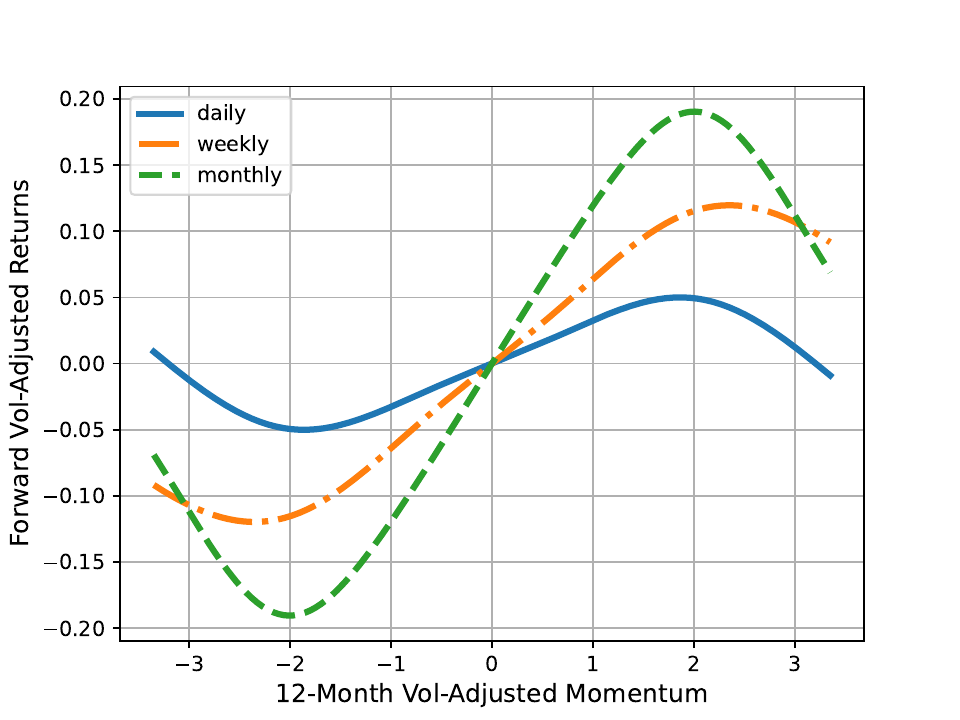}
    \caption{$12$-Month Momentum}
  \end{subfigure}
  
  \caption{Estimated forward vol-adjusted returns as a function of the $1$, $6$ and $12$-month time series momentum factor for KC Wheat Nonlinear models.
  Models are fit using all available training data from January $1950$ through June $2026$. }
\end{figure}

\noindent \textbf{Canola}
\begin{figure}[H]
  \centering
  \begin{subfigure}[b]{0.30\linewidth}
    \includegraphics[width=\linewidth ]{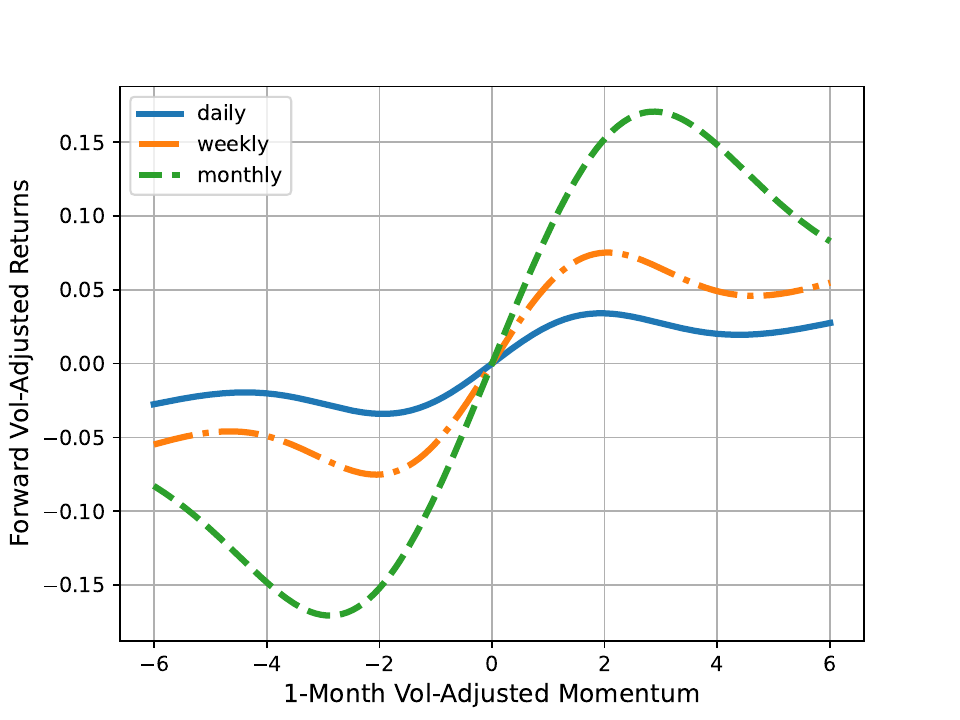}
    \caption{$1$-Month Momentum}
  \end{subfigure}
  \begin{subfigure}[b]{0.30\linewidth}
    \includegraphics[width=\linewidth ]{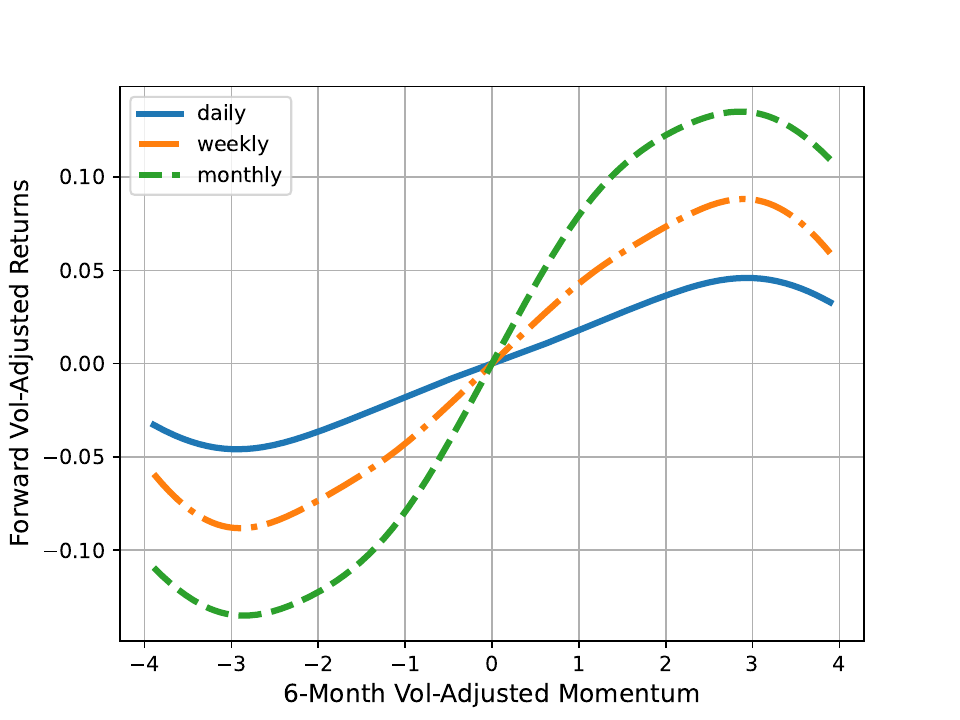}
    \caption{$6$-Month Momentum}
  \end{subfigure}
  \begin{subfigure}[b]{0.30\linewidth}
    \includegraphics[width=\linewidth ]{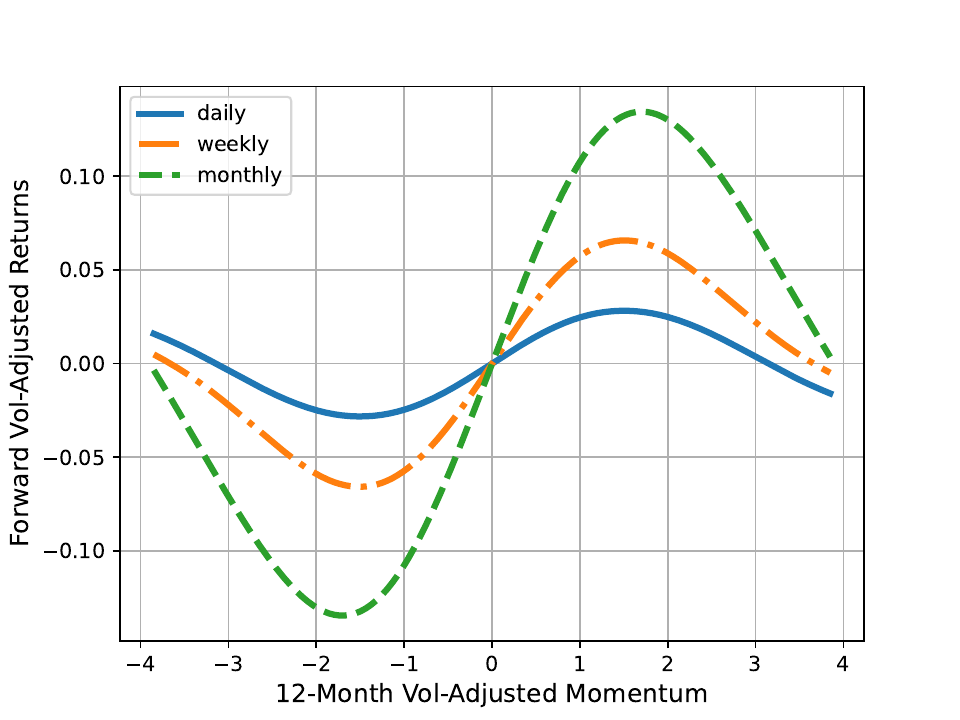}
    \caption{$12$-Month Momentum}
  \end{subfigure}
  
  \caption{Estimated forward vol-adjusted returns as a function of the $1$, $6$ and $12$-month time series momentum factor for Canola Nonlinear models.
  Models are fit using all available training data from January $1950$ through June $2026$. }
\end{figure}

\noindent \textbf{Soybean}
\begin{figure}[H]
  \centering
  \begin{subfigure}[b]{0.30\linewidth}
    \includegraphics[width=\linewidth ]{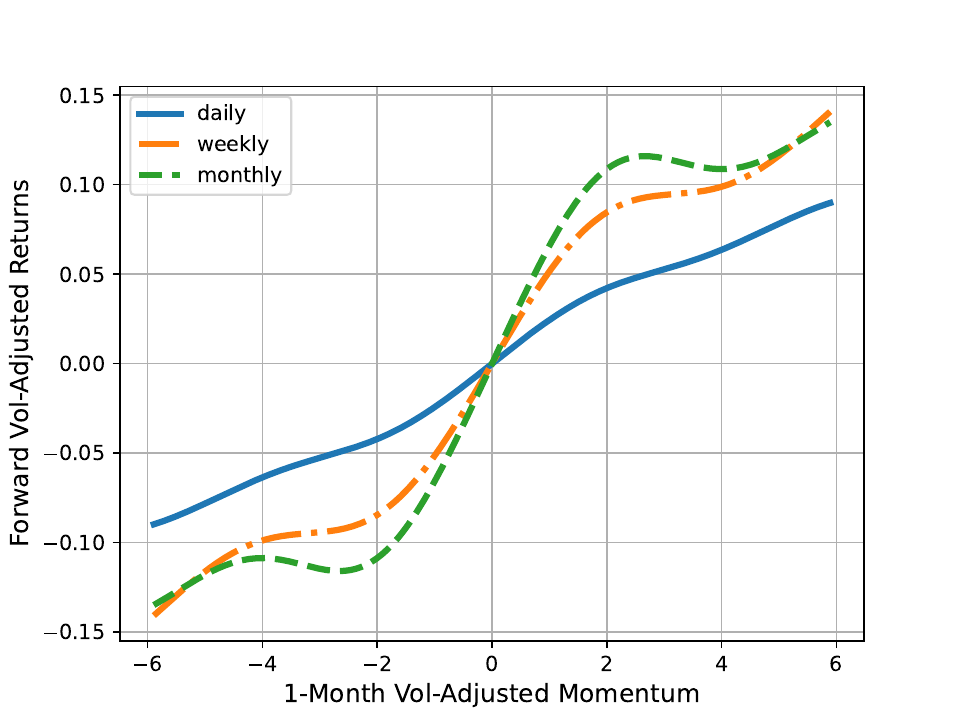}
    \caption{$1$-Month Momentum}
  \end{subfigure}
  \begin{subfigure}[b]{0.30\linewidth}
    \includegraphics[width=\linewidth ]{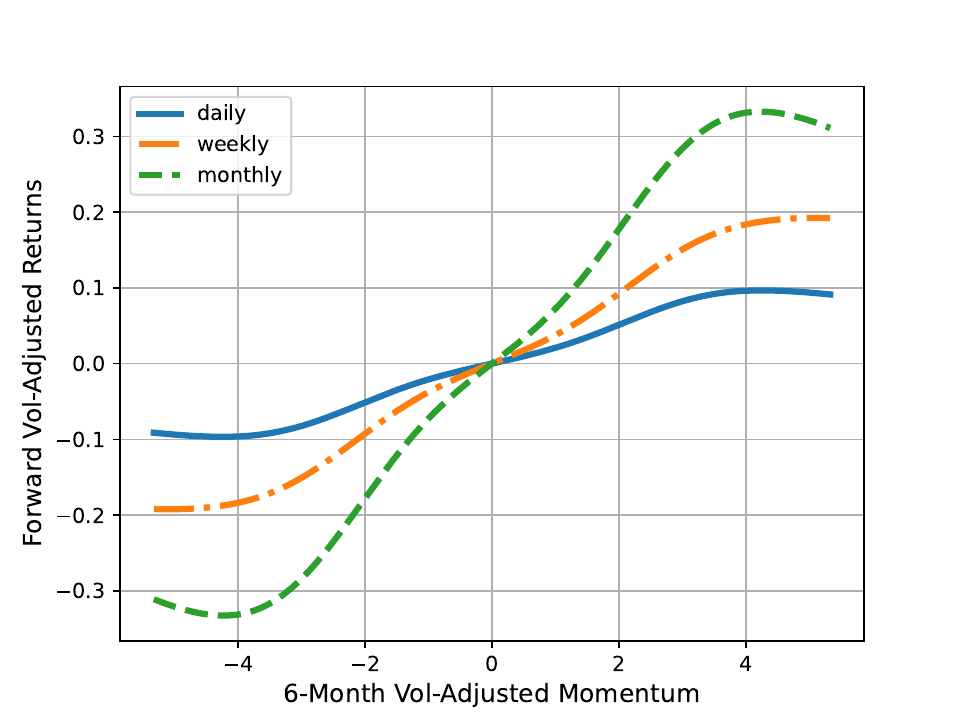}
    \caption{$6$-Month Momentum}
  \end{subfigure}
  \begin{subfigure}[b]{0.30\linewidth}
    \includegraphics[width=\linewidth ]{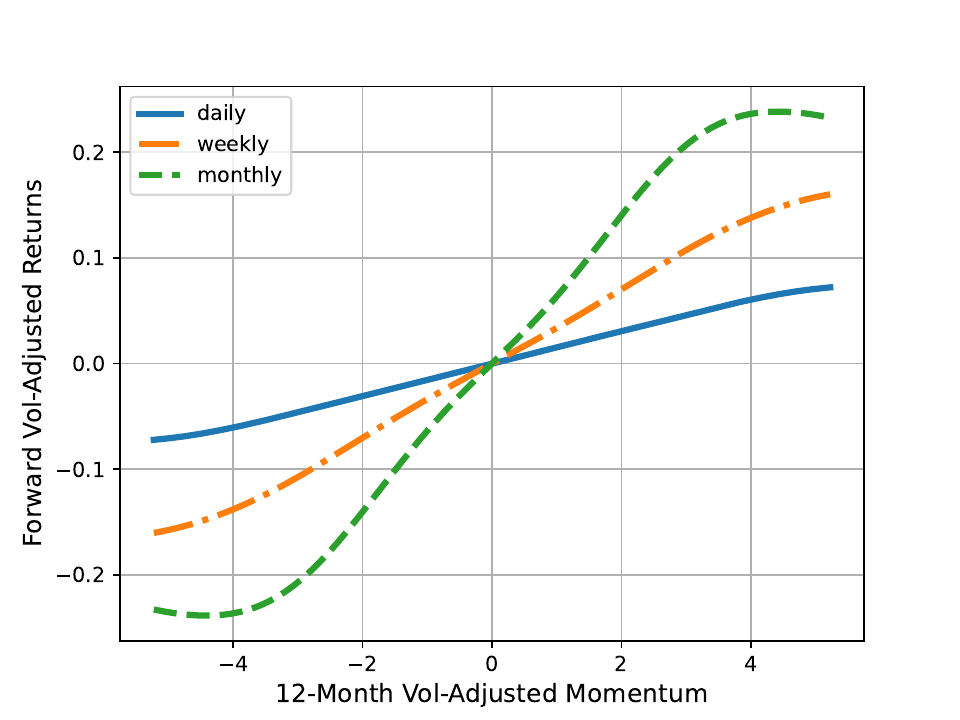}
    \caption{$12$-Month Momentum}
  \end{subfigure}
  
  \caption{Estimated forward vol-adjusted returns as a function of the $1$, $6$ and $12$-month time series momentum factor for Soybean Nonlinear models.
  Models are fit using all available training data from January $1950$ through June $2026$. }
\end{figure}

\noindent \textbf{Soy Meal}
\begin{figure}[H]
  \centering
  \begin{subfigure}[b]{0.30\linewidth}
    \includegraphics[width=\linewidth ]{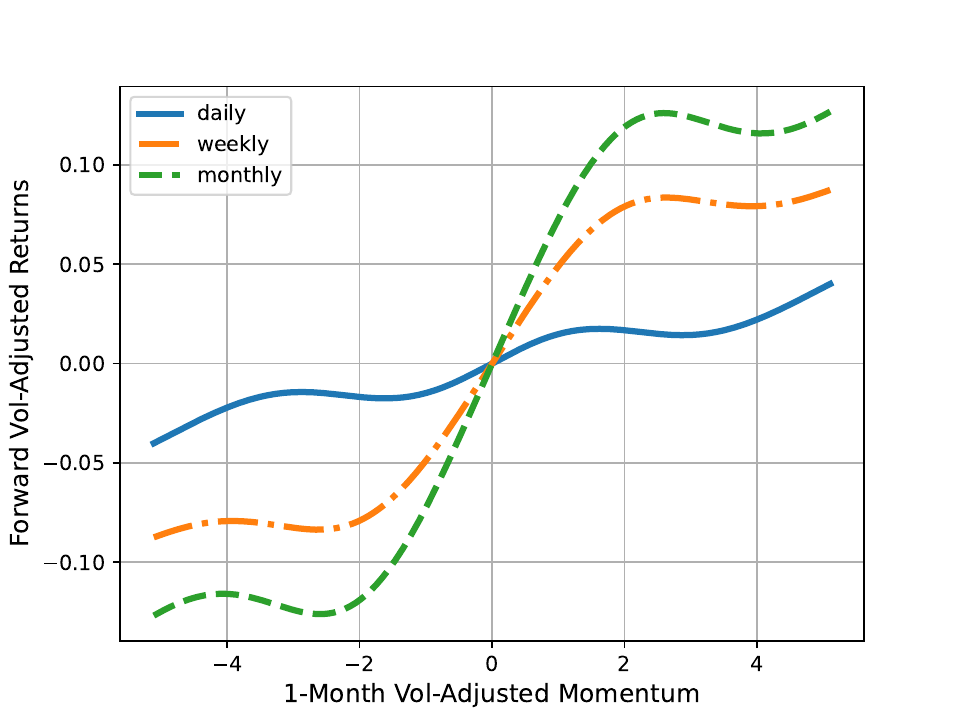}
    \caption{$1$-Month Momentum}
  \end{subfigure}
  \begin{subfigure}[b]{0.30\linewidth}
    \includegraphics[width=\linewidth ]{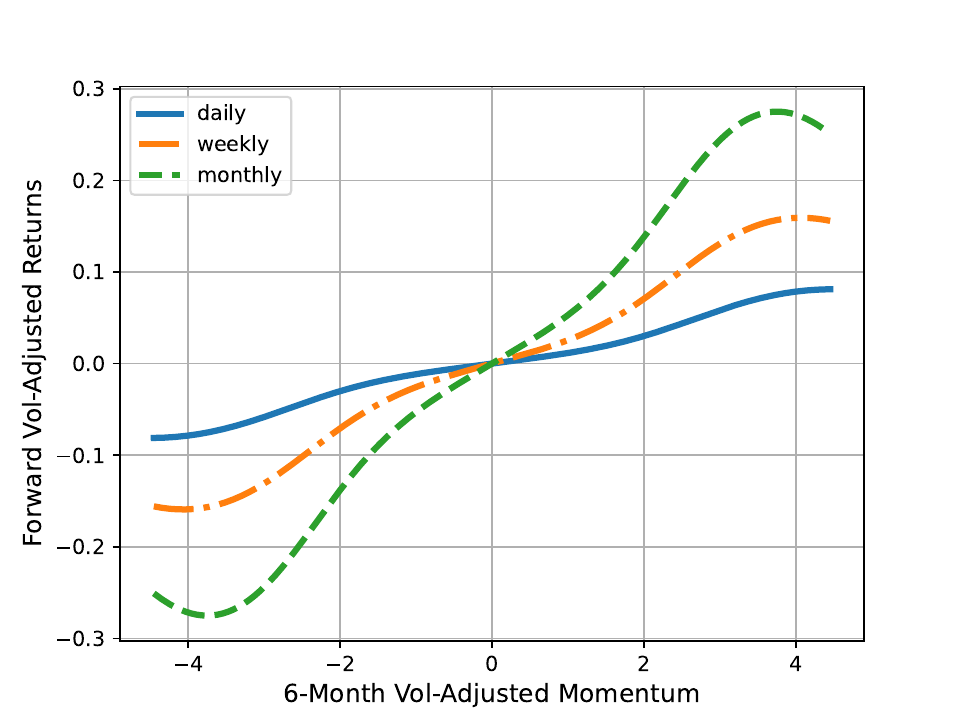}
    \caption{$6$-Month Momentum}
  \end{subfigure}
  \begin{subfigure}[b]{0.30\linewidth}
    \includegraphics[width=\linewidth ]{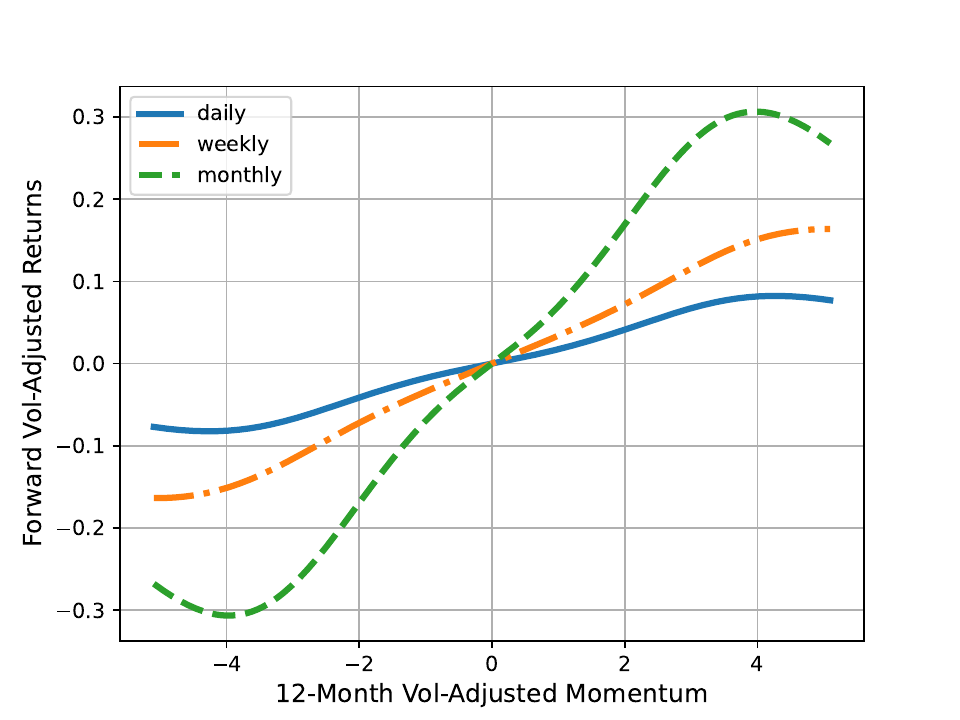}
    \caption{$12$-Month Momentum}
  \end{subfigure}
  
  \caption{Estimated forward vol-adjusted returns as a function of the $1$, $6$ and $12$-month time series momentum factor for Soy Meal Nonlinear models.
  Models are fit using all available training data from January $1950$ through June $2026$. }
\end{figure}

\noindent \textbf{Wheat}
\begin{figure}[H]
  \centering
  \begin{subfigure}[b]{0.30\linewidth}
    \includegraphics[width=\linewidth ]{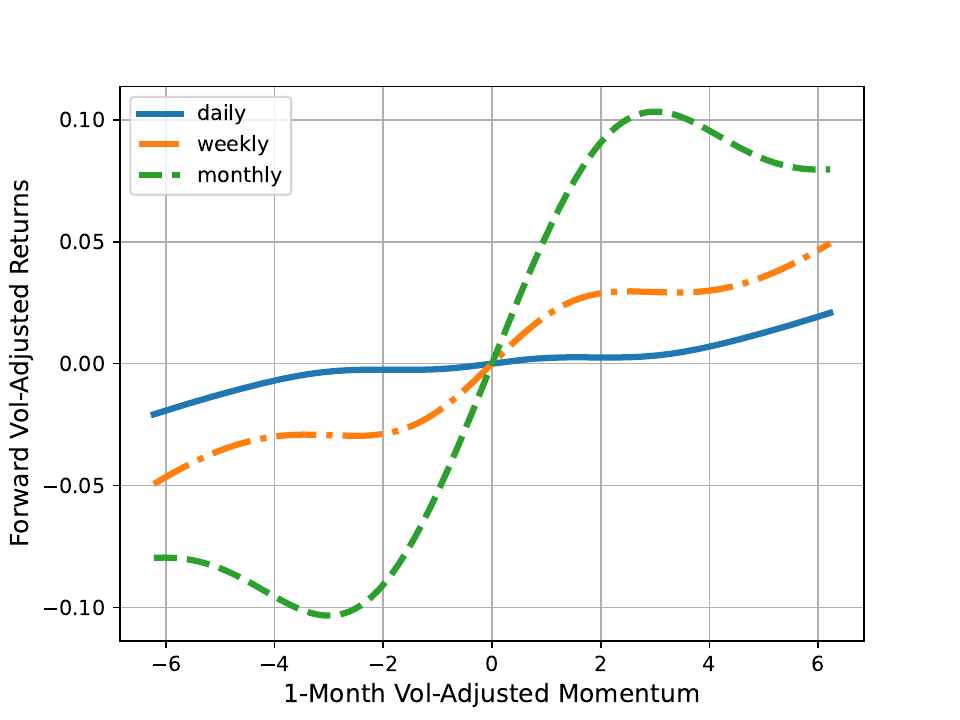}
    \caption{$1$-Month Momentum}
  \end{subfigure}
  \begin{subfigure}[b]{0.30\linewidth}
    \includegraphics[width=\linewidth ]{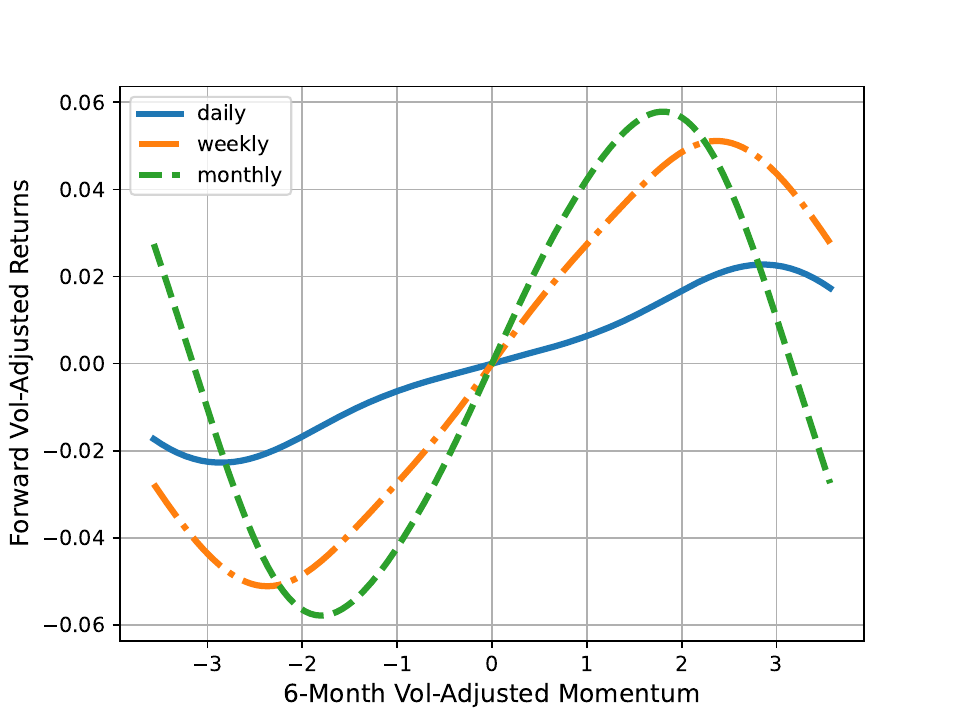}
    \caption{$6$-Month Momentum}
  \end{subfigure}
  \begin{subfigure}[b]{0.30\linewidth}
    \includegraphics[width=\linewidth ]{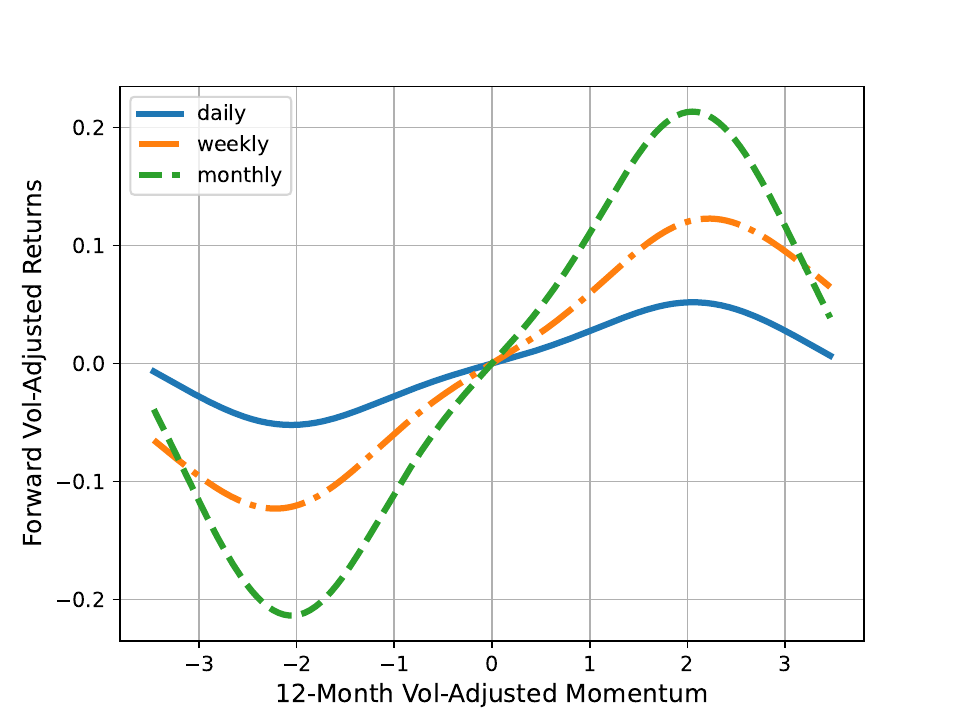}
    \caption{$12$-Month Momentum}
  \end{subfigure}
  
  \caption{Estimated forward vol-adjusted returns as a function of the $1$, $6$ and $12$-month time series momentum factor for Wheat Nonlinear models.
  Models are fit using all available training data from January $1950$ through June $2026$. }
\end{figure}

\subsection{Metal Sector Nonlinear Models}
\begin{figure}[H]
  \centering
  \begin{subfigure}[b]{0.30\linewidth}
    \includegraphics[width=\linewidth ]{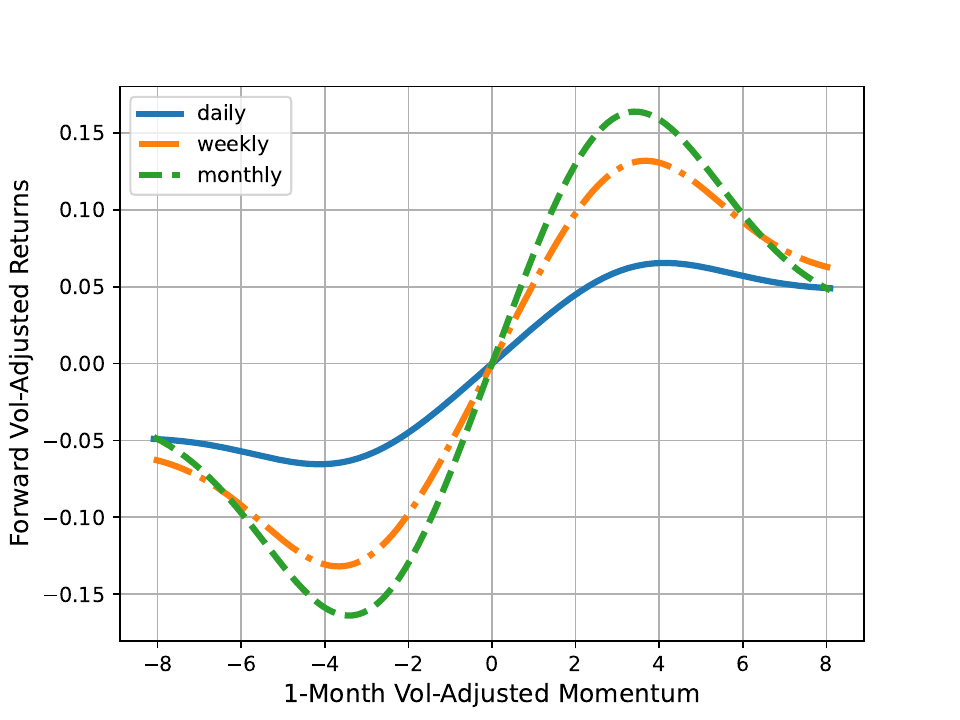}
    \caption{$1$-Month Momentum}
  \end{subfigure}
  \begin{subfigure}[b]{0.30\linewidth}
    \includegraphics[width=\linewidth ]{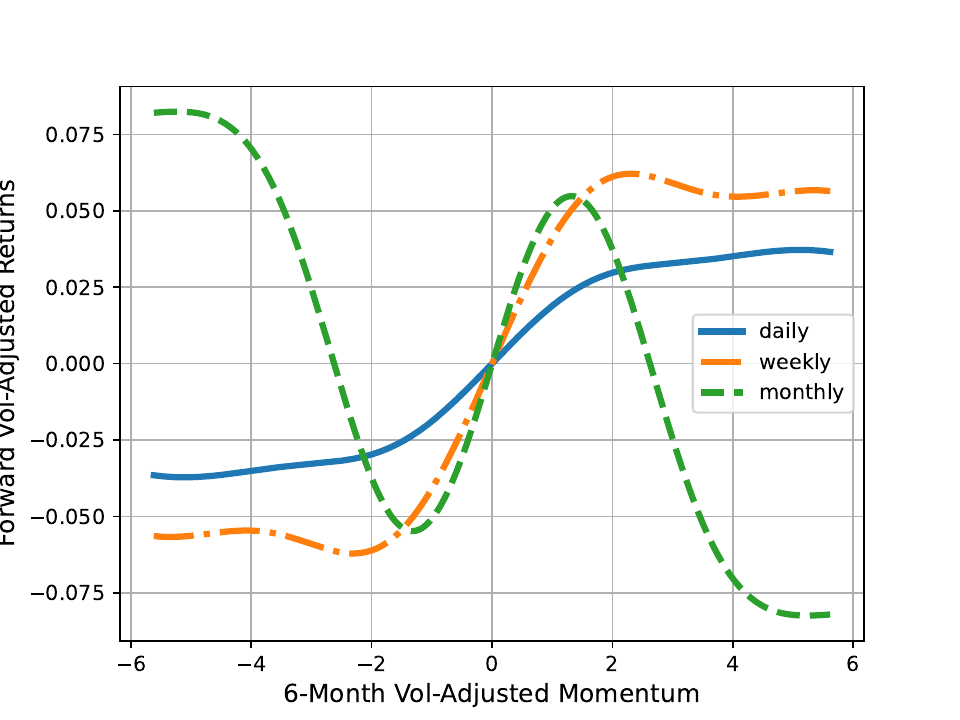}
    \caption{$6$-Month Momentum}
  \end{subfigure}
  \begin{subfigure}[b]{0.30\linewidth}
    \includegraphics[width=\linewidth ]{figs/model_metal_momo_12_month.pdf}
    \caption{$12$-Month Momentum}
  \end{subfigure}
  
  \caption{Estimated forward vol-adjusted returns as a function of the $1$, $6$ and $12$-month time series momentum factor for Metal Sector Nonlinear models.
  Models are fit using all available training data from January $1950$ through June $2026$. }
\end{figure}

\noindent \textbf{Gold}
\begin{figure}[H]
  \centering
  \begin{subfigure}[b]{0.30\linewidth}
    \includegraphics[width=\linewidth ]{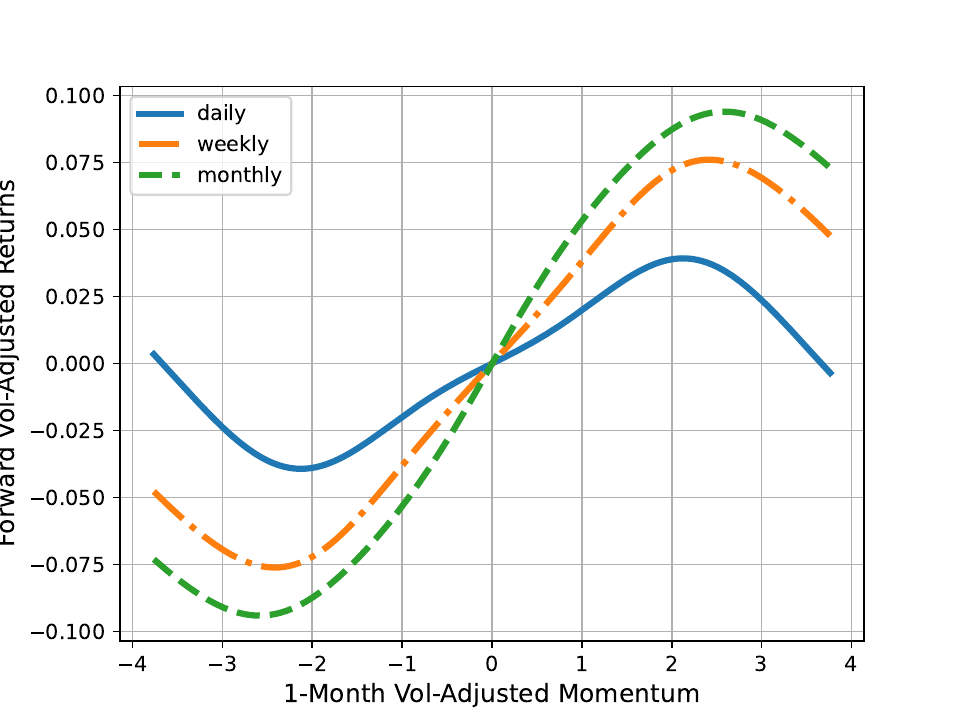}
    \caption{$1$-Month Momentum}
  \end{subfigure}
  \begin{subfigure}[b]{0.30\linewidth}
    \includegraphics[width=\linewidth ]{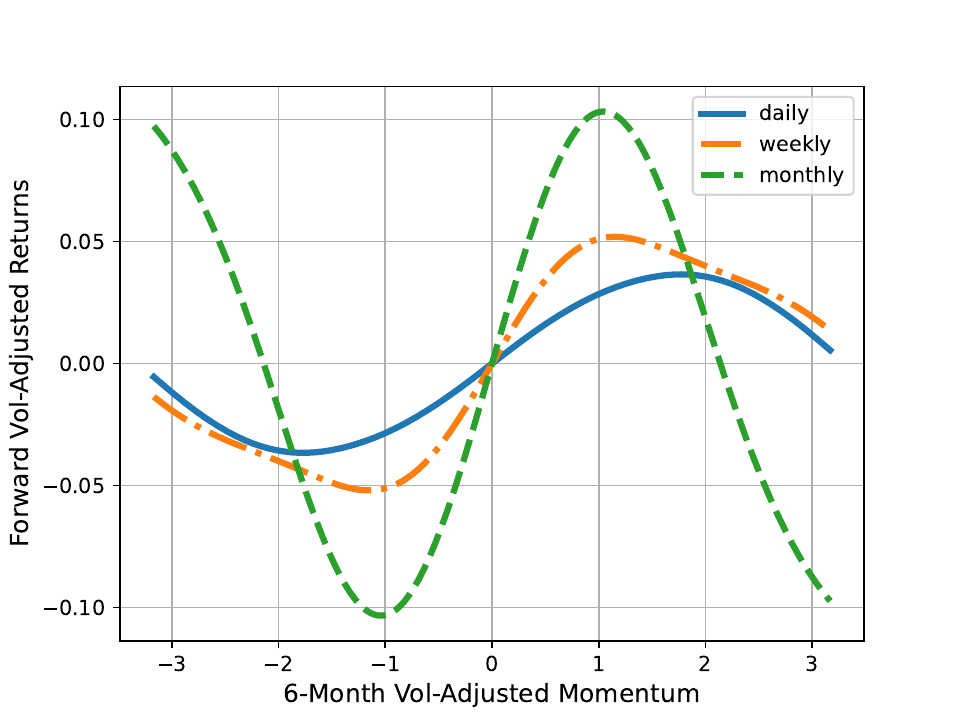}
    \caption{$6$-Month Momentum}
  \end{subfigure}
  \begin{subfigure}[b]{0.30\linewidth}
    \includegraphics[width=\linewidth ]{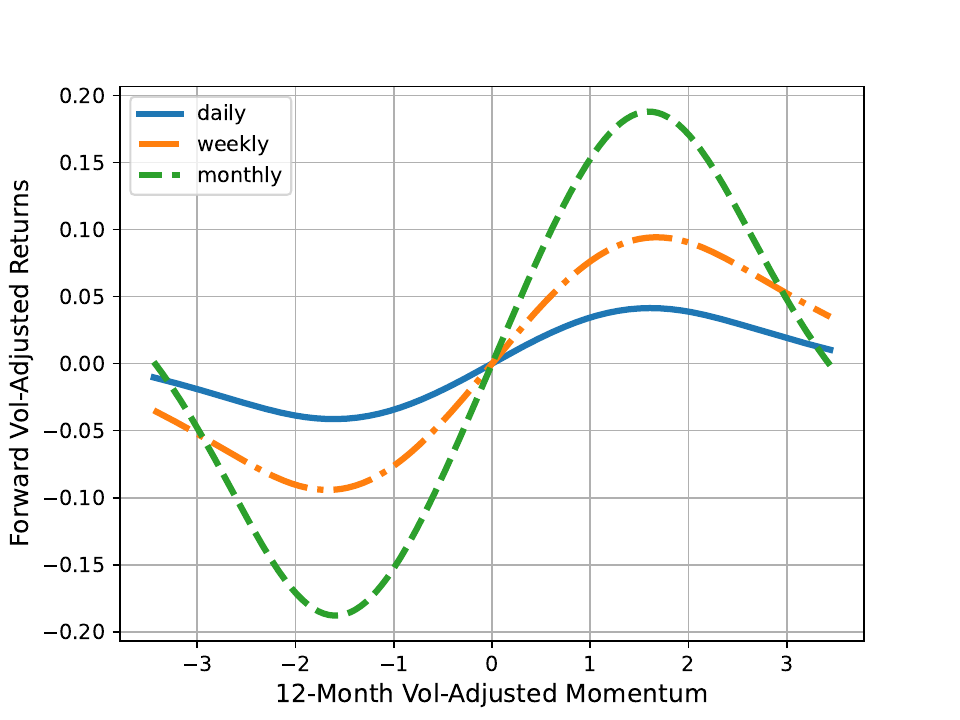}
    \caption{$12$-Month Momentum}
  \end{subfigure}
  
  \caption{Estimated forward vol-adjusted returns as a function of the $1$, $6$ and $12$-month time series momentum factor for Gold Nonlinear models.
  Models are fit using all available training data from January $1950$ through June $2026$. }
\end{figure}

\noindent \textbf{Copper}
\begin{figure}[H]
  \centering
  \begin{subfigure}[b]{0.30\linewidth}
    \includegraphics[width=\linewidth ]{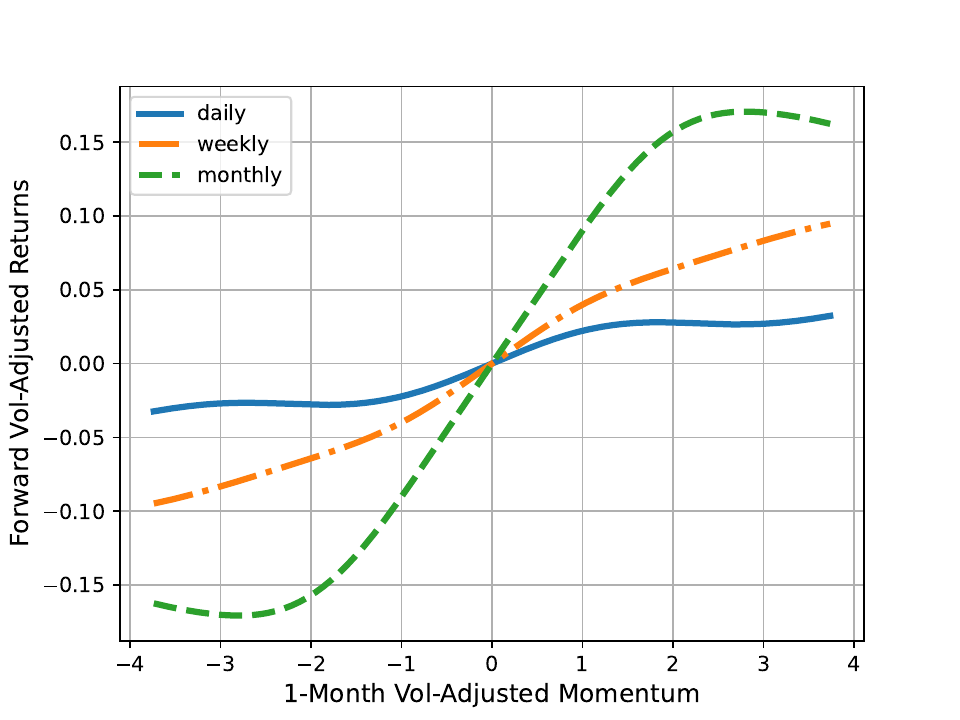}
    \caption{$1$-Month Momentum}
  \end{subfigure}
  \begin{subfigure}[b]{0.30\linewidth}
    \includegraphics[width=\linewidth ]{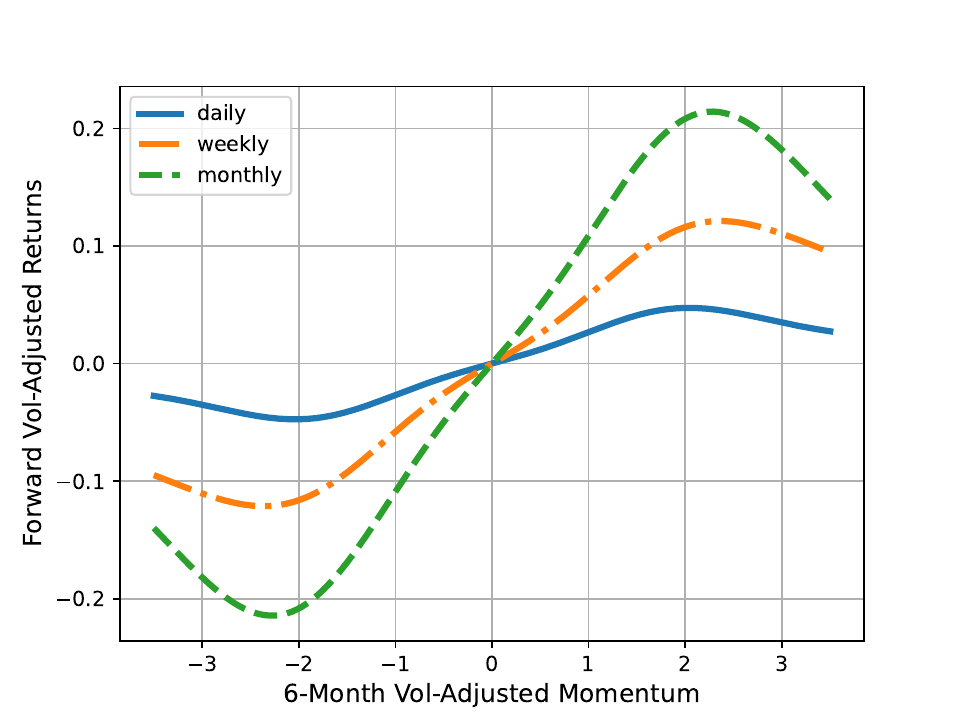}
    \caption{$6$-Month Momentum}
  \end{subfigure}
  \begin{subfigure}[b]{0.30\linewidth}
    \includegraphics[width=\linewidth ]{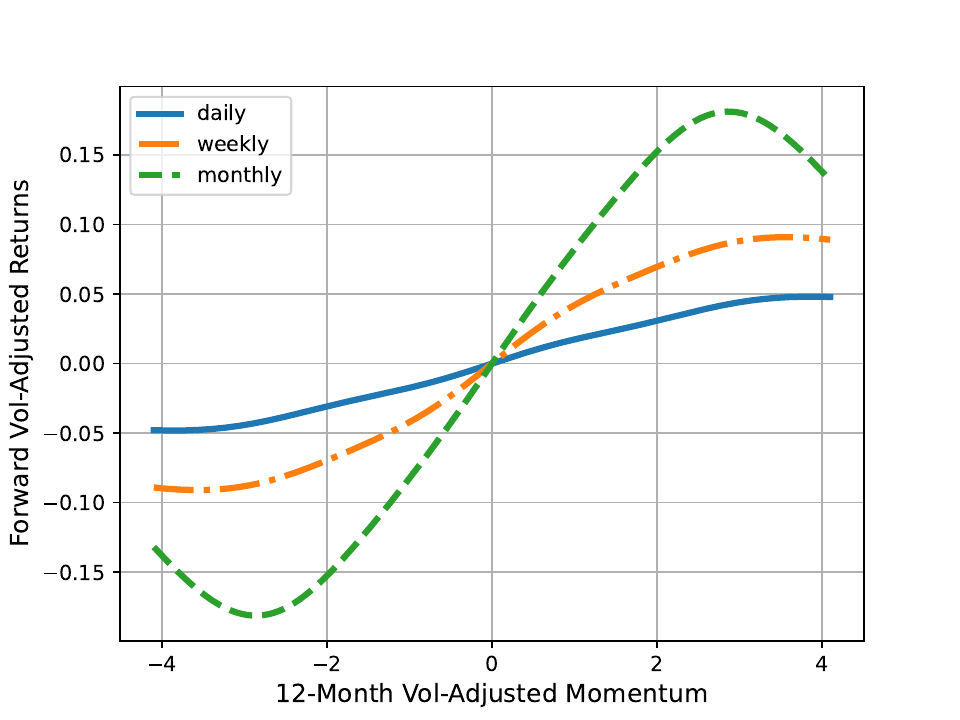}
    \caption{$12$-Month Momentum}
  \end{subfigure}
  
  \caption{Estimated forward vol-adjusted returns as a function of the $1$, $6$ and $12$-month time series momentum factor for Copper Nonlinear models.
  Models are fit using all available training data from January $1950$ through June $2026$. }
\end{figure}

\noindent \textbf{Palladium}
\begin{figure}[H]
  \centering
  \begin{subfigure}[b]{0.30\linewidth}
    \includegraphics[width=\linewidth ]{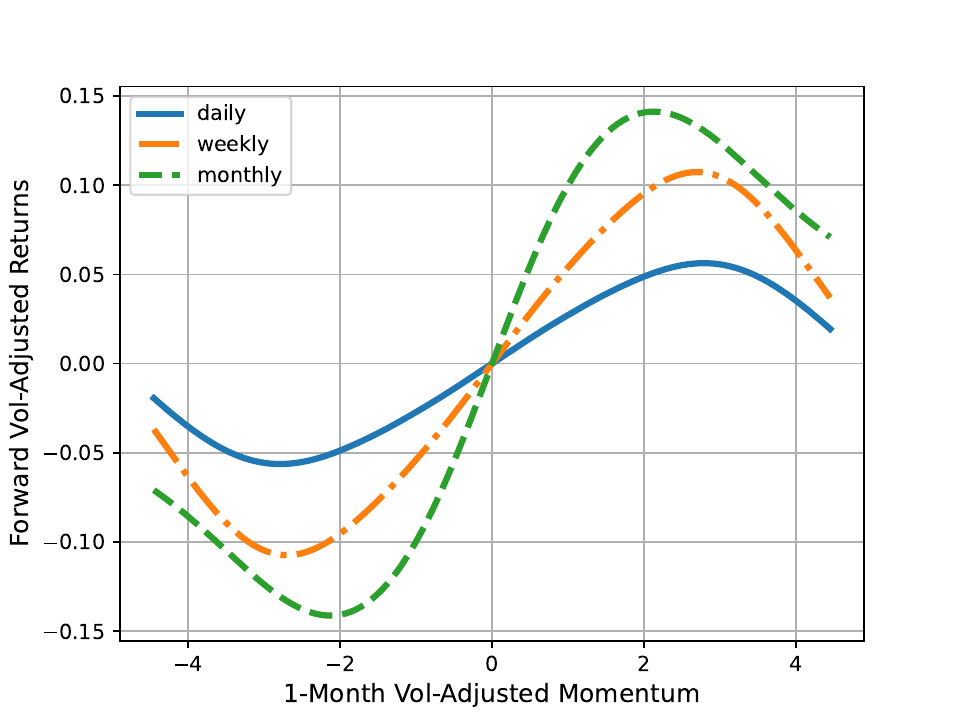}
    \caption{$1$-Month Momentum}
  \end{subfigure}
  \begin{subfigure}[b]{0.30\linewidth}
    \includegraphics[width=\linewidth ]{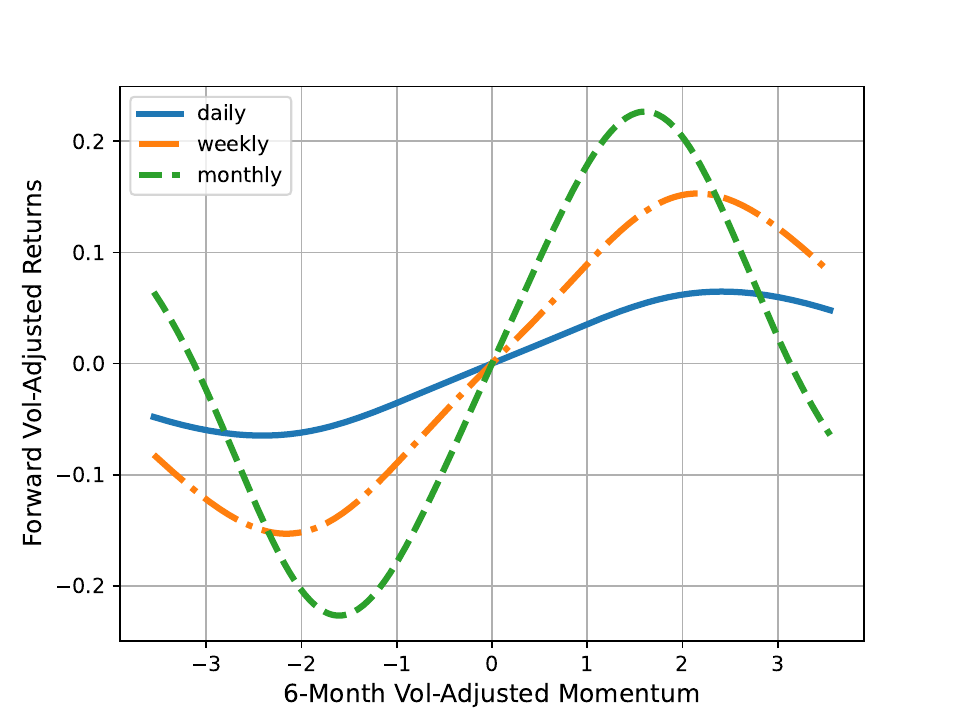}
    \caption{$6$-Month Momentum}
  \end{subfigure}
  \begin{subfigure}[b]{0.30\linewidth}
    \includegraphics[width=\linewidth ]{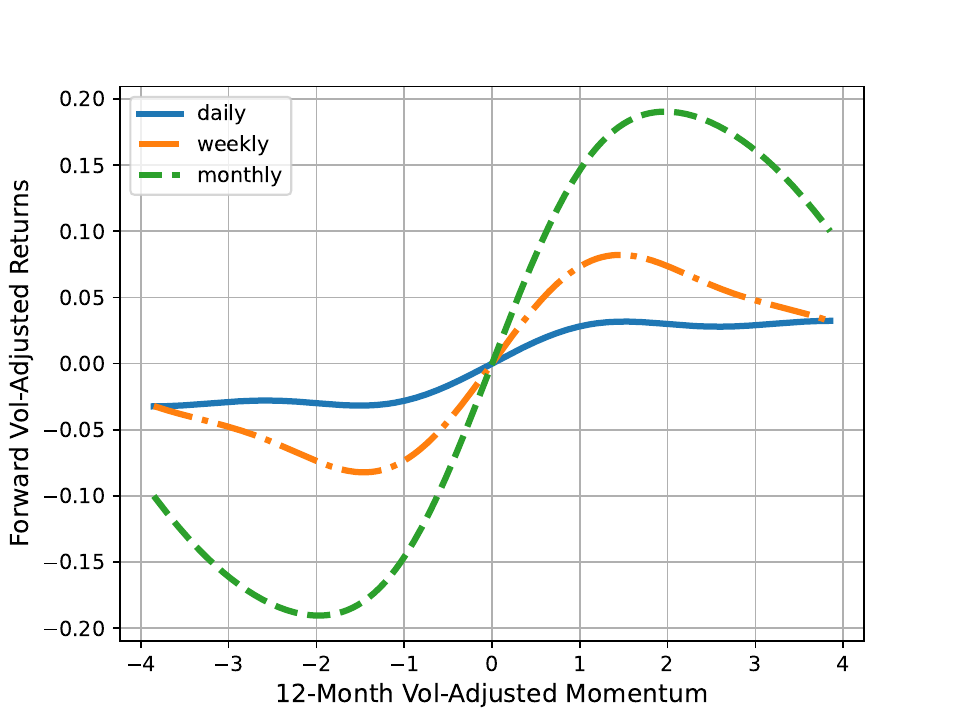}
    \caption{$12$-Month Momentum}
  \end{subfigure}
  
  \caption{Estimated forward vol-adjusted returns as a function of the $1$, $6$ and $12$-month time series momentum factor for Palladium Nonlinear models.
  Models are fit using all available training data from January $1950$ through June $2026$. }
\end{figure}

\noindent \textbf{Platinum}
\begin{figure}[H]
  \centering
  \begin{subfigure}[b]{0.30\linewidth}
    \includegraphics[width=\linewidth ]{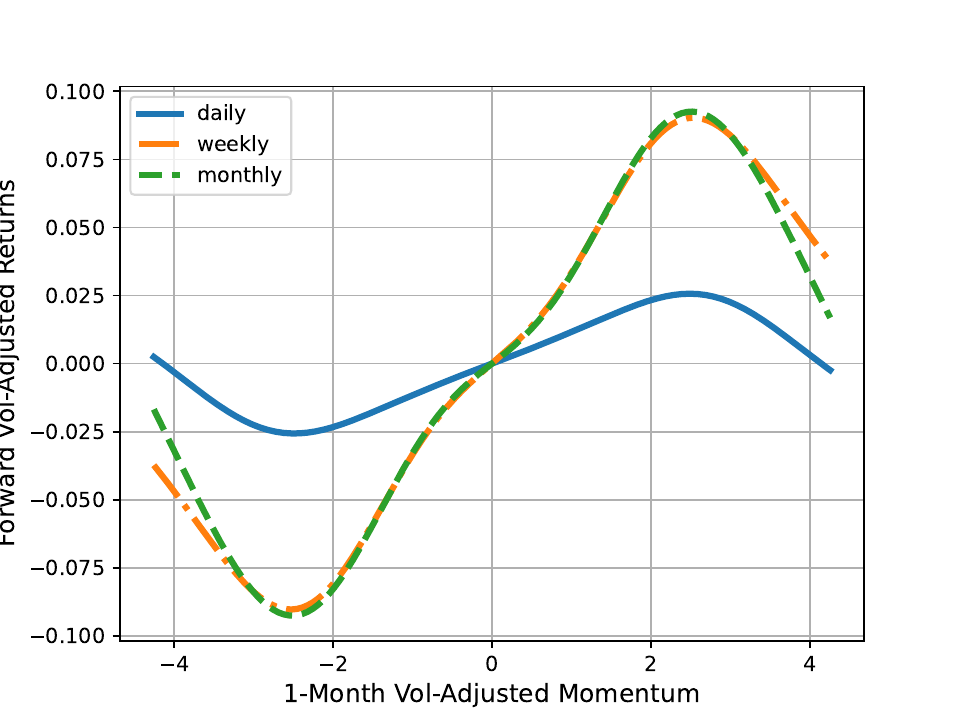}
    \caption{$1$-Month Momentum}
  \end{subfigure}
  \begin{subfigure}[b]{0.30\linewidth}
    \includegraphics[width=\linewidth ]{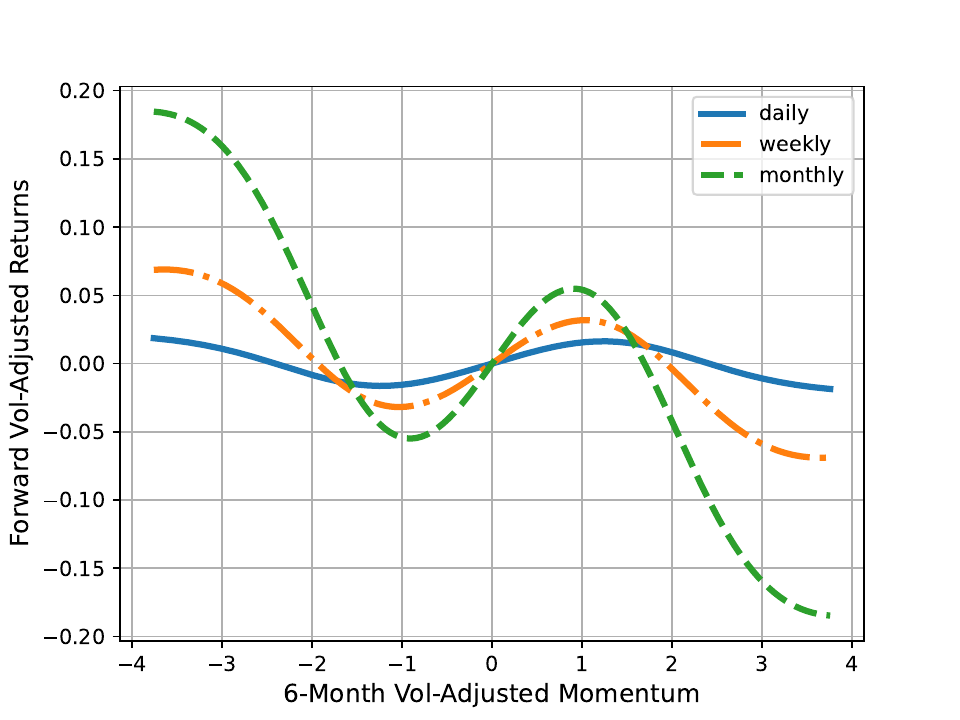}
    \caption{$6$-Month Momentum}
  \end{subfigure}
  \begin{subfigure}[b]{0.30\linewidth}
    \includegraphics[width=\linewidth ]{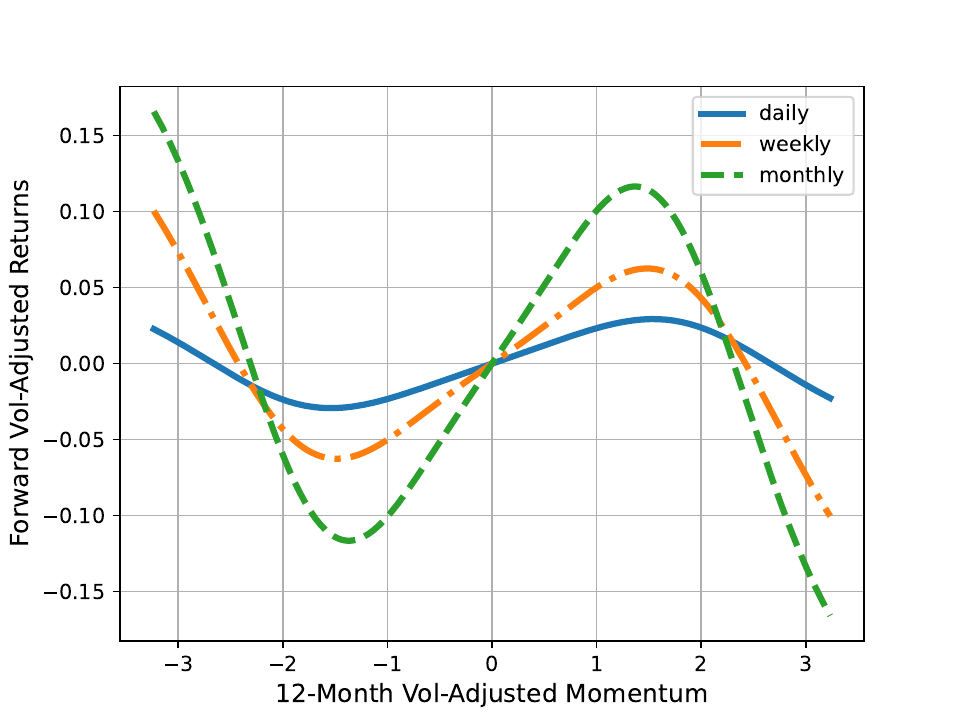}
    \caption{$12$-Month Momentum}
  \end{subfigure}
  
  \caption{Estimated forward vol-adjusted returns as a function of the $1$, $6$ and $12$-month time series momentum factor for Platinum Nonlinear models.
  Models are fit using all available training data from January $1950$ through June $2026$. }
\end{figure}

\noindent \textbf{Silver}
\begin{figure}[H]
  \centering
  \begin{subfigure}[b]{0.30\linewidth}
    \includegraphics[width=\linewidth ]{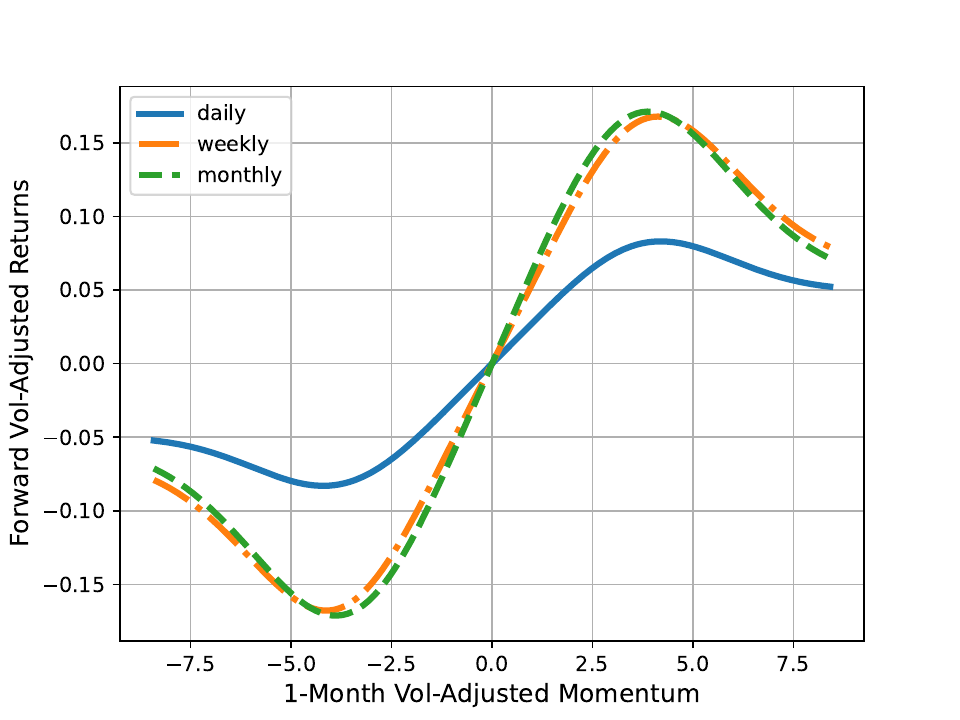}
    \caption{$1$-Month Momentum}
  \end{subfigure}
  \begin{subfigure}[b]{0.30\linewidth}
    \includegraphics[width=\linewidth ]{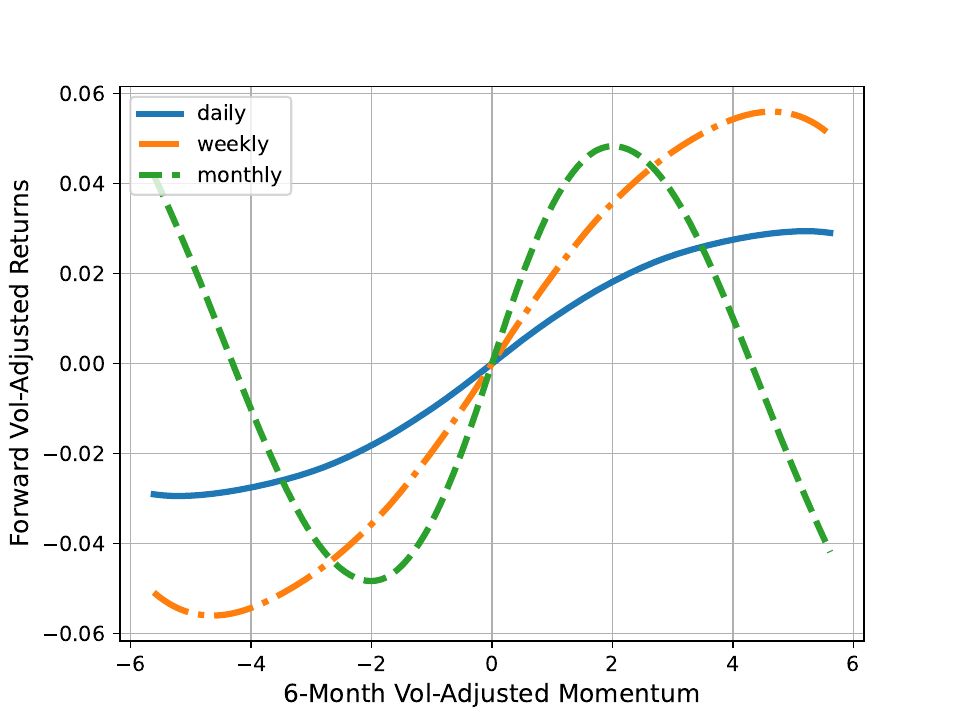}
    \caption{$6$-Month Momentum}
  \end{subfigure}
  \begin{subfigure}[b]{0.30\linewidth}
    \includegraphics[width=\linewidth ]{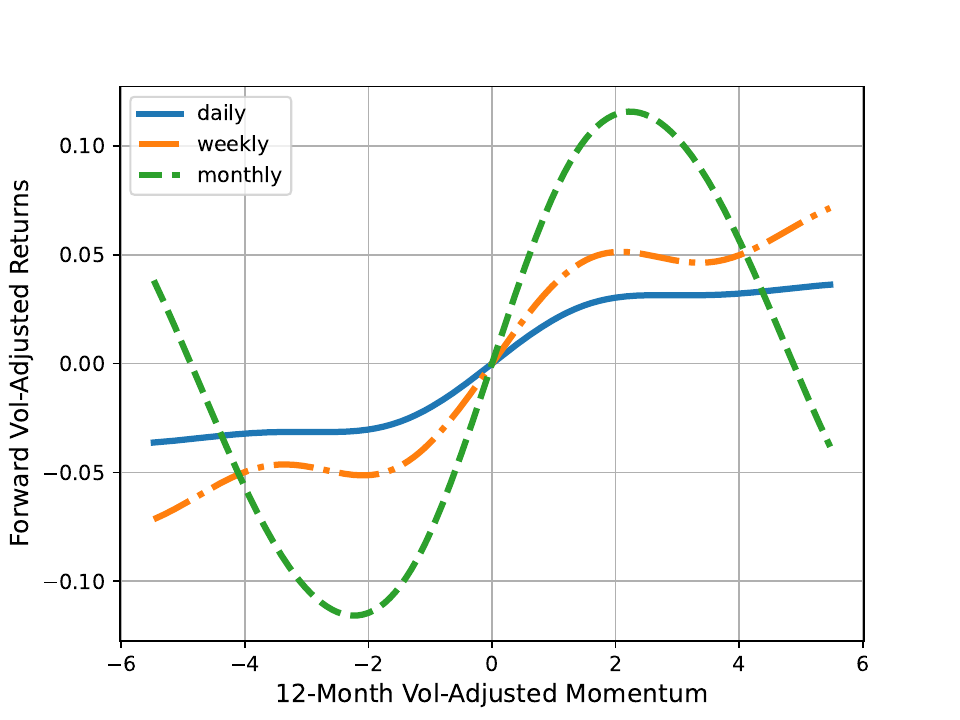}
    \caption{$12$-Month Momentum}
  \end{subfigure}
  
  \caption{Estimated forward vol-adjusted returns as a function of the $1$, $6$ and $12$-month time series momentum factor for Silver Nonlinear models.
  Models are fit using all available training data from January $1950$ through June $2026$. }
\end{figure}

\subsection{Soft Sector Nonlinear Models}
\begin{figure}[H]
  \centering
  \begin{subfigure}[b]{0.30\linewidth}
    \includegraphics[width=\linewidth ]{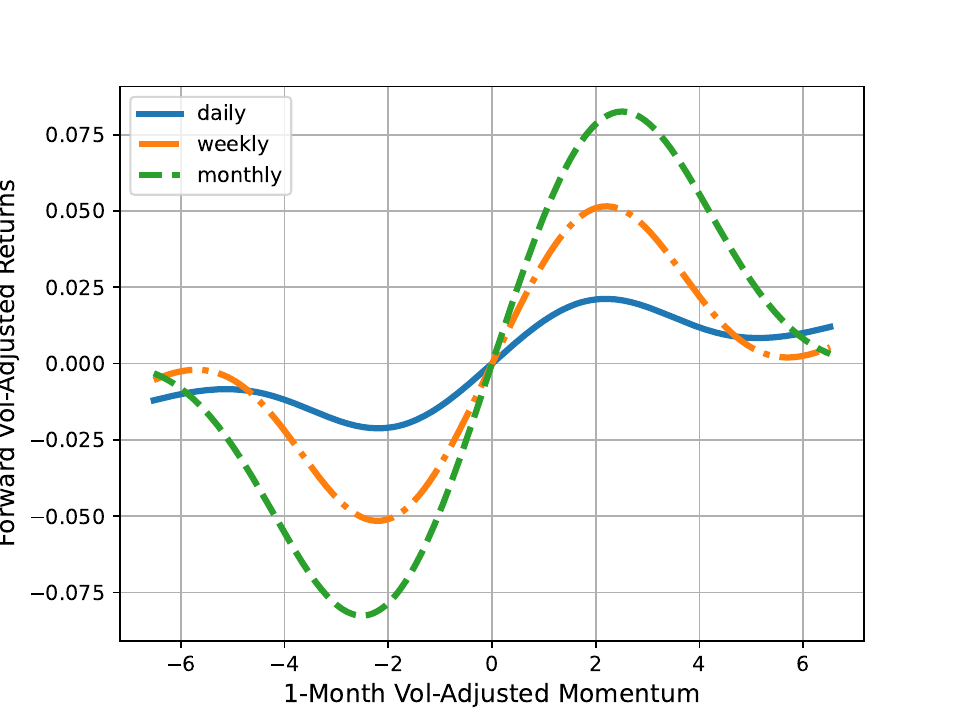}
    \caption{$1$-Month Momentum}
  \end{subfigure}
  \begin{subfigure}[b]{0.30\linewidth}
    \includegraphics[width=\linewidth ]{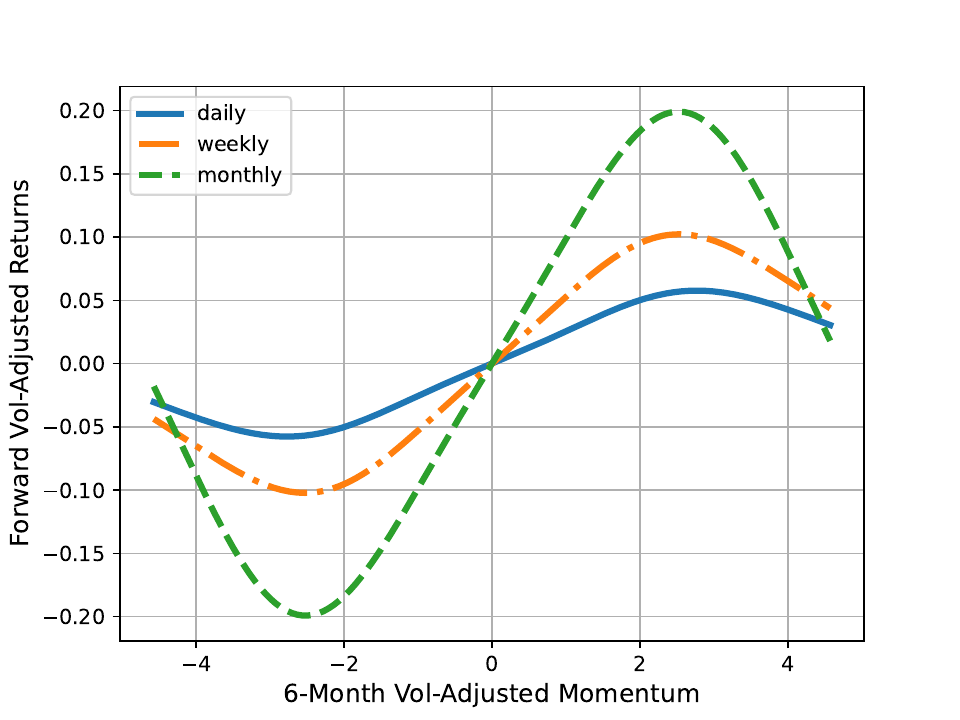}
    \caption{$6$-Month Momentum}
  \end{subfigure}
  \begin{subfigure}[b]{0.30\linewidth}
    \includegraphics[width=\linewidth ]{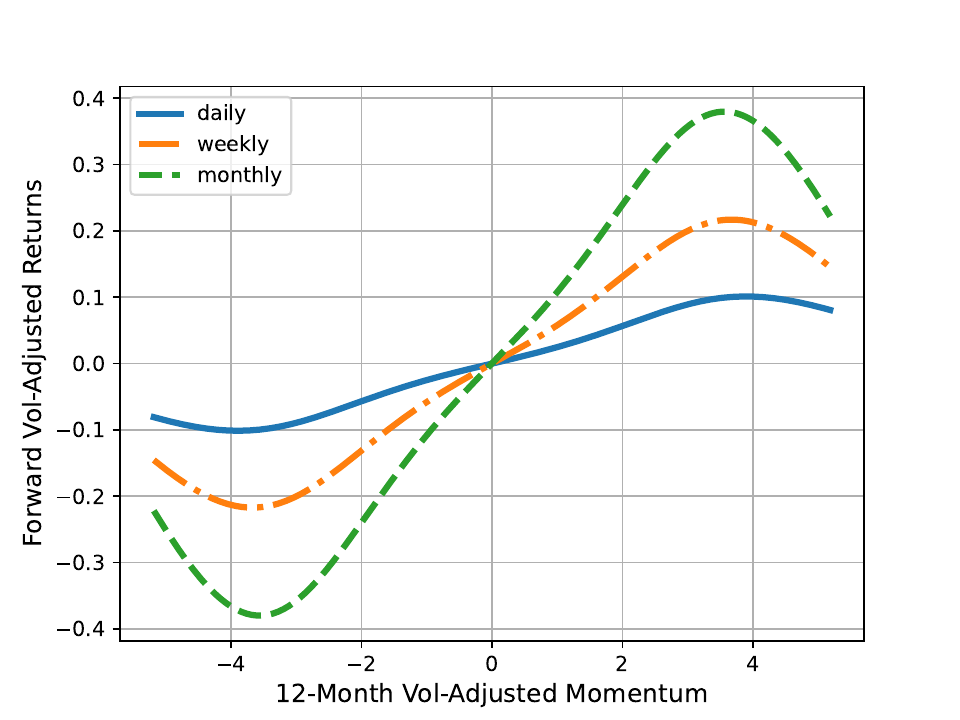}
    \caption{$12$-Month Momentum}
  \end{subfigure}
  
  \caption{Estimated forward vol-adjusted returns as a function of the $1$, $6$ and $12$-month time series momentum factor for Soft Sector Nonlinear models.
  Models are fit using all available training data from January $1950$ through June $2026$. }
\end{figure}

\noindent \textbf{Cocoa}
\begin{figure}[H]
  \centering
  \begin{subfigure}[b]{0.30\linewidth}
    \includegraphics[width=\linewidth ]{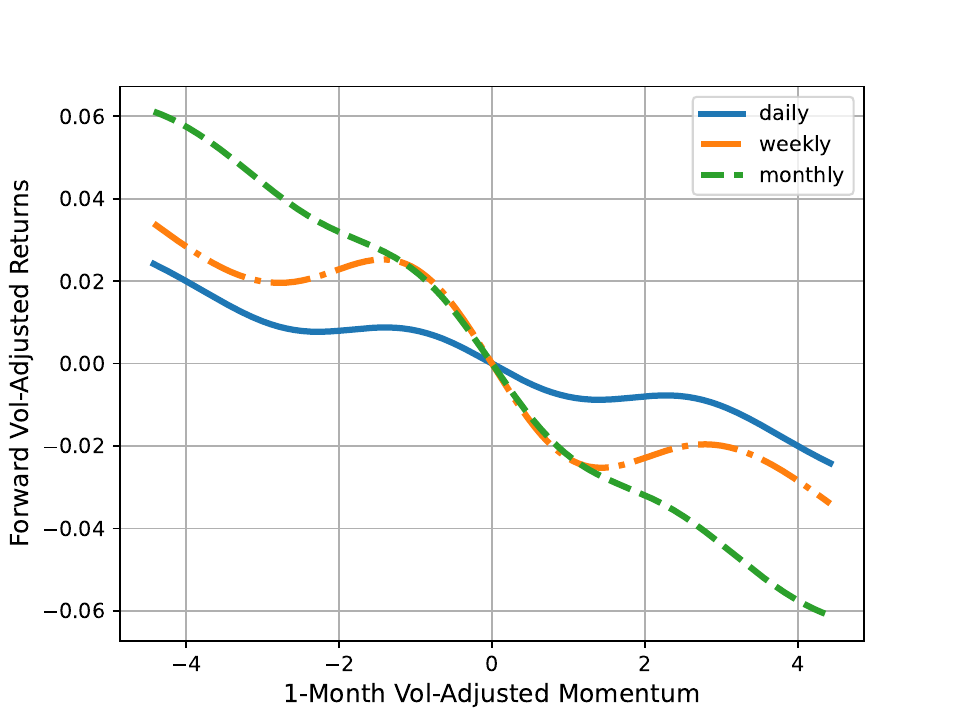}
    \caption{$1$-Month Momentum}
  \end{subfigure}
  \begin{subfigure}[b]{0.30\linewidth}
    \includegraphics[width=\linewidth ]{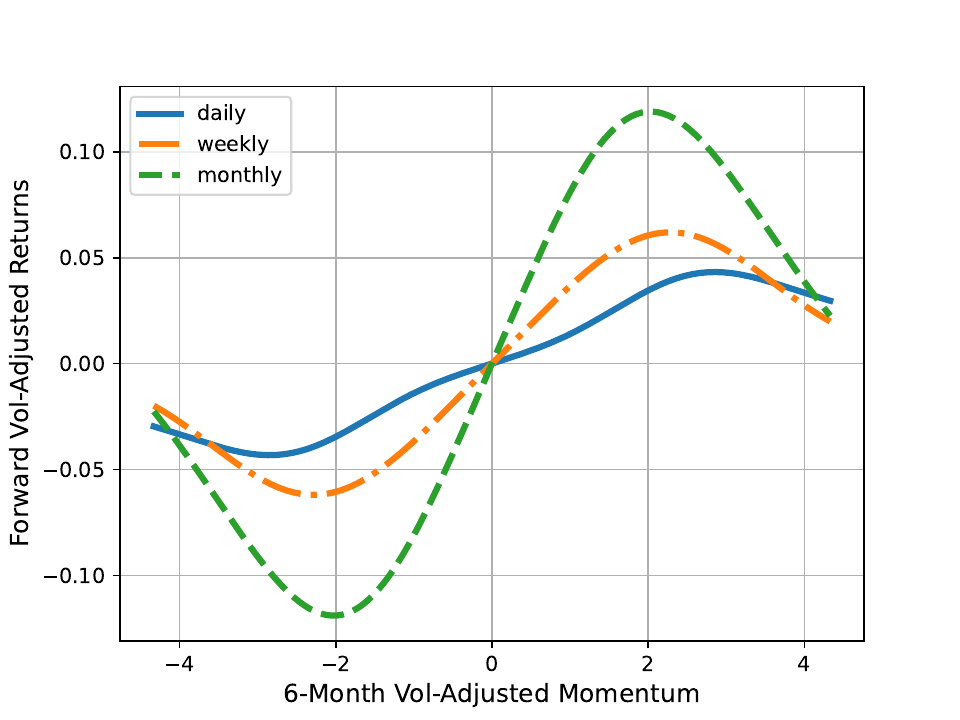}
    \caption{$6$-Month Momentum}
  \end{subfigure}
  \begin{subfigure}[b]{0.30\linewidth}
    \includegraphics[width=\linewidth ]{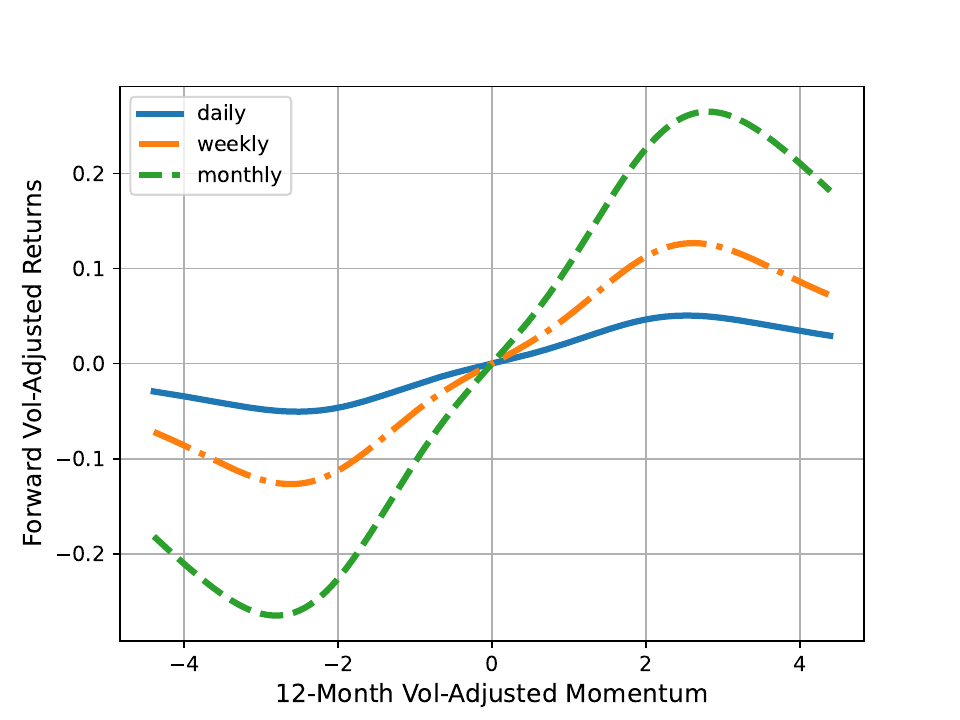}
    \caption{$12$-Month Momentum}
  \end{subfigure}
  
  \caption{Estimated forward vol-adjusted returns as a function of the $1$, $6$ and $12$-month time series momentum factor for Cocoa Nonlinear models.
  Models are fit using all available training data from January $1950$ through June $2026$. }
\end{figure}

\noindent \textbf{Cotton}
\begin{figure}[H]
  \centering
  \begin{subfigure}[b]{0.30\linewidth}
    \includegraphics[width=\linewidth ]{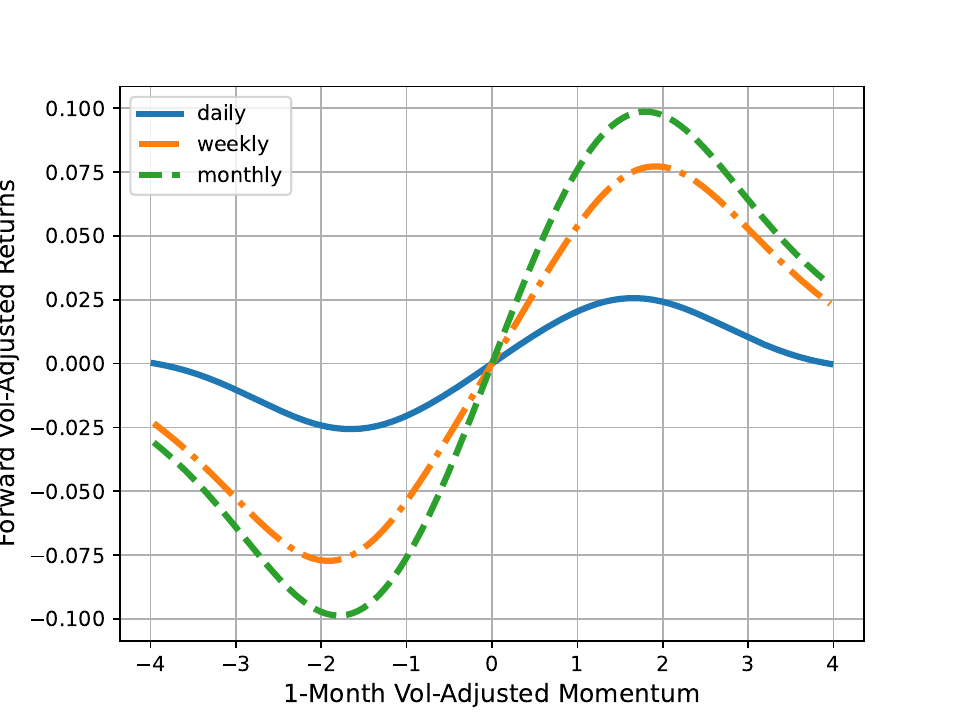}
    \caption{$1$-Month Momentum}
  \end{subfigure}
  \begin{subfigure}[b]{0.30\linewidth}
    \includegraphics[width=\linewidth ]{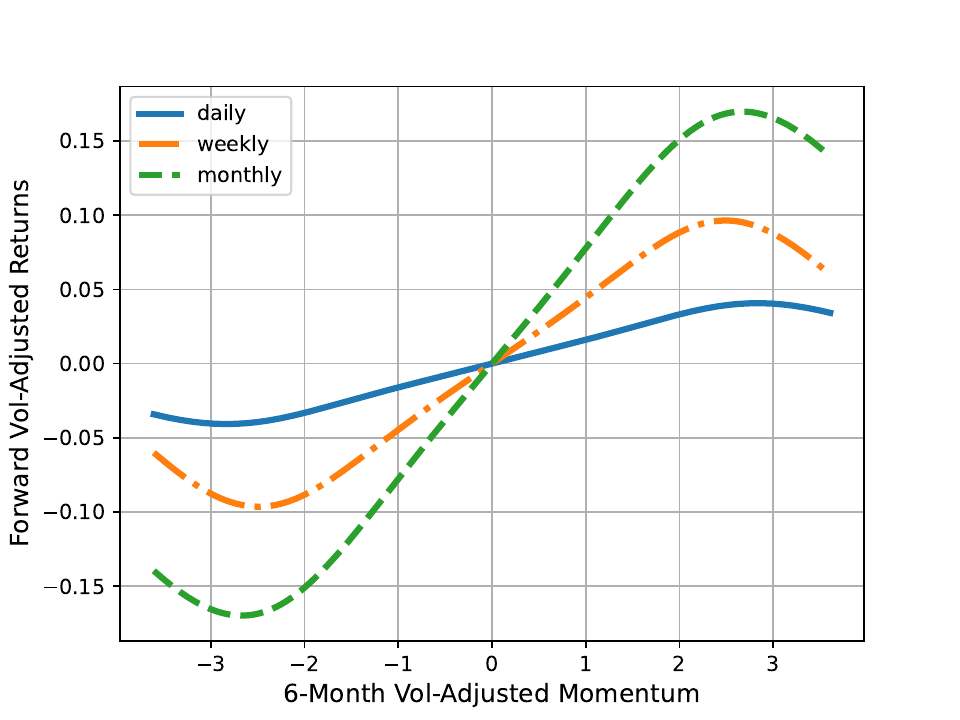}
    \caption{$6$-Month Momentum}
  \end{subfigure}
  \begin{subfigure}[b]{0.30\linewidth}
    \includegraphics[width=\linewidth ]{figs/model_soft_CT_momo_12_month.pdf}
    \caption{$12$-Month Momentum}
  \end{subfigure}
  
  \caption{Estimated forward vol-adjusted returns as a function of the $1$, $6$ and $12$-month time series momentum factor for Cotton Nonlinear models.
  Models are fit using all available training data from January $1950$ through June $2026$. }
\end{figure}

\noindent \textbf{Robusta Coffee}
\begin{figure}[H]
  \centering
  \begin{subfigure}[b]{0.30\linewidth}
    \includegraphics[width=\linewidth ]{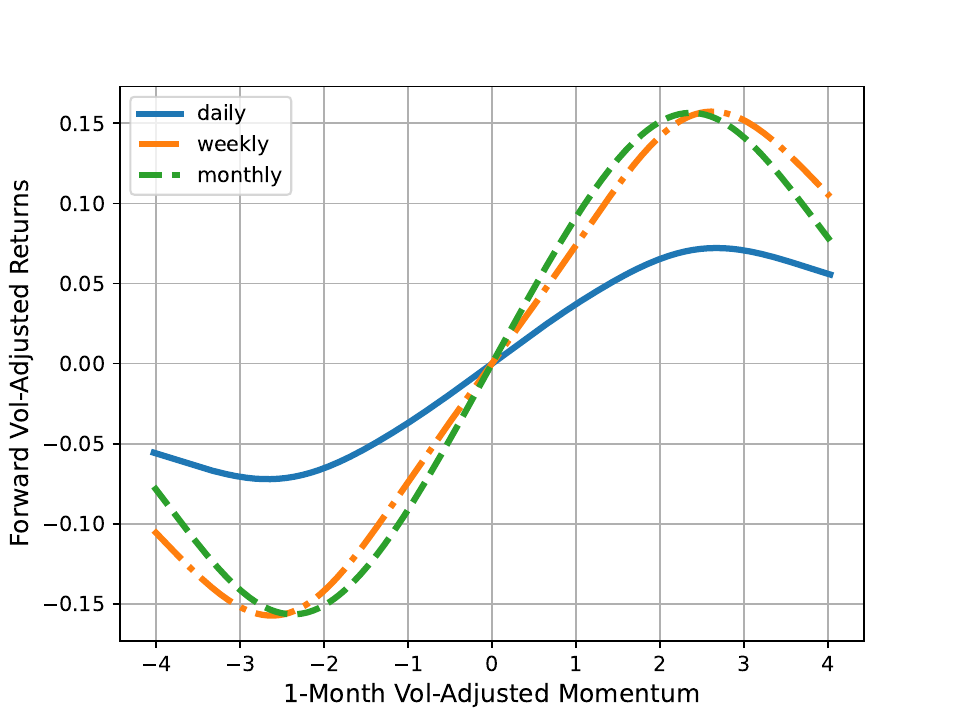}
    \caption{$1$-Month Momentum}
  \end{subfigure}
  \begin{subfigure}[b]{0.30\linewidth}
    \includegraphics[width=\linewidth ]{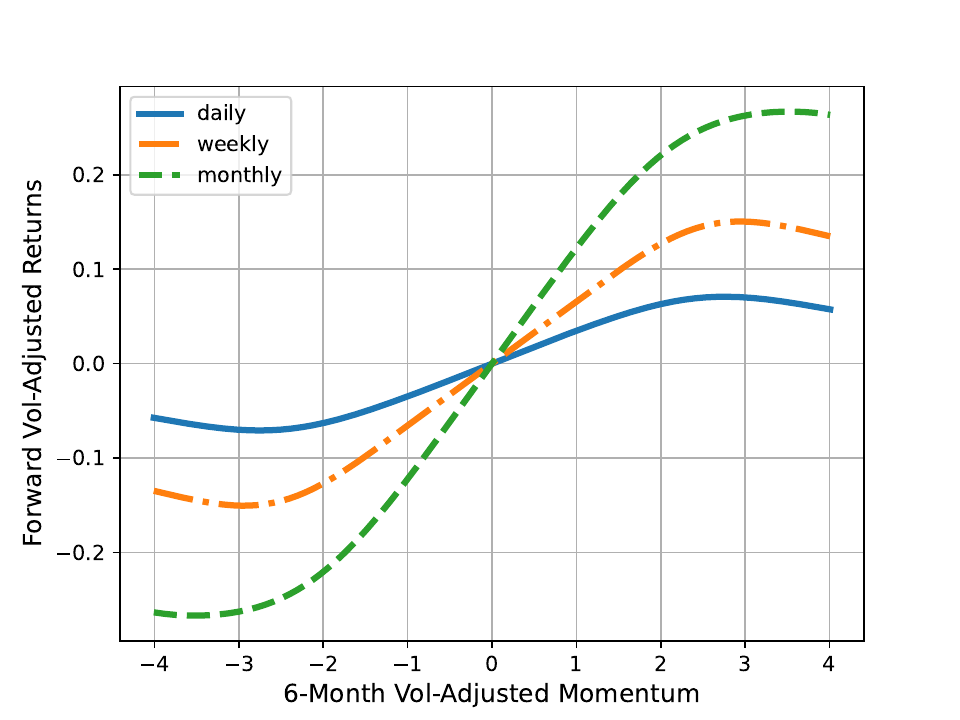}
    \caption{$6$-Month Momentum}
  \end{subfigure}
  \begin{subfigure}[b]{0.30\linewidth}
    \includegraphics[width=\linewidth ]{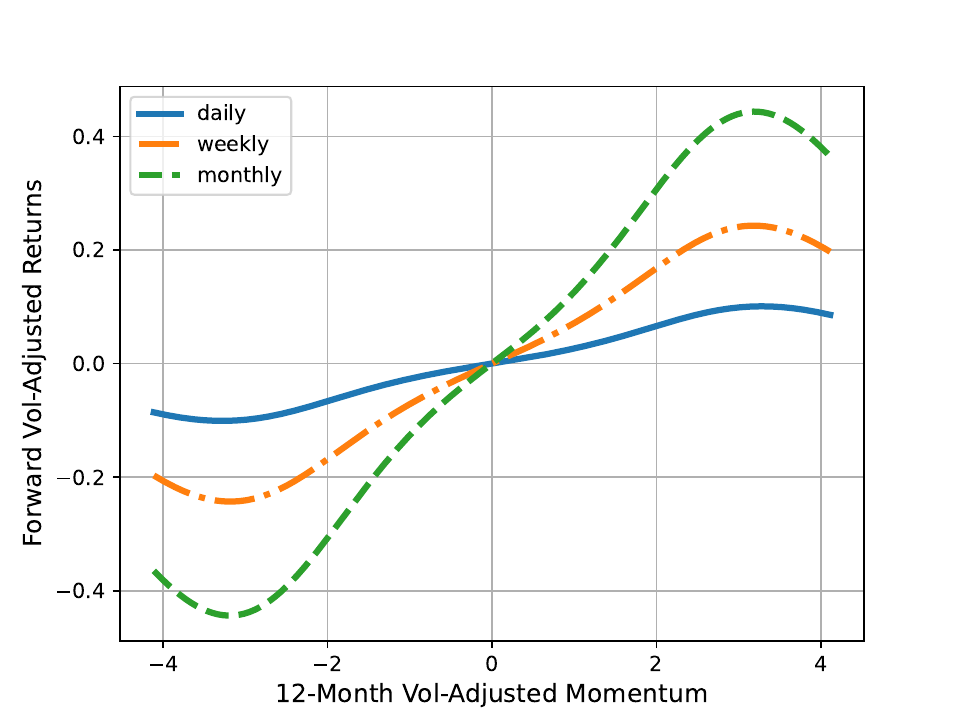}
    \caption{$12$-Month Momentum}
  \end{subfigure}
  
  \caption{Estimated forward vol-adjusted returns as a function of the $1$, $6$ and $12$-month time series momentum factor for Robusta Coffee Nonlinear models.
  Models are fit using all available training data from January $1950$ through June $2026$. }
\end{figure}

\noindent \textbf{Coffee}
\begin{figure}[H]
  \centering
  \begin{subfigure}[b]{0.30\linewidth}
    \includegraphics[width=\linewidth ]{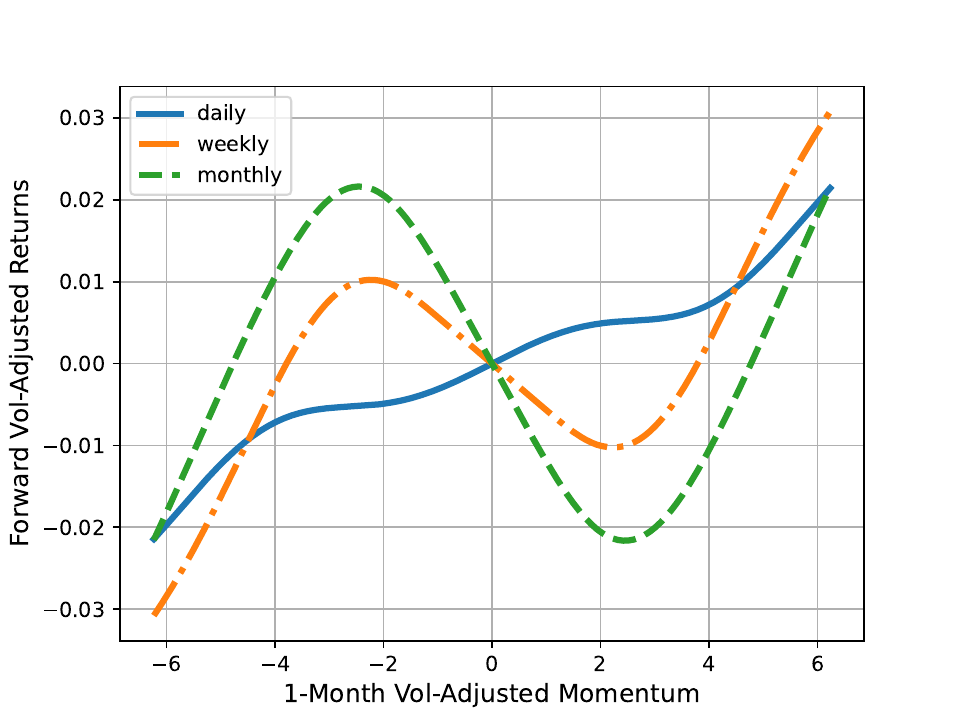}
    \caption{$1$-Month Momentum}
  \end{subfigure}
  \begin{subfigure}[b]{0.30\linewidth}
    \includegraphics[width=\linewidth ]{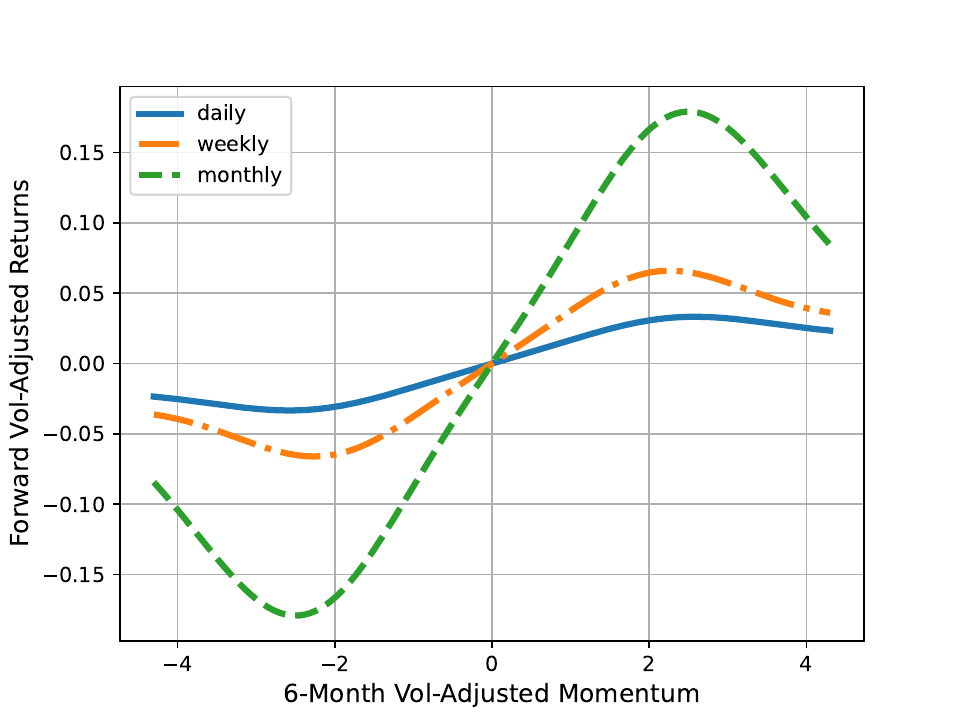}
    \caption{$6$-Month Momentum}
  \end{subfigure}
  \begin{subfigure}[b]{0.30\linewidth}
    \includegraphics[width=\linewidth ]{figs/model_soft_KC_momo_12_month.pdf}
    \caption{$12$-Month Momentum}
  \end{subfigure}
  
  \caption{Estimated forward vol-adjusted returns as a function of the $1$, $6$ and $12$-month time series momentum factor for Coffee Nonlinear models.
  Models are fit using all available training data from January $1950$ through June $2026$. }
\end{figure}

\noindent \textbf{UK Cocoa}
\begin{figure}[H]
  \centering
  \begin{subfigure}[b]{0.30\linewidth}
    \includegraphics[width=\linewidth ]{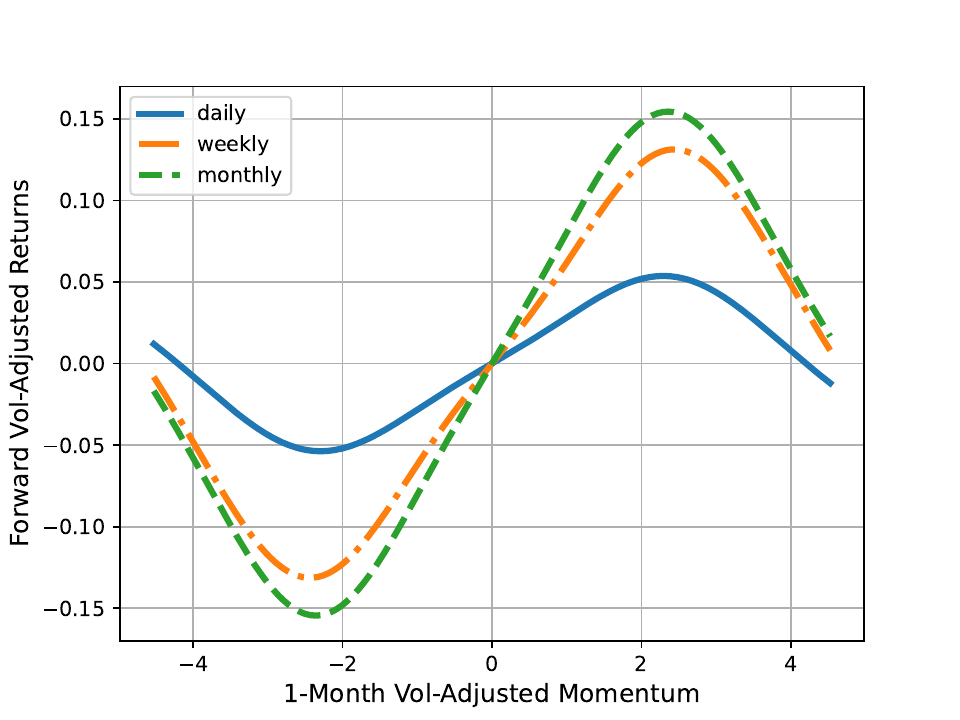}
    \caption{$1$-Month Momentum}
  \end{subfigure}
  \begin{subfigure}[b]{0.30\linewidth}
    \includegraphics[width=\linewidth ]{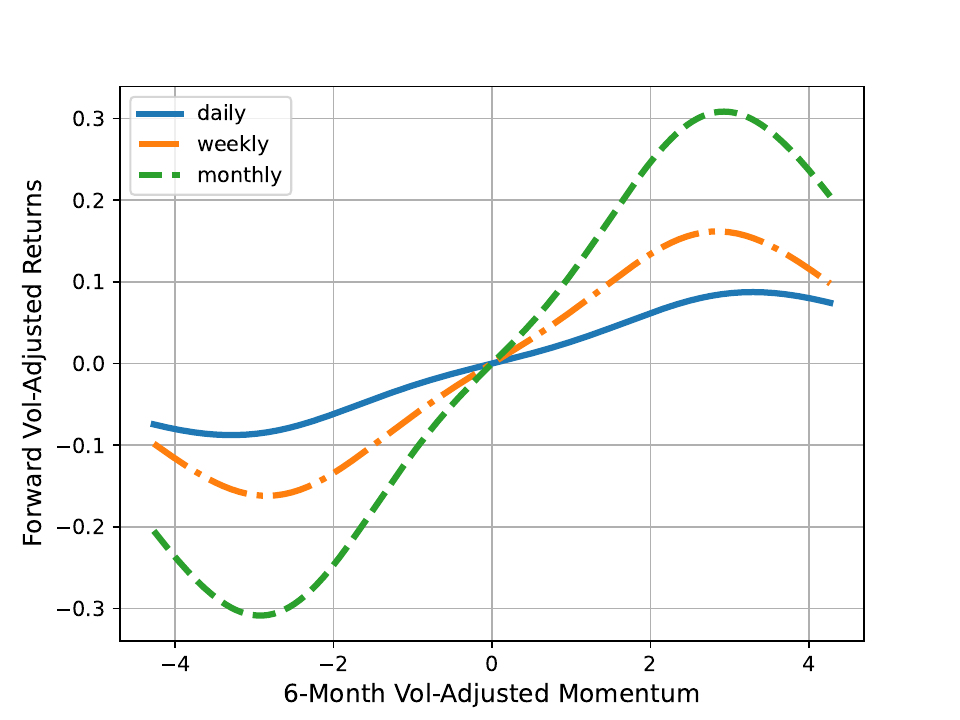}
    \caption{$6$-Month Momentum}
  \end{subfigure}
  \begin{subfigure}[b]{0.30\linewidth}
    \includegraphics[width=\linewidth ]{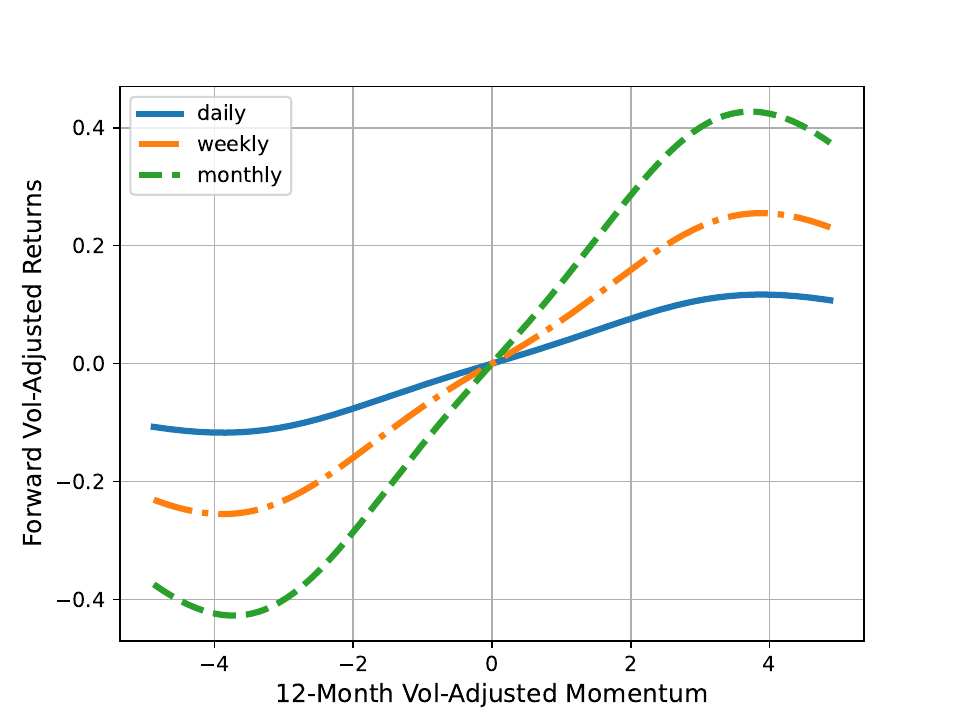}
    \caption{$12$-Month Momentum}
  \end{subfigure}
  
  \caption{Estimated forward vol-adjusted returns as a function of the $1$, $6$ and $12$-month time series momentum factor for UK Cocoa Nonlinear models.
  Models are fit using all available training data from January $1950$ through June $2026$. }
\end{figure}

\noindent \textbf{UK Sugar}
\begin{figure}[H]
  \centering
  \begin{subfigure}[b]{0.30\linewidth}
    \includegraphics[width=\linewidth ]{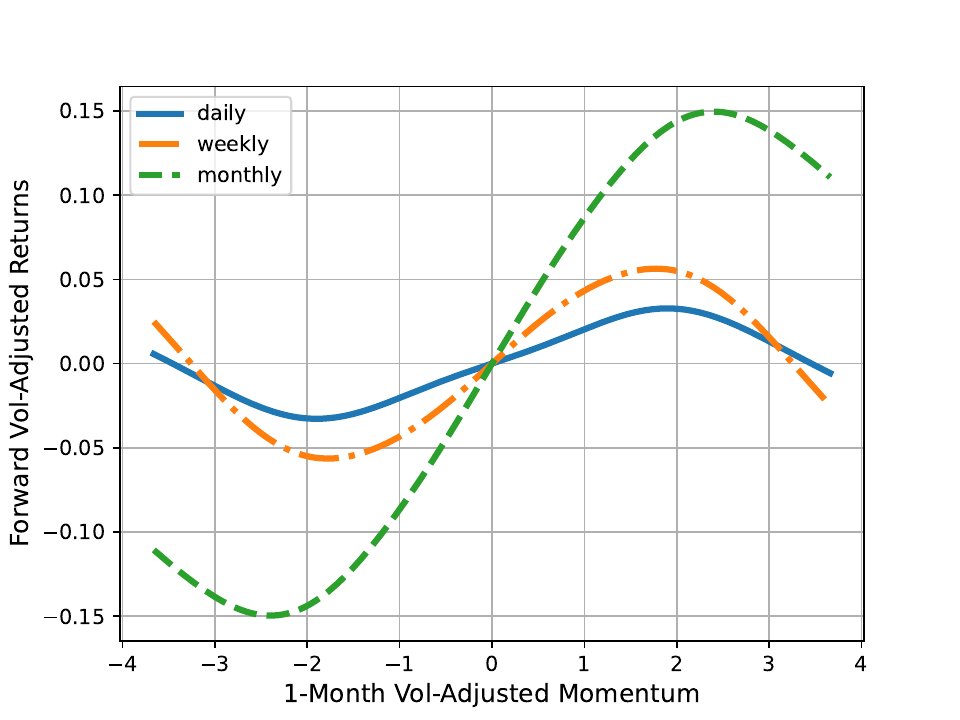}
    \caption{$1$-Month Momentum}
  \end{subfigure}
  \begin{subfigure}[b]{0.30\linewidth}
    \includegraphics[width=\linewidth ]{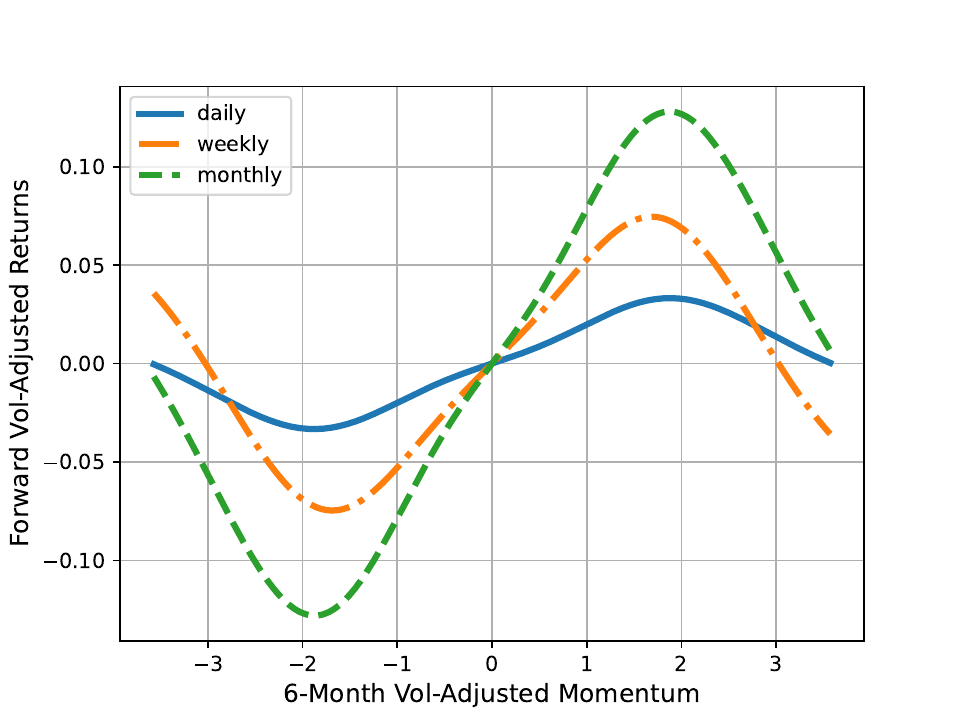}
    \caption{$6$-Month Momentum}
  \end{subfigure}
  \begin{subfigure}[b]{0.30\linewidth}
    \includegraphics[width=\linewidth ]{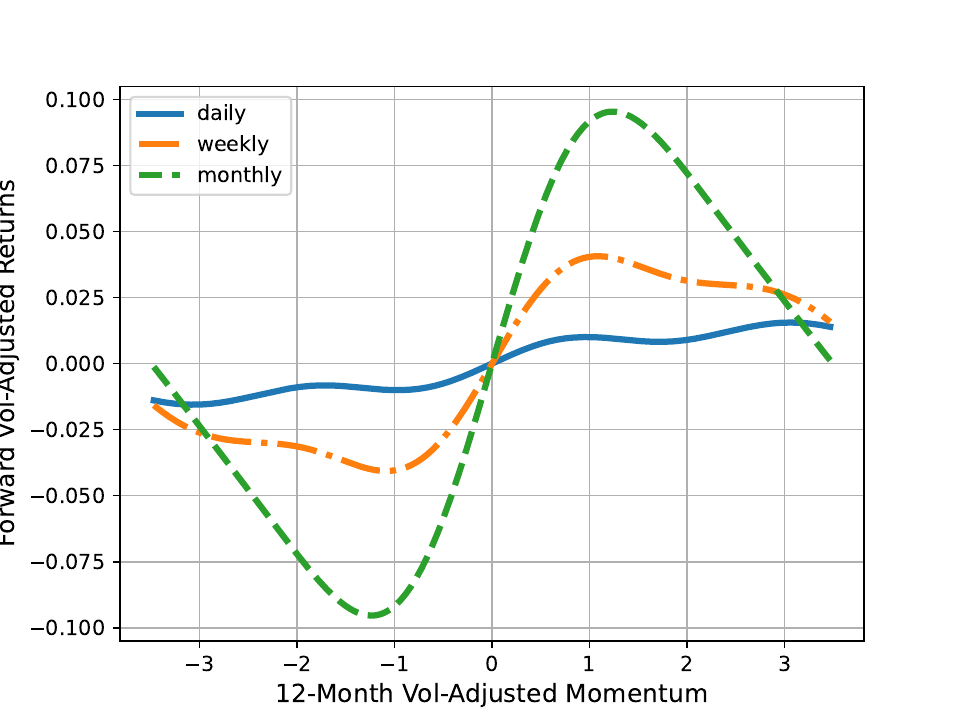}
    \caption{$12$-Month Momentum}
  \end{subfigure}
  
  \caption{Estimated forward vol-adjusted returns as a function of the $1$, $6$ and $12$-month time series momentum factor for UK Sugar Nonlinear models.
  Models are fit using all available training data from January $1950$ through June $2026$. }
\end{figure}

\noindent \textbf{Sugar}
\begin{figure}[H]
  \centering
  \begin{subfigure}[b]{0.30\linewidth}
    \includegraphics[width=\linewidth ]{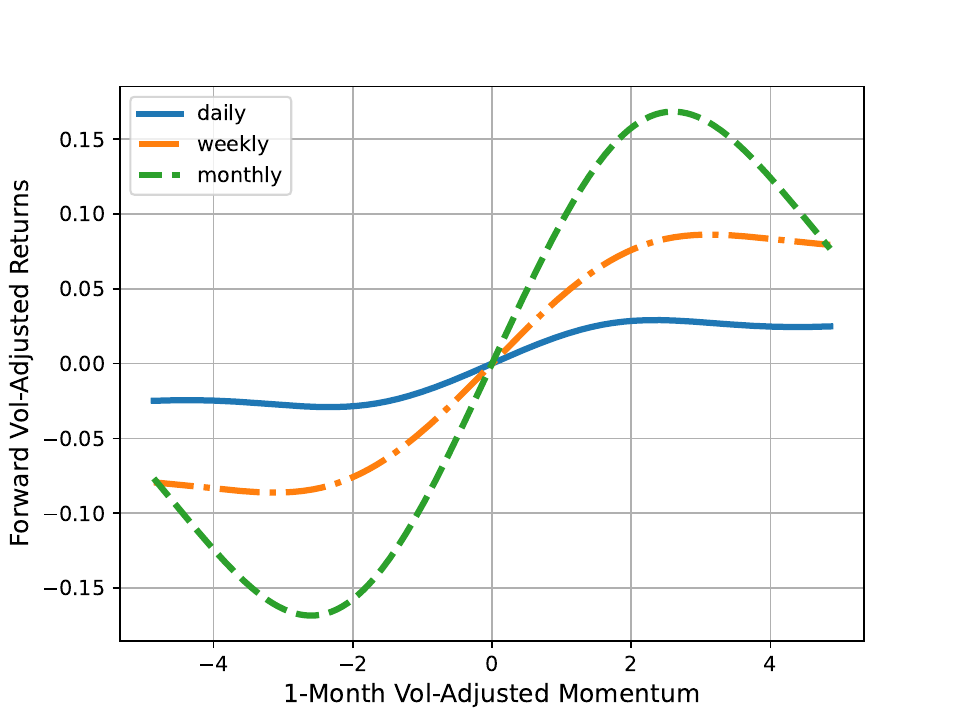}
    \caption{$1$-Month Momentum}
  \end{subfigure}
  \begin{subfigure}[b]{0.30\linewidth}
    \includegraphics[width=\linewidth ]{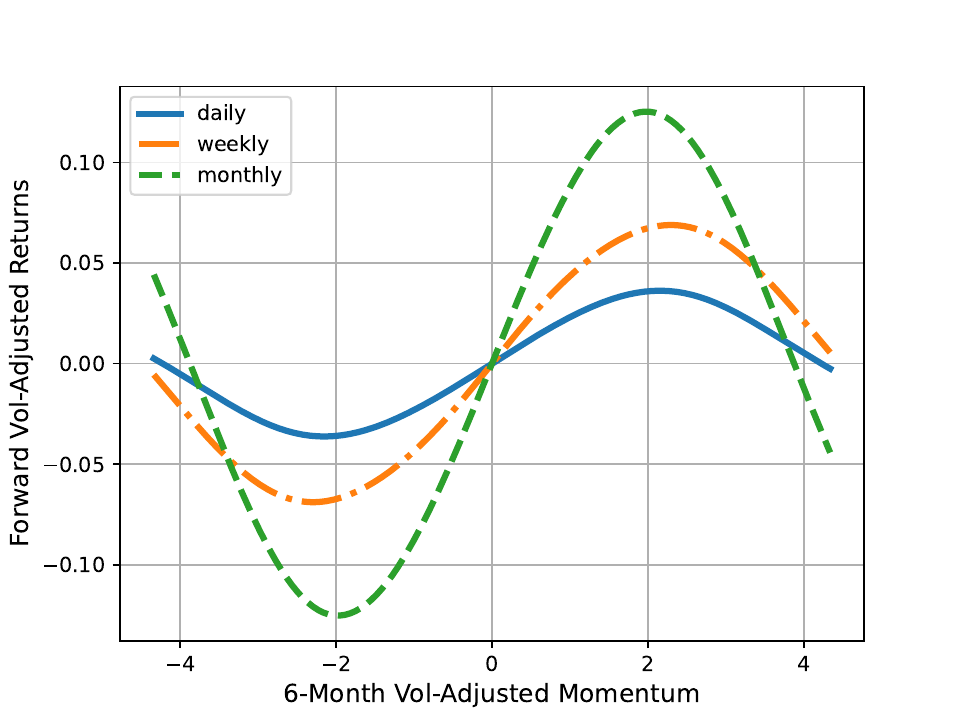}
    \caption{$6$-Month Momentum}
  \end{subfigure}
  \begin{subfigure}[b]{0.30\linewidth}
    \includegraphics[width=\linewidth ]{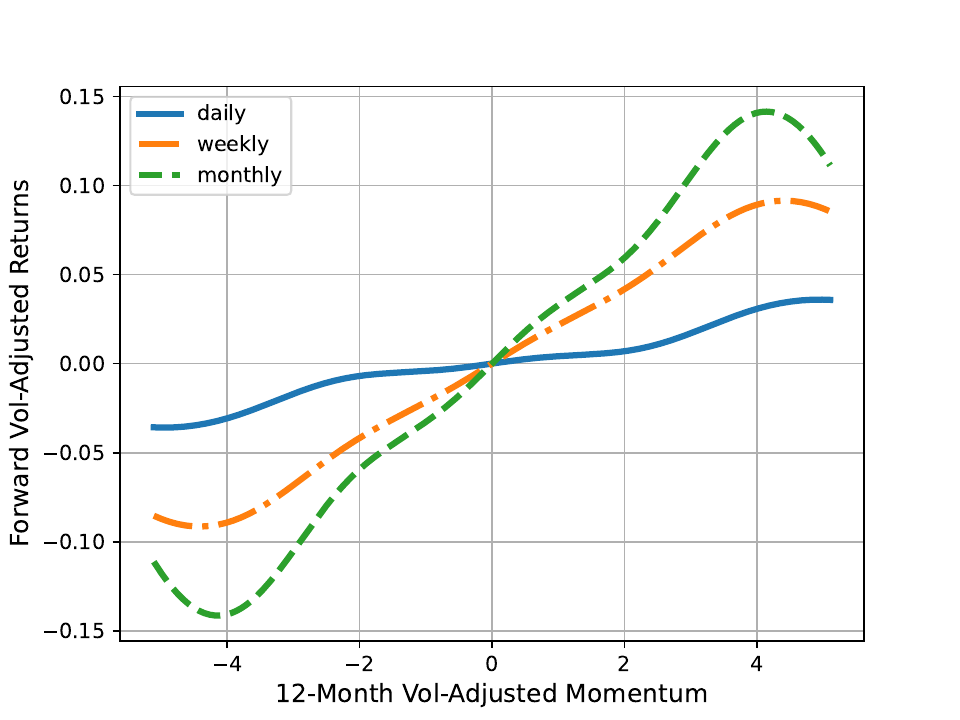}
    \caption{$12$-Month Momentum}
  \end{subfigure}
  
  \caption{Estimated forward vol-adjusted returns as a function of the $1$, $6$ and $12$-month time series momentum factor for Sugar Nonlinear models.
  Models are fit using all available training data from January $1950$ through June $2026$. }
\end{figure}

\section{Futures Market Data}
\begin{table}[H]
\centering
\scriptsize
\begin{tabular}{l l l l l}
\hline
\textbf{Symbol} & \textbf{Market} &  \textbf{Tick Size} & \textbf{Trade Slippage Cost} & \textbf{Commission Cost}\\
\hline
\textbf{Bond} & & & & \\
CN & Cdn 10Yr & 0.01 & 1.0 & 4.94\\
FV & US 5Yr & 0.0078125 & 1.0 & 4.34\\
G & Gilts & 0.01 & 1.0 & 4.18\\
JB & JGB & 0.01 & 1.0 & 10.58\\
OE & Bobl & 0.01 & 1.0 & 3.70\\
RX & Bund & 0.01 & 1.0 & 3.70\\
TY & US 10Yr & 0.015625 & 1.0 & 4.54\\
US & US 30Yr & 0.03125 & 1.0 & 4.64\\
\textbf{Currency} & & & & \\
AD & AUD & 0.005 & 1.0 & 6.24\\
BP & GBP & 0.01 & 1.0 & 6.24\\
CD & CAD & 0.005 & 2.0 & 6.24\\
DX & USD & 0.005 & 1.0 & 6.24\\
EC & EUR & 0.005 & 2.0 & 6.24\\
JY & JPY & 0.005 & 1.0 & 6.24\\
NV & NZD & 0.01 & 1.0 & 6.24\\
SF & CHF & 0.01 & 3.0 & 6.24\\
\textbf{Energy} & & & & \\
CL & Crude & 0.01 & 2.0 & 6.04\\
CO & Brent Crude & 0.01 & 3.0 & 4.64\\
HO & Heating Oil & 0.01 & 3.0 & 6.04\\
NG & Nat Gas & 0.001 & 3.0 & 6.04\\
QS & Gasoil & 0.25 & 2.0 & 4.64\\
XB & RBOB & 0.01 & 7.0 & 6.04\\
\textbf{Equity} & & & & \\
CF & CAC 40 & 0.50 & 5.0 & 3.96\\
ES & S\&P 500 & 0.25 & 1.0 & 5.50\\
GX & DAX & 1.0 & 2.0 & 5.42\\
HI & Hang Seng & 1.0 & 5.0 & 9.14\\
NI & Nikkei & 5.0 & 1.0 & 4.00\\
PT & TSX 60 & 0.10 & 3.0 & 5.30\\
XP & ASX 200 & 1.0 & 1.0 & 3.78\\
Z & FTSE 100 & 0.50 & 2.0 & 4.24\\
\textbf{Grain} & & & & \\
BO & Bean Oil & 0.01 & 3.0 & 6.94\\
C & Corn & 0.25 & 1.0 & 6.94\\
KO & Palm Oil & 1.0 & 1.0 & 14.48\\
KW & KC Wheat & 0.25 & 2.0 & 6.94\\
RS & Canola & 0.10 & 5.0 & 4.94\\
S & Soybean & 0.25 & 3.0 & 6.94\\
SM & Soy Meal & 0.10 & 2.0 & 6.94\\
W & Wheat & 0.25 & 1.0 & 6.94\\
\textbf{Metal} & & & & \\
GC & Gold & 0.10 & 2.0 & 6.14\\
HG & Copper & 0.05 & 4.0 & 6.14\\
PA & Palladium & 0.50 & 5.0 & 6.14\\
PL & Platinum & 0.10 & 9.0 & 6.14\\
SI & Silver & 0.005 & 2.0 & 6.14\\
\textbf{Soft} & & & & \\
CC & Cocoa & 1.0 & 4.0 & 7.24\\
CT & Cotton & 0.01 & 6.0 & 7.24\\
DF & Robusta Coffee & 1.0 & 3.0 & 1.56\\
KC & Coffee & 0.05 & 5.0 & 7.24\\
QC & UK Cocoa & 1.0 & 2.0 & 4.24\\
QW & UK Sugar & 0.10 & 7.0 & 4.24\\
SB & Sugar & 0.01 & 2.0 & 7.24\\
\hline
\end{tabular}
\caption{Futures market universe and trade cost assumptions. Symbols follow Bloomberg market symbology. Data is provided by Commodity Systems Inc (CSI).
Trade slippage cost is expressed in number of ticks per contract traded whereas commission cost is expressed in USD per contract traded.
Costs are estimated from live trading results on a global futures strategy with an  AUM of approximately $\$100$ million USD and an 
annualized volatility of approximately $20\%$.}
\label{table:universe}
\end{table}

\bibliographystyle{plainnat}
\bibliography{Bibliography}

\end{document}